\documentclass[aps,superscriptaddress,onecolumn,nofootinbib,10pt,prx]{revtex4-2}

\usepackage{silence}
\usepackage{amsmath,mathtools,amsthm,amssymb,amsfonts,amstext}
\usepackage{tabularx}
\usepackage{tabularray}
\usepackage{tikz}
\usepackage{tabularx}
\usepackage{tabularray}

\usepackage{graphicx}

\usepackage{color}
\usepackage{bbm}
\usepackage{dsfont}
\usepackage{enumitem}
\usepackage{centernot}
\usepackage{comment}
\usepackage{array}
\usepackage{wrapfig}
\usepackage{fontsize}
\graphicspath{{figures/}}
\usepackage{biolinum}
\usepackage{mathrsfs}
\usepackage{mathdots}
\usepackage{enumitem}
\usepackage[table]{xcolor}
\usepackage{booktabs}
\usepackage{esvect}
\usepackage{enumitem}
\usepackage{thmtools}
\usepackage{tikz}

\DeclareMathOperator{\supp}{supp}

\newcommand{\be}{\begin{equation}\begin{aligned}\hspace{0pt}}
\newcommand{\bbb}{\begin{equation*}\begin{aligned}}
\newcommand{\ee}{\end{aligned}\end{equation}}
\newcommand{\eee}{\end{aligned}\end{equation*}}

\newcommand{\negl}[0]{\operatorname{negl}}
\definecolor{lightred}{RGB}{243,229,231}
\definecolor{lightgreen}{RGB}{241,255,239}
\definecolor{lightblue}{RGB}{232,240,244}
\definecolor{RoyalBlue}{RGB}{65,105,225}
\definecolor{ForestGreen}{RGB}{34,139,34} 
\definecolor{Maroon}{RGB}{135,0,0}
\definecolor{myrefcolor}{rgb}{0.067,0.5,0.5}
\definecolor{myurlcolor}{rgb}{0.1,0,0.9}

\usepackage[linktocpage,colorlinks,hypertexnames=false]{hyperref}
\hypersetup{
    colorlinks = true,
    citecolor  = ForestGreen,
    linkcolor  = Crimson,
    urlcolor   = NavyBlue,
}
\usepackage[capitalize]{cleveref}

\crefname{section}{Sec.}{Secs.}
\Crefname{section}{Sec.}{Secs.}
\crefname{appendix}{Appendix}{Appendices}
\Crefname{appendix}{Appendix}{Appendices}
\crefname{subappendix}{Appendix}{Appendices}
\Crefname{subappendix}{Appendix}{Appendices}
\crefname{subsubappendix}{Appendix}{Appendices}
\Crefname{subsubappendix}{Appendix}{Appendices}

\newtheorem{theorem}{Theorem}
\newtheorem{proposition}[theorem]{Proposition}
\newtheorem{lemma}[theorem]{Lemma}
\newtheorem{corollary}[theorem]{Corollary}
\theoremstyle{definition}
\newtheorem{definition}[theorem]{Definition}

\newtheorem{assumption}[theorem]{Assumption}

\usepackage{macro}

\begin{document}

\title{Computational entanglement distillation beyond pure states: efficient protocols and fundamental limitations}

\author{Takeru~Utsumi}
\thanks{These authors contributed equally. \\ Contact: takeru-utsumi@g.ecc.u-tokyo.ac.jp, jacopo.rizzo@fu-berlin.de}
\affiliation{Graduate School of Arts and Sciences, The University of Tokyo, 3-8-1 Komaba, Meguro-ku, Tokyo 153-8902, Japan}
\author{Jacopo~Rizzo}
\thanks{These authors contributed equally. \\ Contact: takeru-utsumi@g.ecc.u-tokyo.ac.jp, jacopo.rizzo@fu-berlin.de}
\affiliation{Dahlem Center for Complex Quantum Systems, Freie Universit\"{a}t Berlin, 14195 Berlin, Germany}
\author{Ryuji~Takagi}
\affiliation{Graduate School of Arts and Sciences, The University of Tokyo, 3-8-1 Komaba, Meguro-ku, Tokyo 153-8902, Japan}
\author{Lorenzo~Leone}
\affiliation{Dipartimento di Ingegneria Industriale, Universit\`{a} degli Studi di Salerno, Via Giovanni Paolo II, 132, 84084 Fisciano (SA), Italy}
\affiliation{Istituto Nazionale di Fisica Nucleare (INFN) Sezione di Napoli, Gruppo Collegato di Salerno, Italy}

\date{\today}

\begin{abstract}

The information-theoretic approach to entanglement typically relies on arbitrarily many input copies and unboundedly complex operations; however, the available resources are constrained in reality. This motivates the study of entanglement theory with limited resources. One of its central questions is which quantities characterize entanglement manipulations, and so far, only the idealized case of pure states has been well explored. In this work, we address entanglement distillation from mixed states and identify the R\'{e}nyi coherent information as the quantity determining the optimal rate in the worst case under two distinct resource constraints. We first characterize the sample-efficient regime, in which unbounded computational power is allowed. Our universal protocol achieves this rate without knowledge of the input state under certain polynomial rank constraints. This rate is optimal among universal protocols: we construct states whose distillable entanglement is nearly extensive, yet from which no universal sample-efficient protocol attains a rate above their R\'{e}nyi coherent information. We then push this characterization to the computationally efficient frontier: the same bound holds for any efficient protocol even when the input state is known, and another universal protocol achieves this rate efficiently under additional polynomial constraints on the marginal ranks. Overall, our results give the R\'{e}nyi coherent information a novel operational meaning in two computational settings: universal distillation with limited copies and efficient distillation even with knowledge of the input state.

\end{abstract}

\maketitle

\tableofcontents

\section{Introduction}
\label{sec:introduction}

Quantum information theory characterizes the ultimate limits of information processing.
Entanglement theory is a prime example: it studies how entanglement, a central resource of quantum information processing, can be quantified and manipulated under local operations and classical communication (LOCC)~\cite{Horodecki2009quantumentanglement}.
One of its most fundamental tasks is entanglement distillation, which converts multiple copies of a noisy entangled state into nearly perfect Bell pairs~\cite{Bennett1996PurificationNoisyEntanglement,Bennett1996Mixedstateentanglementandqec}.
Distillation supplies resources for teleportation~\cite{bennett1993teleportingunknown}, quantum repeaters~\cite{Briegel1998QuantumRepeaters}, and quantum cryptography~\cite{Shor2000simpleproofsecurity}, and lies at the heart of quantum Shannon theory~\cite{wilde2013QItheory}, closely tied to quantum communication over noisy channels~\cite{Bennett1996Mixedstateentanglementandqec}.
Which quantities characterize the optimal asymptotic distillation rate, known as the distillable entanglement, is a central question in entanglement theory.
For pure states, this rate is given by the entanglement entropy~\cite{BennettConcentratingpartialentanglement}.
Noisy states, however, are generally mixed, and their distillable entanglement is harder to characterize: a general closed-form expression is unknown, and some entangled states cannot be distilled at all~\cite{Horodecki1998BoundEntanglementinNature}.
The optimal distillation rate can nevertheless be characterized in terms of the coherent information~\cite{Horodecki2000UnifiedApproachtoQuantumCapacities,devetak2005distillation}.

These entropic characterizations, however, describe what is possible in principle rather than what is feasible in practice. Since physical resources are limited, accounting for the cost of manipulating entanglement is crucial. Specifically, what matters is how the number of copies and the complexity of operations scale with the number $n$ of qubits. Once efficiency is imposed by restricting both to grow polynomially with $n$, the picture changes drastically~\cite{Leone2025Entanglementtheorywithlimited,meyer2026computationalrelativeentropy}. This is illustrated by pseudoentanglement~\cite{aaronson2023quantumpseudoentanglement}, which extends quantum pseudorandomness~\cite{Ji2018PseudorandomQuantumStates} to entanglement and provides two ensembles with very different amounts of entanglement that no efficient observer can tell apart.
Since efficient protocols cannot distinguish these ensembles, there exist states with large distillable entanglement from which they distill only a small amount.
Such observations, together with the advance of quantum technologies toward practical use, have motivated computational entanglement theory, which seeks to characterize the limits attainable by efficient protocols and identifies computational counterparts of entanglement measures and entropic quantities~\cite{arnonfriedman2023computationalentanglementtheory,Leone2025Entanglementtheorywithlimited,Avidan2026FullyQuantumComputationalEntropies,meyer2026computationalrelativeentropy,yanguez2026accessiblequantumcorrelationscomplexity}.

For pure states, Leone et al.~\cite{Leone2025Entanglementtheorywithlimited} recently showed that, with resources polynomial in $n$, the optimal distillation rate is characterized in the worst case by the min-entropy of the reduced state rather than its von Neumann entropy.
Yet pure states are highly idealized---the natural setting for distillation involves mixed states produced by noise. In the distillation protocol for pure states~\cite{Matsumoto2007universaldistortionfree} on which Ref.~\cite{Leone2025Entanglementtheorywithlimited} relies, simply measuring the Schur labels of the copies already leaves a maximally entangled state. In contrast, for mixed states we must deal with correlations with the environment, and an additional decoding step is needed to recover Bell pairs.
The standard decoupling approach~\cite{hayden2008decoupling,dupuis2010decoupling,dupuis2014one} to mixed-state distillation uses Uhlmann's theorem~\cite{uhlmann1976transition} to guarantee the existence of a suitable decoder, known as the Uhlmann transformation, but this does not by itself yield an efficient implementation~\cite{bostanci2023unicompuhlmann}. From the perspective of fundamental limitations, existing converse bounds indicate large gaps between information-theoretic and efficient distillation, but do not identify the quantity governing the efficient rate~\cite{Bansal2025PseudorandomDensityMatrices,meyer2026computationaltwowayquantumcapacity}. These facts leave open questions: how much entanglement can be distilled from mixed states under resource constraints, and which quantity characterizes the optimal rate?

We answer these questions by showing that the sandwiched R\'{e}nyi coherent information~\cite{muller2013quantumrenyinew,khatri2024principlemodern} of an intermediate order $\beta \in (1/2, 1)$ determines, in the worst case, the optimal distillation rate in two distinct settings with resource constraints. In the first setting, the number of copies is polynomial in the number $n$ of qubits, while the circuit size is unrestricted. Here we consider universal one-way LOCC distillation, which uses no description of the input state. 
The universal entanglement distillation was initially considered for pure-state inputs~\cite{Matsumoto2007universaldistortionfree} and has recently been extended to general mixed states~\cite{UQC,Takagi2026Universal,Tirone2026UniversalEntanglementDistillation}.
In particular, Rizzo et al.~\cite{UQC} generalized the protocol for pure states~\cite{Matsumoto2007universaldistortionfree} to universal Schur coding for mixed states. For polynomial-rank states, their protocol achieves rates given by the R\'{e}nyi coherent information capped at $C\log{n}$ for any constant $C > 0$. We further extend this guarantee to any state close to a polynomial-rank state, including a low-rank state subject to weak noise, by making the decoder robust to deviations in a small fraction of the copies. The rate is then determined by the R\'{e}nyi coherent information of this nearby state. While this protocol is sample-efficient, its efficient implementation remains open.
We next turn to the second setting, in which the circuit size is also polynomial in $n$.\footnote{Computationally efficient protocols are automatically sample-efficient, since a circuit of polynomial size can act on only polynomially many copies.} Under the additional condition that the nearby state has polynomial marginal ranks, we obtain another universal one-way LOCC protocol based on decoupling that achieves the R\'{e}nyi coherent information. Since known implementations of the Uhlmann transformation~\cite{metger2023stateqipspace,utsumi2025algorithmsUhlmanntrans} become costly in high dimensions, our key idea is universal collective compression, which reduces the local dimension per copy to a polynomial. Under the marginal-rank condition, the compression loses almost no entanglement. The receiving party then learns the compressed state from some of the copies and efficiently implements the Uhlmann transformation on the others.

To establish the optimality of these rates in the worst case, we construct a family of $n$-qubit mixed states of polynomial rank whose information-theoretic distillable entanglement is as large as $\widetilde{\Omega}(n)$. For universal LOCC protocols using $k = \poly(n)$ copies, the rate is capped by the smaller of the R\'{e}nyi coherent information and $\log{k}$, even when the circuit size is unrestricted. Its main ingredient is statistical indistinguishability: given polynomially many copies, ensembles constructed from Haar-random states cannot be distinguished from ensembles carrying little entanglement. The Schur-coding protocol nearly matches this bound, attaining the R\'{e}nyi coherent information up to $o(1)$ when it is $\cO(\log{n})$ and otherwise a constant fraction of $\log{k}$ at the number $k$ of copies it uses. We next turn to the computationally efficient setting, now allowing the protocol to be tailored to the input state, and we show that the same bound holds even in this setting. The states in this construction, however, need not be efficiently preparable, so the bound may be attributed to the complexity of the states themselves~\cite{arnonfriedman2023computationalentanglementtheory}.
Computational indistinguishability based on public-key pseudoentanglement~\cite {bouland2024PublicKeyPseudoentanglement} closes this loophole. Under a post-quantum hardness assumption for learning with errors (LWE)~\cite{regev2009Onlatticeslearningwitherrors}, we construct efficiently preparable states for which a similar bound holds even against protocols efficiently compiled from a description of their preparation circuit, with $\log{k}$ replaced by a freely chosen function of order $\omega(\log{n})$. When the R\'{e}nyi coherent information is $\cO(\log{n})$, our efficient protocol attains the bound up to $o(1)$ without a description of the state.
It is therefore optimal even against protocols tailored to the input state or efficiently compiled from a description of its preparation circuit.

Our results reveal some features of efficient distillation that are specific to mixed states.
First, intermediate R\'{e}nyi orders are essential for characterizing the optimal rate of efficient distillation from mixed states. The R\'{e}nyi coherent information at the endpoint $\beta = 1/2$ equals the min-entropy conditioned on the purifying system by the duality relation~\cite{muller2013quantumrenyinew}.
One might expect this endpoint to suffice for mixed states, as the min-entropy does for pure states~\cite{Leone2025Entanglementtheorywithlimited}. For pure states, the trivial purifying system makes the values at all fixed orders $\beta \in [1/2, 1)$ agree up to a constant factor, whereas for mixed states these values can separate by an amount growing with the rank. Indeed, we construct polynomial-rank states whose conditional min-entropy is $-\Theta(\log{n})$, yet whose optimal rate of efficient distillation is $\Theta(\log{n})$ and attained by our efficient protocol up to $o(1)$. More strongly, we find that every order in $(1/2, 1)$ captures the optimal rate for some family of our states but misses it for another; hence, no single fixed order characterizes efficient distillation from mixed states.
Second, the optimal rate of efficient distillation undergoes a sharp transition as the rank grows from polynomial to superpolynomial. At superpolynomial rank, our construction yields states from which no efficient protocol distills entanglement at all, even though their information-theoretic distillable entanglement remains $\widetilde{\Omega}(n)$. Such a transition was indicated for special classes of states in earlier work~\cite{Bansal2025PseudorandomDensityMatrices,meyer2026computationaltwowayquantumcapacity}; our results establish it for general mixed states in the worst case.

Finally, we mention that our results have implications across a broad range of areas in quantum information science.  In quantum networks, decoding complexity is a practical bottleneck for distillation~\cite{Gu2026ConstantOverheadEntanglementDistillation}, and noise on shared states is often not fully characterized. Our efficient protocol, which requires no description of the state, is a step toward addressing these problems.
Beyond distillation, implementing the Uhlmann transformation in a compressed space may also facilitate decoding in quantum communication, state merging, and quantum information recovery~\cite{bostanci2023unicompuhlmann,meyer2026computationaltwowayquantumcapacity}. 
Pseudoentanglement, in turn, connects our mixed-state constructions to quantum cryptography~\cite{goulao2024pseudoentanglementnecessaryefipairs} and to questions of computational complexity in quantum gravity~\cite{aaronson2023quantumpseudoentanglement,akers2024holographicpseudoentanglementcomplexityadscft}, where mixed states play a central role~\cite{arnonfriedman2023computationalentanglementtheory,Engelhardt2026SpoofingEntanglementHolography}. 
Together, our protocols and constructions provide tools for investigating how resource constraints shape entanglement manipulation and its applications.

\section{Overview of main results}
\label{sec:overview of main results}

We consider families of bipartite states $\rho^{\sA\sB}$ on $\Theta(n)$ qubits,\footnote{Considering $\Theta(n)$-qubit families loses no generality, since any family on $n^{\Theta(1)}$ qubits can be re-indexed by its qubit number, under which circuit sizes and numbers of copies remain polynomial.} indexed by a parameter $n$, where $\sA$ and $\sB$ denote Alice's and Bob's systems, respectively.
The bipartition $\sA|\sB$ is extensive, with the two sides containing comparable numbers of qubits.
Alice and Bob are given $k$ identical copies of $\rho^{\sA\sB}$, and their goal is to turn them into as many approximate ebits as possible using only LOCC.
A $k$-shot distillation protocol distills $m$ ebits at error $\varepsilon$ and rate $R = m/k$ if its output has fidelity at least $1 - \varepsilon$ with $m$ Bell pairs (see also \Cref{def:protocol}). Here, the fidelity is defined by $\rF(\rho,\sigma) \coloneqq \|\sqrt{\rho}\sqrt{\sigma}\|_1^2$, where $\|A\|_1 \coloneqq \tr\big[\sqrt{A^\dag A}\big]$ denotes the trace norm.
The information-theoretic distillable entanglement $E_\rD(\sA:\sB)_\rho$ is the largest rate achievable asymptotically, as the number of copies tends to infinity and the error tends to zero, without restricting the parties' computational power.

We study how this rate changes under two resource constraints.
In the first setting, the number of copies is polynomial in $n$, while the circuit size is unrestricted; we call such protocols \emph{sample-efficient}.
In the second, we also require the LOCC channel, including its classical communication, to be implemented by a circuit of polynomial size; we call such protocols \emph{computationally efficient}. See \Cref{sec: General preliminary} for their formal definitions.

To provide our results, we use the (sandwiched) R\'{e}nyi coherent information $\widetilde{I}_\beta(\sA\rangle\sB)_\rho \coloneqq -\widetilde{H}_\beta(\sA|\sB)_\rho$ for the order $\beta \in [1/2, 1)$, where 
\begin{align}
    \widetilde{H}_\beta(\sA|\sB)_\rho
    \coloneqq \max_{\sigma^\sB}\f{1}{1-\beta}\log\tr\Big[\big((\bI^\sA\otimes\sigma^\sB)^{\f{1-\beta}{2\beta}}\rho^{\sA\sB}(\bI^\sA\otimes\sigma^\sB)^{\f{1-\beta}{2\beta}}\big)^\beta\Big]
\end{align}
is the sandwiched conditional R\'{e}nyi entropy for a state $\rho^{\sA\sB}$, and the maximization is taken over all states on $\sB$. The quantity $\widetilde{I}_\beta(\sA\rangle\sB)_\rho$ is non-decreasing in $\beta$ and tends to the standard coherent information as $\beta \to 1$~\cite{muller2013quantumrenyinew,khatri2024principlemodern}.


\subsection{Achievability: Sample-efficient and computationally efficient protocols}

Our two distillation protocols apply to states $\rho^{\sA\sB}$ that are close to a state $\hat{\rho}^{\sA\sB}$ of polynomial rank, in the sense that the trace distance $\f{1}{2}\|\rho^{\sA\sB}-\hat{\rho}^{\sA\sB}\|_1$ is small. Their rates are determined by the R\'{e}nyi coherent information $\widetilde{I}_\beta(\sA\rangle\sB)_{\hat{\rho}}$ of the nearby state. The first protocol requires only a global-rank condition and is sample-efficient. The second is computationally efficient under an additional condition on the rank of Alice's marginal. Both are universal: they use the promised rank bounds, a target rate, and the target error, but no description of either the input state $\rho^{\sA\sB}$ or the nearby state $\hat{\rho}^{\sA\sB}$.

\begin{theorem}[Achievability for mixed-state distillation; formal versions in \Cref{thm:achievability by schur coding,thm:achievability by efficient protocol formal}]
\label{thm:achievability by efficient protocol}
Let $\beta \in [1/2, 1)$ and $c, C > 0$ be any constants, let $\varepsilon \in (0, 1)$ satisfy $\log(1/\varepsilon) = \poly(n)$, and let $\hat{\rho}^{\sA\sB}$ be a state with $\widetilde{I}_\beta(\sA\rangle\sB)_{\hat{\rho}} = \Omega(1)$. In each of the following two cases, there exists a universal one-way LOCC protocol such that, for every state $\rho^{\sA\sB}$ satisfying $\f{1}{2}\|\rho^{\sA\sB}-\hat\rho^{\sA\sB}\|_1\leq 1/\log^{1+c}n$, it distills Bell pairs from $k = \poly(n)$ copies of $\rho^{\sA\sB}$ with error $\varepsilon$ at a rate $R$ bounded as follows.
\begin{enumerate}[label=\textnormal{(\Roman*)}]
    \item If $\rank(\hat{\rho}^{\sA\sB}) = \poly(n)$, then $R \geq \min\big\{\widetilde{I}_\beta(\sA\rangle\sB)_{\hat{\rho}}, \, C\log{n}\big\} - o(1)$, with no restriction on the circuit size of the protocol.
    \item If $\rank(\hat{\rho}^{\sA\sB}), \, \rank(\hat{\rho}^{\sA}) = \poly(n)$, then $R \geq \widetilde{I}_\beta(\sA\rangle\sB)_{\hat{\rho}} - o(1)$, and the protocol is efficiently implemented by a circuit of size $\poly(n)$.
\end{enumerate}
\end{theorem}

Since the rank conditions concern $\hat{\rho}^{\sA\sB}$ alone, the input $\rho^{\sA\sB}$ itself may have large rank, as long as its weight is concentrated near a state of polynomial rank; a typical example is a low-rank state subject to weak noise. Such an input is distilled at the rate determined by the nearby state.
Moreover, the protocols do not need to know which input they receive. A single protocol is determined by the rank bounds, a target rate, and the target error, and it serves every input close to a state that meets these bounds and has R\'{e}nyi coherent information at least the target rate. This universality makes the protocols applicable even when the noise affecting a shared state is not fully characterized.

The two cases differ in their rank requirements and computational guarantees. In~(II), the protocol is computationally efficient, but the marginal-rank condition it requires bounds the R\'{e}nyi coherent information by the logarithm of Alice's marginal rank, $\widetilde{I}_\beta(\sA\rangle\sB)_{\hat{\rho}} \leq \log{\rank(\hat{\rho}^{\sA})} = \cO(\log{n})$.\footnote{Both marginal ranks are in fact polynomially bounded: the bound on Bob's marginal follows from the global-rank bound and the bound on Alice's marginal, since $\rank(\hat{\rho}^{\sB}) \leq \rank(\hat{\rho}^{\sA\sB})\rank(\hat{\rho}^{\sA})$.}
Nearby states with R\'{e}nyi coherent information of order $\omega(\log{n})$ thus have superpolynomial marginal rank and are left to~(I). Dropping the marginal-rank condition, (I) reaches the rate $C\log{n}$ for any constant $C > 0$ even in this regime, at the cost of an unrestricted circuit size.

Below, we outline how the two protocols are constructed: (I) extends universal Schur coding to perturbed inputs, while (II) makes the decoupling approach efficient by first compressing the local systems of the copies. For details, see \Cref{sec:schur coding} for~(I) and \Cref{sec:achievability} for~(II).

For~(I), we start from the universal Schur coding of Ref.~\cite{UQC}.
Alice and Bob apply the Schur transform to their copies and measure the local Schur labels. Alice then measures her symmetric-group register, keeping a register of smaller dimension, and sends her label and the outcome to Bob, who applies a decoder determined by the two labels and this outcome, independent of the input state. The labels select subspaces, the Schur supports, whose permutation symmetry makes the local marginals flat, allowing one decoder to work uniformly on each support; the R\'{e}nyi coherent information controls the weight outside them. From polynomially many copies of polynomial-rank states, this protocol distills at the R\'{e}nyi coherent information, capped at order $\log{n}$~\cite{UQC}.

Our contribution to~(I) is to extend this guarantee to inputs that are only close to a polynomial-rank state and may themselves have large rank. The guarantee for the nearby state does not transfer directly: a small deviation in each copy accumulates, and the tensor powers $\rho^{\otimes k}$ and $\hat{\rho}^{\otimes k}$ can be far apart for polynomially many copies. We instead enlarge the Schur supports to include deviations on a small fraction of the copies, and bound both the size of the enlargement and the weight remaining outside the enlarged supports (\Cref{lem:enlarged subspaces}). A single decoder then distills from every such input at the rate of the nearby state, still from polynomially many copies (\Cref{lem:uniform support estimate}). Finally, a concatenated stabilizer code suppresses the remaining error (\Cref{lem:error reduction}). The decoder, however, is specified through Kronecker coefficients on spaces whose dimension grows with the number of copies, and an efficient implementation remains open.

For~(II), we start from a different approach based on decoupling~\cite{hayden2008decoupling,dupuis2010decoupling,dupuis2014one}.
In this approach, Alice applies a random measurement, retains a register $\sS$, and sends the outcome to Bob.
The decoupling theorem~\cite{dupuis2014one} guarantees that $\sS$ is approximately completely mixed and decoupled from the purifying system when its size is chosen according to the R\'{e}nyi coherent information.
Uhlmann's theorem~\cite{uhlmann1976transition} then guarantees a decoder on Bob's systems, the Uhlmann transformation, that recovers Bell pairs shared with $\sS$. Known algorithms for this transformation~\cite{metger2023stateqipspace,utsumi2025algorithmsUhlmanntrans}, however, become exponentially costly when applied directly to the original high-dimensional systems.

To resolve this problem, our first step is universal collective compression (\Cref{lem:compression}). Alice and Bob reduce the local dimension per copy to a polynomial without learning the original eigenbases.
The marginal-rank condition ensures that this compression preserves the nearby state, up to local isometries and a small error. The parties perform local tomography on some of these compressed copies, and Alice sends her outcomes to Bob so that he can estimate the joint compressed state. However, compression alone does not yet make the decoding efficient: decoupling many compressed copies jointly would produce exponentially many outcomes, and the cost of Bob's decoder grows with their number.
Alice thus measures in two stages (\Cref{lem:first instrument,lem:second instrument}): she first shrinks each compressed copy separately, retaining its conditional entropy, and then decouples a block of the shrunken copies with only polynomially many outcomes. For the reported outcome, Bob computes the Uhlmann transformation from his estimate as the unitary factor of a polar decomposition. The tensor-product structure of the estimate allows this transformation to be implemented at a cost polynomial in the dimension of one compressed copy and the block length, without constructing a matrix on the full block explicitly (\Cref{lem:polar}).
Finally, a concatenated stabilizer code again suppresses the remaining error.
Together, these steps incur only $o(1)$ rate loss, and the protocol thus attains the rate in~(II) with polynomially many copies and a polynomial-size circuit. All communication goes from Alice to Bob, and the circuit is computed from the promised rank bounds, the target rate, and the target error.


\subsection{Converse: No-go results and optimality}

We now ask whether a different protocol can substantially improve these rates. Our converse results construct a family of states with large information-theoretic distillable entanglement but much lower rates under the resource constraints.
To state what these bounds rule out, we also need to specify how a protocol may depend on the input state.
We distinguish the following three models; see \Cref{def:computational model} for the formal definition.
\begin{itemize}
    \item \emph{State-agnostic.} A single protocol is required to succeed on every state in a prescribed set, without information about which member it receives. Its circuit size is unrestricted.
    \item \emph{State-dependent.} The protocol has polynomial circuit size and may depend on the state in any way; no account is taken of how the circuit is found.
    \item \emph{Description-aware.} An efficient compiler receives a classical description of a circuit $C^{\sA\sB\sE}$ preparing a purification of $\rho^{\sA\sB}$, $C^{\sA\sB\sE}\ket{0}^{\sA\sB\sE} = \ket{\rho}^{\sA\sB\sE}$, and outputs an efficient LOCC protocol. Here $\sE$ is the purifying system. A single compiler is required to succeed on all but a negligible fraction of preparation-circuit descriptions drawn from a given distribution.
\end{itemize}
Thus, moving from state-agnostic to state-dependent models restricts computation while allowing arbitrary dependence on the state; the description-aware model further accounts for the cost of constructing the protocol.

Our states consist of two components with separate roles: a tunable \emph{spike state} (\Cref{lem:spikestate}) controls the R\'{e}nyi coherent information, while a weakly weighted, highly entangled \emph{entanglement amplifier} keeps $E_\rD(\sA:\sB)_\rho$ nearly extensive. Following the pseudoentanglement-based approach of Refs.~\cite{aaronson2023quantumpseudoentanglement,arnonfriedman2023computationalentanglementtheory,bouland2024PublicKeyPseudoentanglement,bouland2024hardnesslearninggroundstate,Leone2025Entanglementtheorywithlimited}, we make the amplifier indistinguishable from a weakly entangled counterpart. A resource-limited protocol cannot distill more entanglement from a hidden component than its counterpart holds, since its output would otherwise tell the two apart. The rate is thus capped by the R\'{e}nyi coherent information set by the spike.

The spike state provides R\'{e}nyi coherent information only up to order $\cO(\log{n})$; for larger values, we replace the spike with a highly entangled state that is also indistinguishable from a counterpart with less entanglement. We choose these counterparts so that they remain indistinguishable from the high-entanglement states given any polynomial number $k$ of copies. How little entanglement such a counterpart may carry depends on the rank of our states, a free parameter that can be chosen without changing $E_\rD(\sA:\sB)_\rho$ or the R\'{e}nyi coherent information. At polynomial rank, the counterpart must still have entanglement of order $\omega(\log{n})$, which appears in the description-aware model as the second entry $\theta$ of the cap in the following theorem. In the other two models, the second entry is instead $\theta = \log{k}$, obtained by a separate argument. At superpolynomial rank, the whole state can be hidden behind one with no distillable entanglement, and the rate drops to zero.\footnote{The converse bounds impose $\varepsilon < 1/2$ and $k = \omega(1)$ for the following reasons. Since a separable state can have fidelity $1/2$ with a Bell pair, a protocol at $\varepsilon \geq 1/2$ can declare $1$ ebit without distilling any, and an upper bound of zero on the rate requires $\varepsilon < 1/2$. The number of ebits declared in this way is at most $\log(1/(1-\varepsilon))$, and bounds of order $o(1)$ nevertheless hold for every constant $\varepsilon < 1$. More generally, $m - t$ Bell pairs together with a product state can have fidelity $2^{-t}$ with $m$ Bell pairs; a fixed error thus allows a constant excess $t$ in the declared number of ebits, whose effect $t/k$ on the rate vanishes for $k = \omega(1)$.}

\begin{theorem}[Converse for mixed-state distillation; formal versions in \Cref{thm:agnostic unbounded,thm:agnostic no go,thm:lwe mixed state no go}]
\label{thm:no go informal}
There are a constant $\beta_\rc \in (1/2, 1)$ and families of $\Theta(n)$-qubit states $\{\rho^{\sA\sB}\}_n$ such that, for every constant $\beta \in (\beta_\rc, 1)$, every constant $\varepsilon < 1/2$, and $\omega(1) \leq k \leq \poly(n)$, the rate $R$ of distillation from $k$ copies at error $\varepsilon$ in the models specified below is bounded as
\begin{align}
\label{eq:no go informal cap}
    R
    \leq
    \begin{cases}
        \min\big\{\widetilde{I}_\beta(\sA\rangle\sB)_\rho, \ \theta\big\} + o(1),
        & \text{if } \rank(\rho^{\sA\sB}) = \poly(n), \\
        0,
        & \text{if } \rank(\rho^{\sA\sB}) = 2^{\omega(\log{n})},
    \end{cases}
\end{align}
while $E_\rD(\sA:\sB)_\rho = \widetilde{\Omega}(n)$, and the rank can be varied without changing $E_\rD(\sA:\sB)_\rho$ or $\widetilde{I}_\beta(\sA\rangle\sB)_\rho$.
Here, the bound holds with $\theta = \log{k}$ in the state-agnostic and state-dependent models.
Under the superpolynomial hardness of LWE with binary secrets (\Cref{assumpt:lwe assumptions}), for any prescribed efficiently computable polylogarithmic function $\theta = \omega(\log{n})$, the families can be chosen to be efficiently preparable, and the bound then holds in the description-aware model.
\end{theorem}

A basic ingredient of the construction is a mixture of two branches marked by local flags:
\begin{align}
\label{eq:overview separation state}
    \rho_b^{\sI_\sA\sI_\sB}
    = (1-\eta)\ketbra{00}{00}^{\sF_\sA\sF_\sB} \otimes \xi_{\mathrm{sp}}^{\sG_\sA\sG_\sB} \otimes \ketbra{00}{00}^{\sH_\sA\sH_\sB}
    + \eta\ketbra{11}{11}^{\sF_\sA\sF_\sB} \otimes \ketbra{00}{00}^{\sG_\sA\sG_\sB} \otimes \phi_b^{\sH_\sA\sH_\sB},
\end{align}
where $\sI_\sA = \sF_\sA\sG_\sA\sH_\sA$ and $\sI_\sB = \sF_\sB\sG_\sB\sH_\sB$. The states $\xi_{\mathrm{sp}}$ and $\phi_b$ are the spike state and the pure entanglement amplifier, respectively, and the amplifier is drawn at $b = 0$ from a high-entanglement ensemble and at $b = 1$ from an indistinguishable low-entanglement one.
Our families are built from the states at $b = 0$, with the weight $\eta = 1/\polylog(n)$ chosen small enough that, at $b = 1$, the low-entanglement amplifier adds only $o(1)$ to the cap set by the spike. This weight still leaves $E_\rD(\sI_\sA:\sI_\sB)_{\rho_0} \geq \eta\cdot\Omega(n) = \widetilde{\Omega}(n)$ at $b = 0$. The states of \Cref{thm:no go informal} are then mixtures, under orthogonal labels $j$ held by both parties, of one realization of $\rho_0^{\sI_\sA\sI_\sB}$ conjugated by local unitaries that depend on the label:
\begin{align}
\label{eq:overview labeled mixture}
    \rho^{\sA\sB}
    = \sum_{j} p_j \big(U_{1,j}^{\sI_\sA}\otimes U_{2,j}^{\sI_\sB}\big)\rho_0^{\sI_\sA\sI_\sB}\big(U_{1,j}^{\sI_\sA}\otimes U_{2,j}^{\sI_\sB}\big)^\dag \otimes \ketbra{jj}{jj}^{\sK_\sA\sK_\sB},
\end{align}
where $p_j > 0$ is the probability of the label $j$, $\sA = \sI_\sA\sK_\sA$ and $\sB = \sI_\sB\sK_\sB$, and the local unitaries $U_{1,j}^{\sI_\sA}$ and $U_{2,j}^{\sI_\sB}$ scramble the realization differently for each label (see also Eqs.~\eqref{eq:agnostic mixed} and~\eqref{eq:master mixed}).
By construction, $E_\rD(\sA:\sB)_\rho$ and the R\'{e}nyi coherent information equal those of $\rho_0^{\sI_\sA\sI_\sB}$, while $\rank(\rho^{\sA\sB})$ is the number of labels times $\rank(\rho_0^{\sI_\sA\sI_\sB})$.

The bounds in the three models rest on different types of indistinguishability: statistical in the first two, discussed in \Cref{sec:Converse: State-agnostic setting}, and computational in the third, discussed in \Cref{sec:Converse Description-aware setting}.
In the first two models, the amplifier and its counterpart are drawn from Haar-random states and Haar-subsystem states, respectively, which cannot be distinguished from polynomially many copies, regardless of computational power~\cite{Leone2025Entanglementtheorywithlimited}. Hence, the bounds in these models require no hardness assumption.
In the state-agnostic model, where a single protocol must work throughout the state set, this indistinguishability alone gives the first entry of the cap, with no restriction on the circuit size.
In the state-dependent model, the protocol may instead be tailored to a single state, and the bound then requires one state that defeats every such protocol. We select such a hard instance from the same set by concentration of measure and a union bound over polynomial-size circuits, which is where the circuit-size restriction enters.
In the description-aware model, the compiler also receives a preparation-circuit description, so statistical indistinguishability of the copies alone does not suffice.
Under the LWE hardness assumptions, public-key pseudoentanglement~\cite{bouland2024PublicKeyPseudoentanglement,bouland2024hardnesslearninggroundstate} instead supplies computational indistinguishability even given these descriptions.\footnote{The LWE assumptions are imposed at superpolynomial strength, and for binary secrets, to cover the widest parameter ranges. Under polynomial hardness, or with standard secrets, the smallest attainable $\theta$ rises and the zero-rate case recedes to a larger rank, leaving the ranks in between uncovered. The bound $E_\rD(\sA:\sB)_\rho=\widetilde{\Omega}(n)$ would also weaken under polynomial hardness. The gap between $E_\rD(\sA:\sB)_\rho$ and the rate of any efficient protocol remains in every case; we will discuss these points at the end of \Cref{sec:description-aware converse result}.}

The bound of \Cref{thm:no go informal} also reveals a phenomenon that has no counterpart for pure states: a sharp transition in the rank. As the second case in Eq.~\eqref{eq:no go informal cap} shows, the rate of resource-limited protocols drops from the cap of the first case to zero once the rank becomes superpolynomial, while $E_\rD(\sA:\sB)_\rho$ and $\widetilde{I}_\beta(\sA\rangle\sB)_\rho$ remain unchanged. Since $\rho_0^{\sI_\sA\sI_\sB}$ has polynomial rank, superpolynomial rank means superpolynomially many labels in Eq.~\eqref{eq:overview labeled mixture}. Polynomially many copies then almost never share a label, and each copy is scrambled by its own local unitary. In the statistical construction, these unitaries are independent and Haar random, and the mixture becomes statistically indistinguishable from a separable state that is completely mixed on $\sI_\sA\sI_\sB$ (\Cref{lem:maxmix}). In the computational construction, they are Heisenberg--Weyl displacements that look independent and uniformly random to efficient observers, and the same indistinguishability holds computationally (\Cref{lem:labeled generator}). No entanglement can be distilled from this separable state. This indistinguishability requires superpolynomial rank: at polynomial rank, the purity of the state is at least $1/\poly(n)$, and a SWAP test efficiently distinguishes it from this separable state of purity $2^{-\Theta(n)}$.

In fact, the bound of \Cref{thm:no go informal} holds in every model even if each party is additionally given $l = \poly(n)$ copies of a purification $\ket{\rho}^{\sA\sB\sE}$ of $\rho^{\sA\sB}$ in its own local laboratory, which carry no entanglement across the cut $\sA|\sB$. In the state-agnostic and state-dependent models, \Cref{thm:agnostic unbounded,thm:agnostic no go} include these local copies, with the second entry $\theta = \log{k}$ replaced by $\log{(k+2l)}$, since our arguments extend to the purifications. In the description-aware model, such copies add nothing, as the compiler can prepare them efficiently from the given circuit description.

We now show that our protocols are optimal on suitable families of \Cref{thm:no go informal}. When the R\'{e}nyi coherent information of $\rho^{\sA\sB}$ is $\omega(1)$ and $\cO(\log{n})$, both protocols attain the first entry of the cap up to $o(1)$. To see this, note that removing the amplifier branch from $\rho_0^{\sI_\sA\sI_\sB}$ in every term of Eq.~\eqref{eq:overview labeled mixture} leaves a state $\hat{\rho}^{\sA\sB}$ within trace distance $\eta$ of $\rho^{\sA\sB}$. If $\rho^{\sA\sB}$ has polynomial rank, Eq.~\eqref{eq:overview labeled mixture} has polynomially many labels. As only the spike state remains in each term, the global and marginal ranks of $\hat{\rho}^{\sA\sB}$ are at most the number of labels times those of the spike state, and are hence polynomial, whereas the amplifier makes the marginal ranks of $\rho^{\sA\sB}$ exponential. The R\'{e}nyi coherent information of $\hat{\rho}^{\sA\sB}$ equals that of $\rho^{\sA\sB}$ up to $o(1)$. With $\eta \leq 1/\log^{1+c}{n}$ for a constant $c > 0$, $\hat{\rho}^{\sA\sB}$ serves as the nearby state in both cases of \Cref{thm:achievability by efficient protocol}. In~(I), we take $C$ large enough that $C\log{n}$ exceeds the R\'{e}nyi coherent information of $\hat{\rho}^{\sA\sB}$. For larger R\'{e}nyi coherent information, only~(I) applies, and an estimate of its number of copies shows that its rate reaches any fixed fraction $B < 1/2$ of the second entry $\log{k}$ of the cap. We thus obtain the following corollary (see \Cref{sec:achievability on converse families} for a detailed discussion).

\begin{corollary}[Worst-case optimality]
\label{cor:worst case optimality}
There exist polynomial-rank families $\{\rho^{\sA\sB}\}_n$ of \Cref{thm:no go informal}, with $E_\rD(\sA:\sB)_\rho = \widetilde{\Omega}(n)$, for which the following statements on the universal protocols (I) and (II) of \Cref{thm:achievability by efficient protocol} hold at every fixed order $\beta \in (\beta_\rc, 1)$ and every constant error $\varepsilon \in (0, 1/2)$, where optimality is taken among LOCC protocols acting on $k$ copies with $\omega(1) \leq k \leq \poly(n)$.
\begin{enumerate}[label=\textnormal{(\roman*)}]
    \item If $\widetilde{I}_\beta(\sA\rangle\sB)_\rho$ is $\omega(1)$ and $\cO(\log{n})$, the rate of~(I) is optimal up to $o(1)$ in the state-agnostic model, and that of~(II) in all three models. In the description-aware model, the optimality of~(II) holds for efficiently preparable families under the LWE assumptions of \Cref{thm:no go informal}.
    \item If $\widetilde{I}_\beta(\sA\rangle\sB)_\rho$ is $\omega(\log{n})$, then, for every constant $B \in (0, 1/2)$, protocol~(I) attains $R \geq B\log{k}$ at the number of copies $k = \poly(n)$ it uses, while the optimal rate in the state-agnostic model at this $k$ is at most $\log{k} + o(1)$.
\end{enumerate}
\end{corollary}

As the spike state is tunable, the R\'{e}nyi coherent information of the families in \Cref{cor:worst case optimality} can be set to any value between $\omega(1)$ and $\cO(\log{n})$ up to $o(1)$, and the guarantee $R \geq \widetilde{I}_\beta(\sA\rangle\sB)_{\hat{\rho}} - o(1)$ of \Cref{thm:achievability by efficient protocol} cannot be improved at any value in this range. The protocols of \Cref{thm:achievability by efficient protocol} use only the rank bounds, the target rate, and the target error, which are common to all states of the prescribed set from which these families are selected. Hence both~(I) and~(II) work throughout the set, as the state-agnostic model requires. Protocol~(II) has a polynomial-size circuit and thus also qualifies in the state-dependent model; as it uses no description of the state, it also qualifies in the description-aware model, where a compiler simply outputs our protocol regardless of the given description.

The second entry $\theta$ in the polynomial-rank case of Eq.~\eqref{eq:no go informal cap} matters when the R\'{e}nyi coherent information exceeds it. In the state-agnostic and state-dependent models, $\theta = \log{k}$ comes from the local unitaries in Eq.~\eqref{eq:overview labeled mixture}, which are Haar random in the statistical construction. Averaging over them upper bounds the rate by $\log{k} + o(1)$ at every $k$, for a family fixed independently of $k$ (\Cref{lem:maxmix}). By (ii) of \Cref{cor:worst case optimality}, the optimal rate in the state-agnostic model is $\Theta(\log{k})$ at the number $k$ of copies used by~(I), as for pure states~\cite{Leone2025Entanglementtheorywithlimited}.
In the description-aware model, $\theta$ is instead the entanglement of the weakly entangled states behind which the highly entangled components are hidden. It can be any prescribed function of order $\omega(\log{n})$, and the families are chosen for this $\theta$.
The order $\omega(\log{n})$ cannot be reduced to $\cO(\log{n})$ by comparison with a computationally indistinguishable ensemble of polynomial rank whose distillation rate admits such an information-theoretic cap uniformly over polynomial numbers of copies: a polynomial-rank ensemble indistinguishable from states with R\'{e}nyi coherent information $\omega(\log{n})$ still allows~(I) to distill at rate $C\log{n}$ for every constant $C > 0$ (\Cref{prop:indistinguishability rank}). Whether computationally efficient protocols match the cap in this regime remains open, as we obtain no efficient protocol without a nearby state of polynomial marginal rank.

The other restriction in the converse concerns the R\'{e}nyi order: the cap holds only above a threshold $\beta_\rc > 1/2$, leaving a range of lower orders uncovered. Whether this exclusion is an artifact of our construction or is genuine for mixed states, and more broadly whether a single fixed order suffices to characterize efficient distillation from mixed states, are the questions of the next subsection.


\subsection{Separation: Role of the R\'{e}nyi order and nearly maximal gaps}

Our rates are expressed by the R\'{e}nyi coherent information $\widetilde{I}_\beta(\sA\rangle\sB)_\rho$ at a fixed order $\beta < 1$. The guarantee of \Cref{thm:achievability by efficient protocol} holds at every order $\beta \in [1/2, 1)$, while the cap of \Cref{thm:no go informal} holds only above the threshold $\beta_\rc$. We now ask which R\'{e}nyi orders $\beta$ are singled out by efficient distillation from mixed states. The two end points of the range have clear meanings: as $\beta \to 1$, the R\'{e}nyi coherent information tends to the coherent information, which is achievable without any resource constraint~\cite{devetak2005distillation}; at $\beta = 1/2$, it becomes the conditional min-entropy given the purifying system $\sE$ of $\rho^{\sA\sB}$, i.e., $\widetilde{I}_{1/2}(\sA\rangle\sB)_\rho = H_{\rm min}(\sA|\sE)_\rho$~\cite{muller2013quantumrenyinew}, which for pure states reduces to the min-entropy characterizing distillation with limited resources~\cite{Leone2025Entanglementtheorywithlimited}.

For pure states, this question does not arise, since the R\'{e}nyi coherent information at every fixed order agrees with the min-entropy up to a constant factor. To see this, we provide the following relation that bounds how far the R\'{e}nyi coherent information can lie above its value at $\beta = 1/2$.
While this relation follows from a direct calculation, for completeness we derive it in \Cref{Separations for conditional entropies and converse bounds}.

\begin{restatable}[Rank bound on the separation of R\'{e}nyi orders]{proposition}{proprenyiminentropies}
\label{prop:renyi min relation}
For any state $\varrho^{\sA\sB}$ and $\beta \in (1/2, 1)$,
\begin{align}
    \label{eq:entropy has same order}
    \widetilde{I}_{1/2}(\sA\rangle\sB)_\varrho
    \leq \widetilde{I}_\beta(\sA\rangle\sB)_\varrho
    \leq \frac{\beta}{1-\beta}\big(\widetilde{I}_{1/2}(\sA\rangle\sB)_\varrho + 2\log{\rank(\varrho^{\sA\sB})}\big).
\end{align}
\end{restatable}

From this proposition, we see that, for pure states, where $\rank(\rho^{\sA\sB}) = 1$, all fixed orders $\beta \in [1/2, 1)$ lie within a constant factor of the min-entropy $H_{\rm min}(\sA)_\rho$. For mixed states, the extra term $2\log{\rank(\varrho^{\sA\sB})}$ leaves room for different orders to separate.

We next show that efficient distillation from mixed states reflects this separation. To this end, we again use the construction of the families of \Cref{thm:no go informal}, in which the R\'{e}nyi coherent information of the spike state is tuned by two parameters $\beta_\rc, \beta_\rs$ such that $1/2 < \beta_\rs < \beta_\rc < 1$ (\Cref{lem:spikestate}). To leading order in the logarithm of the rank of the spike state, this quantity is constant for $\beta_\rc \leq \beta < 1$, decreases as $\beta$ decreases below $\beta_\rc$, and changes sign at $\beta = \beta_\rs$. We place a spike state of rank $r - 1$ in $\rho_0^{\sI_\sA\sI_\sB}$ of Eq.~\eqref{eq:overview separation state} and take Eq.~\eqref{eq:overview labeled mixture} with a single label; the resulting state $\rho^{\sA\sB}$ has rank $r$. The following theorem shows that the efficient cap is then given by the R\'{e}nyi coherent information at every order above $\beta_\rc$. Below $\beta_\rc$, the R\'{e}nyi coherent information is smaller than the cap, and below $\beta_\rs$ it is even negative. Since $\beta_\rc$ and $\beta_\rs$ can be chosen freely, the orders at which these changes occur can be placed anywhere in $(1/2, 1)$.

\begin{restatable}[Separation]{theorem}{threegapsseparations}
\label{prop:separation}
For every pair of constants $1/2 < \beta_\rs < \beta_\rc < 1$ and every $r$ satisfying $\omega(1) \leq r \leq \poly(n)$, there is a family of mixed states $\{\rho^{\sA\sB}\}_n$ of \Cref{thm:no go informal} with $\rank(\rho^{\sA\sB}) = r$ that exhibits the following two gaps:
\begin{enumerate}[label=\textnormal{(\arabic*)}]
    \item the R\'{e}nyi coherent information $\widetilde{I}_\beta(\sA\rangle\sB)_{\rho}$ changes sign at $\beta = \beta_\rs$: it is $+\Theta(\log{r})$ for every fixed $\beta \in (\beta_\rs, 1)$ and $-\Theta(\log{r})$ for every fixed $\beta \in [1/2, \beta_\rs)$;
    \item $E_\rD(\sA:\sB)_{\rho} = n^{\Omega(1)}$, yet, for every constant $\varepsilon < 1/2$ and $\omega(1) \leq k \leq \poly(n)$, the rate $R$ of distillation from $k$ copies at error $\varepsilon$ in the models specified below is at most $\widetilde{I}_\beta(\sA\rangle\sB)_{\rho} + o(1)$ for any fixed $\beta \in (\beta_\rc, 1)$.
\end{enumerate}
Here, the rate bound holds in the state-agnostic and state-dependent models.
Under the superpolynomial hardness of LWE with binary secrets (\Cref{assumpt:lwe assumptions}), the family can be chosen to be efficiently preparable, and the rate bound then holds in the description-aware model.
\end{restatable}

Each state of the family in \Cref{prop:separation} indeed admits a nearby state of polynomial global and marginal ranks whose R\'{e}nyi coherent information agrees with the cap in~(2) up to $o(1)$. Protocol~(II) of \Cref{thm:achievability by efficient protocol} then attains this cap at any fixed order $\beta \in (\beta_\rc, 1)$, and the optimal efficient rate is thus $\widetilde{I}_\beta(\sA\rangle\sB)_\rho + o(1) = \Theta(\log{r})$ at these orders. As the R\'{e}nyi coherent information is smaller than this rate below $\beta_\rc$, the exclusion of the orders below $\beta_\rc$ in \Cref{thm:no go informal} is genuine: for these states, a cap at any such order, including the endpoint $\beta = 1/2$, would be violated. Since $\rho^{\sA\sB}$ has rank $r$, the gap $\Theta(\log{r})$ between the R\'{e}nyi coherent information at orders above $\beta_\rc$ and the conditional min-entropy is the largest, up to constant factors, that Eq.~\eqref{eq:entropy has same order} permits at that rank. In particular, at $r = n^{\Theta(1)}$, the optimal efficient rate and the conditional min-entropy are $\Theta(\log{n})$ and $-\Theta(\log{n})$, respectively.

Since $\beta_\rc$ and $\beta_\rs$ can be chosen freely, any two fixed orders can be separated in this way. For any $1/2 \leq \beta_2 < \beta_1 < 1$, choosing $\beta_2 < \beta_\rs < \beta_\rc < \beta_1$ yields a family of states $\{\rho^{\sA\sB}\}_n$ with
\begin{align}
    \widetilde{I}_{\beta_1}(\sA\rangle\sB)_\rho = \Theta(\log{r}),
    \qquad
    \widetilde{I}_{\beta_2}(\sA\rangle\sB)_\rho = -\Theta(\log{r}),
\end{align}
where the former equals the optimal efficient rate up to $o(1)$. Hence, every order $\beta$ in $(1/2, 1)$ captures the optimal efficient rate for some family but misses it for another, even up to constant factors. We can conclude that no single fixed order characterizes efficient distillation from mixed states. This cannot occur for pure states, where all fixed orders agree up to constant factors by \Cref{prop:renyi min relation}, and is thus specific to mixed states. The negative conditional min-entropy above is the case $\beta_2 = 1/2$.

Finally, we focus on the gap between information-theoretic and efficient distillation. We can tune the spike state to make this gap nearly maximal: at $\beta_\rc - 1/2 = o(1/\log{n})$, the spike state becomes separable. Its R\'{e}nyi coherent information tends to zero at every fixed order $\beta \in [1/2, 1)$, closing the separation between different orders. The efficient rate then becomes $o(1)$, while the entanglement amplifier keeps the information-theoretic distillable entanglement nearly extensive. The rank is again set by the number of labels in Eq.~\eqref{eq:overview labeled mixture}. It can be taken down to $r = 1$, for which we omit the spike and obtain a pure state (see \Cref{Separations for conditional entropies and converse bounds} for details).

\begin{proposition}[Nearly maximal separation; formal version in \Cref{prop:degenerate spike}]
\label{prop:nearly maximal separation}
For every $r$ with $1 \leq r \leq \poly(n)$, there is a family of $\Theta(n)$-qubit states $\{\rho^{\sA\sB}\}_n$ with $\rank(\rho^{\sA\sB}) = r$, such that $E_\rD(\sA:\sB)_{\rho} = \widetilde{\Omega}(n)$, yet, for every constant $\varepsilon < 1$ and $\omega(1) \leq k \leq \poly(n)$, the rate $R$ of distillation from $k$ copies at error $\varepsilon$ in the models specified below is $o(1)$. Here, the bound holds in the state-agnostic and state-dependent models. Under the superpolynomial hardness of LWE with binary secrets (\Cref{assumpt:lwe assumptions}), the family can be chosen to be efficiently preparable, and the bound then holds in the description-aware model.
\end{proposition}

The same construction as in \Cref{prop:separation} thus also reproduces the nearly maximal separation known for pure states~\cite{Leone2025Entanglementtheorywithlimited}, at every rank up to $\poly(n)$.
At superpolynomial rank, \Cref{thm:no go informal} strengthens the worst-case conclusion to zero distilled ebits, while $E_\rD(\sA:\sB)_\rho$ remains $\widetilde{\Omega}(n)$, in line with related separations shown for special classes of states in earlier work~\cite{Bansal2025PseudorandomDensityMatrices,meyer2026computationaltwowayquantumcapacity}.
Hence, at every rank, efficiency can reduce a distillable entanglement of $\widetilde{\Omega}(n)$ to a vanishing rate in the worst case. 
Note that these are limitations exhibited by the constructed families: rank alone does not preclude efficient distillation from an arbitrary mixed state.

\section{Discussion}
\label{sec:discussion}

We have shown that, under rank conditions, entanglement can be distilled from mixed states at a rate determined by the R\'{e}nyi coherent information, without using a description of the state. With polynomially many copies, the Schur-coding protocol~\cite{UQC} attains this rate for polynomial-rank states, capped at any rate that grows logarithmically with the number of qubits, and we have extended this guarantee to states close to a polynomial-rank state. When the nearby state also has polynomial marginal ranks, our decoupling-based protocol attains the rate by a polynomial-size circuit. Conversely, we have constructed states from which every protocol using polynomially many copies in the state-agnostic model distills essentially at most this rate, whatever its circuit size. The Schur-coding protocol is thus nearly optimal in this model.
The same bound holds for efficient protocols tailored to the state and, under hardness assumptions on LWE, a similar one for those efficiently compiled from a description of the preparation circuit. When the R\'{e}nyi coherent information is at most of order $\log{n}$, our decoupling-based protocol attains this bound without any description of the state, and is therefore optimal even against such protocols. Once the rank of these states is superpolynomial, no efficient protocol distills any entanglement, although their information-theoretic distillable entanglement stays almost extensive at every rank. The rank thus marks a sharp transition in the computational distillable entanglement, one that is invisible from an information-theoretic perspective.

Conceptually, our results show that efficient distillation still admits an entropic characterization, with the coherent information, which governs distillation without resource constraints, replaced by its R\'{e}nyi generalization. Two further implications emerge from these results. First, in the worst case, knowledge of the state brings no advantage to efficient distillation: a protocol that uses no description of the state does as well as one tailored to the state or efficiently compiled from a description of its preparation circuit.
The observation that protocols tailored to the state have no advantage over universal ones has been obtained in information-theoretic distillation~\cite{Matsumoto2007universaldistortionfree,Takagi2026Universal,Tirone2026UniversalEntanglementDistillation} and in computational distillation from pure states~\cite{Leone2025Entanglementtheorywithlimited}; our results extend it to computational distillation from mixed states.
Second, mixed states reveal a structure that pure states hide. For pure states, where the R\'{e}nyi coherent information at all fixed orders $\beta \in [1/2, 1)$ agrees up to a constant factor, efficiency turns the min-entropy from a one-shot quantity into the rate of distillation from many copies~\cite{Leone2025Entanglementtheorywithlimited}. For mixed states, this role passes to the R\'{e}nyi coherent information of an intermediate order, which can be positive and attainable even when the conditional min-entropy is negative. Moreover, the relevant order depends on the state, and no single fixed order characterizes the rate for all mixed states. This change in the quantity characterizing the rate appears only beyond pure states.

Our results point to several directions for future work. The most immediate question is whether efficient protocols can achieve a rate determined by the R\'{e}nyi coherent information without the condition on the marginal ranks.
 This condition stems from compressing the local systems to polynomial dimension in our efficient protocol, whereas the Schur-coding protocol attains the rate without it from polynomially many copies. The remaining obstacle is thus purely computational, and it is unclear whether it reflects a genuine barrier or only a limitation of our approach. Removing the condition would extend our results to settings such as communication over quantum channels, where the sender's marginal typically has high rank. Another question concerns the upper bounds for protocols efficiently compiled from a description of the preparation circuit, which rely on the hardness of LWE. Whether similar bounds and parameter ranges can be obtained under weaker assumptions, and which kind of computational hardness is necessary to keep entanglement inaccessible to such protocols, remain open. More broadly, one may ask which structures beyond rank make entanglement efficiently accessible: our converse is a worst-case statement, and a tensor product of many two-qubit Werner states, for instance, has exponentially large rank yet can be distilled efficiently at a constant rate per pair~\cite{Gu2026ConstantOverheadEntanglementDistillation}.


\begin{acknowledgments}
The authors thank Kaito Watanabe for helpful discussions. This work was supported by JST SPRING Grant Number JPMJSP2108, JSPS KAKENHI Grant Nos.\ 24K16975, 25K00924, 26H02015, the Japan Science and Technology Agency (JST) CREST Grant No.\ JPMJCR23I3, NEXUS Grant Number JPMJNX26C2, RIKEN iTHEMS, and RIKEN Pioneering Project `Mathematical foundation of quantum information' (PI Yasuyuki Kawahigashi). J.R. was supported by the BMFTR (QR.N, QuSol, PraktiQOM), the Clusters of Excellence (ML4Q, MATH+), QuantERA (SDPCode), the Munich Quantum Valley, Berlin Quantum, the Quantum Flagship (Millenion, PASQuanS2), the DFG (CRC 183, SPP 2514), and the European Research Council (DebuQC).
\end{acknowledgments}

\bibliographystyle{bibsty}
\bibliography{ref}

\begin{thebibliography}{101}%
\makeatletter
\providecommand \@ifxundefined [1]{%
 \@ifx{#1\undefined}
}%
\providecommand \@ifnum [1]{%
 \ifnum #1\expandafter \@firstoftwo
 \else \expandafter \@secondoftwo
 \fi
}%
\providecommand \@ifx [1]{%
 \ifx #1\expandafter \@firstoftwo
 \else \expandafter \@secondoftwo
 \fi
}%
\providecommand \natexlab [1]{#1}%
\providecommand \emph  [1]{``#1''}%
\providecommand \bibnamefont  [1]{#1}%
\providecommand \bibfnamefont [1]{#1}%
\providecommand \citenamefont [1]{#1}%
\providecommand \href@noop [0]{\@secondoftwo}%
\providecommand \href [0]{\begingroup \@sanitize@url \@href}%
\providecommand \@href[1]{\@@startlink{#1}\@@href}%
\providecommand \@@href[1]{\endgroup#1\@@endlink}%
\providecommand \@sanitize@url [0]{\catcode `\\12\catcode `\$12\catcode `\&12\catcode `\#12\catcode `\^12\catcode `\_12\catcode `\%12\relax}%
\providecommand \@@startlink[1]{}%
\providecommand \@@endlink[0]{}%
\providecommand \url  [0]{\begingroup\@sanitize@url \@url }%
\providecommand \@url [1]{\endgroup\@href {#1}{\urlprefix }}%
\providecommand \urlprefix  [0]{URL }%
\providecommand \Eprint [0]{\href }%
\providecommand \doibase [0]{http://dx.doi.org/}%
\providecommand \bibinfo  [0]{\@secondoftwo}%
\providecommand \bibfield  [0]{\@secondoftwo}%
\providecommand \translation [1]{[#1]}%
\providecommand \BibitemOpen [0]{}%
\providecommand \bibitemStop [0]{}%
\providecommand \bibitemNoStop [0]{.\EOS\space}%
\providecommand \EOS [0]{\spacefactor3000\relax}%
\providecommand \BibitemShut  [1]{\csname bibitem#1\endcsname}%
\let\auto@bib@innerbib\@empty
\bibitem [{\citenamefont {Horodecki}\ \emph {et~al.}(2009)\citenamefont {Horodecki}, \citenamefont {Horodecki}, \citenamefont {Horodecki},\ and\ \citenamefont {Horodecki}}]{Horodecki2009quantumentanglement}%
  \BibitemOpen
  \bibfield  {author} {\bibinfo {author} {\bibfnamefont {R.}~\bibnamefont {Horodecki}}, \bibinfo {author} {\bibfnamefont {P.}~\bibnamefont {Horodecki}}, \bibinfo {author} {\bibfnamefont {M.}~\bibnamefont {Horodecki}}, \ and\ \bibinfo {author} {\bibfnamefont {K.}~\bibnamefont {Horodecki}},\ }\bibfield  {title} {Quantum entanglement,\ }\href {\doibase 10.1103/RevModPhys.81.865} {\bibfield  {journal} {\bibinfo  {journal} {Rev. Mod. Phys.}\ }\textbf {\bibinfo {volume} {81}},\ \bibinfo {pages} {865} (\bibinfo {year} {2009})}\BibitemShut {NoStop}%
\bibitem [{\citenamefont {Bennett}\ \emph {et~al.}(1996{\natexlab{a}})\citenamefont {Bennett}, \citenamefont {Brassard}, \citenamefont {Popescu}, \citenamefont {Schumacher}, \citenamefont {Smolin},\ and\ \citenamefont {Wootters}}]{Bennett1996PurificationNoisyEntanglement}%
  \BibitemOpen
  \bibfield  {author} {\bibinfo {author} {\bibfnamefont {C.~H.}\ \bibnamefont {Bennett}}, \bibinfo {author} {\bibfnamefont {G.}~\bibnamefont {Brassard}}, \bibinfo {author} {\bibfnamefont {S.}~\bibnamefont {Popescu}}, \bibinfo {author} {\bibfnamefont {B.}~\bibnamefont {Schumacher}}, \bibinfo {author} {\bibfnamefont {J.~A.}\ \bibnamefont {Smolin}}, \ and\ \bibinfo {author} {\bibfnamefont {W.~K.}\ \bibnamefont {Wootters}},\ }\bibfield  {title} {Purification of noisy entanglement and faithful teleportation via noisy channels,\ }\href {\doibase 10.1103/PhysRevLett.76.722} {\bibfield  {journal} {\bibinfo  {journal} {Phys. Rev. Lett.}\ }\textbf {\bibinfo {volume} {76}},\ \bibinfo {pages} {722} (\bibinfo {year} {1996}{\natexlab{a}})}\BibitemShut {NoStop}%
\bibitem [{\citenamefont {Bennett}\ \emph {et~al.}(1996{\natexlab{b}})\citenamefont {Bennett}, \citenamefont {DiVincenzo}, \citenamefont {Smolin},\ and\ \citenamefont {Wootters}}]{Bennett1996Mixedstateentanglementandqec}%
  \BibitemOpen
  \bibfield  {author} {\bibinfo {author} {\bibfnamefont {C.~H.}\ \bibnamefont {Bennett}}, \bibinfo {author} {\bibfnamefont {D.~P.}\ \bibnamefont {DiVincenzo}}, \bibinfo {author} {\bibfnamefont {J.~A.}\ \bibnamefont {Smolin}}, \ and\ \bibinfo {author} {\bibfnamefont {W.~K.}\ \bibnamefont {Wootters}},\ }\bibfield  {title} {Mixed-state entanglement and quantum error correction,\ }\href {\doibase 10.1103/PhysRevA.54.3824} {\bibfield  {journal} {\bibinfo  {journal} {Phys. Rev. A}\ }\textbf {\bibinfo {volume} {54}},\ \bibinfo {pages} {3824} (\bibinfo {year} {1996}{\natexlab{b}})}\BibitemShut {NoStop}%
\bibitem [{\citenamefont {Bennett}\ \emph {et~al.}(1993)\citenamefont {Bennett}, \citenamefont {Brassard}, \citenamefont {Cr\'epeau}, \citenamefont {Jozsa}, \citenamefont {Peres},\ and\ \citenamefont {Wootters}}]{bennett1993teleportingunknown}%
  \BibitemOpen
  \bibfield  {author} {\bibinfo {author} {\bibfnamefont {C.~H.}\ \bibnamefont {Bennett}}, \bibinfo {author} {\bibfnamefont {G.}~\bibnamefont {Brassard}}, \bibinfo {author} {\bibfnamefont {C.}~\bibnamefont {Cr\'epeau}}, \bibinfo {author} {\bibfnamefont {R.}~\bibnamefont {Jozsa}}, \bibinfo {author} {\bibfnamefont {A.}~\bibnamefont {Peres}}, \ and\ \bibinfo {author} {\bibfnamefont {W.~K.}\ \bibnamefont {Wootters}},\ }\bibfield  {title} {Teleporting an unknown quantum state via dual classical and {E}instein-{P}odolsky-{R}osen channels,\ }\href {\doibase 10.1103/PhysRevLett.70.1895} {\bibfield  {journal} {\bibinfo  {journal} {Phys. Rev. Lett.}\ }\textbf {\bibinfo {volume} {70}},\ \bibinfo {pages} {1895} (\bibinfo {year} {1993})}\BibitemShut {NoStop}%
\bibitem [{\citenamefont {Briegel}\ \emph {et~al.}(1998)\citenamefont {Briegel}, \citenamefont {D\"ur}, \citenamefont {Cirac},\ and\ \citenamefont {Zoller}}]{Briegel1998QuantumRepeaters}%
  \BibitemOpen
  \bibfield  {author} {\bibinfo {author} {\bibfnamefont {H.-J.}\ \bibnamefont {Briegel}}, \bibinfo {author} {\bibfnamefont {W.}~\bibnamefont {D\"ur}}, \bibinfo {author} {\bibfnamefont {J.~I.}\ \bibnamefont {Cirac}}, \ and\ \bibinfo {author} {\bibfnamefont {P.}~\bibnamefont {Zoller}},\ }\bibfield  {title} {Quantum repeaters: The role of imperfect local operations in quantum communication,\ }\href {\doibase 10.1103/PhysRevLett.81.5932} {\bibfield  {journal} {\bibinfo  {journal} {Phys. Rev. Lett.}\ }\textbf {\bibinfo {volume} {81}},\ \bibinfo {pages} {5932} (\bibinfo {year} {1998})}\BibitemShut {NoStop}%
\bibitem [{\citenamefont {Shor}\ and\ \citenamefont {Preskill}(2000)}]{Shor2000simpleproofsecurity}%
  \BibitemOpen
  \bibfield  {author} {\bibinfo {author} {\bibfnamefont {P.~W.}\ \bibnamefont {Shor}}\ and\ \bibinfo {author} {\bibfnamefont {J.}~\bibnamefont {Preskill}},\ }\bibfield  {title} {Simple proof of security of the {BB84} quantum key distribution protocol,\ }\href {\doibase 10.1103/PhysRevLett.85.441} {\bibfield  {journal} {\bibinfo  {journal} {Phys. Rev. Lett.}\ }\textbf {\bibinfo {volume} {85}},\ \bibinfo {pages} {441} (\bibinfo {year} {2000})}\BibitemShut {NoStop}%
\bibitem [{\citenamefont {Wilde}(2017)}]{wilde2013QItheory}%
  \BibitemOpen
  \bibfield  {author} {\bibinfo {author} {\bibfnamefont {M.~M.}\ \bibnamefont {Wilde}},\ }\emph {\bibinfo {title} {Quantum Information Theory}},\ \bibinfo {edition} {2nd}\ ed.,\ \href {\doibase 10.1017/9781316809976} {Cambridge University Press (\bibinfo {year} {2017})}\BibitemShut {NoStop}%
\bibitem [{\citenamefont {Bennett}\ \emph {et~al.}(1996{\natexlab{c}})\citenamefont {Bennett}, \citenamefont {Bernstein}, \citenamefont {Popescu},\ and\ \citenamefont {Schumacher}}]{BennettConcentratingpartialentanglement}%
  \BibitemOpen
  \bibfield  {author} {\bibinfo {author} {\bibfnamefont {C.~H.}\ \bibnamefont {Bennett}}, \bibinfo {author} {\bibfnamefont {H.~J.}\ \bibnamefont {Bernstein}}, \bibinfo {author} {\bibfnamefont {S.}~\bibnamefont {Popescu}}, \ and\ \bibinfo {author} {\bibfnamefont {B.}~\bibnamefont {Schumacher}},\ }\bibfield  {title} {Concentrating partial entanglement by local operations,\ }\href {\doibase 10.1103/PhysRevA.53.2046} {\bibfield  {journal} {\bibinfo  {journal} {Phys. Rev. A}\ }\textbf {\bibinfo {volume} {53}},\ \bibinfo {pages} {2046} (\bibinfo {year} {1996}{\natexlab{c}})}\BibitemShut {NoStop}%
\bibitem [{\citenamefont {Horodecki}\ \emph {et~al.}(1998)\citenamefont {Horodecki}, \citenamefont {Horodecki},\ and\ \citenamefont {Horodecki}}]{Horodecki1998BoundEntanglementinNature}%
  \BibitemOpen
  \bibfield  {author} {\bibinfo {author} {\bibfnamefont {M.}~\bibnamefont {Horodecki}}, \bibinfo {author} {\bibfnamefont {P.}~\bibnamefont {Horodecki}}, \ and\ \bibinfo {author} {\bibfnamefont {R.}~\bibnamefont {Horodecki}},\ }\bibfield  {title} {Mixed-state entanglement and distillation: Is there a ``bound'' entanglement in nature?,\ }\href {\doibase 10.1103/PhysRevLett.80.5239} {\bibfield  {journal} {\bibinfo  {journal} {Phys. Rev. Lett.}\ }\textbf {\bibinfo {volume} {80}},\ \bibinfo {pages} {5239} (\bibinfo {year} {1998})}\BibitemShut {NoStop}%
\bibitem [{\citenamefont {Horodecki}\ \emph {et~al.}(2000)\citenamefont {Horodecki}, \citenamefont {Horodecki},\ and\ \citenamefont {Horodecki}}]{Horodecki2000UnifiedApproachtoQuantumCapacities}%
  \BibitemOpen
  \bibfield  {author} {\bibinfo {author} {\bibfnamefont {M.}~\bibnamefont {Horodecki}}, \bibinfo {author} {\bibfnamefont {P.}~\bibnamefont {Horodecki}}, \ and\ \bibinfo {author} {\bibfnamefont {R.}~\bibnamefont {Horodecki}},\ }\bibfield  {title} {Unified approach to quantum capacities: Towards quantum noisy coding theorem,\ }\href {\doibase 10.1103/PhysRevLett.85.433} {\bibfield  {journal} {\bibinfo  {journal} {Phys. Rev. Lett.}\ }\textbf {\bibinfo {volume} {85}},\ \bibinfo {pages} {433} (\bibinfo {year} {2000})}\BibitemShut {NoStop}%
\bibitem [{\citenamefont {Devetak}\ and\ \citenamefont {Winter}(2005)}]{devetak2005distillation}%
  \BibitemOpen
  \bibfield  {author} {\bibinfo {author} {\bibfnamefont {I.}~\bibnamefont {Devetak}}\ and\ \bibinfo {author} {\bibfnamefont {A.}~\bibnamefont {Winter}},\ }\bibfield  {title} {Distillation of secret key and entanglement from quantum states,\ }\href {\doibase 10.1098/rspa.2004.1372} {\bibfield  {journal} {\bibinfo  {journal} {Proc. R. Soc. A Math. Phys. Eng. Sci.}\ }\textbf {\bibinfo {volume} {461}},\ \bibinfo {pages} {207} (\bibinfo {year} {2005})}\BibitemShut {NoStop}%
\bibitem [{\citenamefont {Leone}\ \emph {et~al.}(2025)\citenamefont {Leone}, \citenamefont {Rizzo}, \citenamefont {Eisert},\ and\ \citenamefont {Jerbi}}]{Leone2025Entanglementtheorywithlimited}%
  \BibitemOpen
  \bibfield  {author} {\bibinfo {author} {\bibfnamefont {L.}~\bibnamefont {Leone}}, \bibinfo {author} {\bibfnamefont {J.}~\bibnamefont {Rizzo}}, \bibinfo {author} {\bibfnamefont {J.}~\bibnamefont {Eisert}}, \ and\ \bibinfo {author} {\bibfnamefont {S.}~\bibnamefont {Jerbi}},\ }\bibfield  {title} {Entanglement theory with limited computational resources,\ }\href {\doibase 10.1038/s41567-025-03048-8} {\bibfield  {journal} {\bibinfo  {journal} {Nat. Phys.}\ }\textbf {\bibinfo {volume} {21}},\ \bibinfo {pages} {1847} (\bibinfo {year} {2025})},\ \bibinfo {note} {{S}ee also the {Supplementary Information}}\BibitemShut {NoStop}%
\bibitem [{\citenamefont {Meyer}\ \emph {et~al.}(2026{\natexlab{a}})\citenamefont {Meyer}, \citenamefont {Raza}, \citenamefont {Rizzo}, \citenamefont {Leone}, \citenamefont {Jerbi},\ and\ \citenamefont {Eisert}}]{meyer2026computationalrelativeentropy}%
  \BibitemOpen
  \bibfield  {author} {\bibinfo {author} {\bibfnamefont {J.~J.}\ \bibnamefont {Meyer}}, \bibinfo {author} {\bibfnamefont {A.}~\bibnamefont {Raza}}, \bibinfo {author} {\bibfnamefont {J.}~\bibnamefont {Rizzo}}, \bibinfo {author} {\bibfnamefont {L.}~\bibnamefont {Leone}}, \bibinfo {author} {\bibfnamefont {S.}~\bibnamefont {Jerbi}}, \ and\ \bibinfo {author} {\bibfnamefont {J.}~\bibnamefont {Eisert}},\ }\bibfield  {title} {Computational relative entropy,\ }\href {https://arxiv.org/abs/2509.20472} {\bibfield  {journal} {\bibinfo  {journal} {arXiv:2509.20472}\ } (\bibinfo {year} {2026}{\natexlab{a}})}\BibitemShut {NoStop}%
\bibitem [{\citenamefont {Aaronson}\ \emph {et~al.}(2024)\citenamefont {Aaronson}, \citenamefont {Bouland}, \citenamefont {Fefferman}, \citenamefont {Ghosh}, \citenamefont {Vazirani}, \citenamefont {Zhang},\ and\ \citenamefont {Zhou}}]{aaronson2023quantumpseudoentanglement}%
  \BibitemOpen
  \bibfield  {author} {\bibinfo {author} {\bibfnamefont {S.}~\bibnamefont {Aaronson}}, \bibinfo {author} {\bibfnamefont {A.}~\bibnamefont {Bouland}}, \bibinfo {author} {\bibfnamefont {B.}~\bibnamefont {Fefferman}}, \bibinfo {author} {\bibfnamefont {S.}~\bibnamefont {Ghosh}}, \bibinfo {author} {\bibfnamefont {U.}~\bibnamefont {Vazirani}}, \bibinfo {author} {\bibfnamefont {C.}~\bibnamefont {Zhang}}, \ and\ \bibinfo {author} {\bibfnamefont {Z.}~\bibnamefont {Zhou}},\ }\bibfield  {title} {Quantum pseudoentanglement,\ }\href {\doibase 10.4230/LIPIcs.ITCS.2024.2} {in\ \emph {\bibinfo {booktitle} {Proc. 15th {ITCS}}},\ \bibinfo {series} {LIPIcs}, Vol.\ \bibinfo {volume} {287} (\bibinfo {year} {2024}),\ pp.\ \bibinfo {pages} {2:1--2:21}}\BibitemShut {NoStop}%
\bibitem [{\citenamefont {Ji}\ \emph {et~al.}(2018)\citenamefont {Ji}, \citenamefont {Liu},\ and\ \citenamefont {Song}}]{Ji2018PseudorandomQuantumStates}%
  \BibitemOpen
  \bibfield  {author} {\bibinfo {author} {\bibfnamefont {Z.}~\bibnamefont {Ji}}, \bibinfo {author} {\bibfnamefont {Y.-K.}\ \bibnamefont {Liu}}, \ and\ \bibinfo {author} {\bibfnamefont {F.}~\bibnamefont {Song}},\ }\bibfield  {title} {Pseudorandom quantum states,\ }\href {\doibase 10.1007/978-3-319-96878-0_5} {in\ \emph {\bibinfo {booktitle} {Adv. Cryptol. -- CRYPTO 2018}},\ \bibinfo {series} {Lect. Notes Comput. Sci.}, Vol.\ \bibinfo {volume} {10993} (\bibinfo {year} {2018}),\ pp.\ \bibinfo {pages} {126--152}}\BibitemShut {NoStop}%
\bibitem [{\citenamefont {Arnon-Friedman}\ \emph {et~al.}(2023)\citenamefont {Arnon-Friedman}, \citenamefont {Brakerski},\ and\ \citenamefont {Vidick}}]{arnonfriedman2023computationalentanglementtheory}%
  \BibitemOpen
  \bibfield  {author} {\bibinfo {author} {\bibfnamefont {R.}~\bibnamefont {Arnon-Friedman}}, \bibinfo {author} {\bibfnamefont {Z.}~\bibnamefont {Brakerski}}, \ and\ \bibinfo {author} {\bibfnamefont {T.}~\bibnamefont {Vidick}},\ }\bibfield  {title} {Computational entanglement theory,\ }\href {https://arxiv.org/abs/2310.02783} {\bibfield  {journal} {\bibinfo  {journal} {arXiv:2310.02783}\ } (\bibinfo {year} {2023})}\BibitemShut {NoStop}%
\bibitem [{\citenamefont {Avidan}\ \emph {et~al.}(2026)\citenamefont {Avidan}, \citenamefont {Hahn}, \citenamefont {Renes},\ and\ \citenamefont {Arnon}}]{Avidan2026FullyQuantumComputationalEntropies}%
  \BibitemOpen
  \bibfield  {author} {\bibinfo {author} {\bibfnamefont {N.}~\bibnamefont {Avidan}}, \bibinfo {author} {\bibfnamefont {T.~A.}\ \bibnamefont {Hahn}}, \bibinfo {author} {\bibfnamefont {J.~M.}\ \bibnamefont {Renes}}, \ and\ \bibinfo {author} {\bibfnamefont {R.}~\bibnamefont {Arnon}},\ }\bibfield  {title} {Fully quantum computational entropies,\ }\href {\doibase 10.4230/LIPICS.ITCS.2026.13} {in\ \emph {\bibinfo {booktitle} {Proc. 17th {ITCS}}},\ \bibinfo {series} {LIPIcs}, Vol.\ \bibinfo {volume} {362} (\bibinfo {year} {2026}),\ pp.\ \bibinfo {pages} {13:1--13:3}}\BibitemShut {NoStop}%
\bibitem [{\citenamefont {Y{\'a}ng{\"u}ez}\ \emph {et~al.}(2026)\citenamefont {Y{\'a}ng{\"u}ez}, \citenamefont {Avidan}, \citenamefont {Kochanowski},\ and\ \citenamefont {Hahn}}]{yanguez2026accessiblequantumcorrelationscomplexity}%
  \BibitemOpen
  \bibfield  {author} {\bibinfo {author} {\bibfnamefont {{\'A}.}~\bibnamefont {Y{\'a}ng{\"u}ez}}, \bibinfo {author} {\bibfnamefont {N.}~\bibnamefont {Avidan}}, \bibinfo {author} {\bibfnamefont {J.}~\bibnamefont {Kochanowski}}, \ and\ \bibinfo {author} {\bibfnamefont {T.~A.}\ \bibnamefont {Hahn}},\ }\bibfield  {title} {Accessible quantum correlations under complexity constraints,\ }\href {https://arxiv.org/abs/2604.15540} {\bibfield  {journal} {\bibinfo  {journal} {arXiv:2604.15540}\ } (\bibinfo {year} {2026})}\BibitemShut {NoStop}%
\bibitem [{\citenamefont {Matsumoto}\ and\ \citenamefont {Hayashi}(2007)}]{Matsumoto2007universaldistortionfree}%
  \BibitemOpen
  \bibfield  {author} {\bibinfo {author} {\bibfnamefont {K.}~\bibnamefont {Matsumoto}}\ and\ \bibinfo {author} {\bibfnamefont {M.}~\bibnamefont {Hayashi}},\ }\bibfield  {title} {Universal distortion-free entanglement concentration,\ }\href {\doibase 10.1103/PhysRevA.75.062338} {\bibfield  {journal} {\bibinfo  {journal} {Phys. Rev. A}\ }\textbf {\bibinfo {volume} {75}},\ \bibinfo {pages} {062338} (\bibinfo {year} {2007})}\BibitemShut {NoStop}%
\bibitem [{\citenamefont {Hayden}\ \emph {et~al.}(2008)\citenamefont {Hayden}, \citenamefont {Horodecki}, \citenamefont {Winter},\ and\ \citenamefont {Yard}}]{hayden2008decoupling}%
  \BibitemOpen
  \bibfield  {author} {\bibinfo {author} {\bibfnamefont {P.}~\bibnamefont {Hayden}}, \bibinfo {author} {\bibfnamefont {M.}~\bibnamefont {Horodecki}}, \bibinfo {author} {\bibfnamefont {A.}~\bibnamefont {Winter}}, \ and\ \bibinfo {author} {\bibfnamefont {J.}~\bibnamefont {Yard}},\ }\bibfield  {title} {A decoupling approach to the quantum capacity,\ }\href {\doibase 10.1142/S1230161208000043} {\bibfield  {journal} {\bibinfo  {journal} {Open Syst. Inf. Dyn.}\ }\textbf {\bibinfo {volume} {15}},\ \bibinfo {pages} {7} (\bibinfo {year} {2008})}\BibitemShut {NoStop}%
\bibitem [{\citenamefont {Dupuis}(2009)}]{dupuis2010decoupling}%
  \BibitemOpen
  \bibfield  {author} {\bibinfo {author} {\bibfnamefont {F.}~\bibnamefont {Dupuis}},\ }\emph {\bibinfo {title} {The Decoupling Approach to Quantum Information Theory}},\ \href {\doibase 10.71781/10720} {PhD thesis,\ \bibinfo  {school} {University of Montreal} (\bibinfo {year} {2009})}\BibitemShut {NoStop}%
\bibitem [{\citenamefont {Dupuis}\ \emph {et~al.}(2014)\citenamefont {Dupuis}, \citenamefont {Berta}, \citenamefont {Wullschleger},\ and\ \citenamefont {Renner}}]{dupuis2014one}%
  \BibitemOpen
  \bibfield  {author} {\bibinfo {author} {\bibfnamefont {F.}~\bibnamefont {Dupuis}}, \bibinfo {author} {\bibfnamefont {M.}~\bibnamefont {Berta}}, \bibinfo {author} {\bibfnamefont {J.}~\bibnamefont {Wullschleger}}, \ and\ \bibinfo {author} {\bibfnamefont {R.}~\bibnamefont {Renner}},\ }\bibfield  {title} {One-shot decoupling,\ }\href {\doibase 10.1007/s00220-014-1990-4} {\bibfield  {journal} {\bibinfo  {journal} {Commun. Math. Phys.}\ }\textbf {\bibinfo {volume} {328}},\ \bibinfo {pages} {251} (\bibinfo {year} {2014})}\BibitemShut {NoStop}%
\bibitem [{\citenamefont {Uhlmann}(1976)}]{uhlmann1976transition}%
  \BibitemOpen
  \bibfield  {author} {\bibinfo {author} {\bibfnamefont {A.}~\bibnamefont {Uhlmann}},\ }\bibfield  {title} {The ``transition probability'' in the state space of a $*$-algebra,\ }\href {\doibase 10.1016/0034-4877(76)90060-4} {\bibfield  {journal} {\bibinfo  {journal} {Rep. Math. Phys.}\ }\textbf {\bibinfo {volume} {9}},\ \bibinfo {pages} {273} (\bibinfo {year} {1976})}\BibitemShut {NoStop}%
\bibitem [{\citenamefont {Bostanci}\ \emph {et~al.}(2026)\citenamefont {Bostanci}, \citenamefont {Efron}, \citenamefont {Metger}, \citenamefont {Poremba}, \citenamefont {Qian},\ and\ \citenamefont {Yuen}}]{bostanci2023unicompuhlmann}%
  \BibitemOpen
  \bibfield  {author} {\bibinfo {author} {\bibfnamefont {J.}~\bibnamefont {Bostanci}}, \bibinfo {author} {\bibfnamefont {Y.}~\bibnamefont {Efron}}, \bibinfo {author} {\bibfnamefont {T.}~\bibnamefont {Metger}}, \bibinfo {author} {\bibfnamefont {A.}~\bibnamefont {Poremba}}, \bibinfo {author} {\bibfnamefont {L.}~\bibnamefont {Qian}}, \ and\ \bibinfo {author} {\bibfnamefont {H.}~\bibnamefont {Yuen}},\ }\bibfield  {title} {Unitary complexity and the {U}hlmann transformation problem,\ }\href {\doibase 10.4230/LIPIcs.ITCS.2026.24} {in\ \emph {\bibinfo {booktitle} {Proc. 17th {ITCS}}},\ \bibinfo {series} {LIPIcs}, Vol.\ \bibinfo {volume} {362} (\bibinfo {year} {2026}),\ pp.\ \bibinfo {pages} {24:1--24:17}}\BibitemShut {NoStop}%
\bibitem [{\citenamefont {Bansal}\ \emph {et~al.}(2025)\citenamefont {Bansal}, \citenamefont {Mok}, \citenamefont {Bharti}, \citenamefont {Koh},\ and\ \citenamefont {Haug}}]{Bansal2025PseudorandomDensityMatrices}%
  \BibitemOpen
  \bibfield  {author} {\bibinfo {author} {\bibfnamefont {N.}~\bibnamefont {Bansal}}, \bibinfo {author} {\bibfnamefont {W.-K.}\ \bibnamefont {Mok}}, \bibinfo {author} {\bibfnamefont {K.}~\bibnamefont {Bharti}}, \bibinfo {author} {\bibfnamefont {D.~E.}\ \bibnamefont {Koh}}, \ and\ \bibinfo {author} {\bibfnamefont {T.}~\bibnamefont {Haug}},\ }\bibfield  {title} {Pseudorandom density matrices,\ }\href {\doibase 10.1103/PRXQuantum.6.020322} {\bibfield  {journal} {\bibinfo  {journal} {PRX Quantum}\ }\textbf {\bibinfo {volume} {6}},\ \bibinfo {pages} {020322} (\bibinfo {year} {2025})}\BibitemShut {NoStop}%
\bibitem [{\citenamefont {Meyer}\ \emph {et~al.}(2026{\natexlab{b}})\citenamefont {Meyer}, \citenamefont {Rizzo}, \citenamefont {Raza}, \citenamefont {Leone}, \citenamefont {Jerbi},\ and\ \citenamefont {Eisert}}]{meyer2026computationaltwowayquantumcapacity}%
  \BibitemOpen
  \bibfield  {author} {\bibinfo {author} {\bibfnamefont {J.~J.}\ \bibnamefont {Meyer}}, \bibinfo {author} {\bibfnamefont {J.}~\bibnamefont {Rizzo}}, \bibinfo {author} {\bibfnamefont {A.}~\bibnamefont {Raza}}, \bibinfo {author} {\bibfnamefont {L.}~\bibnamefont {Leone}}, \bibinfo {author} {\bibfnamefont {S.}~\bibnamefont {Jerbi}}, \ and\ \bibinfo {author} {\bibfnamefont {J.}~\bibnamefont {Eisert}},\ }\bibfield  {title} {The computational two-way quantum capacity,\ }\href {https://arxiv.org/abs/2601.15393} {\bibfield  {journal} {\bibinfo  {journal} {arXiv:2601.15393}\ } (\bibinfo {year} {2026}{\natexlab{b}})}\BibitemShut {NoStop}%
\bibitem [{\citenamefont {M{\"u}ller-Lennert}\ \emph {et~al.}(2013)\citenamefont {M{\"u}ller-Lennert}, \citenamefont {Dupuis}, \citenamefont {Szehr}, \citenamefont {Fehr},\ and\ \citenamefont {Tomamichel}}]{muller2013quantumrenyinew}%
  \BibitemOpen
  \bibfield  {author} {\bibinfo {author} {\bibfnamefont {M.}~\bibnamefont {M{\"u}ller-Lennert}}, \bibinfo {author} {\bibfnamefont {F.}~\bibnamefont {Dupuis}}, \bibinfo {author} {\bibfnamefont {O.}~\bibnamefont {Szehr}}, \bibinfo {author} {\bibfnamefont {S.}~\bibnamefont {Fehr}}, \ and\ \bibinfo {author} {\bibfnamefont {M.}~\bibnamefont {Tomamichel}},\ }\bibfield  {title} {On quantum {R}{\'e}nyi entropies: a new generalization and some properties,\ }\href {https://doi.org/10.1063/1.4838856} {\bibfield  {journal} {\bibinfo  {journal} {J. Math. Phys.}\ }\textbf {\bibinfo {volume} {54}},\ \bibinfo {pages} {122203} (\bibinfo {year} {2013})}\BibitemShut {NoStop}%
\bibitem [{\citenamefont {Khatri}\ and\ \citenamefont {Wilde}(2024)}]{khatri2024principlemodern}%
  \BibitemOpen
  \bibfield  {author} {\bibinfo {author} {\bibfnamefont {S.}~\bibnamefont {Khatri}}\ and\ \bibinfo {author} {\bibfnamefont {M.~M.}\ \bibnamefont {Wilde}},\ }\bibfield  {title} {Principles of quantum communication theory: a modern approach,\ }\href {https://doi.org/10.48550/arXiv.2011.04672} {\bibfield  {journal} {\bibinfo  {journal} {arXiv:2011.04672}\ } (\bibinfo {year} {2024})}\BibitemShut {NoStop}%
\bibitem [{\citenamefont {Rizzo}\ \emph {et~al.}(2026)\citenamefont {Rizzo}, \citenamefont {Lami}, \citenamefont {Eisert},\ and\ \citenamefont {Leone}}]{UQC}%
  \BibitemOpen
  \bibfield  {author} {\bibinfo {author} {\bibfnamefont {J.}~\bibnamefont {Rizzo}}, \bibinfo {author} {\bibfnamefont {L.}~\bibnamefont {Lami}}, \bibinfo {author} {\bibfnamefont {J.}~\bibnamefont {Eisert}}, \ and\ \bibinfo {author} {\bibfnamefont {L.}~\bibnamefont {Leone}},\ }\bibfield  {title} {Universal quantum coding,\ }\href {https://arxiv.org/abs/2609.38038v1} {\bibfield  {journal} {\bibinfo  {journal} {arXiv:2609.38038}\ } (\bibinfo {year} {2026})},\ \bibinfo {note} {version 1}\BibitemShut {NoStop}%
\bibitem [{\citenamefont {Takagi}\ \emph {et~al.}(2026)\citenamefont {Takagi}, \citenamefont {Watanabe}, \citenamefont {Matsuura}, \citenamefont {Arai},\ and\ \citenamefont {Hayashi}}]{Takagi2026Universal}%
  \BibitemOpen
  \bibfield  {author} {\bibinfo {author} {\bibfnamefont {R.}~\bibnamefont {Takagi}}, \bibinfo {author} {\bibfnamefont {K.}~\bibnamefont {Watanabe}}, \bibinfo {author} {\bibfnamefont {T.}~\bibnamefont {Matsuura}}, \bibinfo {author} {\bibfnamefont {H.}~\bibnamefont {Arai}}, \ and\ \bibinfo {author} {\bibfnamefont {M.}~\bibnamefont {Hayashi}},\ }\bibfield  {title} {Universal distillation of quantum entanglement,\ }\href {https://arxiv.org/abs/2609.40215} {\bibfield  {journal} {\bibinfo  {journal} {arXiv:2609.40215}\ } (\bibinfo {year} {2026})}\BibitemShut {NoStop}%
\bibitem [{\citenamefont {Tirone}\ \emph {et~al.}(2026)\citenamefont {Tirone}, \citenamefont {Mele}, \citenamefont {Giovannetti},\ and\ \citenamefont {Lami}}]{Tirone2026UniversalEntanglementDistillation}%
  \BibitemOpen
  \bibfield  {author} {\bibinfo {author} {\bibfnamefont {S.}~\bibnamefont {Tirone}}, \bibinfo {author} {\bibfnamefont {F.~A.}\ \bibnamefont {Mele}}, \bibinfo {author} {\bibfnamefont {V.}~\bibnamefont {Giovannetti}}, \ and\ \bibinfo {author} {\bibfnamefont {L.}~\bibnamefont {Lami}},\ }\bibfield  {title} {Universal entanglement distillation,\ }\href {https://arxiv.org/abs/2609.40080} {\bibfield  {journal} {\bibinfo  {journal} {arXiv:2609.40080}\ } (\bibinfo {year} {2026})}\BibitemShut {NoStop}%
\bibitem [{\citenamefont {Metger}\ and\ \citenamefont {Yuen}(2023)}]{metger2023stateqipspace}%
  \BibitemOpen
  \bibfield  {author} {\bibinfo {author} {\bibfnamefont {T.}~\bibnamefont {Metger}}\ and\ \bibinfo {author} {\bibfnamefont {H.}~\bibnamefont {Yuen}},\ }\bibfield  {title} {state{QIP} = state{PSPACE},\ }\href {\doibase 10.1109/FOCS57990.2023.00082} {in\ \emph {\bibinfo {booktitle} {Proc. 64th {FOCS}}} (\bibinfo {year} {2023}),\ pp.\ \bibinfo {pages} {1349--1356}}\BibitemShut {NoStop}%
\bibitem [{\citenamefont {Utsumi}\ \emph {et~al.}(2025)\citenamefont {Utsumi}, \citenamefont {Nakata}, \citenamefont {Wang},\ and\ \citenamefont {Takagi}}]{utsumi2025algorithmsUhlmanntrans}%
  \BibitemOpen
  \bibfield  {author} {\bibinfo {author} {\bibfnamefont {T.}~\bibnamefont {Utsumi}}, \bibinfo {author} {\bibfnamefont {Y.}~\bibnamefont {Nakata}}, \bibinfo {author} {\bibfnamefont {Q.}~\bibnamefont {Wang}}, \ and\ \bibinfo {author} {\bibfnamefont {R.}~\bibnamefont {Takagi}},\ }\bibfield  {title} {Quantum algorithms for {U}hlmann transformation,\ }\href {https://arxiv.org/abs/2509.03619} {\bibfield  {journal} {\bibinfo  {journal} {arXiv:2509.03619}\ } (\bibinfo {year} {2025})}\BibitemShut {NoStop}%
\bibitem [{\citenamefont {Bouland}\ \emph {et~al.}(2024{\natexlab{a}})\citenamefont {Bouland}, \citenamefont {Fefferman}, \citenamefont {Ghosh}, \citenamefont {Metger}, \citenamefont {Vazirani}, \citenamefont {Zhang},\ and\ \citenamefont {Zhou}}]{bouland2024PublicKeyPseudoentanglement}%
  \BibitemOpen
  \bibfield  {author} {\bibinfo {author} {\bibfnamefont {A.}~\bibnamefont {Bouland}}, \bibinfo {author} {\bibfnamefont {B.}~\bibnamefont {Fefferman}}, \bibinfo {author} {\bibfnamefont {S.}~\bibnamefont {Ghosh}}, \bibinfo {author} {\bibfnamefont {T.}~\bibnamefont {Metger}}, \bibinfo {author} {\bibfnamefont {U.}~\bibnamefont {Vazirani}}, \bibinfo {author} {\bibfnamefont {C.}~\bibnamefont {Zhang}}, \ and\ \bibinfo {author} {\bibfnamefont {Z.}~\bibnamefont {Zhou}},\ }\bibfield  {title} {Public-key pseudoentanglement and the hardness of learning ground state entanglement structure,\ }\href {\doibase 10.4230/LIPIcs.CCC.2024.21} {in\ \emph {\bibinfo {booktitle} {Proc. 39th {CCC}}},\ \bibinfo {series} {LIPIcs}, Vol.\ \bibinfo {volume} {300} (\bibinfo {year} {2024}{\natexlab{a}}),\ pp.\ \bibinfo {pages} {21:1--21:23}},\ \bibinfo {note} {{S}ee also the full version: \href{https://arxiv.org/abs/2311.12017v1}{arXiv:2311.12017v1}}\BibitemShut {NoStop}%
\bibitem [{\citenamefont {Regev}(2009)}]{regev2009Onlatticeslearningwitherrors}%
  \BibitemOpen
  \bibfield  {author} {\bibinfo {author} {\bibfnamefont {O.}~\bibnamefont {Regev}},\ }\bibfield  {title} {On lattices, learning with errors, random linear codes, and cryptography,\ }\href {\doibase 10.1145/1568318.1568324} {\bibfield  {journal} {\bibinfo  {journal} {J. ACM}\ }\textbf {\bibinfo {volume} {56}},\ \bibinfo {pages} {34} (\bibinfo {year} {2009})}\BibitemShut {NoStop}%
\bibitem [{\citenamefont {Gu}\ \emph {et~al.}(2026)\citenamefont {Gu}, \citenamefont {Leone}, \citenamefont {Goodenough},\ and\ \citenamefont {Khatri}}]{Gu2026ConstantOverheadEntanglementDistillation}%
  \BibitemOpen
  \bibfield  {author} {\bibinfo {author} {\bibfnamefont {A.}~\bibnamefont {Gu}}, \bibinfo {author} {\bibfnamefont {L.}~\bibnamefont {Leone}}, \bibinfo {author} {\bibfnamefont {K.}~\bibnamefont {Goodenough}}, \ and\ \bibinfo {author} {\bibfnamefont {S.}~\bibnamefont {Khatri}},\ }\bibfield  {title} {Constant overhead entanglement distillation via scrambling,\ }\href {\doibase 10.1103/3q4z-llv8} {\bibfield  {journal} {\bibinfo  {journal} {Phys. Rev. Lett.}\ }\textbf {\bibinfo {volume} {136}},\ \bibinfo {pages} {110805} (\bibinfo {year} {2026})}\BibitemShut {NoStop}%
\bibitem [{\citenamefont {Goul{\~a}o}\ and\ \citenamefont {Elkouss}(2024)}]{goulao2024pseudoentanglementnecessaryefipairs}%
  \BibitemOpen
  \bibfield  {author} {\bibinfo {author} {\bibfnamefont {M.}~\bibnamefont {Goul{\~a}o}}\ and\ \bibinfo {author} {\bibfnamefont {D.}~\bibnamefont {Elkouss}},\ }\bibfield  {title} {Pseudo-entanglement is necessary for {EFI} pairs,\ }\href {https://arxiv.org/abs/2406.06881} {\bibfield  {journal} {\bibinfo  {journal} {arXiv:2406.06881}\ } (\bibinfo {year} {2024})}\BibitemShut {NoStop}%
\bibitem [{\citenamefont {Akers}\ \emph {et~al.}(2024)\citenamefont {Akers}, \citenamefont {Bouland}, \citenamefont {Chen}, \citenamefont {Kohler}, \citenamefont {Metger},\ and\ \citenamefont {Vazirani}}]{akers2024holographicpseudoentanglementcomplexityadscft}%
  \BibitemOpen
  \bibfield  {author} {\bibinfo {author} {\bibfnamefont {C.}~\bibnamefont {Akers}}, \bibinfo {author} {\bibfnamefont {A.}~\bibnamefont {Bouland}}, \bibinfo {author} {\bibfnamefont {L.}~\bibnamefont {Chen}}, \bibinfo {author} {\bibfnamefont {T.}~\bibnamefont {Kohler}}, \bibinfo {author} {\bibfnamefont {T.}~\bibnamefont {Metger}}, \ and\ \bibinfo {author} {\bibfnamefont {U.}~\bibnamefont {Vazirani}},\ }\bibfield  {title} {Holographic pseudoentanglement and the complexity of the {AdS/CFT} dictionary,\ }\href {https://arxiv.org/abs/2411.04978} {\bibfield  {journal} {\bibinfo  {journal} {arXiv:2411.04978}\ } (\bibinfo {year} {2024})}\BibitemShut {NoStop}%
\bibitem [{\citenamefont {Engelhardt}\ \emph {et~al.}(2026)\citenamefont {Engelhardt}, \citenamefont {Folkestad}, \citenamefont {Levine}, \citenamefont {Verheijden},\ and\ \citenamefont {Yang}}]{Engelhardt2026SpoofingEntanglementHolography}%
  \BibitemOpen
  \bibfield  {author} {\bibinfo {author} {\bibfnamefont {N.}~\bibnamefont {Engelhardt}}, \bibinfo {author} {\bibfnamefont {{\AA}.}~\bibnamefont {Folkestad}}, \bibinfo {author} {\bibfnamefont {A.}~\bibnamefont {Levine}}, \bibinfo {author} {\bibfnamefont {E.}~\bibnamefont {Verheijden}}, \ and\ \bibinfo {author} {\bibfnamefont {L.}~\bibnamefont {Yang}},\ }\bibfield  {title} {Spoofing entanglement in holography,\ }\href {\doibase 10.1007/JHEP05(2026)107} {\bibfield  {journal} {\bibinfo  {journal} {J. High Energy Phys.}\ }\textbf {\bibinfo {volume} {05}},\ \bibinfo {pages} {107} (\bibinfo {year} {2026})}\BibitemShut {NoStop}%
\bibitem [{\citenamefont {Bouland}\ \emph {et~al.}(2024{\natexlab{b}})\citenamefont {Bouland}, \citenamefont {Zhang},\ and\ \citenamefont {Zhou}}]{bouland2024hardnesslearninggroundstate}%
  \BibitemOpen
  \bibfield  {author} {\bibinfo {author} {\bibfnamefont {A.}~\bibnamefont {Bouland}}, \bibinfo {author} {\bibfnamefont {C.}~\bibnamefont {Zhang}}, \ and\ \bibinfo {author} {\bibfnamefont {Z.}~\bibnamefont {Zhou}},\ }\bibfield  {title} {On the hardness of learning ground state entanglement of geometrically local {H}amiltonians,\ }\href {https://arxiv.org/abs/2411.04353} {\bibfield  {journal} {\bibinfo  {journal} {arXiv:2411.04353}\ } (\bibinfo {year} {2024}{\natexlab{b}})}\BibitemShut {NoStop}%
\bibitem [{\citenamefont {Kitaev}(1997)}]{Kitaev1997QCAlgoEC}%
  \BibitemOpen
  \bibfield  {author} {\bibinfo {author} {\bibfnamefont {A.}~\bibnamefont {Kitaev}},\ }\bibfield  {title} {Quantum computations: algorithms and error correction,\ }\href {\doibase 10.1070/RM1997v052n06ABEH002155} {\bibfield  {journal} {\bibinfo  {journal} {Russian Math. Surveys}\ }\textbf {\bibinfo {volume} {52}},\ \bibinfo {pages} {1191} (\bibinfo {year} {1997})}\BibitemShut {NoStop}%
\bibitem [{\citenamefont {Watrous}(2005)}]{watrous2004notessuperoperatornorm}%
  \BibitemOpen
  \bibfield  {author} {\bibinfo {author} {\bibfnamefont {J.}~\bibnamefont {Watrous}},\ }\bibfield  {title} {Notes on super-operator norms induced by {S}chatten norms,\ }\href {https://dl.acm.org/doi/10.5555/2011608.2011614} {\bibfield  {journal} {\bibinfo  {journal} {Quantum Inf. Comput.}\ }\textbf {\bibinfo {volume} {5}},\ \bibinfo {pages} {58} (\bibinfo {year} {2005})}\BibitemShut {NoStop}%
\bibitem [{\citenamefont {Watrous}(2018)}]{watrous2018TheoryQI}%
  \BibitemOpen
  \bibfield  {author} {\bibinfo {author} {\bibfnamefont {J.}~\bibnamefont {Watrous}},\ }\emph {\bibinfo {title} {The Theory of Quantum Information}},\ \href {\doibase 10.1017/9781316848142} {Cambridge University Press (\bibinfo {year} {2018})}\BibitemShut {NoStop}%
\bibitem [{\citenamefont {Tomamichel}\ \emph {et~al.}(2010)\citenamefont {Tomamichel}, \citenamefont {Colbeck},\ and\ \citenamefont {Renner}}]{tomamichel2010duality}%
  \BibitemOpen
  \bibfield  {author} {\bibinfo {author} {\bibfnamefont {M.}~\bibnamefont {Tomamichel}}, \bibinfo {author} {\bibfnamefont {R.}~\bibnamefont {Colbeck}}, \ and\ \bibinfo {author} {\bibfnamefont {R.}~\bibnamefont {Renner}},\ }\bibfield  {title} {Duality between smooth min- and max-entropies,\ }\href {\doibase 10.1109/TIT.2010.2054130} {\bibfield  {journal} {\bibinfo  {journal} {IEEE Trans. Inf. Theory}\ }\textbf {\bibinfo {volume} {56}},\ \bibinfo {pages} {4674} (\bibinfo {year} {2010})}\BibitemShut {NoStop}%
\bibitem [{\citenamefont {Fuchs}\ and\ \citenamefont {van~de Graaf}(1999)}]{Fuchs1999fuchsvandegraaf}%
  \BibitemOpen
  \bibfield  {author} {\bibinfo {author} {\bibfnamefont {C.~A.}\ \bibnamefont {Fuchs}}\ and\ \bibinfo {author} {\bibfnamefont {J.}~\bibnamefont {van~de Graaf}},\ }\bibfield  {title} {Cryptographic distinguishability measures for quantum-mechanical states,\ }\href {\doibase 10.1109/18.761271} {\bibfield  {journal} {\bibinfo  {journal} {IEEE Trans. Inf. Theory}\ }\textbf {\bibinfo {volume} {45}},\ \bibinfo {pages} {1216} (\bibinfo {year} {1999})}\BibitemShut {NoStop}%
\bibitem [{\citenamefont {Horn}\ and\ \citenamefont {Johnson}(1991)}]{Horn1991TopicsinMatrixAnalysis}%
  \BibitemOpen
  \bibfield  {author} {\bibinfo {author} {\bibfnamefont {R.~A.}\ \bibnamefont {Horn}}\ and\ \bibinfo {author} {\bibfnamefont {C.~R.}\ \bibnamefont {Johnson}},\ }\emph {\bibinfo {title} {Topics in Matrix Analysis}},\ \href {\doibase 10.1017/CBO9780511840371} {Cambridge University Press (\bibinfo {year} {1991})}\BibitemShut {NoStop}%
\bibitem [{\citenamefont {Gilchrist}\ \emph {et~al.}(2011)\citenamefont {Gilchrist}, \citenamefont {Terno},\ and\ \citenamefont {Wood}}]{gilchrist2011vectorizationquantumsuse}%
  \BibitemOpen
  \bibfield  {author} {\bibinfo {author} {\bibfnamefont {A.}~\bibnamefont {Gilchrist}}, \bibinfo {author} {\bibfnamefont {D.~R.}\ \bibnamefont {Terno}}, \ and\ \bibinfo {author} {\bibfnamefont {C.~J.}\ \bibnamefont {Wood}},\ }\bibfield  {title} {Vectorization of quantum operations and its use,\ }\href {https://arxiv.org/abs/0911.2539} {\bibfield  {journal} {\bibinfo  {journal} {arXiv:0911.2539}\ } (\bibinfo {year} {2011})}\BibitemShut {NoStop}%
\bibitem [{\citenamefont {Miszczak}(2011)}]{Miszczak2011SINGULARVALUEDECOMPOSITION}%
  \BibitemOpen
  \bibfield  {author} {\bibinfo {author} {\bibfnamefont {J.~A.}\ \bibnamefont {Miszczak}},\ }\bibfield  {title} {Singular value decomposition and matrix reorderings in quantum information theory,\ }\href {\doibase 10.1142/s0129183111016683} {\bibfield  {journal} {\bibinfo  {journal} {Int. J. Mod. Phys. C}\ }\textbf {\bibinfo {volume} {22}},\ \bibinfo {pages} {897} (\bibinfo {year} {2011})}\BibitemShut {NoStop}%
\bibitem [{\citenamefont {Winter}(2004)}]{Winter2004ExtrinsicIntrinsic}%
  \BibitemOpen
  \bibfield  {author} {\bibinfo {author} {\bibfnamefont {A.}~\bibnamefont {Winter}},\ }\bibfield  {title} {``{E}xtrinsic'' and ``intrinsic'' data in quantum measurements: Asymptotic convex decomposition of positive operator valued measures,\ }\href {\doibase 10.1007/s00220-003-0989-z} {\bibfield  {journal} {\bibinfo  {journal} {Commun. Math. Phys.}\ }\textbf {\bibinfo {volume} {244}},\ \bibinfo {pages} {157} (\bibinfo {year} {2004})}\BibitemShut {NoStop}%
\bibitem [{\citenamefont {Moore}\ \emph {et~al.}(2009)\citenamefont {Moore}, \citenamefont {Kearfott},\ and\ \citenamefont {Cloud}}]{Moore2009IntroductionIntervalAnalysis}%
  \BibitemOpen
  \bibfield  {author} {\bibinfo {author} {\bibfnamefont {R.~E.}\ \bibnamefont {Moore}}, \bibinfo {author} {\bibfnamefont {R.~B.}\ \bibnamefont {Kearfott}}, \ and\ \bibinfo {author} {\bibfnamefont {M.~J.}\ \bibnamefont {Cloud}},\ }\emph {\bibinfo {title} {Introduction to Interval Analysis}},\ \href {\doibase 10.1137/1.9780898717716} {Society for Industrial and Applied Mathematics (\bibinfo {year} {2009})}\BibitemShut {NoStop}%
\bibitem [{\citenamefont {Dawson}\ and\ \citenamefont {Nielsen}(2006)}]{Dawson2006SolovayKitaevalgorithm}%
  \BibitemOpen
  \bibfield  {author} {\bibinfo {author} {\bibfnamefont {C.~M.}\ \bibnamefont {Dawson}}\ and\ \bibinfo {author} {\bibfnamefont {M.~A.}\ \bibnamefont {Nielsen}},\ }\bibfield  {title} {The {S}olovay--{K}itaev algorithm,\ }\href {https://dl.acm.org/doi/10.5555/2011679.2011685} {\bibfield  {journal} {\bibinfo  {journal} {Quantum Inf. Comput.}\ }\textbf {\bibinfo {volume} {6}},\ \bibinfo {pages} {81} (\bibinfo {year} {2006})}\BibitemShut {NoStop}%
\bibitem [{\citenamefont {Shoup}(2009)}]{Shoup2012Computationalintroductionnumbertheoryalgebra}%
  \BibitemOpen
  \bibfield  {author} {\bibinfo {author} {\bibfnamefont {V.}~\bibnamefont {Shoup}},\ }\emph {\bibinfo {title} {A Computational Introduction to Number Theory and Algebra}},\ \bibinfo {edition} {2nd}\ ed.,\ \href {\doibase 10.1017/CBO9780511814549} {Cambridge University Press (\bibinfo {year} {2009})}\BibitemShut {NoStop}%
\bibitem [{\citenamefont {Nielsen}\ and\ \citenamefont {Chuang}(2010)}]{nielsen2010quantum}%
  \BibitemOpen
  \bibfield  {author} {\bibinfo {author} {\bibfnamefont {M.~A.}\ \bibnamefont {Nielsen}}\ and\ \bibinfo {author} {\bibfnamefont {I.~L.}\ \bibnamefont {Chuang}},\ }\emph {\bibinfo {title} {Quantum Computation and Quantum Information}},\ \href {\doibase 10.1017/CBO9780511976667} {Cambridge University Press (\bibinfo {year} {2010})}\BibitemShut {NoStop}%
\bibitem [{\citenamefont {Barenco}\ \emph {et~al.}(1995)\citenamefont {Barenco}, \citenamefont {Bennett}, \citenamefont {Cleve}, \citenamefont {DiVincenzo}, \citenamefont {Margolus}, \citenamefont {Shor}, \citenamefont {Sleator}, \citenamefont {Smolin},\ and\ \citenamefont {Weinfurter}}]{Barenco1995Elementarygatesquantumcomputation}%
  \BibitemOpen
  \bibfield  {author} {\bibinfo {author} {\bibfnamefont {A.}~\bibnamefont {Barenco}}, \bibinfo {author} {\bibfnamefont {C.~H.}\ \bibnamefont {Bennett}}, \bibinfo {author} {\bibfnamefont {R.}~\bibnamefont {Cleve}}, \bibinfo {author} {\bibfnamefont {D.~P.}\ \bibnamefont {DiVincenzo}}, \bibinfo {author} {\bibfnamefont {N.}~\bibnamefont {Margolus}}, \bibinfo {author} {\bibfnamefont {P.}~\bibnamefont {Shor}}, \bibinfo {author} {\bibfnamefont {T.}~\bibnamefont {Sleator}}, \bibinfo {author} {\bibfnamefont {J.~A.}\ \bibnamefont {Smolin}}, \ and\ \bibinfo {author} {\bibfnamefont {H.}~\bibnamefont {Weinfurter}},\ }\bibfield  {title} {Elementary gates for quantum computation,\ }\href {\doibase 10.1103/PhysRevA.52.3457} {\bibfield  {journal} {\bibinfo  {journal} {Phys. Rev. A}\ }\textbf {\bibinfo {volume} {52}},\ \bibinfo {pages} {3457} (\bibinfo {year} {1995})}\BibitemShut {NoStop}%
\bibitem [{\citenamefont {Mezzadri}(2007)}]{Mezzadri2007Generaterandommatrices}%
  \BibitemOpen
  \bibfield  {author} {\bibinfo {author} {\bibfnamefont {F.}~\bibnamefont {Mezzadri}},\ }\bibfield  {title} {How to generate random matrices from the classical compact groups,\ }\href {https://www.ams.org/notices/200705/fea-mezzadri-web.pdf} {\bibfield  {journal} {\bibinfo  {journal} {Notices Am. Math. Soc.}\ }\textbf {\bibinfo {volume} {54}},\ \bibinfo {pages} {592} (\bibinfo {year} {2007})}\BibitemShut {NoStop}%
\bibitem [{\citenamefont {Brent}\ and\ \citenamefont {Zimmermann}(2010)}]{Brent2010Moderncomputerarithmetic}%
  \BibitemOpen
  \bibfield  {author} {\bibinfo {author} {\bibfnamefont {R.~P.}\ \bibnamefont {Brent}}\ and\ \bibinfo {author} {\bibfnamefont {P.}~\bibnamefont {Zimmermann}},\ }\emph {\bibinfo {title} {Modern Computer Arithmetic}},\ \bibinfo {series} {Cambridge Monographs on Applied and Computational Mathematics}, Vol.~\bibinfo {volume} {18},\ \href {\doibase 10.1017/CBO9780511921698} {Cambridge University Press (\bibinfo {year} {2010})}\BibitemShut {NoStop}%
\bibitem [{\citenamefont {Goodman}\ and\ \citenamefont {Wallach}(1998)}]{goodman1998representations}%
  \BibitemOpen
  \bibfield  {author} {\bibinfo {author} {\bibfnamefont {R.}~\bibnamefont {Goodman}}\ and\ \bibinfo {author} {\bibfnamefont {N.~R.}\ \bibnamefont {Wallach}},\ }\emph {\bibinfo {title} {Representations and Invariants of the Classical Groups}},\ \bibinfo {series} {Encyclopedia of Mathematics and its Applications}, Vol.~\bibinfo {volume} {68},\ \href {\doibase 10.1007/978-0-387-79852-3} {Cambridge University Press (\bibinfo {year} {1998})},\ \bibinfo {note} {revised edition: \emph{Symmetry, Representations, and Invariants}, Graduate Texts in Mathematics, Vol.~255, Springer (2009)}\BibitemShut {NoStop}%
\bibitem [{\citenamefont {Harrow}(2005)}]{harrow2005applications}%
  \BibitemOpen
  \bibfield  {author} {\bibinfo {author} {\bibfnamefont {A.~W.}\ \bibnamefont {Harrow}},\ }\emph {\bibinfo {title} {Applications of Coherent Classical Communication and the {S}chur Transform to Quantum Information Theory}},\ \href {http://hdl.handle.net/1721.1/34973} {PhD thesis,\ \bibinfo  {school} {Massachusetts Institute of Technology} (\bibinfo {year} {2005})}\BibitemShut {NoStop}%
\bibitem [{\citenamefont {Knill}\ and\ \citenamefont {Laflamme}(1996)}]{knill1996concatenatedquantumcodes}%
  \BibitemOpen
  \bibfield  {author} {\bibinfo {author} {\bibfnamefont {E.}~\bibnamefont {Knill}}\ and\ \bibinfo {author} {\bibfnamefont {R.}~\bibnamefont {Laflamme}},\ }\bibfield  {title} {Concatenated quantum codes,\ }\href {https://arxiv.org/abs/quant-ph/9608012} {\bibfield  {journal} {\bibinfo  {journal} {arXiv:quant-ph/9608012}\ } (\bibinfo {year} {1996})}\BibitemShut {NoStop}%
\bibitem [{\citenamefont {Gottesman}(1996)}]{Gottesman1996Classofquantumerror}%
  \BibitemOpen
  \bibfield  {author} {\bibinfo {author} {\bibfnamefont {D.}~\bibnamefont {Gottesman}},\ }\bibfield  {title} {Class of quantum error-correcting codes saturating the quantum {H}amming bound,\ }\href {\doibase 10.1103/PhysRevA.54.1862} {\bibfield  {journal} {\bibinfo  {journal} {Phys. Rev. A}\ }\textbf {\bibinfo {volume} {54}},\ \bibinfo {pages} {1862} (\bibinfo {year} {1996})}\BibitemShut {NoStop}%
\bibitem [{\citenamefont {Matsumoto}(2003)}]{Matsumoto2003Conversionquantumstabilizercodeentanglementdistillation}%
  \BibitemOpen
  \bibfield  {author} {\bibinfo {author} {\bibfnamefont {R.}~\bibnamefont {Matsumoto}},\ }\bibfield  {title} {Conversion of a general quantum stabilizer code to an entanglement distillation protocol,\ }\href {\doibase 10.1088/0305-4470/36/29/316} {\bibfield  {journal} {\bibinfo  {journal} {J. Phys. A: Math. Gen.}\ }\textbf {\bibinfo {volume} {36}},\ \bibinfo {pages} {8113} (\bibinfo {year} {2003})}\BibitemShut {NoStop}%
\bibitem [{\citenamefont {Gottesman}(1997)}]{gottesman1997stabilizercodesquantumerror}%
  \BibitemOpen
  \bibfield  {author} {\bibinfo {author} {\bibfnamefont {D.}~\bibnamefont {Gottesman}},\ }\bibfield  {title} {Stabilizer codes and quantum error correction,\ }\href {https://arxiv.org/abs/quant-ph/9705052} {\bibfield  {journal} {\bibinfo  {journal} {arXiv:quant-ph/9705052}\ } (\bibinfo {year} {1997})}\BibitemShut {NoStop}%
\bibitem [{\citenamefont {Koenig}\ and\ \citenamefont {Smolin}(2014)}]{Koenig2014EfficientlyselectarbitraryClifford}%
  \BibitemOpen
  \bibfield  {author} {\bibinfo {author} {\bibfnamefont {R.}~\bibnamefont {Koenig}}\ and\ \bibinfo {author} {\bibfnamefont {J.~A.}\ \bibnamefont {Smolin}},\ }\bibfield  {title} {How to efficiently select an arbitrary {C}lifford group element,\ }\href {\doibase 10.1063/1.4903507} {\bibfield  {journal} {\bibinfo  {journal} {J. Math. Phys.}\ }\textbf {\bibinfo {volume} {55}},\ \bibinfo {pages} {122202} (\bibinfo {year} {2014})}\BibitemShut {NoStop}%
\bibitem [{\citenamefont {von~zur Gathen}\ and\ \citenamefont {Gerhard}(2013)}]{vonzurGathen2013Moderncomputeralgebra}%
  \BibitemOpen
  \bibfield  {author} {\bibinfo {author} {\bibfnamefont {J.}~\bibnamefont {von~zur Gathen}}\ and\ \bibinfo {author} {\bibfnamefont {J.}~\bibnamefont {Gerhard}},\ }\emph {\bibinfo {title} {Modern Computer Algebra}},\ \bibinfo {edition} {3rd}\ ed.,\ \href {\doibase 10.1017/CBO9781139856065} {Cambridge University Press (\bibinfo {year} {2013})}\BibitemShut {NoStop}%
\bibitem [{\citenamefont {Guruswami}\ \emph {et~al.}(2025)\citenamefont {Guruswami}, \citenamefont {Rudra},\ and\ \citenamefont {Sudan}}]{Guruswami2025Essentialcodingtheory}%
  \BibitemOpen
  \bibfield  {author} {\bibinfo {author} {\bibfnamefont {V.}~\bibnamefont {Guruswami}}, \bibinfo {author} {\bibfnamefont {A.}~\bibnamefont {Rudra}}, \ and\ \bibinfo {author} {\bibfnamefont {M.}~\bibnamefont {Sudan}},\ }\emph {\bibinfo {title} {Essential Coding Theory}},\ \href {https://cse.buffalo.edu/faculty/atri/courses/coding-theory/book/web-coding-book.pdf} {\bibinfo  {school} {University at Buffalo} (\bibinfo {year} {2025})},\ \bibinfo {note} {draft of {A}ugust 26, 2025}\BibitemShut {NoStop}%
\bibitem [{\citenamefont {Marwah}\ and\ \citenamefont {Dupuis}(2022)}]{Marwah2022UniformContinuitysandwichedrenyi}%
  \BibitemOpen
  \bibfield  {author} {\bibinfo {author} {\bibfnamefont {A.}~\bibnamefont {Marwah}}\ and\ \bibinfo {author} {\bibfnamefont {F.}~\bibnamefont {Dupuis}},\ }\bibfield  {title} {Uniform continuity bound for sandwiched {R}\'{e}nyi conditional entropy,\ }\href {\doibase 10.1063/5.0088507} {\bibfield  {journal} {\bibinfo  {journal} {J. Math. Phys.}\ }\textbf {\bibinfo {volume} {63}},\ \bibinfo {pages} {052201} (\bibinfo {year} {2022})}\BibitemShut {NoStop}%
\bibitem [{\citenamefont {Jozsa}\ \emph {et~al.}(1998)\citenamefont {Jozsa}, \citenamefont {Horodecki}, \citenamefont {Horodecki},\ and\ \citenamefont {Horodecki}}]{Jozsa1998UniversalQuantumInformationCompression}%
  \BibitemOpen
  \bibfield  {author} {\bibinfo {author} {\bibfnamefont {R.}~\bibnamefont {Jozsa}}, \bibinfo {author} {\bibfnamefont {M.}~\bibnamefont {Horodecki}}, \bibinfo {author} {\bibfnamefont {P.}~\bibnamefont {Horodecki}}, \ and\ \bibinfo {author} {\bibfnamefont {R.}~\bibnamefont {Horodecki}},\ }\bibfield  {title} {Universal quantum information compression,\ }\href {\doibase 10.1103/PhysRevLett.81.1714} {\bibfield  {journal} {\bibinfo  {journal} {Phys. Rev. Lett.}\ }\textbf {\bibinfo {volume} {81}},\ \bibinfo {pages} {1714} (\bibinfo {year} {1998})}\BibitemShut {NoStop}%
\bibitem [{\citenamefont {Hayashi}\ and\ \citenamefont {Matsumoto}(2002)}]{hayashi2002universalvariablelengthsourcecoding}%
  \BibitemOpen
  \bibfield  {author} {\bibinfo {author} {\bibfnamefont {M.}~\bibnamefont {Hayashi}}\ and\ \bibinfo {author} {\bibfnamefont {K.}~\bibnamefont {Matsumoto}},\ }\bibfield  {title} {Quantum universal variable-length source coding,\ }\href {\doibase 10.1103/PhysRevA.66.022311} {\bibfield  {journal} {\bibinfo  {journal} {Phys. Rev. A}\ }\textbf {\bibinfo {volume} {66}},\ \bibinfo {pages} {022311} (\bibinfo {year} {2002})}\BibitemShut {NoStop}%
\bibitem [{\citenamefont {Bennett}\ \emph {et~al.}(2006)\citenamefont {Bennett}, \citenamefont {Harrow},\ and\ \citenamefont {Lloyd}}]{Bennett2006Universalquantumdatacompressionnondestructive}%
  \BibitemOpen
  \bibfield  {author} {\bibinfo {author} {\bibfnamefont {C.~H.}\ \bibnamefont {Bennett}}, \bibinfo {author} {\bibfnamefont {A.~W.}\ \bibnamefont {Harrow}}, \ and\ \bibinfo {author} {\bibfnamefont {S.}~\bibnamefont {Lloyd}},\ }\bibfield  {title} {Universal quantum data compression via nondestructive tomography,\ }\href {\doibase 10.1103/PhysRevA.73.032336} {\bibfield  {journal} {\bibinfo  {journal} {Phys. Rev. A}\ }\textbf {\bibinfo {volume} {73}},\ \bibinfo {pages} {032336} (\bibinfo {year} {2006})}\BibitemShut {NoStop}%
\bibitem [{\citenamefont {Lowe}\ and\ \citenamefont {Tan}(2026)}]{lowe2026randomdimensionreductionlearning}%
  \BibitemOpen
  \bibfield  {author} {\bibinfo {author} {\bibfnamefont {A.}~\bibnamefont {Lowe}}\ and\ \bibinfo {author} {\bibfnamefont {X.}~\bibnamefont {Tan}},\ }\bibfield  {title} {Random dimension reduction and learning symmetric properties of quantum states,\ }\href {https://arxiv.org/abs/2606.23592} {\bibfield  {journal} {\bibinfo  {journal} {arXiv:2606.23592}\ } (\bibinfo {year} {2026})}\BibitemShut {NoStop}%
\bibitem [{\citenamefont {Burchardt}\ \emph {et~al.}(2025)\citenamefont {Burchardt}, \citenamefont {Fei}, \citenamefont {Grinko}, \citenamefont {Larocca}, \citenamefont {Ozols}, \citenamefont {Timmerman},\ and\ \citenamefont {Visnevskyi}}]{burchardt2025highdimensionalschur}%
  \BibitemOpen
  \bibfield  {author} {\bibinfo {author} {\bibfnamefont {A.}~\bibnamefont {Burchardt}}, \bibinfo {author} {\bibfnamefont {J.}~\bibnamefont {Fei}}, \bibinfo {author} {\bibfnamefont {D.}~\bibnamefont {Grinko}}, \bibinfo {author} {\bibfnamefont {M.}~\bibnamefont {Larocca}}, \bibinfo {author} {\bibfnamefont {M.}~\bibnamefont {Ozols}}, \bibinfo {author} {\bibfnamefont {S.}~\bibnamefont {Timmerman}}, \ and\ \bibinfo {author} {\bibfnamefont {V.}~\bibnamefont {Visnevskyi}},\ }\bibfield  {title} {High-dimensional quantum {S}chur transforms,\ }\href {https://arxiv.org/abs/2509.22640} {\bibfield  {journal} {\bibinfo  {journal} {arXiv:2509.22640}\ } (\bibinfo {year} {2025})}\BibitemShut {NoStop}%
\bibitem [{\citenamefont {Tomamichel}\ \emph {et~al.}(2014)\citenamefont {Tomamichel}, \citenamefont {Berta},\ and\ \citenamefont {Hayashi}}]{Tomamichel2014Relatingdifferentquantumgeneralizations}%
  \BibitemOpen
  \bibfield  {author} {\bibinfo {author} {\bibfnamefont {M.}~\bibnamefont {Tomamichel}}, \bibinfo {author} {\bibfnamefont {M.}~\bibnamefont {Berta}}, \ and\ \bibinfo {author} {\bibfnamefont {M.}~\bibnamefont {Hayashi}},\ }\bibfield  {title} {Relating different quantum generalizations of the conditional {R}{\'e}nyi entropy,\ }\href {\doibase 10.1063/1.4892761} {\bibfield  {journal} {\bibinfo  {journal} {J. Math. Phys.}\ }\textbf {\bibinfo {volume} {55}},\ \bibinfo {pages} {082206} (\bibinfo {year} {2014})}\BibitemShut {NoStop}%
\bibitem [{\citenamefont {Dinh}\ \emph {et~al.}(2019)\citenamefont {Dinh}, \citenamefont {Ho}, \citenamefont {Le},\ and\ \citenamefont {Vo}}]{Dinh2019Twotraceinequalities}%
  \BibitemOpen
  \bibfield  {author} {\bibinfo {author} {\bibfnamefont {T.~H.}\ \bibnamefont {Dinh}}, \bibinfo {author} {\bibfnamefont {M.~T.}\ \bibnamefont {Ho}}, \bibinfo {author} {\bibfnamefont {C.~T.}\ \bibnamefont {Le}}, \ and\ \bibinfo {author} {\bibfnamefont {B.~K.}\ \bibnamefont {Vo}},\ }\bibfield  {title} {Two trace inequalities for operator functions,\ }\href {\doibase 10.7153/mia-2019-22-69} {\bibfield  {journal} {\bibinfo  {journal} {Math. Inequal. Appl.}\ }\textbf {\bibinfo {volume} {22}},\ \bibinfo {pages} {1021} (\bibinfo {year} {2019})}\BibitemShut {NoStop}%
\bibitem [{\citenamefont {Cleve}\ \emph {et~al.}(2016)\citenamefont {Cleve}, \citenamefont {Leung}, \citenamefont {Liu},\ and\ \citenamefont {Wang}}]{Cleve2016Nearlinearconstructionsexactunitary2designs}%
  \BibitemOpen
  \bibfield  {author} {\bibinfo {author} {\bibfnamefont {R.}~\bibnamefont {Cleve}}, \bibinfo {author} {\bibfnamefont {D.}~\bibnamefont {Leung}}, \bibinfo {author} {\bibfnamefont {L.}~\bibnamefont {Liu}}, \ and\ \bibinfo {author} {\bibfnamefont {C.}~\bibnamefont {Wang}},\ }\bibfield  {title} {Near-linear constructions of exact unitary 2-designs,\ }\href {\doibase 10.26421/QIC16.9-10-1} {\bibfield  {journal} {\bibinfo  {journal} {Quantum Inf. Comput.}\ }\textbf {\bibinfo {volume} {16}},\ \bibinfo {pages} {721} (\bibinfo {year} {2016})}\BibitemShut {NoStop}%
\bibitem [{\citenamefont {Beigi}(2013)}]{Beigi2013Sandwicheddataprocessing}%
  \BibitemOpen
  \bibfield  {author} {\bibinfo {author} {\bibfnamefont {S.}~\bibnamefont {Beigi}},\ }\bibfield  {title} {Sandwiched {R}{\'e}nyi divergence satisfies data processing inequality,\ }\href {\doibase 10.1063/1.4838855} {\bibfield  {journal} {\bibinfo  {journal} {J. Math. Phys.}\ }\textbf {\bibinfo {volume} {54}},\ \bibinfo {pages} {122202} (\bibinfo {year} {2013})}\BibitemShut {NoStop}%
\bibitem [{\citenamefont {Powers}\ and\ \citenamefont {St{\o}rmer}(1970)}]{powers1970free}%
  \BibitemOpen
  \bibfield  {author} {\bibinfo {author} {\bibfnamefont {R.~T.}\ \bibnamefont {Powers}}\ and\ \bibinfo {author} {\bibfnamefont {E.}~\bibnamefont {St{\o}rmer}},\ }\bibfield  {title} {Free states of the canonical anticommutation relations,\ }\href {\doibase 10.1007/BF01645492} {\bibfield  {journal} {\bibinfo  {journal} {Commun. Math. Phys.}\ }\textbf {\bibinfo {volume} {16}},\ \bibinfo {pages} {1} (\bibinfo {year} {1970})}\BibitemShut {NoStop}%
\bibitem [{\citenamefont {Higham}\ \emph {et~al.}(2004)\citenamefont {Higham}, \citenamefont {Mackey}, \citenamefont {Mackey},\ and\ \citenamefont {Tisseur}}]{Higham2004Computingpolardecompositionmatrixsign}%
  \BibitemOpen
  \bibfield  {author} {\bibinfo {author} {\bibfnamefont {N.~J.}\ \bibnamefont {Higham}}, \bibinfo {author} {\bibfnamefont {D.~S.}\ \bibnamefont {Mackey}}, \bibinfo {author} {\bibfnamefont {N.}~\bibnamefont {Mackey}}, \ and\ \bibinfo {author} {\bibfnamefont {F.}~\bibnamefont {Tisseur}},\ }\bibfield  {title} {Computing the polar decomposition and the matrix sign decomposition in matrix groups,\ }\href {\doibase 10.1137/S0895479803426644} {\bibfield  {journal} {\bibinfo  {journal} {SIAM J. Matrix Anal. Appl.}\ }\textbf {\bibinfo {volume} {25}},\ \bibinfo {pages} {1178} (\bibinfo {year} {2004})}\BibitemShut {NoStop}%
\bibitem [{\citenamefont {Lin}(2022)}]{Lin2022LectureNotes}%
  \BibitemOpen
  \bibfield  {author} {\bibinfo {author} {\bibfnamefont {L.}~\bibnamefont {Lin}},\ }\emph {\bibinfo {title} {Lecture Notes on Quantum Algorithms for Scientific Computation}},\ \href {https://math.berkeley.edu/~linlin/qasc/} {\bibinfo {howpublished} {Lecture notes},\ \bibinfo  {school} {University of California, Berkeley} (\bibinfo {year} {2022})}\BibitemShut {NoStop}%
\bibitem [{\citenamefont {Hiai}\ and\ \citenamefont {Petz}(2014)}]{Hiai2014Introductionmatrixanalysisapplications}%
  \BibitemOpen
  \bibfield  {author} {\bibinfo {author} {\bibfnamefont {F.}~\bibnamefont {Hiai}}\ and\ \bibinfo {author} {\bibfnamefont {D.}~\bibnamefont {Petz}},\ }\emph {\bibinfo {title} {Introduction to Matrix Analysis and Applications}},\ Universitext,\ \href {\doibase 10.1007/978-3-319-04150-6} {Springer (\bibinfo {year} {2014})}\BibitemShut {NoStop}%
\bibitem [{\citenamefont {Gily\'{e}n}\ \emph {et~al.}(2019)\citenamefont {Gily\'{e}n}, \citenamefont {Su}, \citenamefont {Low},\ and\ \citenamefont {Wiebe}}]{gilyen2019qsvt}%
  \BibitemOpen
  \bibfield  {author} {\bibinfo {author} {\bibfnamefont {A.}~\bibnamefont {Gily\'{e}n}}, \bibinfo {author} {\bibfnamefont {Y.}~\bibnamefont {Su}}, \bibinfo {author} {\bibfnamefont {G.~H.}\ \bibnamefont {Low}}, \ and\ \bibinfo {author} {\bibfnamefont {N.}~\bibnamefont {Wiebe}},\ }\bibfield  {title} {Quantum singular value transformation and beyond: exponential improvements for quantum matrix arithmetics,\ }\href {\doibase 10.1145/3313276.3316366} {in\ \emph {\bibinfo {booktitle} {Proc. 51st {STOC}}} (\bibinfo {year} {2019}),\ pp.\ \bibinfo {pages} {193--204}}\BibitemShut {NoStop}%
\bibitem [{\citenamefont {Golub}\ and\ \citenamefont {Van~Loan}(2013)}]{Golub2013Matrixcomputations}%
  \BibitemOpen
  \bibfield  {author} {\bibinfo {author} {\bibfnamefont {G.~H.}\ \bibnamefont {Golub}}\ and\ \bibinfo {author} {\bibfnamefont {C.~F.}\ \bibnamefont {Van~Loan}},\ }\emph {\bibinfo {title} {Matrix Computations}},\ \bibinfo {edition} {4th}\ ed.,\ \href {\doibase 10.1137/1.9781421407944} {Johns Hopkins University Press (\bibinfo {year} {2013})}\BibitemShut {NoStop}%
\bibitem [{\citenamefont {Ok{\v{s}}a}\ \emph {et~al.}(2017)\citenamefont {Ok{\v{s}}a}, \citenamefont {Yamamoto},\ and\ \citenamefont {Vajter{\v{s}}ic}}]{Oksa2017AsymptoticquadraticconvergenceblockJacobi}%
  \BibitemOpen
  \bibfield  {author} {\bibinfo {author} {\bibfnamefont {G.}~\bibnamefont {Ok{\v{s}}a}}, \bibinfo {author} {\bibfnamefont {Y.}~\bibnamefont {Yamamoto}}, \ and\ \bibinfo {author} {\bibfnamefont {M.}~\bibnamefont {Vajter{\v{s}}ic}},\ }\bibfield  {title} {Asymptotic quadratic convergence of the serial block-{Jacobi} {EVD} algorithm for {Hermitian} matrices,\ }\href {\doibase 10.1007/s00211-016-0863-5} {\bibfield  {journal} {\bibinfo  {journal} {Numer. Math.}\ }\textbf {\bibinfo {volume} {136}},\ \bibinfo {pages} {1071} (\bibinfo {year} {2017})}\BibitemShut {NoStop}%
\bibitem [{\citenamefont {Guţă}\ \emph {et~al.}(2020)\citenamefont {Guţă}, \citenamefont {Kahn}, \citenamefont {Kueng},\ and\ \citenamefont {Tropp}}]{Guta2020Faststatetomography}%
  \BibitemOpen
  \bibfield  {author} {\bibinfo {author} {\bibfnamefont {M.}~\bibnamefont {Guţă}}, \bibinfo {author} {\bibfnamefont {J.}~\bibnamefont {Kahn}}, \bibinfo {author} {\bibfnamefont {R.}~\bibnamefont {Kueng}}, \ and\ \bibinfo {author} {\bibfnamefont {J.~A.}\ \bibnamefont {Tropp}},\ }\bibfield  {title} {Fast state tomography with optimal error bounds,\ }\href {\doibase 10.1088/1751-8121/ab8111} {\bibfield  {journal} {\bibinfo  {journal} {J. Phys. A: Math. Theor.}\ }\textbf {\bibinfo {volume} {53}},\ \bibinfo {pages} {204001} (\bibinfo {year} {2020})}\BibitemShut {NoStop}%
\bibitem [{\citenamefont {Smolin}\ \emph {et~al.}(2012)\citenamefont {Smolin}, \citenamefont {Gambetta},\ and\ \citenamefont {Smith}}]{Smolin2012Efficientmethodcomputingmaximumlikelihood}%
  \BibitemOpen
  \bibfield  {author} {\bibinfo {author} {\bibfnamefont {J.~A.}\ \bibnamefont {Smolin}}, \bibinfo {author} {\bibfnamefont {J.~M.}\ \bibnamefont {Gambetta}}, \ and\ \bibinfo {author} {\bibfnamefont {G.}~\bibnamefont {Smith}},\ }\bibfield  {title} {Efficient method for computing the maximum-likelihood quantum state from measurements with additive {G}aussian noise,\ }\href {\doibase 10.1103/PhysRevLett.108.070502} {\bibfield  {journal} {\bibinfo  {journal} {Phys. Rev. Lett.}\ }\textbf {\bibinfo {volume} {108}},\ \bibinfo {pages} {070502} (\bibinfo {year} {2012})}\BibitemShut {NoStop}%
\bibitem [{\citenamefont {Wang}\ and\ \citenamefont {Carreira-Perpi{\~n}{\'a}n}(2013)}]{wang2013projectionprobabilitysimplexefficient}%
  \BibitemOpen
  \bibfield  {author} {\bibinfo {author} {\bibfnamefont {W.}~\bibnamefont {Wang}}\ and\ \bibinfo {author} {\bibfnamefont {M.~{\'A}.}\ \bibnamefont {Carreira-Perpi{\~n}{\'a}n}},\ }\bibfield  {title} {Projection onto the probability simplex: An efficient algorithm with a simple proof, and an application,\ }\href {https://arxiv.org/abs/1309.1541} {\bibfield  {journal} {\bibinfo  {journal} {arXiv:1309.1541}\ } (\bibinfo {year} {2013})}\BibitemShut {NoStop}%
\bibitem [{\citenamefont {Tomamichel}(2016)}]{Tomamichel2016QIPwithFiniteResources}%
  \BibitemOpen
  \bibfield  {author} {\bibinfo {author} {\bibfnamefont {M.}~\bibnamefont {Tomamichel}},\ }\emph {\bibinfo {title} {Quantum Information Processing with Finite Resources: Mathematical Foundations}},\ SpringerBriefs in Mathematical Physics,\ \href {\doibase 10.1007/978-3-319-21891-5} {Springer (\bibinfo {year} {2016})}\BibitemShut {NoStop}%
\bibitem [{\citenamefont {Lubkin}\ and\ \citenamefont {Lubkin}(1993)}]{Lubkin1993AverageQuantalBehavior}%
  \BibitemOpen
  \bibfield  {author} {\bibinfo {author} {\bibfnamefont {E.}~\bibnamefont {Lubkin}}\ and\ \bibinfo {author} {\bibfnamefont {T.}~\bibnamefont {Lubkin}},\ }\bibfield  {title} {Average quantal behavior and thermodynamic isolation,\ }\href {\doibase 10.1007/BF01215300} {\bibfield  {journal} {\bibinfo  {journal} {Int. J. Theor. Phys.}\ }\textbf {\bibinfo {volume} {32}},\ \bibinfo {pages} {933} (\bibinfo {year} {1993})}\BibitemShut {NoStop}%
\bibitem [{\citenamefont {Mele}(2024)}]{Mele2024IntroductionHaarMeasure}%
  \BibitemOpen
  \bibfield  {author} {\bibinfo {author} {\bibfnamefont {A.~A.}\ \bibnamefont {Mele}},\ }\bibfield  {title} {Introduction to {H}aar measure tools in quantum information: A beginner's tutorial,\ }\href {\doibase 10.22331/q-2024-05-08-1340} {\bibfield  {journal} {\bibinfo  {journal} {Quantum}\ }\textbf {\bibinfo {volume} {8}},\ \bibinfo {pages} {1340} (\bibinfo {year} {2024})}\BibitemShut {NoStop}%
\bibitem [{\citenamefont {Tang}\ \emph {et~al.}(2025)\citenamefont {Tang}, \citenamefont {Wright},\ and\ \citenamefont {Zhandry}}]{tang2025conjugatequerieshelp}%
  \BibitemOpen
  \bibfield  {author} {\bibinfo {author} {\bibfnamefont {E.}~\bibnamefont {Tang}}, \bibinfo {author} {\bibfnamefont {J.}~\bibnamefont {Wright}}, \ and\ \bibinfo {author} {\bibfnamefont {M.}~\bibnamefont {Zhandry}},\ }\bibfield  {title} {Conjugate queries can help,\ }\href {https://arxiv.org/abs/2510.07622} {\bibfield  {journal} {\bibinfo  {journal} {arXiv:2510.07622}\ } (\bibinfo {year} {2025})}\BibitemShut {NoStop}%
\bibitem [{\citenamefont {L{\'e}vy}\ and\ \citenamefont {Pellegrino}(1951)}]{levy1951problemes}%
  \BibitemOpen
  \bibfield  {author} {\bibinfo {author} {\bibfnamefont {P.}~\bibnamefont {L{\'e}vy}}\ and\ \bibinfo {author} {\bibfnamefont {F.}~\bibnamefont {Pellegrino}},\ }\emph {\bibinfo {title} {Probl{\`e}mes concrets d'analyse fonctionnelle}},\ \bibinfo {series} {Collection de monographies sur la th{\'e}orie des fonctions}, Vol.~\bibinfo {volume} {36},\ \href {https://books.google.co.jp/books?id=rzAPAAAAIAAJ} {Gauthier-Villars (\bibinfo {year} {1951})},\ p.~\bibinfo {pages} {68}\BibitemShut {NoStop}%
\bibitem [{\citenamefont {Ledoux}(2001)}]{ledoux2001concentration}%
  \BibitemOpen
  \bibfield  {author} {\bibinfo {author} {\bibfnamefont {M.}~\bibnamefont {Ledoux}},\ }\emph {\bibinfo {title} {The Concentration of Measure Phenomenon}},\ \bibinfo {series} {Mathematical Surveys and Monographs}, Vol.~\bibinfo {volume} {89},\ \href {\doibase 10.1090/surv/089} {AMS (\bibinfo {year} {2001})}\BibitemShut {NoStop}%
\bibitem [{\citenamefont {Meckes}(2019)}]{Meckes2019RandomMatrixTheoryClassicalCompact}%
  \BibitemOpen
  \bibfield  {author} {\bibinfo {author} {\bibfnamefont {E.~S.}\ \bibnamefont {Meckes}},\ }\emph {\bibinfo {title} {The Random Matrix Theory of the Classical Compact Groups}},\ \bibinfo {series} {Cambridge Tracts in Mathematics}, Vol.\ \bibinfo {volume} {218},\ \href {\doibase 10.1017/9781108303453} {Cambridge University Press (\bibinfo {year} {2019})}\BibitemShut {NoStop}%
\bibitem [{\citenamefont {Micciancio}(2018)}]{Micciancio2018OntheHardnessofLWEwithBinarySecrets}%
  \BibitemOpen
  \bibfield  {author} {\bibinfo {author} {\bibfnamefont {D.}~\bibnamefont {Micciancio}},\ }\bibfield  {title} {On the hardness of learning with errors with binary secrets,\ }\href {\doibase 10.4086/toc.2018.v014a013} {\bibfield  {journal} {\bibinfo  {journal} {Theory Comput.}\ }\textbf {\bibinfo {volume} {14}},\ \bibinfo {pages} {1} (\bibinfo {year} {2018})}\BibitemShut {NoStop}%
\bibitem [{\citenamefont {Regev}(2010)}]{regev2010Thelearningwitherrorsproblem}%
  \BibitemOpen
  \bibfield  {author} {\bibinfo {author} {\bibfnamefont {O.}~\bibnamefont {Regev}},\ }\bibfield  {title} {The learning with errors problem (invited survey),\ }\href {\doibase 10.1109/CCC.2010.26} {in\ \emph {\bibinfo {booktitle} {Proc. 25th {CCC}}} (\bibinfo {year} {2010}),\ pp.\ \bibinfo {pages} {191--204}}\BibitemShut {NoStop}%
\bibitem [{\citenamefont {Kirchner}\ and\ \citenamefont {Fouque}(2015)}]{Kirchner2015AnImprovedBKWAlgorithmforLWE}%
  \BibitemOpen
  \bibfield  {author} {\bibinfo {author} {\bibfnamefont {P.}~\bibnamefont {Kirchner}}\ and\ \bibinfo {author} {\bibfnamefont {P.-A.}\ \bibnamefont {Fouque}},\ }\bibfield  {title} {An improved {BKW} algorithm for {LWE} with applications to cryptography and lattices,\ }\href {\doibase 10.1007/978-3-662-47989-6_3} {in\ \emph {\bibinfo {booktitle} {Adv. Cryptol. -- CRYPTO 2015}},\ \bibinfo {series} {Lect. Notes Comput. Sci.}, Vol.\ \bibinfo {volume} {9215} (\bibinfo {year} {2015}),\ pp.\ \bibinfo {pages} {43--62}}\BibitemShut {NoStop}%
\bibitem [{\citenamefont {Herold}\ \emph {et~al.}(2018)\citenamefont {Herold}, \citenamefont {Kirshanova},\ and\ \citenamefont {May}}]{Herold2018OnasymptoticcomplexitysolvingLWE}%
  \BibitemOpen
  \bibfield  {author} {\bibinfo {author} {\bibfnamefont {G.}~\bibnamefont {Herold}}, \bibinfo {author} {\bibfnamefont {E.}~\bibnamefont {Kirshanova}}, \ and\ \bibinfo {author} {\bibfnamefont {A.}~\bibnamefont {May}},\ }\bibfield  {title} {On the asymptotic complexity of solving {LWE},\ }\href {\doibase 10.1007/s10623-016-0326-0} {\bibfield  {journal} {\bibinfo  {journal} {Des. Codes Cryptogr.}\ }\textbf {\bibinfo {volume} {86}},\ \bibinfo {pages} {55} (\bibinfo {year} {2018})}\BibitemShut {NoStop}%
\bibitem [{\citenamefont {Peikert}\ and\ \citenamefont {Waters}(2011)}]{Peikert2011LossyTrapdoorFunctions}%
  \BibitemOpen
  \bibfield  {author} {\bibinfo {author} {\bibfnamefont {C.}~\bibnamefont {Peikert}}\ and\ \bibinfo {author} {\bibfnamefont {B.}~\bibnamefont {Waters}},\ }\bibfield  {title} {Lossy trapdoor functions and their applications,\ }\href {\doibase 10.1137/080733954} {\bibfield  {journal} {\bibinfo  {journal} {SIAM J. Comput.}\ }\textbf {\bibinfo {volume} {40}},\ \bibinfo {pages} {1803} (\bibinfo {year} {2011})}\BibitemShut {NoStop}%
\bibitem [{\citenamefont {Hardy}\ and\ \citenamefont {Wright}(2008)}]{Hardy2008AnIntroductiontotheTheoryofNumbers}%
  \BibitemOpen
  \bibfield  {author} {\bibinfo {author} {\bibfnamefont {G.~H.}\ \bibnamefont {Hardy}}\ and\ \bibinfo {author} {\bibfnamefont {E.~M.}\ \bibnamefont {Wright}},\ }\emph {\bibinfo {title} {An Introduction to the Theory of Numbers}},\ \href {\doibase 10.1093/oso/9780199219858.001.0001} {Oxford University Press (\bibinfo {year} {2008})}\BibitemShut {NoStop}%
\bibitem [{\citenamefont {Vadhan}(2012)}]{Vadhan2012Pseudorandomness}%
  \BibitemOpen
  \bibfield  {author} {\bibinfo {author} {\bibfnamefont {S.~P.}\ \bibnamefont {Vadhan}},\ }\bibfield  {title} {Pseudorandomness,\ }\href {\doibase 10.1561/0400000010} {\bibfield  {journal} {\bibinfo  {journal} {Found. Trends Theor. Comput. Sci.}\ }\textbf {\bibinfo {volume} {7}},\ \bibinfo {pages} {1} (\bibinfo {year} {2012})}\BibitemShut {NoStop}%
\bibitem [{\citenamefont {Tarabunga}\ and\ \citenamefont {Haug}(2026)}]{Tarabunga2026quantifyingmixed}%
  \BibitemOpen
  \bibfield  {author} {\bibinfo {author} {\bibfnamefont {P.~S.}\ \bibnamefont {Tarabunga}}\ and\ \bibinfo {author} {\bibfnamefont {T.}~\bibnamefont {Haug}},\ }\bibfield  {title} {Quantifying mixed-state entanglement via partial transpose and realignment moments,\ }\href {\doibase 10.22331/q-2026-08-19-2194} {\bibfield  {journal} {\bibinfo  {journal} {{Quantum}}\ }\textbf {\bibinfo {volume} {10}},\ \bibinfo {pages} {2194} (\bibinfo {year} {2026})}\BibitemShut {NoStop}%
\bibitem [{\citenamefont {Zhandry}(2019)}]{Zhandry2019TheMagicofELFs}%
  \BibitemOpen
  \bibfield  {author} {\bibinfo {author} {\bibfnamefont {M.}~\bibnamefont {Zhandry}},\ }\bibfield  {title} {The magic of {ELFs},\ }\href {\doibase 10.1007/s00145-018-9289-9} {\bibfield  {journal} {\bibinfo  {journal} {J. Cryptol.}\ }\textbf {\bibinfo {volume} {32}},\ \bibinfo {pages} {825} (\bibinfo {year} {2019})}\BibitemShut {NoStop}%
\end{thebibliography}%

\clearpage
\appendix
\crefalias{section}{appendix}
\crefalias{subsection}{subappendix}
\crefalias{subsubsection}{subsubappendix}

\section{Preliminaries}
\label{sec: General preliminary}

We collect the notation used throughout this paper in \Cref{sec:notation}. We define entanglement distillation protocols, the computational models, the corresponding distillable entanglements, and relate these measures in \Cref{sec:Information-theoretic and computational entanglement measures}. In \Cref{sec:numerics}, we describe the conventions on precision for implementing operations with finite computational resources.

\subsection{Notation}
\label{sec:notation}

In this work, we exclusively consider finite-dimensional Hilbert spaces $\cH$, equipped with the standard inner product. We denote a physical system, also called a register, by a sans-serif letter such as $\sA$, and its associated Hilbert space by $\cH^\sA$. We denote the dimension of a Hilbert space $\cH$ by $\dim \cH$, abbreviated as $d_\sA \coloneqq \dim\cH^\sA$ for a system $\sA$. We mostly consider bipartite systems consisting of subsystems $\sA$ and $\sB$, with Hilbert space $\cH^{\sA\sB}=\cH^\sA\otimes\cH^\sB$.
In particular, we focus on extensive bipartitions, where the numbers of qubits in $\sA$ and $\sB$ are of the same order.
For an integer $k \geq 1$, we write $\sA^k$ for $k$ copies of $\sA$, whose Hilbert space is $(\cH^\sA)^{\otimes k}$.

We write $\cB(\cH)$ for the set of linear operators on $\cH$. For any operator $M$ on $\cH$, we denote the complex conjugate and the transpose in a given orthonormal basis by $M^*$ and $M^\top$, respectively, and denote the Hermitian conjugate by $M^\dag$.
A positive semidefinite operator $\rho$ with unit trace is called a quantum state.
We say that a quantum state $\psi$ is pure if $\rank(\psi) = 1$. Pure states are also denoted by $\ketbra{\psi}{\psi}$, or by the corresponding unit vector $\ket{\psi}$.
We denote a bipartite state of systems $\sA\sB$ by $\rho^{\sA\sB}$ and its reduced density matrices of systems $\sA$ and $\sB$ by $\rho^\sA=\tr_\sB[\rho^{\sA\sB}]$ and $\rho^\sB=\tr_\sA[\rho^{\sA\sB}]$, respectively.
A special class of bipartite states is given by maximally entangled states. We denote the maximally entangled state of Schmidt rank $d$ by
\begin{align}
\label{eq:maximally entangled state}
    \ket{\Phi_d}^{\sA\sB} \coloneqq \frac{1}{\sqrt{d}}\sum_{i=1}^{d}\ket{i}^\sA\ket{i}^\sB,
\end{align}
where $\{\ket{i}\}_i$ is the computational basis.
The state $\ket{\Phi_2}^{\sA\sB}$ represents a Bell pair.
We sometimes drop the subscript $d$ when it is clear from the context.
Denoting the identity operator by $\bI$, we write $\pi^\sA \coloneqq \bI^\sA/d_\sA$ for the completely mixed state on system $\sA$.

A quantum channel from $\sA$ to $\sA'$ is denoted by $\cT^{\sA\to\sA'}$, which is a completely positive trace-preserving (CPTP) map acting on operators in the corresponding Hilbert spaces. We denote by $\rid$ the identity channel.
In this manuscript, we primarily focus on \emph{local operations and classical communication (LOCC)} channels defined with respect to a fixed bipartition.
These channels are composed of sequential applications of local quantum channels on subsystems, interleaved with rounds of classical communication.

We omit the superscripts indicating systems or registers when they are clear from the context. Likewise, when no confusion arises, we omit the identity operator and the identity channel on the remaining systems, e.g., writing $\cT^{\sB\to\sC}(\sigma^\sB\rho^{\sA\sB})$ for $(\rid^\sA\otimes\cT^{\sB\to\sC})\big((\bI^\sA\otimes\sigma^\sB)\rho^{\sA\sB}\big)$.

Other fundamental quantities in quantum information are distance measures. For an operator $M$, $\|M\|_p$ with $p \in (0,\infty)$ denotes the Schatten $p$-norm, defined as $\|M\|_p\coloneqq(\tr |M|^p)^{1/p}$.
In particular, $\|M\|_1$ and $\|M\|_2$ are the trace norm and the Hilbert--Schmidt norm, respectively, and $\|M\|_\infty \coloneqq \lim_{p\to\infty}\|M\|_p$ is the operator norm.
The Schatten $p$-norm is nonincreasing in $p$: $\|M\|_p \leq \|M\|_q$, for $p \geq q$.
For a vector $v$, the Euclidean norm is defined by $\|v\| \coloneqq \sqrt{v^\dag v}$.
Given two quantum states $\rho$ and $\sigma$, their trace distance is defined as $\frac{1}{2}\|\rho-\sigma\|_1$.
The distance admits the variational form $\f{1}{2}\|\rho - \sigma\|_1 = \max_{0 \leq \Pi \leq \bI}\tr[\Pi(\rho-\sigma)]$.
The diamond norm~\cite{Kitaev1997QCAlgoEC, watrous2004notessuperoperatornorm, watrous2018TheoryQI} for a linear map $\cT^{\sA\rarr\sB}$ is 
\begin{align}
\label{eq:diamond norm}
    \big\|\cT^{\sA\rarr\sB}\big\|_\diamond 
    \coloneqq \max_{\sR}\max_{M^{\sR\sA}; \|M^{\sR\sA}\|_1 \leq 1}\big\|\rid^\sR \otimes \cT^{\sA\rarr\sB}(M^{\sR\sA})\big\|_1, 
\end{align}
where the maximization is taken over all reference systems $\sR$ of arbitrary size and all operators $M^{\sR\sA}$ with trace norm at most one.
For a difference of two quantum channels, it suffices to take the maximization over all pure states on $\sR\sA$, where $\sR$ can be chosen such that $d_\sR = d_\sA$.
The diamond distance between two channels $\cT^{\sA\rarr\sB}$ and $\cT'^{\sA\rarr\sB}$ is $\f{1}{2}\|\cT - \cT'\|_\diamond$, which is the largest trace distance between their outputs on a common input.
For isometry channels $\til{\cV}$ and $\cV$, the diamond norm and the operator norm are related as~\cite{Kitaev1997QCAlgoEC} 
\begin{align}
\label{eq:rel of operator and diamond norm}
    \f{1}{2}\big\|\til{\cV} - \cV\big\|_\diamond \leq \big\|\til{V} - V\big\|_\infty.
\end{align}

Another common distance measure is the fidelity $\rF(\rho,\sigma) \coloneqq \| \sqrt{\rho}\sqrt{\sigma}\|_1^2$, which reduces to $\rF(\rho, \psi) = \bra{\psi}\rho\ket{\psi}$ when $\psi$ is pure. We call $1 - \rF(\rho,\sigma)$ the infidelity.
The square root of the infidelity, $\sqrt{1 - \rF(\rho,\sigma)}$, is called the purified distance, also known as the sine distance. It satisfies the triangle inequality and does not increase under quantum channels~\cite{tomamichel2010duality}. As the fidelity is multiplicative under tensor products, $\rF(\rho^{\otimes t},\sigma^{\otimes t}) = \rF(\rho,\sigma)^t$, the purified distance between $t$ copies is at most $\sqrt{t(1-\rF(\rho,\sigma))}$.
The trace distance and the fidelity are equivalent distance measures as stated by Fuchs--van de Graaf inequalities~\cite{Fuchs1999fuchsvandegraaf}: 
\begin{equation}
\label{eq:fuchsvandegraaf}
1 - \sqrt{\rF(\rho,\sigma)} \leq \frac{1}{2}\|\rho-\sigma\|_1 \leq \sqrt{1 - \rF(\rho,\sigma)}. 
\end{equation}

For $\alpha \in (0, 1) \cup (1, \infty)$, the \emph{sandwiched quantum $\alpha$-R\'{e}nyi divergence} of a state $\rho$ and a positive semidefinite operator $\sigma$ is defined as 
\begin{align}
    \widetilde{D}_\alpha(\rho \| \sigma) \coloneqq \f{1}{\alpha - 1} \log \Big(\tr\Big[\big(\sigma^{\f{1-\alpha}{2\alpha}}\rho\sigma^{\f{1-\alpha}{2\alpha}}\big)^\alpha\Big]\Big),
\end{align}
if $\alpha \in (0, 1)$ and $\tr(\rho\sigma) \neq 0$, or if $\alpha \in (1, \infty)$ and $\supp[\rho] \subseteq \supp[\sigma]$; otherwise, $\widetilde{D}_\alpha(\rho \| \sigma) \coloneqq \infty$~\cite{muller2013quantumrenyinew}.
The \emph{conditional quantum $\alpha$-R\'{e}nyi entropy} of a bipartite state $\rho^{\sA\sB}$ is then given by
\begin{align}
\label{eq:conditional renyi entropy}
    \widetilde{H}_\alpha(\sA|\sB)_\rho \coloneqq -\min_{\sigma^{\sB}}\widetilde{D}_\alpha\big(\rho^{\sA\sB}\big\|\bI^{\sA}\otimes\sigma^{\sB}\big),
\end{align}
where the minimization is taken over all states $\sigma^{\sB}$. 
The limits $\alpha \to 1$ and $\alpha \to \infty$ yield the conditional von Neumann entropy $H_1(\sA|\sB)_\rho$ and the min-entropy $H_{\rm min}(\sA|\sB)_\rho$, respectively.
When $\sB$ is a trivial system, it reduces to the (unconditional) quantum $\alpha$-R\'{e}nyi entropy $H_\alpha(\sA)_\rho = \f{1}{1-\alpha}\log{\tr[(\rho^{\sA})^\alpha]}$, whose limits at $\alpha \to 1, \infty$ are the von Neumann entropy $H_1(\sA)_\rho = -\tr[\rho^\sA\log{\rho^\sA}]$ and the min-entropy $H_{\rm min}(\sA)_\rho = -\log{\|\rho^{\sA}\|_\infty}$.
The \emph{sandwiched R\'{e}nyi coherent information} is the negative of Eq.~\eqref{eq:conditional renyi entropy}: $\widetilde{I}_\beta(\sA\rangle\sB)_\rho \coloneqq -\widetilde{H}_\beta(\sA|\sB)_\rho$.
Using the duality~\cite{muller2013quantumrenyinew}, for a pure state $\ket{\rho}^{\sA\sB\sE}$, it is equal to the conditional R\'{e}nyi entropy given the purifying system as
\begin{align}
\label{eq:duality}
    \widetilde{H}_\alpha(\sA|\sE)_\rho = \widetilde{I}_\beta(\sA\rangle\sB)_\rho, \qquad
    \f{1}{\alpha} + \f{1}{\beta} = 2,
\end{align}
which pairs $\alpha \in (1,\infty]$ with $\beta \in [1/2,1)$, with $\alpha=\infty$ corresponding to $\beta=1/2$.

We use a technique known as \emph{vectorization}~\cite{Horn1991TopicsinMatrixAnalysis, gilchrist2011vectorizationquantumsuse, Miszczak2011SINGULARVALUEDECOMPOSITION}.
Although a few notational conventions can be found in the literature, we define it as a linear map from $\bC^{d \times d}$ to $\bC^{d} \otimes \bC^{d}$, given by 
\begin{align}
\label{eq:definition of vectorization}
    \rvec: \ketbra{\psi}{\phi} \mapsto \ket{\psi}\ket{\phi^*},
\end{align}
where the complex conjugate is taken in the computational basis. The vectorization in this way is a bijection between matrices and vectors; thus, its inverse $\rvec^{-1}$ is well-defined. The vectorization preserves the Hilbert--Schmidt norm: $\|A - B\|_2 = \|\rvec(A) - \rvec(B)\|$, and satisfies $\rvec(ABC) = (A \otimes C^\top)\rvec(B)$ for any operators $A$, $B$, and $C$.
For a state $\rho^\sA$, the vector $\rvec\big(\sqrt{\rho^*}\big)^{\hat{\sA}\sA}$ has the marginal $\rho^\sA$ on $\sA$, where $\hat{\sA}$ is a copy of $\sA$, and we call it the canonical purification of $\rho^\sA$~\cite{Winter2004ExtrinsicIntrinsic}.

For asymptotic comparisons, we adopt the standard \emph{Landau notation}. We write $f = \cO(g)$ if $f(n) \leq cg(n)$ for some constant $c$ and all large enough $n$, $f = \Omega(g)$ if $f(n) \geq cg(n)$ under the same convention, and $f = \Theta(g)$ if both hold. We write $f = o(g)$ if $f(n)/g(n) \to 0$ and $f = \omega(g)$ if $g(n)/f(n) \to 0$. We also use $\poly(a, b, \ldots)$ to represent $a^{\cO(1)}b^{\cO(1)}\cdots$, and use $\negl(\lambda)$ to indicate a function decaying faster than any inverse polynomial in $\lambda$. The tilde notation hides polylogarithmic factors in the argument: $\widetilde{\cO}(g) = \cO(g \polylog{(g)})$, $\widetilde{\Omega}(g) = \Omega(g/\polylog{(g)})$, and $\widetilde{\Theta}(g)$ means both. Throughout, $\log$ denotes the logarithm to base two, and $\ln$ the natural logarithm.


\subsection{Information-theoretic and computational entanglement measures}
\label{sec:Information-theoretic and computational entanglement measures}

Throughout, states are understood as members of a family indexed by $n$. The bipartite states $\rho^{\sA\sB}$ under consideration are defined on $\Theta(n)$ qubits, and the symbols $o(1)$, $\poly(n)$, and $\negl(n)$ refer to this parameter.

\begin{definition}[$k$-shot entanglement distillation protocol and its rate]
\label{def:protocol}
Let $k \geq 1$ and $m \geq 0$ be integers. A \emph{$k$-shot entanglement distillation protocol} is an LOCC channel $\Lambda^{\sA^k\sB^k\to\sS\hat{\sS}}$ taking $k$ copies of a bipartite state on $\sA\sB$, together with local ancillas in a fixed pure product state, to registers $\sS\hat{\sS}$. For $\varepsilon \in [0, 1)$, we say that $\Lambda^{\sA^k\sB^k\to\sS\hat{\sS}}$ distills $m$ ebits from $\rho^{\sA\sB}$ at error $\varepsilon$ if $\rF\big(\Lambda\big((\rho^{\sA\sB})^{\otimes k}\big), \Phi_2^{\otimes m}\big) \geq 1 - \varepsilon$. We call $R \coloneqq m/k$ the \emph{rate} of the protocol.
\end{definition}

The registers $\sS\hat{\sS}$ carry $m$ pairs, and we say that the protocol \emph{declares} $m$ pairs. In the following, we also compare a protocol with a number of pairs different from the number it declares, which is needed when the protocol is not fixed in advance but varies over a class or is produced from an input. For an integer $m \geq 0$ and an LOCC protocol $\Lambda$, we accordingly define
\begin{align}
\label{eq:fidelity functional}
    \rF_\Lambda(\cdot) \coloneqq \tr\big[\big(\Phi_2^{\otimes m}\otimes\bI\big)\Lambda(\cdot)\big],
\end{align}
where the Bell pairs $\Phi_2^{\otimes m}$ are supported on the first $m$ declared pairs, and $\rF_\Lambda \coloneqq 0$ when $\Lambda$ declares fewer than $m$ pairs. A protocol declaring $m$ pairs distills $m$ ebits at error $\varepsilon$ exactly when $\rF_\Lambda\big((\rho^{\sA\sB})^{\otimes k}\big) \geq 1 - \varepsilon$.

The task is sometimes formulated using a heralded protocol, as in Ref.~\cite{Leone2025Entanglementtheorywithlimited}, and such protocols are covered by \Cref{def:protocol} as follows. The output includes a classical flag register $\sX$, available to both parties, that indicates success or failure. The protocol succeeds with probability $p \geq 1 - \nu$, and, conditioned on success, the output state $\omega^{\sS\hat{\sS}}$ satisfies $\rF\big(\omega, \Phi_2^{\otimes m}\big) \geq 1 - \varepsilon'$. Discarding $\sX$ gives a channel of the form in \Cref{def:protocol}, whose output has fidelity at least $p \rF\big(\omega, \Phi_2^{\otimes m}\big) \geq 1 - \varepsilon' - \nu$. Thus, the heralded protocol can be viewed as a protocol of \Cref{def:protocol} with error $\varepsilon' + \nu$, and all results in this paper apply.

\begin{definition}[Information-theoretic distillable entanglement]
\label{def:it distillable}
For $\varepsilon \in [0,1)$ and $k \geq 1$, let $E_\rD^{(k, \varepsilon)}(\sA:\sB)_\rho$ be the supremum of the rates of all $k$-shot distillation protocols that distill from $\rho^{\sA\sB}$ at error $\varepsilon$. If no such protocol exists, set it to $0$. 
We define the (asymptotic) \emph{information-theoretic distillable entanglement} as
\begin{align}
    E_\rD(\sA:\sB)_\rho \coloneqq \lim_{\varepsilon \to 0}\ \liminf_{k\to\infty}\ E_\rD^{(k, \varepsilon)}(\sA:\sB)_\rho.
\end{align}
\end{definition}

The above definition captures the information-theoretic notion of distillable entanglement, where computational resources are unrestricted.
To study distillation with limited resources, we need to specify both the resource constraints and how a protocol may depend on the input state. We first introduce the resource constraints.
Efficiency is a property of a family indexed by $n$, so Refs.~\cite{arnonfriedman2023computationalentanglementtheory, Leone2025Entanglementtheorywithlimited} specify computational measures not by a value but by the function bounds that efficient families do and do not achieve.
We instead carry the size bound $s$ as a parameter, as Ref.~\cite{Avidan2026FullyQuantumComputationalEntropies} does for the computational entropies: at a fixed $s$, the class of protocols defined below is finite.
Quantifying over $s = \poly(n)$ outside the definition recovers the standard asymptotic notion of efficiency.

Under resource constraints, additional local access to the state may help the parties, and our converse bounds hold even when they are given such access. To formalize this, for a purification $\ket{\rho}^{\sA\sB\sE}$ on a fixed register $\sE$ of $\cO(n)$ qubits and an integer $l \geq 0$, we also allow each party to receive $l$ copies of $\ket{\rho}^{\sA\sB\sE}$, each held entirely in that party's laboratory, which the protocol may use in any way. These copies carry no entanglement between the two laboratories, and thus contribute no distillable entanglement by themselves. The input is then $(\rho^{\sA\sB})^{\otimes k}\otimes(\ketbra{\rho}{\rho}^{\sA\sB\sE})^{\otimes 2l}$, the error is defined by the same fidelity condition in Eq.~\eqref{eq:fidelity functional} evaluated on this input, and the rate is still $R = m/k$.

\begin{definition}[Protocol classes]
\label{def:protocol classes}
Fix a finite universal gate set. For integers $s, k \geq 1$ and $l \geq 0$, let $\mathfrak{L}(s, k, l)$ denote the set of distillation protocols acting on $k$ shared copies and $l$ additional local copies of a purification per party, implemented by LOCC circuits of at most $s$ gates from that set, including the classical communication they use.
A circuit with $s$ gates uses at most $\cO(s)$ ancillas in addition to the $\Theta((k+2l)n)$ input qubits. Each gate can be specified by $\cO(\log{(s(k+2l)n)})$ bits, and hence each protocol by $\cO(s\log{(s(k+2l)n)})$ bits, giving
\begin{align}
\label{eq:protocol count}
    |\mathfrak{L}(s, k, l)| \leq 2^{\cO(s\log{(s(k+2l)n)})}.
\end{align}
\end{definition}

We omit $l$ when $l = 0$ in this notation and in the state-agnostic and computational distillable entanglements below.
We call a family of protocols indexed by $n$ \emph{sample-efficient} if it acts on $k = \poly(n)$ shared copies and $l = \poly(n)$ additional local copies per party, with no restriction on its circuit size, and \emph{computationally efficient} if, in addition, each member lies in $\mathfrak{L}(s, k, l)$ with $s = \poly(n)$.

We next introduce the computational models considered in this work.

\begin{definition}[Computational models]
\label{def:computational model}
A computational model specifies how a $k$-shot distillation protocol may depend on the state it acts on. We consider the following three.
\begin{enumerate}[label=\textnormal{(M\arabic*)}, leftmargin=4em]
    \item \emph{State-agnostic}: a single protocol, of unrestricted size, acts on every state of a prescribed set and receives no information about which one it is given.
    \item \emph{State-dependent}: the protocol is implemented by an LOCC circuit of bounded size and may otherwise depend on the state in any way whatsoever, no account being taken of how it is found.
    \item \emph{Description-aware}: the state carries a purification prepared by a known circuit, and the protocol is output by a single circuit of bounded size on input a classical description of that circuit.
\end{enumerate}
\end{definition}

Model (M1) is the state-agnostic setting of Ref.~\cite{Leone2025Entanglementtheorywithlimited} and formalizes universal distillation; its unrestricted size isolates the limitation due to the number of copies.
Model (M2), defined for nonuniform circuit families, underlies the computational distillable entanglement of Refs.~\cite{arnonfriedman2023computationalentanglementtheory, Leone2025Entanglementtheorywithlimited}, whose fixed-error regime is recovered here at a constant error.
Model (M3) is adopted in the uniform computational measure of Ref.~\cite{arnonfriedman2023computationalentanglementtheory}, where the key indexing the family is passed as input to the protocol; in Ref.~\cite{bouland2024PublicKeyPseudoentanglement} the description is passed to the distinguisher instead.

We first define the distillable entanglement in model (M1), where a single protocol must work for every state in a prescribed set, with no restriction on its circuit size.

\begin{definition}[State-agnostic distillable entanglement]
\label{def:agnostic distillable}
In model (M1), let $\scS$ be a set of states on $\sA\sB$, and fix a register $\sE$ of $\cO(n)$ qubits large enough to purify every state in $\scS$. For integers $k \geq 1$, $l \geq 0$ and an error $\varepsilon \in [0,1)$, let $E_{\rD, \ag}^{(k, l, \varepsilon)}(\sA:\sB)_\scS$ be the supremum of the rates $R$ achievable by a single protocol of unrestricted size that distills $kR$ ebits at error $\varepsilon$ from $k$ shared copies of every $\rho^{\sA\sB} \in \scS$ and $l$ additional local copies per party of every purification of it on $\sA\sB\sE$.
\end{definition}

We next turn to model (M2), which bounds the circuit size while allowing the protocol to depend on the state in any way. Since $\mathfrak{L}(s, k, l)$ is finite at a fixed $s$, the largest rate in the following definition is attained.

\begin{definition}[Computational distillable entanglement]
\label{def:comp distillable}
In model (M2), let $\ket{\rho}^{\sA\sB\sE}$ be a fixed purification. For integers $s, k \geq 1$, $l \geq 0$ and an error $\varepsilon \in [0,1)$, we define the \emph{computational distillable entanglement} $\widehat{E}_\rD^{(k, l, \varepsilon, s)}(\sA:\sB)_\rho$ as the largest rate among the protocols of $\mathfrak{L}(s, k, l)$ distilling from $k$ shared copies of $\rho^{\sA\sB}$ and $l$ additional local copies of $\ket{\rho}^{\sA\sB\sE}$ per party at error $\varepsilon$, and $0$ if there is none.
\end{definition}

The quantity $\widehat{E}_\rD^{(k, l, \varepsilon, s)}$ is non-decreasing in $\varepsilon$, and a bound at a constant $\varepsilon$ also caps the rate at every smaller error.
It places no constraint on how a protocol is found: a circuit of size $s$ may depend on the state and its supplied purification in any way whatsoever, subject only to being describable by $\cO(s\log{(s(k+2l)n)})$ bits.
Model (M3) additionally accounts for the cost of constructing the protocol from a classical description of the state preparation.

\begin{definition}[Description-aware distillable entanglement]
\label{def:aware distillable}
In model (M3), let $\scD$ be a distribution over classical descriptions $d$ of quantum circuits $C_d$, where each circuit prepares a purification $\ket{\rho_d}^{\sA\sB\sE} = C_d\ket{0}$. For $s, k \geq 1$ and $\varepsilon, \nu \in [0,1)$, let $\widehat{E}_{\rD, \aw}^{(k, \varepsilon, \nu, s)}(\sA:\sB)_\scD$ be the supremum of the rates $R$ achievable as follows. A single circuit of at most $s$ gates takes $d$ as input and outputs a protocol in $\mathfrak{L}(s, k)$. This protocol distills $kR$ ebits from $\rho_d^{\sA\sB}$ at error $\varepsilon$ with probability at least $1 - \nu$ over $d$ drawn from $\scD$.
\end{definition}

In this model, we do not supply additional copies of a purification, since each party can prepare them from the given description with polynomial overhead.
At a single fixed state, the description-aware distillable entanglement carries no content beyond \Cref{def:comp distillable}, since a circuit of polynomial size may hardwire the description.
It becomes nontrivial only for a distribution of descriptions, where the compiler cannot depend on which description is sampled; this is why \Cref{def:aware distillable} is a function of $\scD$ and admits a fraction $\nu$ of failures.

\begin{proposition}[Relations among the distillable entanglements]
\label{prop:relations}
Let $s, k \geq 1$ and $l \geq 0$ be integers, and let $\varepsilon, \nu \in [0,1)$. Let $\scS$ be a set of states on $\sA\sB$ and $\scD$ be a distribution over descriptions as in \Cref{def:aware distillable}. Then, the following hold.
\begin{enumerate}[label=\textnormal{(\arabic*)}]
    \item Both $\widehat{E}_\rD^{(k,l,\varepsilon,s)}(\sA:\sB)_\rho$ and $\widehat{E}_{\rD, \aw}^{(k,\varepsilon,\nu,s)}(\sA:\sB)_\scD$ are non-decreasing in $s$; all of $E_\rD^{(k,\varepsilon)}$, $E_{\rD, \ag}^{(k,l,\varepsilon)}$, $\widehat{E}_\rD^{(k,l,\varepsilon,s)}$, and $\widehat{E}_{\rD, \aw}^{(k,\varepsilon,\nu,s)}$ are non-decreasing in $\varepsilon$; both $E_{\rD, \ag}^{(k,l,\varepsilon)}(\sA:\sB)_\scS$ and $\widehat{E}_\rD^{(k,l,\varepsilon,s)}(\sA:\sB)_\rho$ are non-decreasing in $l$; $\widehat{E}_{\rD, \aw}^{(k,\varepsilon,\nu,s)}(\sA:\sB)_\scD$ is non-decreasing in $\nu$; and $E_{\rD, \ag}^{(k,l,\varepsilon)}(\sA:\sB)_\scS$ is non-increasing under enlarging $\scS$.
    \item $E_{\rD, \ag}^{(k,l,\varepsilon)}(\sA:\sB)_{\{\rho\}} = E_\rD^{(k,\varepsilon)}(\sA:\sB)_\rho$.
    \item $E_{\rD, \ag}^{(k,l,\varepsilon)}(\sA:\sB)_\scS \leq \inf_{\rho\in\scS}E_\rD^{(k,\varepsilon)}(\sA:\sB)_\rho$ and $\displaystyle \widehat{E}_\rD^{(k,l,\varepsilon,s)}(\sA:\sB)_\rho \leq E_\rD^{(k,\varepsilon)}(\sA:\sB)_\rho$.
    \item With probability at least $1 - \nu$ over $d$ drawn from $\scD$, $\widehat{E}_{\rD, \aw}^{(k,\varepsilon,\nu,s)}(\sA:\sB)_\scD \leq \widehat{E}_\rD^{(k,l,\varepsilon,s)}(\sA:\sB)_{\rho_d}$ for every $l \geq 0$.
\end{enumerate}
\end{proposition}

\begin{proof}[Proof of \Cref{prop:relations}]

The monotonicity statements in (1) all follow from comparing the sets over which the optimization is performed. A larger $s$ enlarges $\mathfrak{L}(s,k,l)$, a larger $\varepsilon$ weakens the requirement of \Cref{def:protocol} at fixed rate, a larger $\nu$ weakens that of \Cref{def:aware distillable}, and a larger $\scS$ imposes more constraints on the single protocol of \Cref{def:agnostic distillable}. For $l$, a protocol may discard the additional copies, which is a local operation; the set $\scS$ of shared states is unchanged, and only the input to the protocol is enlarged.
For (2), discarding the additional copies gives the inequality $\geq$. Conversely, for a single state and with no size restriction, each party can prepare $l$ copies of a fixed purification of $\rho^{\sA\sB}$ in its own laboratory, so a protocol using these copies yields a $k$-shot distillation protocol at the same rate and error, which gives $\leq$.
The inequalities in (3) follow in the same way: on the left, a protocol with the additional copies is converted into an admissible $k$-shot distillation protocol at error $\varepsilon$ for the state on the right by preparing those copies locally; in the first inequality, this holds for every $\rho \in \scS$.
For (4), the protocol produced from $d$ lies in $\mathfrak{L}(s,k)$ and distills from $\rho_d$ at error $\varepsilon$ for every $d$ outside the excluded fraction, which gives the bound at $l = 0$; the monotonicity in $l$ in (1) extends it to every $l$.

\end{proof}

Comparing $l = 0$ with general $l$ in (1), we have $E_{\rD, \ag}^{(k,\varepsilon)}(\sA:\sB)_\scS \leq E_{\rD, \ag}^{(k,l,\varepsilon)}(\sA:\sB)_\scS$ and $\widehat{E}_\rD^{(k,\varepsilon,s)}(\sA:\sB)_\rho \leq \widehat{E}_\rD^{(k,l,\varepsilon,s)}(\sA:\sB)_\rho$ for every $l \geq 0$, and hence the converse bounds that we prove with additional copies of a purification also apply without them. Across different measures, by contrast, no inequality holds between $E_{\rD, \ag}$ and either of the two hatted measures in either direction. For a single state, (2) reduces the former to the information-theoretic value, which by (3) is at least as large as both, while on a set on which no single protocol succeeds, it vanishes even though the individual states retain positive computational rates. A comparison does hold at matched size and $l = 0$: an agnostic protocol of at most $s$ gates lies in $\mathfrak{L}(s,k)$ at every state in $\scS$ and, at $s = \poly(n)$, is also the output of a compiler of polynomial size that ignores its input, so its rate is bounded by the right-hand sides of both (3) and (4). Only (4) relates two of the computational quantities, and in one direction only: an upper bound on $\widehat{E}_\rD$ at almost every description caps $\widehat{E}_{\rD, \aw}$, but not conversely, since a nonuniform circuit may depend on the sampled description.

Our achievability result in \Cref{sec:achievability} provides a single explicit polynomial-size protocol that uses neither a description of the state nor additional copies of a purification, and therefore lower-bounds the three quantities associated with models (M1)--(M3).
The converse bounds in \Cref{sec:Converse: State-agnostic setting,sec:Converse Description-aware setting} hold even when each party is additionally given $\poly(n)$ copies of a purification $\ket{\rho}^{\sA\sB\sE}$ of the input state, with each copy held entirely in that party's laboratory. In model (M3), each party can prepare these copies locally from the given circuit description with polynomial overhead, so they do not increase the distillation rates achievable by polynomial-size circuits.

\begin{table}[t]
\centering
\caption{The computational model under which each of our results is stated.}
\label{tab:models}
\begin{tabular}{ccc}
\hline
Result & Model & Role \\
\hline
(I) of \Cref{thm:achievability by efficient protocol}, \Cref{thm:achievability by schur coding}
  & (M1) & achievability, sample-efficient \\
(II) of \Cref{thm:achievability by efficient protocol}, \Cref{thm:achievability by efficient protocol formal}
  & (M1)--(M3) & achievability, computationally efficient \\
\hline
\Cref{thm:no go informal}
  & (M1)--(M3) & converse; (M3) under LWE assumptions \\
\Cref{thm:agnostic unbounded,cor:agnostic no go informal}
  & (M1) & converse \\
\Cref{thm:agnostic no go,cor:agnostic no go informal}
  & (M2) & converse \\
\Cref{thm:lwe mixed state no go,thm:lwe mixed state no go informal}
  & (M3) & converse under LWE assumptions \\
\Cref{cor:worst case optimality}
  & (M1) for (I), (M1)--(M3) for (II) & optimality \\
\hline
\Cref{prop:separation}, \Cref{prop:nearly maximal separation}
  & (M1)--(M3) & separation \\
\Cref{prop:degenerate spike} & (M1)--(M2) in (I), (M3) in (II) & separation \\
\hline
\end{tabular}
\end{table}


\subsection{Finite precision and circuit compilation}
\label{sec:numerics}

The circuit size in \Cref{def:protocol classes} counts gates from a finite set, while our constructions are described in ideal terms, such as an exact isometry, a unitary drawn from the Haar measure, or a tolerance given as a real number. None of these is realized by finitely many gates and bits as it stands. We hence provide in this subsection the prescriptions by which an ideal description is turned into a finite circuit, together with the accounting of the resulting errors. A reader interested only in the ideal protocol and familiar with this circuit compilation may proceed to \Cref{sec:common preparation}.

Numerical data are represented by dyadic rationals, with positive tolerances rounded down to within a factor of two and loop bounds rounded up using certified bounds on logarithms and roots~\cite{Moore2009IntroductionIntervalAnalysis}. We call the errors of evaluating numerical quantities under these conventions arithmetic errors.

For quantum operations, we take the finite gate set of \Cref{def:protocol classes} to be inverse-closed and to contain the standard Clifford generators exactly, together with determinant-one one-qubit generators dense in $\mathrm{SU}(2)$. The constructive Solovay--Kitaev algorithm~\cite{Dawson2006SolovayKitaevalgorithm} synthesizes a circuit containing $N_0$ gates outside this set to total operator error at most $\delta$ using $\cO(N_0\log^4{(N_0/\delta)})$ gates and polynomial classical overhead.
Since a controlled call of a unitary turns its global phase into a relative one, we approximate it together with its phase. Before synthesis, we extend it to a determinant-one unitary on the original register and initialized work registers, which fixes the phase and is preserved by the gate set above. The ideal extension returns the work registers to zero.

To bound the total approximation error, we mainly measure the error of each implementation by the diamond distance from the ideal operation. The distance is subadditive under composition~\cite{watrous2018TheoryQI}, so a circuit making at most $G_0$ implementation calls stays within $\delta$ of the ideal one if each call is accurate to $\delta/(1+G_0)$. For adaptive circuits, the composition bound applies after the classical outcome registers are included in the channel description. Finite-gate and arithmetic errors are counted within the error budget of each construction, alongside its deliberate approximations.

For uniform sampling, finitely many random bits produce an exact uniform distribution only when the size of the set is a power of two. We hence use rejection sampling with a fixed trial limit~\cite[Sec.~9.2]{Shoup2012Computationalintroductionnumbertheoryalgebra}. For a set of size $m \geq 2$, we draw $\lceil\log{m}\rceil$ random bits and reject values outside the set, so that each trial succeeds with probability at least $1/2$. For $\zeta \in (0, 1)$, a limit of $\lceil\log{(1/\zeta)}\rceil$ trials bounds the failure probability by $\zeta$, and returning a fixed element on failure changes the output distribution by total variation at most $\zeta$. For $m=1$, we return the sole element. Over at most $R_0$ sampling calls, a failure probability of at most $\delta/(1+R_0)$ per call contributes at most $\delta$ to the diamond distance, even when subsequent operations depend on the sampled labels.

A finite-precision construction works with a matrix near the ideal one rather than the ideal matrix itself. The following lemma bounds the resulting displacement of the isometric factor of the thin QR factorization, with an amplification governed by the smallest singular value. This is used in the following two lemmas to implement a given isometry and a Haar unitary with finite precision.

\begin{lemma}[Perturbation of the isometric factor]
\label{lem:qr perturbation}
Let $X$ be a $d \times q$ matrix with $q \leq d$, and suppose that its minimum singular value $\sigma_{\min}(X)$ satisfies $\sigma_{\min}(X) \geq g > 0$. Let $Q(X)$ denote the isometric factor of its positive-diagonal thin QR factorization. Then, every matrix $E$ of the same size with $\|E\|_\infty \leq g/2$ satisfies
\begin{align}
\label{eq:qr perturbation}
    \|Q(X+E) - Q(X)\|_2
    \leq \f{2\sqrt{2}}{g}\|E\|_2.
\end{align}
\end{lemma}

\begin{proof}[Proof of \Cref{lem:qr perturbation}]

Since $\sigma_{\min}(X+vE) \geq g/2$ for $0 \leq v \leq 1$, we may write $X + vE = Q_v R_v$ with $Q_v$ and $R_v$ differentiable in $v$, and denote the derivatives by $\dot{Q}_v$ and $\dot{R}_v$. 
Differentiation gives
\begin{align}
    Q_v^\dag E R_v^{-1}
    &= Q_v^\dag\dot{Q}_v+\dot{R}_vR_v^{-1}, \\
    (\bI-Q_vQ_v^\dag)\dot{Q}_v
    &= (\bI-Q_vQ_v^\dag)ER_v^{-1}.
\end{align}
The matrix $Q_v^\dag\dot{Q}_v$ is skew-Hermitian, whereas $\dot{R}_vR_v^{-1}$ is upper triangular with real diagonal. 
Comparison of the strict lower triangular and imaginary diagonal parts therefore yields
\begin{align}
    \|Q_v^\dag\dot{Q}_v\|_2^2
    \leq 2\|Q_v^\dag ER_v^{-1}\|_2^2.
\end{align}
The projections $Q_vQ_v^\dag\dot{Q}_v$ and $(\bI-Q_vQ_v^\dag)\dot{Q}_v$ are orthogonal in Hilbert--Schmidt norm, so $\|\dot{Q}_v\|_2\leq\sqrt{2}\|ER_v^{-1}\|_2$.
Since $R_v$ has the same singular values as $X+vE$, we have $\|R_v^{-1}\|_\infty \leq 2/g$, and hence
\begin{align}
    \|Q(X+E) - Q(X)\|_2
    = \Big\|\int_0^1 \dot{Q}_v \,\rd v\Big\|_2
    \leq \sqrt{2}\int_0^1 \|E R_v^{-1}\|_2 \,\rd v
    \leq \f{2\sqrt{2}}{g}\|E\|_2,
\end{align}
where we used $\|ER_v^{-1}\|_2 \leq \|E\|_2\|R_v^{-1}\|_\infty$ in the last inequality.

\end{proof}

For operations specified by an isometry matrix, the following lemma gives a finite circuit with explicit error and gate bounds. The construction eliminates the subdiagonal entries by two-level unitaries, and the error analysis accounts for both numerical rounding and finite-gate synthesis.

\begin{lemma}[Finite-precision isometry synthesis]
\label{lem:isometry synthesis}
Let $V:\bC^q\to\bC^d$ be an isometry, where $1 \leq q \leq d$, and let $\delta \in(0, 1/2)$. Given a rational matrix $A_0$ satisfying $\|A_0-V\|_2 \leq \delta/128$, there is a classical algorithm that constructs a unitary circuit whose first $q$ columns, denoted by $\widehat{V}$, satisfy $\big\|\widehat{V}-V\big\|_\infty \leq \delta$, where the circuit acts on a register of dimension at most $2d$ and $V$ is extended by zero rows to that dimension. The gate count is $\cO\big(d^2\log^3{(2d)}\log^4{(d/\delta)}\big)$, and the classical running time is polynomial in $d$, $\log{(1/\delta)}$, and the bit length of $(A_0,\delta)$.
\end{lemma}

\begin{proof}[Proof of \Cref{lem:isometry synthesis}]

Given a description of $A_0$, we apply to it the standard elimination by two-level unitaries~\cite[Sec.~4.5.1]{nielsen2010quantum}.
To this end, we choose the precision parameters as follows:
\begin{align}
    z \coloneqq \f{\delta}{256d^3}, \qquad 
    b \coloneqq \f{\delta}{256d^2}.
\end{align}
We eliminate the entries of $A_0$ below the diagonal, column by column, by rotations acting on a pivot pair $(\alpha,\beta)$ consisting of the diagonal entry and the entry to be eliminated. At each such pair, we compare the rational number $|\alpha|^2+|\beta|^2$ with $z^2$. If the value is below $z^2$, we set the stored entry $\beta$ to zero without applying a rotation. Otherwise, we store a description of the exact two-level rotation
\begin{align}
    G(\alpha,\beta)
    = \f{1}{\sqrt{|\alpha|^2+|\beta|^2}}
    \begin{pmatrix}
    \alpha^* & \beta^* \\
    -\beta & \alpha
    \end{pmatrix},
\end{align}
which maps the pair $(\alpha, \beta)$ to $(\sqrt{|\alpha|^2+|\beta|^2}, \, 0)$. We compute the corresponding row update with error at most $ b$ in the Hilbert--Schmidt norm and preserve structural zeros exactly.

Let $A_j$ be the working matrix after $j$ steps. Let $Q_j$ be the product of the exact rotations selected through step $j$, where a rotation is extended by the identity on the rows it does not act on, and a skipped pair contributes the identity. We write $N_{\mathrm{skip}}$ and $N_{\mathrm{rot}}$ for the numbers of skipped and rotated pivot pairs. At step $j$, the working matrix $A_j$ differs from $A_{j-1}$ by at most $z$ if the pair is skipped, and from $G_jA_{j-1}$ by at most $b$ if a rotation $G_j$ is applied. Since the subsequent exact rotations preserve the Hilbert--Schmidt norm, the accumulated error is bounded as
\begin{align}
    \|A_j-Q_jV\|_2
    \leq \|A_0 - V\|_2 + N_{\mathrm{skip}}z+N_{\mathrm{rot}} b
    \leq \f{\delta}{128} + d^2 z + d^2 b
    \leq \f{\delta}{64},
\end{align}
where the number of pivot pairs is at most $d^2$. The working entries remain bounded by two, and every nontrivial pivot denominator is at least $z$. The row updates thus require $\cO(\log{(d/\delta)})$ bits of precision.

We convert this bound into an error of the isometry. Let $\epsilon \coloneqq \f{\delta}{64}$, let $J$ be the final step, and write $A_J = \binom{T}{0}$ with $T$ upper triangular of size $q$. Setting $E \coloneqq Q_J^\dag A_J - V$, the bound above reads $\|E\|_2 \leq \epsilon$, and the minimum singular value $\sigma_{\min}(V) = 1$ gives $\sigma_{\min}(T) = \sigma_{\min}(A_J) \geq 1 - \epsilon$. 
The inverse $T^{-1}$ is upper triangular with the diagonal entries $T_{jj}^{-1}$, and hence $|T_{jj}|^{-1} \leq \|T^{-1}\|_\infty \leq (1-\epsilon)^{-1}$. Every diagonal entry of $T$ consequently satisfies $|T_{jj}| \geq 1-\epsilon > 1/2$. An error in $T_{jj}$ is amplified by $1/|T_{jj}|$ in its phase $T_{jj}/|T_{jj}|$, so this bound keeps the phase evaluation stable and requires polynomially many bits.
We choose phases on the first $q$ rows that make the diagonal of $T$ positive, and let $P$ be the corresponding diagonal unitary, completed by the identity. Writing $U \coloneqq (PQ_J)^\dag$, we have
\begin{align}
    V + E
    = Q_J^\dag A_J
    = U \binom{PT}{0},
\end{align}
where $PT$ is upper triangular with positive diagonal. The first $q$ columns of $U$ are therefore the isometric factor $Q(V+E)$. Since $Q(V) = V$ and $\sigma_{\min}(V) = 1$, \Cref{lem:qr perturbation} gives $\|Q(V+E) - V\|_2 \leq 2\sqrt{2}\epsilon$.

For finite-gate synthesis, we fix the phase of the implementation by the convention above. We reserve one unused coordinate on which to place the compensating phase, and pad the matrix register to dimension $2^{\lceil\log{(d+1)}\rceil}\leq2d$. We extend $U$ to $U\oplus(\det U)^*\oplus\bI$, which agrees with $U$ on the first $q$ columns. Its determinant satisfies
\begin{align}
    \det U \cdot (\det U)^* \cdot 1
    = |\det U|^2
    = 1,
\end{align}
where the last equality comes from the unitarity of $U$. The extension is hence a determinant-one unitary, and its determinant phase is computed from the selected rotations and row phases.

The unitary $U$ is a product of two-level factors, and the compensating phase can be absorbed into one of them, so the extension is a product of two-level special unitaries. Following Ref.~\cite[Secs.~7--8]{Barenco1995Elementarygatesquantumcomputation}, these are decomposed into $\cO(d^2\log^3{(2d)})$ elementary gates acting on the matrix register alone. We allocate total operator error at most $\delta/4$ to the evaluation of their angles and to their finite-gate compilation. Since $\|\cdot\|_\infty \leq \|\cdot\|_2$, the resulting circuit satisfies
\begin{align}
    \big\|\widehat{V}-V\big\|_\infty
    \leq 2\sqrt{2}\epsilon + \f{\delta}{4}
    < \delta.
\end{align}
The elimination uses $\cO(d^3)$ scalar operations, and Solovay--Kitaev compilation gives the stated gate count and classical running time.

\end{proof}

Our distillation protocol presented later draws a unitary according to the Haar measure. The following lemma implements one with finite precision. Its guarantee is not that the output circuit is drawn from a distribution close to the Haar measure: the circuit and an exact Haar unitary are defined on a common probability space, and their actions on inputs with initialized work registers are close on each realization, except with a prescribed failure probability.

\begin{lemma}[Finite-precision Haar unitary synthesis]
\label{lem:haar implementation}
For an integer $q \geq 1$ and $\delta, \zeta_\rH \in (0,1/2)$, set $d_\sC \coloneqq 2^{\lceil\log{(q+1)}\rceil}$. There is a randomized classical algorithm that outputs a unitary circuit acting on a register $\sC$ of dimension $d_\sC$ and a one-qubit work register $\sW$, whose implemented unitary is denoted by $\widehat{V}^{\sC\sW}$. On the probability space of its random bits, there is a Haar unitary $V^\sC$ on the span of $q$ fixed vectors of the computational basis in $\sC$, extended by the identity on its complement, such that
\begin{align}
\label{eq:haar implementation}
    \Big\|\widehat{V}^{\sC\sW}\big(\bI^\sC\otimes\ket{0}^\sW\big)
    - V^\sC\otimes\ket{0}^\sW\Big\|_\infty
    \leq \delta,
\end{align}
except with probability at most $\zeta_\rH$. The gate count and the classical running time are $\poly\big(q, \log(1/\delta), \log(1/\zeta_\rH)\big)$.
\end{lemma}

\begin{proof}[Proof of \Cref{lem:haar implementation}]

The positive-diagonal unitary factor $Q(Z)$ of a square complex standard Gaussian matrix $Z$ is Haar distributed~\cite{Mezzadri2007Generaterandommatrices}. Here the entries of $Z$ are independent with $\bE[|Z_{ij}|^2] = 1$.
We couple a finite-bit sample and its numerical QR calculation to this exact matrix. For $Z$ of size $q \times q$, it holds that
\begin{align}
\label{eq:gaussian smallest}
    \Pr[\sigma_{\min}(Z) < b_\rH]\leq q^2b_\rH^2.
\end{align}
To see this, suppose that $\|Zx\| < b_\rH$ for a unit vector $x$, and take an index $j$ with $|x_j| \geq q^{-1/2}$.
Dividing $Zx$ by $x_j$ shows that the $j$-th column of $Z$ is within distance $\sqrt{q}b_\rH$ of the span of the other columns. The index $j$ depends on $x$, so this holds for at least one of the $q$ columns.
Conditional on the other columns, the component of the $j$-th column normal to their span is $y \in \bC$ with the standard complex Gaussian distribution, and $\Pr[|y|\leq s]=1-e^{-s^2}\leq s^2$ at $s = \sqrt{q}b_\rH$ bounds the probability that the $j$-th column is that close to the span by $qb_\rH^2$. A union bound over the $q$ columns then gives Eq.~\eqref{eq:gaussian smallest}.

Choose dyadic rationals $b_\rH$ and $M_\rH$ in the ranges
\begin{align}
    \f{\sqrt{\zeta_\rH}}{8q}\leq b_\rH\leq\f{\sqrt{\zeta_\rH}}{4q}, \qquad
    \f{4q}{\sqrt{\zeta_\rH}}\leq M_\rH\leq\f{8q}{\sqrt{\zeta_\rH}},
\end{align}
and define $\nu_\rH \coloneqq b_\rH\delta/64$.
Since $\bE[\|Z\|_2^2] = q^2$, Eq.~\eqref{eq:gaussian smallest} and Markov's inequality applied to $\|Z\|_2^2$ show that $\sigma_{\min}(Z)\geq b_\rH$ and $\|Z\|_2\leq M_\rH$ except with probability at most $\zeta_\rH/8$.

We generate a finite-bit Gaussian matrix efficiently as follows. For independent uniform $u_1, u_2 \in (0,1)$, the variable $\sqrt{-\ln{u_1}} e^{2\pi iu_2}$ follows the standard complex Gaussian distribution.
We apply this construction independently to the $q^2$ entries. Set $\tau_\rH \coloneqq 2^{-\lceil\log{(64q^2/\zeta_\rH)}\rceil}$.
The event that every $u_1$ lies in $[2\tau_\rH, 1-2\tau_\rH]$ fails with probability at most $\zeta_\rH/16$, and the algorithm returns the identity in this case.
On $[\tau_\rH,1-\tau_\rH]\times[0,1]$, the sum of the two partial derivative bounds of $(u_1,u_2)\mapsto\sqrt{-\ln u_1}e^{2\pi iu_2}$ is at most $L_\rH\coloneqq4\pi\tau_\rH^{-2}$. 
Couple each of the $2q^2$ uniform variables to the midpoint of the interval determined by its first $B$ binary digits, where $B$ satisfies
\begin{align}
    2^{-B}\leq\min\Big\{\tau_\rH,\f{\nu_\rH}{4qL_\rH}\Big\}.
\end{align}
Evaluate each complex entry to absolute accuracy $\nu_\rH/(4q)$, using interval evaluation of the logarithm, square root, sine, and cosine~\cite{Moore2009IntroductionIntervalAnalysis}. 
The rational output $Z_{\mathrm{rat}}$ then obeys $\|Z_{\mathrm{rat}}-Z\|_2 \leq \nu_\rH$ whenever every $u_1$ lies in $[2\tau_\rH, 1-2\tau_\rH]$.
Since $u_1$ stays in $[\tau_\rH, 1-\tau_\rH]$, the logarithm and the square root avoid their singularities; then, the bit count and evaluation time are polynomial in $q$, $\log(1/\delta)$, and $\log(1/\zeta_\rH)$~\cite{Brent2010Moderncomputerarithmetic}.

We compute the QR factor of $Z_{\mathrm{rat}}$ by the same thresholded elimination as in the proof of \Cref{lem:isometry synthesis}, taking the threshold for skipping a pair and the row-update error to be at most $\nu_\rH/(16q^2)$ each. The procedure returns the identity in two cases: if $\|Z_{\mathrm{rat}}\|_2>2M_\rH$ at the outset, or if a final diagonal magnitude is below $b_\rH/4$. Otherwise, it records the phase on each row that makes the diagonal entry positive and outputs a unitary $Q_{\mathrm{num}}$ and a triangular $R_{\mathrm{num}}$ with positive diagonal. The phases and $R_{\mathrm{num}}$ are represented by the rational triangular entries and their positive square roots.

On the good events above, $R_{\mathrm{num}}$ has smallest singular value at least $b_\rH-2\nu_\rH>b_\rH/2$, so every diagonal magnitude is at least $b_\rH/2$, and neither of the two tests returns the identity. After at most $q^2$ steps, the elimination gives
\begin{align}
    \|Z_{\mathrm{rat}} - Q_{\mathrm{num}}R_{\mathrm{num}}\|_2
    \leq \nu_\rH.
\end{align}
The working entries are bounded by $2M_\rH+1$ and all nontrivial pivot denominators by the cutoff, so the precision and the loop bound are polynomial.

We now bound the resulting unitary error. Apply \Cref{lem:qr perturbation} with $X = Z$, $g = b_\rH$, and $E=Q_{\mathrm{num}}R_{\mathrm{num}}-Z$, whose Hilbert--Schmidt norm is at most $2\nu_\rH$ by the triangle inequality. This gives $\|Q_{\mathrm{num}}-Q(Z)\|_2\leq4\sqrt2\nu_\rH/b_\rH<\delta/8$. Extend $Q(Z)$ and $Q_{\mathrm{num}}$ by the identity on the complement of the chosen $q$-dimensional subspace of $\sC$, and denote the resulting unitaries by $V^\sC$ and $V_{\mathrm{num}}^\sC$, respectively.

Evaluate $V_{\mathrm{num}}^\sC$ to a rational matrix $A_0$ with $\|A_0-V_{\mathrm{num}}^\sC\|_2\leq\delta/256$. We apply \Cref{lem:isometry synthesis} with $V=V_{\mathrm{num}}^\sC$, input and output dimension $d_\sC$, and tolerance $\delta/2$. Since $d_\sC$ is a power of two, its construction acts on a register of dimension $2d_\sC$, which we identify with $\sC$ and one work qubit $\sW$, with the original input subspace corresponding to $\sW$ initialized to $\ket{0}^\sW$. The resulting circuit satisfies
\begin{align}
    \Big\|\widehat{V}^{\sC\sW}\big(\bI^\sC\otimes\ket{0}^\sW\big)
    - V_{\mathrm{num}}^\sC\otimes\ket{0}^\sW\Big\|_\infty
    \leq \delta/2.
\end{align}
Combining this with $\|V_{\mathrm{num}}^\sC-V^\sC\|_\infty<\delta/8$ proves Eq.~\eqref{eq:haar implementation}. The selected rotations and row phases can be evaluated to the required precision in polynomial time, and $d_\sC\leq2q$, so \Cref{lem:isometry synthesis} gives the stated gate count and classical running time. The two failure events above have total probability at most $3\zeta_\rH/16<\zeta_\rH$.

\end{proof}

\section{Preparatory lemmas for achievability}
\label{sec:common preparation}

The two appendices that follow present two distillation protocols, and we collect here the material they share. We briefly explain the Schur--Weyl duality in \Cref{sec:schur weyl}, and reduce the error of the distilled pairs by a concatenated stabilizer code in \Cref{sec:error reduction}.

\subsection{Schur--Weyl duality}
\label{sec:schur weyl}

Let $\cH$ be a Hilbert space of dimension $d$, and let $l \geq 1$ be an integer. The Schur--Weyl decomposition~\cite{goodman1998representations, harrow2005applications} of $l$ copies of $\cH$ reads
\begin{align}
\label{eq:schur weyl}
    \cH^{\otimes l} \cong \bigoplus_{\lambda \vdash l}\cW_\lambda(d)\otimes\cS_\lambda,
\end{align}
where $\lambda \vdash l$ means that $\lambda$ is a partition of $l$, $\cW_\lambda(d)$ is the space of the irreducible representation of the unitary group on $\cH$, which is zero when the number $\ell(\lambda)$ of rows of $\lambda$ exceeds $d$, and $\cS_\lambda$ is that of the symmetric group $S_l$ of degree $l$. We write $d_\lambda \coloneqq \dim\cS_\lambda$, and $\Pi_\lambda$ for the projector onto the $\lambda$ sector, carrying the register of $\cH$ as a superscript where several registers appear. The unitaries $U^{\otimes l}$ act as the identity on $\cS_\lambda$, and more generally we write $\pi_\lambda(X)$ for the action of $X^{\otimes l}$ on $\cW_\lambda(d)$, so that $\Pi_\lambda X^{\otimes l}\Pi_\lambda = \pi_\lambda(X)\otimes\bI^{\cS_\lambda}$. The permutations of the $l$ copies act as the identity on $\cW_\lambda(d)$ and irreducibly on $\cS_\lambda$. By Schur--Weyl duality, the operators commuting with every $U^{\otimes l}$ are those of the form $\bigoplus_\lambda\bI^{\cW_\lambda(d)}\otimes X_\lambda$, and they are spanned by the permutations.

We pad a Hilbert space with unused basis states whenever a construction requires its dimension to take a particular value. Such a padding is a local isometric embedding, and it leaves the nonzero eigenvalues of every state, its rank, and its entanglement unchanged. At $d \geq l$, every $\lambda \vdash l$ appears in Eq.~\eqref{eq:schur weyl}, since $\ell(\lambda) \leq l$.


\subsection{Error reduction by a concatenated code}
\label{sec:error reduction}

Each of the two protocols produces polynomially many pairs of qubits with a small infidelity, and the infidelity accumulates over the pairs. We here reduce the error by a QECC, which allows the final error of each protocol to be made exponentially small.
The QECC we use is a concatenated stabilizer code~\cite{knill1996concatenatedquantumcodes, Gottesman1996Classofquantumerror}. A short inner code is decoded by a table listing a correction for each syndrome, and an algebraic outer code repairs the inner blocks on which that table fails.
We choose the inner code as a Clifford code sampled approximately uniformly at random, which requires no code search or distance certification and whose table is built in polynomial time in both distillation protocols.

Both codes act as one-way distillation protocols. To see this, we recall the transpose trick of the maximally entangled state $\ket{\Phi}^{\sA\sB}$, which gives
\begin{align}
\label{eq:bilateral code}
    \big(\big(V^\top\big)^\sA\otimes\big(V^\dag\big)^\sB\big)\big(\bI^\sA\otimes E^\sB\big)\ket{\Phi}^{\sA\sB}
    = \big(\bI^\sA\otimes\big(V^\dag EV\big)^\sB\big)\ket{\Phi}^{\sA\sB}.
\end{align}
Here, $V$ and $V^\dag$ can be regarded as encoding and decoding unitaries, and $E$ as an error operator. 
In particular, when $V$ is Clifford and $E$ is Pauli, the conjugated error $V^\dag EV$ is again a Pauli operator. Alice and Bob apply $V^\top$ and $V^\dag$, respectively, and both measure the check qubits; Alice then sends her outcomes. Their difference is the syndrome of the conjugated error, which Bob corrects~\cite{Bennett1996Mixedstateentanglementandqec, Matsumoto2003Conversionquantumstabilizercodeentanglementdistillation}. The stabilizer formalism and the action of Clifford unitaries on Pauli operators are reviewed, e.g., in Ref.~{\cite[Chaps.~3 and~4]{gottesman1997stabilizercodesquantumerror}}.

In the following, we deal with pairs of qubits, one held by Alice and the other by Bob. The two codes act on such pairs, and the target of each of them is the Bell state $\Phi_2$. All communication goes from Alice to Bob, and neither the input state nor any description of it enters the construction.
The construction is controlled by two parameters, a rate loss $a$ shared by the two codes and the length $N$ of the outer code. The length $L$ of the inner code is then determined by $a$ and $N$, and the error of the output decreases exponentially in $N$.

\begin{lemma}[Error reduction by a concatenated code]
\label{lem:error reduction}
Let $a \in (0, 1/32]$ and $\zeta_\rc \in (0,1)$, and let $N$ and $S$ be integers such that $N \geq 4/a$ and $S \geq 1$. Set
\begin{align}
\label{eq:code parameters}
    \lambda_\rc \coloneqq \lceil\log{(16/a)}\rceil, \qquad 
    u_\rc \coloneqq a/(64\lambda_\rc), \qquad 
    L \coloneqq \max\big\{\lceil 16\lambda_\rc^2/a\rceil, \, 2\lceil\log{(N+1)}\rceil\big\},
\end{align}
\begin{align}
\label{eq:code lengths}
    k_\rc \coloneqq LN, \qquad 
    m_\rc \coloneqq \big(L - \lceil aL\rceil\big)\big(N - 2\lfloor aN\rfloor\big).
\end{align}
Then, there is a one-way LOCC protocol that consumes $S$ code blocks of $k_\rc$ pairs each and outputs $Sm_\rc$ pairs. Its inner code is a random Clifford code drawn at each use of the protocol, and its outer code is computed from $a$ and $N$ by a classical randomized algorithm that succeeds with probability at least $1 - \zeta_\rc$; when this algorithm fails, the protocol returns a fixed state of the same size. The following statements hold.
\begin{enumerate}
    \item Suppose that, within each code block, the state is a product across its $k_\rc$ pairs, each of which has fidelity at least $1 - u_\rc$ with $\Phi_2$. Then, when the algorithm succeeds, the output, averaged over the random inner code, has fidelity at least $1 - S2^{-aN/2}$ with $\Phi_2^{\otimes Sm_\rc}$.
    \item The numbers of input and output pairs obey $m_\rc \geq (1-2a)^2k_\rc$.
    \item The encoding, the syndrome extraction, and the correction use $\poly(S, k_\rc)$ one- and two-qubit quantum gates, and the classical computation, comprising the construction and decoding of the two codes, runs in time
    \begin{align}
    \label{eq:code cost}
        \poly\big(S, N, L, \log{(1/\zeta_\rc)}\big)\exp\big(\cO(\log^2{(1/a)})\big)N^{\cO(a)}.
    \end{align}
\end{enumerate}
\end{lemma}

\begin{proof}[Proof of \Cref{lem:error reduction}]

We first describe the inner code and bound the probability that an inner block carries a logical error, then describe the outer code and bound the failure probability of one code block, and finally collect the lengths and the costs. Throughout we set
\begin{align}
\label{eq:inner code}
    s_\rc \coloneqq \lceil aL\rceil, \qquad 
    w_\rc \coloneqq L - s_\rc, \qquad 
    v_\rc \coloneqq \big\lfloor aL/(8\lambda_\rc)\big\rfloor, \qquad
    r_\rc \coloneqq \lfloor aN\rfloor,
\end{align}
so that $k_\rc = LN$ and $m_\rc = w_\rc(N - 2r_\rc)$.

Divide each code block into $N$ inner blocks of $L$ pairs. For each $i = 1, \ldots, N$, Alice draws a uniform $L$-qubit Clifford $V_\rc$, independently for distinct $i$, and sends the seed that specifies it; the same $V_\rc$ is used at the $i$-th inner block of every code block, so that $N$ seeds are drawn in total. Sampling and circuit compilation use Ref.~\cite{Koenig2014EfficientlyselectarbitraryClifford}, with finite sampling as in \Cref{sec:numerics}. The stabilizer is generated by $V_\rc Z_1V_\rc^\dag,\ldots,V_\rc Z_{s_\rc}V_\rc^\dag$, and the code is applied through Eq.~\eqref{eq:bilateral code}. Bob enumerates the Paulis of weight at most $v_\rc$ and stores one correction for each syndrome, choosing the first in a fixed ordering when several share a syndrome; a syndrome absent from the table receives a fixed correction.

The uniform Clifford also supplies the twirl needed in the analysis. Decompose it as a uniform Pauli factor times an independent representative modulo Paulis, in this order. The transpose and the adjoint in Eq.~\eqref{eq:bilateral code} reverse the order of the two factors, and the Pauli factor therefore acts first in both laboratories, where averaging it is the bilateral Pauli twirl on the inner block. Under this average, each of the $L$ pairs becomes Bell diagonal with weight at least $1 - u_\rc$ on $\Phi_2$, independently of the others. Pauli factors change conjugated stabilizers only by signs, while the syndrome differences and the correction table depend only on the representative's symplectic action. We may thus analyze each inner block as an independent Pauli error of marginal probability at most $u_\rc$ per pair, whatever that action is, and no separate twirling operation or seed is required.

We next bound the size of the table. Putting $y \coloneqq 2v_\rc/L \leq a/(4\lambda_\rc)$,
\begin{align}
\label{eq:enumeration}
    \sum_{j\leq2v_\rc}\binom{L}{j}3^j
    \leq 2^{L[h_2(y)+y\log3]}
    \leq 2^{aL/2},
\end{align}
where the last step uses $h_2(y)+y\log3\leq y\log(3e/y)$ together with $\log(12e\lambda_\rc/a)\leq2\lambda_\rc$, which follows from Eq.~\eqref{eq:code parameters} and $\lambda_\rc \geq 9$. The table is therefore built without enumerating all $4^L$ Pauli errors.

An inner block carries a logical error only if the actual Pauli error has weight above $v_\rc$ or the table returns a different Pauli with the same syndrome. For a fixed actual error of weight at most $v_\rc$, the product of the two Paulis in a collision is a fixed nonidentity Pauli, which for a uniform $s_\rc$-dimensional stabilizer lies in the normalizer with probability $(2^{2L-s_\rc}-1)/(2^{2L}-1)\leq2^{1-s_\rc}$~{\cite[Chap.~24]{wilde2013QItheory}}. A union bound over the table thus bounds the collision probability by $2^{1-s_\rc+aL/2}$, uniformly in the actual error, by Eq.~\eqref{eq:enumeration}. 
The pairs being independent with error probability at most $u_\rc$, the probability of more than $v_\rc$ errors is at most $(eLu_\rc/(v_\rc+1))^{v_\rc+1}$. Here $v_\rc+1>aL/(8\lambda_\rc)\geq2\lambda_\rc$ and $eLu_\rc/(v_\rc+1)<e/8$, so the probability $p_\rb$ that an inner block carries a logical error, averaged over the uniform Clifford $V_\rc$, obeys
\begin{align}
\label{eq:inner failure}
    p_\rb \leq (e/8)^{2\lambda_\rc}+2^{1-s_\rc+aL/2}
    \leq 2^{-2\lambda_\rc}
    \leq \f{a}{16e}.
\end{align}
Indeed the first term is at most $2^{-3\lambda_\rc}$ by $(e/8)^2 < 1/8$, the second at most $2^{1-8\lambda_\rc^2} \leq 2^{-3\lambda_\rc}$ by $aL \geq 16\lambda_\rc^2$, and $2^{-\lambda_\rc} \leq a/16$. The protocol need not identify such an inner block.

We turn to the outer code, which treats each inner block as a symbol of $w_\rc$ qubits. Since $w_\rc \geq L/2$, we have $2^{w_\rc} \geq N+1$, and we may work over $\bF_{2^{w_\rc}}$ and choose distinct $z_1, \ldots, z_N$ there. Setting
\begin{align}
\label{eq:grs}
    \varsigma_i^2 \coloneqq \bigg(\prod_{j \neq i}(z_i - z_j)\bigg)^{-1}, \qquad 
    \scG \coloneqq \big\{(\varsigma_if(z_i))_{i=1}^{N} : \deg{f} < r_\rc\big\},
\end{align}
where square roots are unique in this field, Lagrange interpolation gives $\sum_i\varsigma_i^2g(z_i) = 0$ for every $g$ of degree at most $N-2$; applying this to a product of two polynomials of degree below $r_\rc$ shows $\scG \subseteq \scG^\perp$. The CSS construction with both check spaces equal to $\scG$ therefore has parameters $[[N, N-2r_\rc, \geq r_\rc+1]]_{2^{w_\rc}}$, the distance because $\scG^\perp$ is a generalized Reed--Solomon code of dimension $N-r_\rc$. Its dual multipliers are the same $\varsigma_i$, since $\varsigma_i^{-1}\big(\prod_{j\neq i}(z_i-z_j)\big)^{-1}=\varsigma_i$. This uses the chosen length $N$ only, and not every element of the field as an evaluation point.

A field of degree $w_\rc$ is obtained by testing random monic polynomials for irreducibility~{\cite[Chap.~14]{vonzurGathen2013Moderncomputeralgebra}}, among which the irreducible ones number at least $2^{w_\rc}/(2w_\rc)$. Each trial thus succeeds with probability at least $1/(2w_\rc)$, and at least one of $\lceil2w_\rc\ln(1/\zeta_\rc)\rceil$ independent trials succeeds with probability at least $1 - \zeta_\rc$. This is the only step at which the construction can fail. Alice chooses and sends the field description used jointly.

To decode either syndrome, Bob first finds a vector $\hat{x}$ with that syndrome by Gaussian elimination, so that $\hat{x}-x\in\scG^\perp$ for the true error $x$. After dividing the coordinates of $\hat{x}$ by $\varsigma_i$, he applies the Welch--Berlekamp decoder with dimension $N-r_\rc$ and radius $\lfloor r_\rc/2\rfloor$, restores the multipliers, and subtracts the decoded codeword from $\hat{x}$. This returns the correct error whenever its weight is at most that radius, because $\scG^\perp$ has distance $r_\rc+1$. The algorithm, its correctness, and its $\cO(N^3)$ field-operation bound are Ref.~\cite[Algorithm~12.1.1 and Theorems~12.1.3 and~12.1.6]{Guruswami2025Essentialcodingtheory}. Bob verifies the candidate's syndrome and weight, and uses a fixed correction if decoding or verification fails. Binary expansion in trace-dual field bases converts the checks to commuting qubit Paulis, and symplectic Gaussian elimination produces the encoding Clifford. As $N \geq 4/a$ implies $\lfloor r_\rc/2\rfloor \geq aN/4$, the outer code corrects any pattern of at most $aN/4$ inner blocks carrying an arbitrary logical error on their $w_\rc$ qubits.

We now bound the failure probability of one code block, with the field description fixed. Distinct inner blocks receive disjoint parts of the product state and independent seeds, and the events that they carry a logical error are consequently independent, with probabilities bounded by $p_\rb$ of Eq.~\eqref{eq:inner failure}. The Chernoff bound gives
\begin{align}
\label{eq:outer failure}
    \Pr[\text{outer decoding fails in a code block}]
    \leq \bigg(\f{4ep_\rb}{a}\bigg)^{aN/4}
    \leq 2^{-aN/2},
\end{align}
and on the complementary event the code block outputs $\Phi_2^{\otimes m_\rc}$ exactly. This estimate averages the seeds of that code block alone and holds uniformly in the field description, hence also after averaging it. Code blocks may remain correlated, both through the shared seeds and through their inputs. If $\Pi_j$ is the target Bell projector for the $j$-th code block, then $\bI-\prod_j\Pi_j \leq \sum_j(\bI-\Pi_j)$, because these projectors act on disjoint registers and commute. Taking the output expectation bounds the total infidelity by the sum of the $S$ individual infidelities, without a joint classical-error representation or independence between code blocks.

The seeds are sampled only approximately, and we leave room for this. Equation~\eqref{eq:inner failure} also gives $p_\rb \leq 2^{-2\lambda_\rc} \leq a^2/256$, with which the right-hand side of Eq.~\eqref{eq:outer failure} improves to $2^{-aN}$, and the infidelity with exactly uniform seeds is at most $S2^{-aN}$. We sample each of the $N$ seeds within total variation distance $2^{-aN/2}/(2N)$ of the uniform distribution, with polynomial overhead. As the output fidelity changes by at most the total variation distance of the seeds, the total infidelity is at most $S2^{-aN} + \f{1}{2}2^{-aN/2} \leq S2^{-aN/2}$, which proves the first statement.

The lengths follow from $s_\rc \leq aL + 1 \leq 2aL$, which holds as $aL \geq 16\lambda_\rc^2 \geq 1$, and from $r_\rc \leq aN$. Hence $w_\rc/L \geq 1 - 2a$ and $(N-2r_\rc)/N \geq 1-2a$, which is the second statement.

It remains to collect the costs. By Eq.~\eqref{eq:enumeration}, the table and the Clifford circuit of one inner block are constructed in $\poly(L)2^{\cO(aL)}$ operations, and the $N$ tables serve every code block. From Eq.~\eqref{eq:code parameters} we have $aL = \cO(\lambda_\rc^2 + a\log{N})$ and $\lambda_\rc = \cO(\log{(1/a)})$, and the bound $2^{\cO(aL)} = \exp(\cO(\log^2{(1/a)}))N^{\cO(a)}$ stated in Eq.~\eqref{eq:code cost} follows. The irreducibility tests cost $\poly(w_\rc, \log{(1/\zeta_\rc)})$, each field operation has $\poly(w_\rc)$ bit cost, and the outer decoding uses $\cO(SN^3)$ field operations over the two syndromes of each of the $S$ code blocks. The inner Clifford acts on $L$ qubits and the outer encoding Clifford on $w_\rc N$ qubits, so that the circuits over all $S$ code blocks have $\poly(S, k_\rc)$ gates.

\end{proof}


\begin{figure}
\adjustbox{width=170mm}{%
\begin{tikzpicture}[
    lab/.style={font=\small, align=center},
    seg/.style={draw, line width=1.2pt, fill=white, align=center, font=\small},
    cell/.style={draw, line width=0.6pt, fill=white, minimum size=0.40cm, inner sep=0pt},
    hl/.style={draw, line width=0.6pt, fill=black!15, minimum size=0.40cm, inner sep=0pt},
    arr/.style={-Latex, line width=1.0pt}
]
\foreach \y/\name in {0.9/1, 0.3/2, -0.9/{k_\rc}}{
  \foreach \x in {0,1,2,3,4,5,6,7}{
    \node[cell] at (1.8+\x*0.46,\y) {};
  }
  \node[hl] at (1.8+3*0.46,\y) {};
  \node[lab, anchor=east] at (1.45,\y) {batch $\name$};
}
\node at (1.8+3*0.46,-0.3) {$\vdots$};
\draw[line width=1.0pt, dashed] (1.8+3*0.46-0.28,1.15) rectangle (1.8+3*0.46+0.28,-1.15);
\node[lab, anchor=north] at (3.18,-1.6) {$m_\rb$ pairs per batch};
\draw[arr] (5.6,0) -- (8.2,0);
\node[lab, anchor=south] at (6.8,0.5) {one pair \\ from each batch};
\node[seg, minimum width=1.0cm, minimum height=0.8cm] at (8.8,0) {$L$};
\node[seg, minimum width=1.0cm, minimum height=0.8cm] at (10.0,0) {$L$};
\node at (10.9,0) {$\cdots$};
\node[seg, minimum width=1.0cm, minimum height=0.8cm] at (11.8,0) {$L$};
\node[lab, anchor=north] at (10.3,-1.6) {one code block: $k_\rc = LN$ pairs\\ in a product state,\\ in $N$ inner blocks of $L$ pairs};
\draw[arr] (12.6,0) -- (15.2,0);
\node[lab, anchor=south] at (13.8,0.50) {concatenated code};
\node[seg, minimum width=2.2cm, minimum height=0.8cm] at (16.5,0) {$m_\rc$ pairs};
\node[lab, anchor=north] at (16.6,-1.6) {output of the code block};
\end{tikzpicture}}
    \caption{The code blocks of \Cref{cor:batch distillation}. Each of the $k_\rc$ batches yields $m_\rb$ pairs of qubits, and the pairs with the same index in all the batches form a code block of $k_\rc = LN$ pairs, which is in a product state. The concatenated code of \Cref{lem:error reduction} divides each code block into $N$ inner blocks of $L$ pairs and turns it into $m_\rc$ pairs.}
    \label{fig:code blocks}
\end{figure}
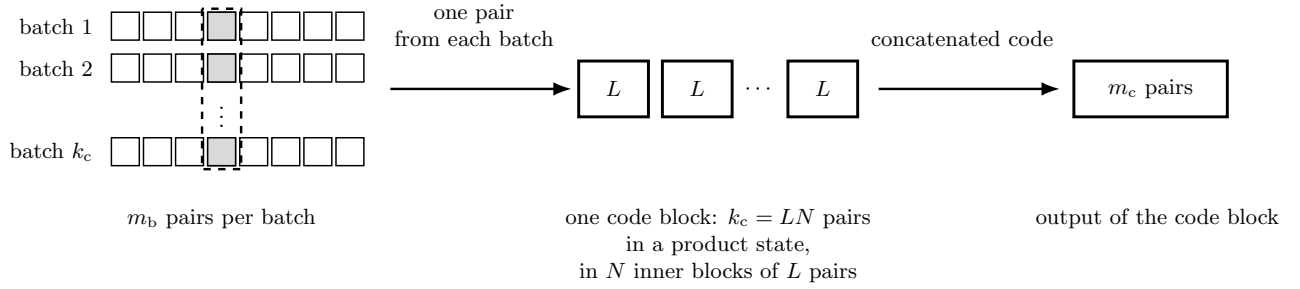


Both protocols, in Secs.~\ref{sec:schur coding} and~\ref{sec:achievability}, apply a channel to each of $k_\rc$ disjoint batches of the shared copies, and every batch produces $m_\rb$ pairs of qubits. The state of the $m_\rb$ pairs of one batch is not a product across them in general, whereas \Cref{lem:error reduction} requires a product state within each code block. A code block is thus formed across the batches, by taking the pair of the same index from each. The batches being independent, the state of such a code block is a product across its pairs. The following corollary collects this step.

\begin{corollary}[Distillation from a batch protocol]
\label{cor:batch distillation}
Let $a \in (0, 1/32]$ and $\varepsilon_\rc \in (0,1)$, and let $K$ and $m_\rb$ be positive integers. Set
\begin{align}
\label{eq:code target}
    N \coloneqq \bigg\lceil\f{2}{a}\bigg\lceil\log{\bigg(\f{4m_\rb}{\varepsilon_\rc}\bigg)}\bigg\rceil\bigg\rceil, \qquad 
    \zeta_\rc \coloneqq \f{\varepsilon_\rc}{2},
\end{align}
and take $u_\rc$, $L$, $k_\rc$, and $m_\rc$ of Eqs.~\eqref{eq:code parameters} and~\eqref{eq:code lengths} at this $a$ and $N$. 
Let $\rho^{\sA\sB}$ be a state, and let $\Lambda^{\sA^K\sB^K\to\sS\hat{\sS}}$ be a one-way LOCC channel, with $\sS$ and $\hat{\sS}$ of $m_\rb$ qubits each, such that every one of the $m_\rb$ pairs of $\Lambda^{\sA^K\sB^K\to\sS\hat{\sS}}\big((\rho^{\sA\sB})^{\otimes K}\big)$ has fidelity at least $1 - u_\rc$ with $\Phi_2$. Consider the one-way LOCC protocol on $k \coloneqq Kk_\rc$ copies that partitions them into $k_\rc$ batches of $K$ copies and applies $\Lambda^{\sA^K\sB^K\to\sS\hat{\sS}}$ to each batch independently. It then applies \Cref{lem:error reduction} at this $a$, $N$, and $\zeta_\rc$ to the $S \coloneqq m_\rb$ code blocks, one for each index of the pairs. Then, the following hold.
\begin{enumerate}
    \item The protocol distills $m \coloneqq m_\rb m_\rc$ ebits from $\rho^{\sA\sB}$ at error $\varepsilon_\rc$ in the sense of \Cref{def:protocol}.
    \item Its rate obeys $\f{m}{k} \geq \f{m_\rb}{K}(1-4a)$.
\end{enumerate}
\end{corollary}

\begin{proof}[Proof of \Cref{cor:batch distillation}]

As $4m_\rb/\varepsilon_\rc \geq 4$, the inner ceiling in Eq.~\eqref{eq:code target} is at least two, and hence $N \geq 4/a$, which is the condition of \Cref{lem:error reduction}. The batches consist of disjoint sets of copies of $\rho^{\sA\sB}$, to each of which $\Lambda^{\sA^K\sB^K\to\sS\hat{\sS}}$ is applied independently. For each $j \in \{1, \ldots, m_\rb\}$, the $k_\rc$ pairs collected into the $j$-th code block come from distinct batches. After the classical records of the batches, which the code does not use, are discarded, these pairs are thus in a product state, and each of them has infidelity at most $u_\rc$ with $\Phi_2$ by the conditions. Every code block therefore meets the conditions of the first statement of \Cref{lem:error reduction}, which at $S = m_\rb$ bounds the infidelity of the output with $\Phi_2^{\otimes m}$ by
\begin{align}
    m_\rb2^{-aN/2} \leq \f{\varepsilon_\rc}{4}
\end{align}
when the construction of the outer code succeeds. This construction fails with probability at most $\zeta_\rc = \varepsilon_\rc/2$, which does not change the number $m$ of output pairs, and the fidelity decreases by at most this probability. The infidelity of the output is thus at most $3\varepsilon_\rc/4 < \varepsilon_\rc$, which is (1). For (2), the second statement of \Cref{lem:error reduction} gives $m_\rc/k_\rc \geq (1-2a)^2 \geq 1-4a$, and $m = m_\rb m_\rc$ and $k = Kk_\rc$ yield the claim.

\end{proof}

\section{Achievability: Universal Schur coding with polynomially many copies}
\label{sec:schur coding}

We prove sample-efficient entanglement distillation for states close to a state of polynomial rank, without restricting the marginal ranks. The coding and entropy estimates are imported from Ref.~\cite{UQC}; we give the additional support-enlargement and perturbation arguments needed for states that are only close to a state of polynomial rank. We present these tools in \Cref{sec:preparatory lemmas schur}, and combine them into the sample-efficient achievability result in \Cref{sec:sample efficient result}.

\subsection{Schur coding tools}
\label{sec:preparatory lemmas schur}

In \Cref{sec:Schur information spectrum and CG channels}, we recall from Ref.~\cite{UQC} the Schur information spectrum, the Clebsch--Gordan channel picture, and the universal coding bound, which bounds the infidelity by the weight of the information spectrum below a threshold. We bound this weight by the R\'{e}nyi coherent information in \Cref{sec:entropy bounds}. In \Cref{sec:Uniform recovery on invariant supports}, we show that a single channel recovers Bell pairs from every permutation-invariant state whose weight lies mostly on an invariant projector of small rank. In \Cref{sec:schur subspaces}, we enlarge the Schur supports to include deviations on a small fraction of the copies, which makes this recovery stable under perturbations of the input state. These tools are combined in \Cref{sec:sample efficient result}, where the remaining error of the distilled pairs is reduced by the concatenated code of \Cref{sec:error reduction}.

\subsubsection{Schur information spectrum and Clebsch--Gordan channels}
\label{sec:Schur information spectrum and CG channels}

We apply the Schur--Weyl decomposition of \Cref{sec:schur weyl} to $\sA^l$, $\sB^l$, and $\sE^l$, and use $\lambda$, $\mu$, and $\nu$ for their labels, respectively.
Let $\ket\rho^{\sA\sB\sE}$ purify $\rho^{\sA\sB}$, with
$d_\sE=\rank(\rho^{\sA\sB})$, and write $g_{\lambda\mu\nu}$ for the
Kronecker coefficient, the multiplicity of $\cS_\nu$ in
$\cS_\lambda\otimes\cS_\mu$.
The projector $\Pi_\eta^{\lambda\mu}$ onto the $\eta$-isotypic component of $\cS_\lambda\otimes\cS_\mu$ under the simultaneous permutations of the copies has rank $g_{\lambda\mu\eta}d_\eta$.
Block operators are extended by the identity on the unitary factors
and zero on other local sectors. For an integer $\hat{r} \geq 1$, put $p_{\lambda\mu\nu} \coloneqq \big\|(\Pi_\lambda^{\sA^l}\otimes\Pi_\mu^{\sB^l} \otimes\Pi_\nu^{\sE^l})(\ket\rho^{\sA\sB\sE})^{\otimes l}\big\|^2$ and 
\begin{align}
\label{eq:information spectrum}
    \cV_{\lambda\mu}^{(\hat{r})}(d) 
    \coloneqq\sum_{\substack{\eta\vdash l:\ell(\eta) \leq \hat{r}\,,\ d_\eta\leq d}} g_{\lambda\mu\eta}d_\eta, \qquad
    Z_l^{(\hat{r})}(\lambda,\mu,\nu)
    \coloneqq \log{d_\mu} - \log\cV_{\lambda\mu}^{(\hat{r})}(d_\nu).
\end{align}
The numbers $p_{\lambda\mu\nu}$ form a probability distribution of the labels $(\lambda,\mu,\nu)$, which is independent of the purification, and we write $\Pr_\rho$ and $\bE_\rho$ for its probabilities and expectations, at the number of copies appearing in the event.
The purification theorem of Ref.~\cite[Theorem~14]{UQC}
implies that positive probability requires $g_{\lambda\mu\nu}>0$ and
$\ell(\nu)\leq\rank(\rho^{\sA\sB})$.  Thus $Z_l^{(\hat{r})}$ is finite on
the support when $\rank(\rho^{\sA\sB}) \leq \hat{r}$.  It is exactly
$J_l^{(\hat{r})}$ in the notation of Ref.~\cite{UQC}.

We here introduce the Clebsch--Gordan (CG) channel and the coding idea. For fixed local labels $(\lambda,\mu)$, the CG
branch isometries and their associated channels are given by
\begin{align}
\label{eq:cg branch channel}
    (V_{\mu,\nu}^{\lambda,\alpha})^\dagger : \cS_\lambda\longrightarrow\cS_\mu\otimes\cS_\nu, \qquad 
    \cN_{\lambda\to\mu}^{(\nu,\alpha)}(X) 
    \coloneqq \tr_{\cS_\nu}\Big[(V_{\mu,\nu}^{\lambda,\alpha})^\dagger X V_{\mu,\nu}^{\lambda,\alpha}\Big], \qquad 
    1\leq\alpha\leq g_{\lambda\mu\nu}.
\end{align}
Here $\cS_\lambda$ is the channel input, identified with a reference
copy of Alice's symmetric-group register.  These are the channels of
Ref.~\cite[Appendix~C.1]{UQC}, written with $\cS$ and $\cW$ in place
of that paper's $\mathcal P$ and $\mathcal Q$ for the permutation and
unitary factors.  Each is trace preserving and sends
$\bI^{\cS_\lambda}/d_\lambda$ to $\bI^{\cS_\mu}/d_\mu$.
After conditioning on $(\lambda,\mu)$ and tracing out the unitary
factors, the conditional permutation state $\omega_{\lambda\mu}$ has the representation
\begin{align}
\label{eq:cg physical state}
 \omega_{\lambda\mu}
   =(\rid\otimes\cN_{\lambda\to\mu})
       (\Phi_{d_\lambda}).
\end{align}
The physical channel $\cN_{\lambda\to\mu}$ contains the unknown
state-dependent weights and coherences between multiplicity indices;
it need not be a classical mixture of the branch channels
in Eq.~\eqref{eq:cg branch channel}~\cite[Eq.~(C5)]{UQC}.

The coding idea is to replace the unknown physical channel by a
reference CG channel that is flat on the retained error space: its
branch weights are proportional to $d_\nu$~\cite[Eq.~(C7)]{UQC}.
After resolving the local Schur labels, Alice measures her symmetric-group register into randomly rotated code blocks of the target dimension and communicates the outcome to Bob, whose recovery is fixed by this reference channel and the outcome. Its performance is controlled by the number $\cV_{\lambda\mu}^{(\hat{r})}(d_\nu)$ of retained error directions relative to $d_\mu/q$, where $q$ is the Schmidt rank of the target maximally entangled state.
We import the resulting guarantee; the encoding and recovery maps are developed in Ref.~\cite[Appendix~C]{UQC}.

\begin{theorem}[Universal Schur coding bound {\cite[Theorem~16]{UQC}}]
\label{thm:schur coding bound}
For integers $\hat{r},l\geq1$ and $q\geq2$, and a real number $\vartheta\geq0$ with $q2^{-l\vartheta}\leq1$, there is a common one-way LOCC channel $\Lambda_{l,q,\vartheta}^{\sA^l\sB^l\to\sS\hat{\sS}}$, with
$\sS$ and $\hat{\sS}$ of dimension $q$, such that every state
$\rho^{\sA\sB}$ with $\rank(\rho^{\sA\sB}) \leq \hat{r}$ satisfies
\begin{align}
\label{eq:schur coding bound}
 1-\rF(\Lambda_{l,q,\vartheta}(\rho^{\otimes l}),\Phi_q)
 \leq\Pr_\rho[Z_l^{(\hat{r})}<l\vartheta]+q2^{-l\vartheta}.
\end{align}
The channel depends only on the local spaces and $(\hat{r},l,q,\vartheta)$.
\end{theorem}
We suppress the channel's register superscripts and those of
$\Phi_q^{\sS\hat{\sS}}$ when unambiguous.
This is the universal distillation theorem of Ref.~\cite{UQC}, with
$(k,K,r,L)=(l,q,\hat{r},2^{l\vartheta}/q)$ and its stronger bound
$p+(1-p)/L$ replaced by $p+1/L$.

\subsubsection{Entropy bounds on the information spectrum}
\label{sec:entropy bounds}

We bound the tail of the information spectrum by the sandwiched R\'{e}nyi coherent information of \Cref{sec:notation}, through the Petz--R\'{e}nyi divergence. For a state $\rho$ and a positive semidefinite operator $\sigma$, the Petz--R\'{e}nyi divergence is
\begin{align}
\label{eq:petz}
    D_\beta^{\rP}(\rho\|\sigma)
    \coloneqq \f{1}{\beta-1}\log\tr[\rho^\beta\sigma^{1-\beta}].
\end{align}
This divergence has value $+\infty$ when the trace vanishes, and its reference $\sigma$ need not be normalized. We abbreviate the regularized quantity by
\begin{align}
\label{eq:shifted petz notation}
 I^{\rP}_{\beta,\varpi}(\rho)
 \coloneqq D_\beta^{\rP}(\rho^{\sA\sB}\|
            \bI^\sA\otimes(\rho^\sB+\varpi\bI^\sB)),
 \qquad\varpi\geq0.
\end{align}
This is $\mathcal{I}_{1-s,\tau}$ of Ref.~\cite{UQC}, with
$s=1-\beta$ and $\tau=\varpi$.  On a purification, its unshifted
value is a Petz conditional entropy~\cite[Eq.~(E2)]{UQC}:
\begin{align}
\label{eq:petz conditional entropy}
 I^{\rP}_{\beta,0}(\rho)
 =-D_{2-\beta}^{\rP}(\rho^{\sA\sE}\|
                         \bI^\sA\otimes\rho^\sE).
\end{align}
For order $2-\beta>1$, the same trace formula is evaluated on the
reference support, with value $+\infty$ unless the state is supported
there.  The next estimates connect this entropy to the spectrum.

\begin{lemma}[Tail bound on the information spectrum {\cite[Lemmas~24 and~26]{UQC}}]
\label{lem:moment bound}
\label{lem:petz comparison}
Let $\hat{r},l\geq1$ be integers, $\beta\in[1/2,1)$, $\varpi>0$, and $\vartheta\geq0$, and let $\rho^{\sA\sB}$ be a state with $\rank(\rho^{\sA\sB}) \leq \hat{r}$.
Set $s=1-\beta$, $v=s/(1+s)$, and
\begin{align}
\label{eq:moment constant}
    A_{\hat{r},\varpi}
    \coloneqq\f{\hat{r}}{\varpi} + \f{\hat{r}^2}{2\varpi^2},\qquad
    C_l \coloneqq |\{\lambda\vdash l\}|\exp(A_{\hat{r},\varpi}).
\end{align}
Then,
\begin{align}
    \label{eq:moment bound}
    \bE_\rho\Big[2^{-vZ_l^{(\hat{r})}}\Big]
    \leq C_l^v2^{-vlI^{\rP}_{\beta,\varpi}(\rho)}, \qquad
    \Pr_\rho[Z_l^{(\hat{r})}<l\vartheta]
    \leq C_l^v2^{-vl(I^{\rP}_{\beta,\varpi}(\rho)-\vartheta)}, \qquad
    2^{-sI^{\rP}_{\beta,\varpi}(\rho)}
    \leq 2^{-s\widetilde I_\beta(\sA\rangle\sB)_\rho} + (\hat{r}\varpi)^s.
\end{align}
\end{lemma}
The first estimate is Ref.~\cite[Lemma~24]{UQC} with
$(k,r,\tau,s)=(l,\hat{r},\varpi,1-\beta)$; Markov's inequality gives
the second. The third is Ref.~\cite[Lemma~26]{UQC}, using the duality $\widetilde H_{\beta/(2\beta-1)}(\sA|\sE)_\rho=\widetilde I_\beta(\sA\rangle\sB)_\rho$ of Eq.~\eqref{eq:duality} for $\beta>1/2$ and its min-entropy endpoint for $\beta=1/2$.
The optimized entropy $\widetilde H_\alpha^\uparrow$ of Ref.~\cite{UQC} is our $\widetilde H_\alpha$: the same variational formula at order $\alpha>1$, with negative powers on the reference support and divergence
$+\infty$ unless the state is supported there.
Here $\exp$ is natural, so
$\log{C_l}=\log|\{\lambda\vdash l\}|+A_{\hat{r},\varpi}/\ln2$.

\subsubsection{Uniform recovery on invariant supports}
\label{sec:Uniform recovery on invariant supports}

The smoothing argument uses the following consequence of the uniform
recovery estimate in Ref.~\cite[Lemma~10 and proof of Theorem~16]{UQC}.

\begin{lemma}[Uniform recovery on an invariant support]
\label{lem:uniform support estimate}
Let $q \geq 2$ be an integer and $c \in [0, 1/q]$, and let $P$ be a projector commuting with every $(U^\sA)^{\otimes l}\otimes(V^\sB)^{\otimes l}$ and with the simultaneous permutations of the copies. Suppose that its blocks $P^{\lambda\mu}$ on $\cS_\lambda\otimes\cS_\mu$ satisfy $\rank(P^{\lambda\mu}) \leq c d_\mu$. There is a common one-way LOCC channel $\Lambda_{P,q}$
such that every simultaneously permutation-invariant state $\omega$
on $\sA^l\sB^l$ obeys
\begin{align}
\label{eq:imported support estimate}
 1-\rF(\Lambda_{P,q}(\omega),\Phi_q)
 \leq\left(\sqrt{\tr[(\bI-P)\omega]}+\sqrt{qc}\right)^2.
\end{align}
The channel depends only on $P,q$ and the local spaces.
\end{lemma}

\begin{proof}[Proof of \Cref{lem:uniform support estimate}]

For a nonzero permutation block, write $d=d_\lambda$, $D=d_\mu$,
and $N=\rank(P^{\lambda\mu})$.  Invariance and Schur's lemma give
\begin{align}
     \tr_{\cS_\mu}[P^{\lambda\mu}] = \f{N}{d}\bI^{\cS_\lambda},\qquad
     \tr_{\cS_\lambda}[P^{\lambda\mu}] = \f{N}{D}\bI^{\cS_\mu}.\nonumber
\end{align}
Consequently, if $\{\ket{f_i}\}_{i=1}^N$ is an orthonormal basis of
its range and $\ket{f_i}=(\bI^{\cS_\lambda}\otimes F_i)\ket{\Phi_d}$, then
\begin{align}
\label{eq:uniformity identities}
    \tr[F_i^\dag F_j] = d \delta_{ij}, \qquad
    \sum_iF_i^\dag F_i = N\bI^{\cS_\lambda}, \qquad
    \sum_iF_iF_i^\dag = \f{dN}{D}\bI^{\cS_\mu}.
\end{align}
These are the conditions of the uniform recovery estimate in
Ref.~\cite[Lemma~10, Eq.~(A35)]{UQC}.  The complete-channel bound
in its proof of Theorem~16, Eq.~(C25), then gives infidelity at most
$qN/D$ on every invariant state supported in this block.
Indeed, that estimate is uniform over every positive coefficient
matrix on the error space; it thus also covers a selected
subspace of a multiplicity space.  The dimension condition $q\leq d$
follows from $D\leq dN$ and $qN\leq D$.
The local blocks share the same Alice operation, as in that proof,
so their bounds combine into one channel with error at most $qc$
on every invariant state supported on $P$.

For a general invariant $\omega$, put
$M=\Lambda_{P,q}^\dag(\bI-\Phi_q)$, so $0\leq M\leq\bI$.
The normalized state $P\omega P/\tr[P\omega]$, when defined, remains
permutation-invariant.
The support guarantee gives $\tr[MP\omega P]\leq qc \tr[P\omega]$, and hence
\begin{align}
 \sqrt{\tr[M\omega]}
 &\leq\|\sqrt{M} P\sqrt{\omega}\|_2
        +\|\sqrt{M}(\bI-P)\sqrt{\omega}\|_2
 \leq\sqrt{qc}+\sqrt{\tr[(\bI-P)\omega]}.
\end{align}
The case $\tr[P\omega]=0$ is immediate.  The argument applies even when
$P$ and $\omega$ do not commute.

\end{proof}

\subsubsection{Enlargement of Schur supports}
\label{sec:schur subspaces}

For $\vartheta\geq0$ define
\begin{align}
\label{eq:schur subspace}
    P_{l,\vartheta}^{\lambda\mu}
     \coloneqq\sum_{\substack{\eta\vdash l:\ell(\eta)\leq\hat{r}\,,\ \cV_{\lambda\mu}^{(\hat{r})}(d_\eta)\leq d_\mu2^{-l\vartheta}}} \Pi_\eta^{\lambda\mu},\qquad \qquad
    P_{l,\vartheta}
    \coloneqq\sum_{\lambda,\mu}P_{l,\vartheta}^{\lambda\mu}.
\end{align}
The cumulative-volume observation in the proof of
Ref.~\cite[Theorem~16]{UQC}, together with its purification theorem,
gives
\begin{align}
\label{eq:schur weight}
 \rank(P_{l,\vartheta}^{\lambda\mu}) \leq d_\mu 2^{-l\vartheta},\qquad
 \tr[(\bI-P_{l,\vartheta})\hat{\rho}^{\otimes l}]
 =\Pr_{\hat{\rho}}[Z_l^{(\hat{r})}<l\vartheta]
 \quad(\rank(\hat{\rho}) \leq \hat{r}).
\end{align}
Indeed, the selected diagrams form an initial segment in $d_\eta$, including ties, and its rank is its largest cumulative volume.
The weight identity uses
$\tr[\Pi_\eta^{\lambda\mu}\hat{\rho}^{\otimes l}]=p_{\lambda\mu\eta}$
from Ref.~\cite[Theorem~14]{UQC}.
For $t \in \{0, \ldots, l-1\}$, let $P_{l,t,\vartheta}$ project onto the sum of the
ranges of $P_{l-j,\vartheta}\otimes\bI^{\cI}$ over
$0\leq j\leq t$ and $\cI\subseteq\{1,\ldots,l\}$ with $|\cI|=j$;
the identity acts on the copies in $\cI$.

\begin{lemma}[Enlarged Schur supports]
\label{lem:enlarged subspaces}
The projector $P_{l,t,\vartheta}$ has both invariances in
\Cref{lem:uniform support estimate}, contains every
$P_{l-j,\vartheta}\otimes\bI^{\cI}$ with $|\cI|=j\leq t$, and satisfies
\begin{align}
\label{eq:enlarged rank}
    \rank(P_{l,t,\vartheta}^{\lambda\mu})
    \leq d_\mu(t+1)l^{2t}2^{-(l-t)\vartheta}.
\end{align}
If $\rank(\hat{\rho}) \leq \hat{r}$ and $\f{1}{2}\|\rho-\hat{\rho}\|_1 \leq \hat{\epsilon}$, then
\begin{align}
\label{eq:transfer}
    \tr[(\bI-P_{l,t,\vartheta})\rho^{\otimes l}]
    \leq(1+\hat{\epsilon})^l\max_{0\leq j\leq t} \Pr_{\hat{\rho}}[Z_{l-j}^{(\hat{r})}<(l-j)\vartheta] 
    + \sum_{j=t+1}^l\binom lj\hat{\epsilon}^j.
\end{align}
\end{lemma}

\begin{proof}[Proof of \Cref{lem:enlarged subspaces}]

Containment and invariance follow from the definition.  Under restriction
to $S_{l-j}$, write
$\cS_\lambda=\bigoplus_{\lambda'}\cS_{\lambda'}\otimes
\bC^{f^{\lambda/\lambda'}}$, where $f^{\lambda/\lambda'}$ counts paths
removing $j$ boxes.  The standard-tableau branching rule gives
$\sum_{\lambda'}f^{\lambda/\lambda'}\leq l^j$ and
$\sum_{\mu'}f^{\mu/\mu'}d_{\mu'}=d_\mu$.
Thus for each fixed $\cI$ its block rank is at most
\begin{align}
 \sum_{\lambda',\mu'}f^{\lambda/\lambda'}f^{\mu/\mu'} \rank(P_{l-j,\vartheta}^{\lambda'\mu'})
 \leq l^jd_\mu2^{-(l-j)\vartheta}.
\end{align}
Summing over $j\leq t$ and $\binom lj\leq l^j$ subsets proves
Eq.~\eqref{eq:enlarged rank}.
For Eq.~\eqref{eq:transfer}, let $\Delta$ be the positive part of $\rho - \hat{\rho}$, for which $\rho \leq \hat{\rho} + \Delta$ and $\tr[\Delta] \leq \hat{\epsilon}$. Then $\rho^{\otimes l} \leq (\hat{\rho} + \Delta)^{\otimes l}$, and we expand the right-hand side into the terms with $\Delta$ on a subset $\cI$ of the copies and $\hat{\rho}$ on the others. For $|\cI| = j \leq t$, the containment gives $\bI - P_{l,t,\vartheta} \leq \bI - P_{l-j,\vartheta}\otimes\bI^{\cI}$, and the trace of the term factorizes into $\tr[(\bI - P_{l-j,\vartheta})\hat{\rho}^{\otimes(l-j)}]\tr[\Delta]^j \leq \hat{\epsilon}^j\Pr_{\hat{\rho}}[Z_{l-j}^{(\hat{r})}<(l-j)\vartheta]$ by Eq.~\eqref{eq:schur weight}. For $j > t$, the term is at most $\hat{\epsilon}^j$. Summing over $\cI$ and using $\sum_{j\leq t}\binom lj\hat{\epsilon}^j \leq (1+\hat{\epsilon})^l$ proves Eq.~\eqref{eq:transfer}.

\end{proof}

\subsection{Sample-efficient achievability result}
\label{sec:sample efficient result}

We now combine these tools for a state $\rho^{\sA\sB}$ close to a nearby state $\hat{\rho}^{\sA\sB}$ of low rank. We retain $\hat{r}$ for the rank bound of the nearby state, $\hat{\epsilon}$ for the perturbation radius, $J$ for the target rate, $\gamma$ for the allowed rate loss, and $\varepsilon$ for the output error. The nearby state is used only in the analysis; the common channel depends on the promised parameters.

For $\hat{\epsilon}\in[0,1)$ and an integer $\hat{r}\geq1$, define
\begin{align}
\label{eq:smooth coherent information global}
    \widetilde{I}_\beta^{\,\hat{\epsilon}, \hat{r}}(\sA\rangle\sB)_\rho \coloneqq \sup\Big\{\widetilde{I}_\beta(\sA\rangle\sB)_{\hat{\rho}} : \f{1}{2}\big\|\rho^{\sA\sB} - \hat{\rho}^{\sA\sB}\big\|_1 \leq \hat{\epsilon}, \ \rank(\hat{\rho}^{\sA\sB}) \leq \hat{r}\Big\}.
\end{align}
The supremum is $-\infty$ if the feasible set is empty. Otherwise, it is attained, since the feasible set is compact and $\widetilde{I}_\beta$ is continuous for $\beta \in [1/2, 1)$~\cite{Marwah2022UniformContinuitysandwichedrenyi}. Only the nearby state $\hat{\rho}$ is subject to the rank constraint. At $\hat{\epsilon}=0$ and $\hat{r}=\rank(\rho)$, this reduces to $\widetilde I_\beta(\sA\rangle\sB)_\rho$.

\begin{theorem}[Sample-efficient distillation for $k$ copies of a mixed state]
\label{thm:achievability by schur coding}
Fix a constant $\beta\in[1/2,1)$ and an integer $\hat{r}=\poly(n)$.
Let $J,\gamma,\varepsilon$ satisfy
\begin{align}
\label{eq:schur rate loss range}
    \gamma \in (0, J/4],\qquad 
    J=\cO(\log{n}),\qquad
    1/\gamma=\poly(n),\qquad
    \varepsilon \in (0,1),\qquad
    \log(1/\varepsilon)=\poly(n).
\end{align}
There is $\epsilon_\star\in(0,1)$, determined by $(\hat{r},J,\gamma)$
at this fixed $\beta$, with
\begin{align}
\label{eq:schur smoothing radius order}
    \epsilon_\star
    = \Omega\bigg(\f{\gamma}{\log\hat{r} + J + \log(1/\gamma)+1}\bigg),
\end{align}
and a single one-way LOCC channel such that, for every $\hat{\epsilon} \in [0, \epsilon_\star]$ and every state $\rho^{\sA\sB}$ satisfying
$\widetilde I_\beta^{\,\hat{\epsilon},\hat{r}}(\sA\rangle\sB)_\rho\geq J$,
the channel distills Bell pairs from $\rho^{\sA\sB}$ at error $\varepsilon$ in the sense of \Cref{def:protocol}, with rate $J\geq R\geq J-\gamma$ from $k = \poly(\hat{r}, 2^J, 1/\gamma, \log(1/\varepsilon)) = \poly(n)$ copies.
For fixed local spaces it depends only on $(n,\hat{r},J,\gamma,\varepsilon)$, independently of $\hat{\epsilon}$, $\rho$, and the nearby state.
\end{theorem}

\begin{proof}[Proof of \Cref{thm:achievability by schur coding}]

We first fix the parameters and bound the number of copies. Set $s=1-\beta$, $v=s/(1+s)$, and
\begin{align}
\label{eq:threshold regularization}
    a \coloneqq \f{\gamma}{8J}, \qquad
    \vartheta \coloneqq J-\gamma/8,\qquad
    \varpi \coloneqq \f{2^{-J}}{\hat{r}} (2^{s\gamma/16}-1)^{1/s}.
\end{align}
Since $\gamma \leq J/4$, we have $a \leq 1/32$, and we take $u_\rc$ of Eq.~\eqref{eq:code parameters} at this $a$. 
We use the constant $C_l$ of Eq.~\eqref{eq:moment constant}.
Let $K\geq2$ be the smallest integer satisfying
\begin{gather}
    \label{eq:batch conditions}
    K \geq 8/\gamma, \qquad 
    K\gamma \geq 64\big(J + 3\log(K+1) + 3\big), \qquad
    K\gamma/64 \geq \log{C_K}, \qquad 
    \f{v\gamma K}{128(J + 2\log{K} + 1)} \geq \log(6/u_\rc).
\end{gather}
These conditions have a polynomial solution.  To check this directly,
write $A=A_{\hat{r},\varpi}$ and $H=\log(6/u_\rc)$.
The partition bound used in Ref.~\cite[proof of Proposition~5]{UQC} gives
\begin{align}
\label{eq:copy bound constant}
    \log{C_K}
    \leq \f{\pi}{\ln{2}} \sqrt{\f{2K}{3}} + \f{A}{\ln{2}}, \qquad
    A 
    \leq 2\hat{r}^4 2^{2J}\bigg(\f{32}{s\gamma}\bigg)^{2/s}.
\end{align}
Here $0<\varpi<1$ and
$\varpi\geq2^{-J}\hat{r}^{-1}(s\gamma/32)^{1/s}$, using
$2^x-1\geq x/2$ for $x\geq0$. For $B = 2 + \gamma^{-2} + [A + (J+1)(H+1)]/\gamma$, every integer $K'$ that exceeds $B\log(2+B)$ by a sufficiently large constant factor, which depends only on $\beta$, satisfies Eq.~\eqref{eq:batch conditions}: $B\geq\gamma^{-2}$ controls the square-root term, and $B\gamma\geq A+(J+1)(H+1)$ controls the constant and logarithmic terms. Thus $K$ is polynomial in $\hat{r},2^J,1/\gamma$, and
\begin{align}
\label{eq:copy logarithm}
    \log{K}
    = \cO(\log\hat{r} + J + \log(1/\gamma) + 1).
\end{align}
If $\gamma>1$, first replace $\log(1/\gamma)$ by zero in the upper
bound; the displayed form follows since $\gamma\leq J/4$ makes
$J+1$ and $J-\log\gamma+1$ comparable.

Define
\begin{align}
\label{eq:schur smoothing radius}
    \epsilon_\star \coloneqq \f{v\gamma}{512(J + 2\log{K} + 1)}, \qquad
    t \coloneqq \lceil8\epsilon_\star K\rceil, \qquad
    m_1 \coloneqq \lfloor K(J-\gamma/4)\rfloor, \qquad 
    q \coloneqq 2^{m_1}.
\end{align}
The conditions imply $K>256$, $\epsilon_\star<1/512$, $t\leq K/2<K$,
and $m_1\geq1$.  Equation~\eqref{eq:copy logarithm} proves the claimed
order of $\epsilon_\star$, which is independent of $\varepsilon$.

We next show that each batch yields pairs of small infidelity, even though the input only lies close to a state of rank at most $\hat{r}$. Fix a nearby state $\hat{\rho}$ attaining
Eq.~\eqref{eq:smooth coherent information global}.
The third inequality of Eq.~\eqref{eq:moment bound} gives $I^{\rP}_{\beta,\varpi}(\hat{\rho})\geq J-\gamma/16
=\vartheta+\gamma/16$, because $2^{-sJ}+(\hat{r}\varpi)^s=2^{-s(J-\gamma/16)}$. Hence, uniformly for $0\leq j\leq t$,
\begin{align}
\label{eq:witness tail}
    \Pr_{\hat{\rho}}[Z_{K-j}^{(\hat{r})}<(K-j)\vartheta]
    \leq C_{K-j}^v2^{-v(K-j)\gamma/16}
    \leq2^{-vK\gamma/64}.
\end{align}
We used $K-j\geq K/2$ and $\log{C_{K-j}} \leq \log{C_K} \leq K\gamma/64$.
Equation~\eqref{eq:transfer} at $\hat{\epsilon} \leq \epsilon_\star$ then yields
\begin{align}
\label{eq:outside weight}
    \tr[(\bI-P_{K,t,\vartheta})\rho^{\otimes K}]
    \leq\delta_1\coloneqq2^{-vK\gamma/128}+2^{-4\epsilon_\star K},
\end{align}
since $(1+\epsilon_\star)^K\leq2^{vK\gamma/128}$ and $\sum_{j>t}\binom Kj\epsilon_\star^j \leq2^{-t}(1+2\epsilon_\star)^K\leq2^{-4\epsilon_\star K}$.
The volume cost satisfies
\begin{align}
\label{eq:decoder error}
    \delta_2\coloneqq q(t+1)K^{2t}2^{-(K-t)\vartheta}
    < 2^{-3K\gamma/32}
    < 1.
\end{align}
Indeed, its logarithm is at most
$-K\gamma/8+t\vartheta+2t\log{K} + \log(t+1)$, and
\begin{align}
    t \vartheta + 2t\log{K} + \log(t+1)
    \leq vK\gamma/64 + J + 3\log(K+1)
    < K\gamma/32.
\end{align}
Applying \Cref{lem:uniform support estimate} to the invariant input $\rho^{\otimes K}$ and $P_{K,t,\vartheta}$ at $c = \delta_2/q$, which Eq.~\eqref{eq:enlarged rank} allows, we obtain a single channel on $K$ copies with infidelity at most
\begin{align}
    (\sqrt{\delta_1} + \sqrt{\delta_2})^2
    \leq 2\delta_1 + 2\delta_2
    \leq 6 \cdot 2^{-4\epsilon_\star K}
    \leq u_\rc.
\end{align}
Here $4\epsilon_\star \leq v\gamma/128 \leq 3\gamma/32$ and the last
inequality is the fourth condition in Eq.~\eqref{eq:batch conditions}.

We finally reduce the error and bound the rate. Each of the $m_1$ pair marginals has infidelity at most $u_\rc$, and the channel on one batch thus meets the conditions of \Cref{cor:batch distillation} at $m_\rb=m_1$ and $\varepsilon_\rc=\varepsilon$. The resulting protocol distills $m=m_1m_\rc$ ebits from $k=Kk_\rc$ copies at error $\varepsilon$, with $m/k\geq(m_1/K)(1-4a)$. Since $1/K\leq\gamma/8$,
\begin{align}
\label{eq:schur rate proof}
    J \geq R = \f{m_1m_\rc}{Kk_\rc}
    \geq (J-3\gamma/8)(1-4a)
    \geq J-7\gamma/8
    \geq J-\gamma.
\end{align}
Finally, Eqs.~\eqref{eq:code target} and~\eqref{eq:code parameters} give $N = \cO(a^{-1}\log(m_1/\varepsilon))$ and $L = \cO(a^{-1}\log^2(1/a)+\log{N})$. Since $a^{-1}=8J/\gamma$ and $m_1\leq KJ$, the number of copies $k=KLN$ is polynomial in the stated parameters, and $k\geq N\geq4/a$. For constant $\varepsilon$ and parameters $J = \Theta(\log{n})$ and $1/\gamma = \Theta(\log\log{n})$, the bound on $B$ above and Eq.~\eqref{eq:copy bound constant} give $K \leq \hat{r}^4 2^{2J}\polylog(n)$, while $N,L = \polylog(n)$. Hence,
\begin{align}
\label{eq:schur copy count}
    \log{k} \leq 4\log{\hat{r}} + 2J + \cO(\log\log{n}).
\end{align}
Tensor products and composition preserve one-way LOCC, and all channels depend only on the promised parameters.

\end{proof}

For the set $\scS$ of states with $\widetilde I_\beta^{\,\hat{\epsilon},\hat{r}}(\sA\rangle\sB)_\rho\geq J$ at some $\hat{\epsilon} \in [0, \epsilon_\star]$, this gives
\begin{align}
 E_{\rD,\ag}^{(k,\varepsilon)}(\sA:\sB)_\scS\geq J-\gamma,
 \qquad E_\rD^{(k,\varepsilon)}(\sA:\sB)_\rho\geq J-\gamma
 \quad(\rho\in\scS).
\end{align}
Here $E_{\rD,\ag}^{(k,\varepsilon)}(\sA:\sB)_\scS$ is the rate of a single protocol that works throughout $\scS$, as in \Cref{def:agnostic distillable}, and the second bound follows from the first by (3) of \Cref{prop:relations}.

Case~(I) of \Cref{thm:achievability by efficient protocol} is obtained by choosing the parameters as follows. For a state $\hat{\rho}^{\sA\sB}$ in that theorem, we take $\hat{r} = \rank(\hat{\rho}^{\sA\sB})$, $1/\gamma = \Theta(\log\log{n})$, $J \coloneqq \min\{\widetilde{I}_\beta(\sA\rangle\sB)_{\hat{\rho}}, \, C\log{n}\}$, and an error $\varepsilon$ as in Eq.~\eqref{eq:schur rate loss range}. The target rate is capped at $C\log{n}$ because the number of copies in \Cref{thm:achievability by schur coding} is polynomial in $2^J$. These parameters obey Eq.~\eqref{eq:schur rate loss range} for large enough $n$, since $\widetilde{I}_\beta(\sA\rangle\sB)_{\hat{\rho}} = \Omega(1)$ implies $J \geq 4\gamma$, and Eq.~\eqref{eq:schur smoothing radius order} yields $\epsilon_\star = \Omega(1/(\log{n}\log\log{n}))$, which exceeds $1/\log^{1+c}{n}$ for every constant $c > 0$ and large enough $n$. Every state $\rho^{\sA\sB}$ within trace distance $1/\log^{1+c}{n}$ of $\hat{\rho}^{\sA\sB}$ thus obeys $\widetilde{I}_\beta^{\,\hat{\epsilon}, \hat{r}}(\sA\rangle\sB)_\rho \geq J$ at $\hat{\epsilon} = 1/\log^{1+c}{n}$, with $\hat{\rho}^{\sA\sB}$ as the nearby state. The single channel of \Cref{thm:achievability by schur coding}, which uses $\hat{\rho}^{\sA\sB}$ only through $\hat{r}$ and $J$, distills from each such state on $\poly(n)$ copies at a rate $R \geq J - \gamma = \min\{\widetilde{I}_\beta(\sA\rangle\sB)_{\hat{\rho}}, \, C\log{n}\} - o(1)$. This provides (I) of \Cref{thm:achievability by efficient protocol}.

\section{Achievability: Collective compression and efficient decoupling}
\label{sec:achievability}

The distillation protocol in \Cref{sec:schur coding} acts on the registers of the shared copies themselves, whose dimension grows with the number of copies. Its recovery is specified through the Kronecker coefficients on these registers, and we do not know how to implement it by a circuit of polynomial size. We now require the nearby state to have polynomial rank on its marginals as well, and under this condition we distill at the same rate by a protocol whose circuit size is polynomial.
Alice and Bob each begin by compressing their systems of the shared copies into registers of polynomial dimension, without any description of the state, and every subsequent operation acts on these registers alone. The finite-precision implementations of these operations follow the compilation conventions in \Cref{sec:numerics}. We present the building blocks of the protocol in \Cref{sec:preparatory lemmas achievability}, combine them into the distillation protocol in \Cref{sec:protocol}, and provide the formal version of Case~(II) of \Cref{thm:achievability by efficient protocol} in \Cref{sec:achievability result}.

\subsection{Compression and decoupling tools}
\label{sec:preparatory lemmas achievability}

We present the primitives of our distillation protocol, which we combine in \Cref{sec:protocol}. In \Cref{sec:collective compression}, we construct the compression channel and characterize its output on selected copies. We then describe Alice's two-stage measurement in \Cref{sec:two instruments}, whose first stage retains most of the entropy and whose second stage achieves decoupling, and construct Bob's decoder in \Cref{sec:decoder}.

\subsubsection{Universal collective compression}
\label{sec:collective compression}

Universal compression of quantum states has long been studied~\cite{Jozsa1998UniversalQuantumInformationCompression, hayashi2002universalvariablelengthsourcecoding, Bennett2006Universalquantumdatacompressionnondestructive}. These schemes jointly map the i.i.d. copies into a subspace whose dimension is characterized by the entropy, and do not retain a register for each copy. A per-copy dimension reduction is given in Ref.~\cite{lowe2026randomdimensionreductionlearning}, which maps each copy into a register of dimension equal to a given upper bound on the rank, with an exact guarantee for inputs obeying that bound. Our channel also reduces each copy separately, but applies to i.i.d. inputs of arbitrary rank together with an arbitrary reference: its output is close to a truncation of each copy at a random threshold, with an error that does not depend on the input.

Let $\cH^\sA$ be a Hilbert space of dimension $d_\sA$. For $s > 0$ and a state $\tau^\sA$, let $P_{\tau,s}^\sA$ be the projector onto the span of the eigenstates of $\tau^\sA$ corresponding to its eigenvalues at least $s$. Let $\cH^\sC$ be a ($Q+1$)-dimensional space spanned by the orthonormal basis $\{\ket{\!\!\perp}^\sC, \ket{1}^\sC, \ldots, \ket{Q}^\sC\}$, where the integer $Q$ satisfies $Q \geq 1/s$, and let $V_{\tau, s}^{\sA\to\sC}$ be an isometry from the range of $P_{\tau,s}^\sA$ into the span of $\{\ket{1}^\sC, \ldots, \ket{Q}^\sC\}$.
We define the truncation channel
\begin{align}
\label{eq:truncation channel}
    \cN_{\tau,s}^{\sA\to\sC}(X^\sA) \coloneqq V_{\tau, s}^{\sA\to\sC}P_{\tau,s}^\sA X^\sA P_{\tau,s}^\sA (V_{\tau, s}^{\sA\to\sC})^\dag 
    + \tr\big[(\bI^\sA - P_{\tau,s}^\sA) X^\sA\big]\ketbra{\!\!\perp}{\perp\!\!}^\sC.
\end{align}
Since the rank of $P_{\tau,s}^\sA$ is at most $1/s \leq Q$, such an isometry exists. 
For a unitary $V^\sC$ on the span of $\{\ket{1}^\sC, \ldots, \ket{Q}^\sC\}$, extended by the identity on $\ket{\!\!\perp}^\sC$, we write $\cV^\sC$ for the channel defined as $\cV^\sC(Y^\sC) \coloneqq V^\sC Y^\sC (V^\sC)^\dag$.

The following lemma shows that a single channel, constructed without any description of the input state, compresses each of the $K$ copies of a register $\sA$ of dimension $d_\sA$ into a register $\sC$ of dimension $Q+1$. Its output on any $h$ of them, together with an arbitrary reference $\sR$, is close to the output of $\cV^\sC \circ \cN_{\tau,s}^{\sA\to\sC}$ applied to every copy, where the threshold $s$ and the unitary $V^\sC$ are common to all the copies.

\begin{lemma}[Universal collective compression]
\label{lem:compression}
Let $w \in (0, 1/4)$, let $Q \geq \lceil 2/w \rceil$ and $G \geq 1$ be integers, and let $K \geq 4Q^2$ and $1 \leq h \leq K$. Then, there is a channel $\cC^{\sA^K\to\sC^K}$ satisfying that, for any register $\sR$, any state $\varrho^{\sA^K\sR}$ such that $\tr_\sR[\varrho^{\sA^K\sR}] = \tau^{\otimes K}$, and any subset $\cI \subseteq \{1, \ldots, K\}$ with $|\cI| = K - h$,
\begin{align}
\label{eq:single compression error}
    \f{1}{2}\Big\|\tr_{\cI}\big[\big(\cC^{\sA^K\to\sC^K}\otimes\rid^\sR\big)\big(\varrho^{\sA^K\sR}\big)\big] - \tr_{\cI}\big[\big(\cD_{\tau}^{\sA^K\to\sC^K}\otimes\rid^\sR\big)\big(\varrho^{\sA^K\sR}\big)\big]\Big\|_1
    \leq \f{4h}{\sqrt{w}\,K^{1/4}} + \f{3h}{\sqrt{G}}.
\end{align}
Here, $\cD_\tau^{\sA^K\to\sC^K}$ divides $[w, 2w]$ into $G$ equal subintervals, draws a threshold $s$ uniformly from their midpoints and a unitary $V^\sC$ from the Haar measure, and applies $\cV^\sC\circ\cN_{\tau,s}^{\sA\to\sC}$ with this single $s$ and $V^\sC$ to every copy.
For any $\delta > 0$, the channel $\cC^{\sA^K\to\sC^K}$ can be implemented within diamond distance $\delta$ by a circuit of $\poly(K, \log{d_\sA}, Q, \log{G}, b_w, \log(1/\delta))$ one- and two-qubit gates, where $w$ is specified as a rational number of bit length $b_w$.
\end{lemma}

The compression channel $\cC^{\sA^K\to\sC^K}$ is built on the Schur--Weyl decomposition~\cite{goodman1998representations, harrow2005applications} of $(\cH^\sA)^{\otimes K}$ into irreducible representations of the unitary group and the symmetric group, labeled by partitions $\lambda$ of $K$. The rows of $\lambda$ estimate the eigenvalues of $\tau$, while its eigenbasis is carried by the unitary-group register, the only one whose dimension grows with $d_\sA$. Discarding this register and keeping the rows above the threshold is thus efficient, but loses the eigenbasis that $\cN_{\tau,s}^{\sA\to\sC}$ retains. The common Haar unitary $V^\sC$ removes this difference, since after averaging over it the outputs no longer depend on the choice of basis within the retained subspace.

\begin{proof}[Proof of \Cref{lem:compression}]

We use the Schur--Weyl decomposition of \Cref{sec:schur weyl} at $l = K$ copies of $\sA$.
Following Ref.~\cite{burchardt2025highdimensionalschur}, whose circuits we will use later, we take the Gelfand--Tsetlin basis of $\cW_\lambda(d_\sA)$ and the Young--Yamanouchi basis of $\cS_\lambda$. The latter basis is constructed along the chain $S_1 \subset \cdots \subset S_K$, and thus fixes the symmetric-group register without reference to $d_\sA$.
We here pad $\cH^\sA$ so that $d_\sA$ is a power of two at least $\max\{K, Q\}$. Every $\lambda \vdash K$ then appears in the decomposition, and the rows retained fit in $\cH^\sC$ when we truncate $\lambda$ to at most $Q$ rows below.

We define $\cC^{\sA^K\to\sC^K}$ as the channel realized by the following procedure.
We first draw a threshold $s$ and a unitary $V^\sC$ as in the statement. We then measure $\lambda$, discard $\cW_\lambda(d_\sA)$ while retaining $\cS_\lambda$, append the highest weight vector $\ket{\mathrm{hw}_\lambda}$ of $\cW_\lambda(d_\sA)$, and perform the inverse Schur transform.
We select the row indices $a$ of $\lambda = (\lambda_1, \lambda_2, \ldots)$ satisfying
\begin{align}
\label{eq:selection rule}
    \f{\lambda_a - a}{K-1} \geq s,
\end{align}
and write $Q_\lambda(s)$ for their number.
We then apply to each copy of $\sA$ a channel that maps $\ket{a}^\sA$ to $\ket{a}^\sC$ for a selected $a$ and replaces the orthogonal complement with $\ket{\!\!\perp}^\sC$, followed by the Haar random unitary $V^\sC$. We finally discard $\lambda$ and $s$.
Since $s \geq w$, every selected index obeys $\lambda_a \geq w(K-1)$, so $Q_\lambda(s) \leq K/[w(K-1)] \leq Q$; the selected indices fit in $\sC$.

Efficient circuits for the high-dimensional Schur transform are given in Ref.~\cite{burchardt2025highdimensionalschur}, in which $\ket{\mathrm{hw}_\lambda}$ is taken as a computational basis state. The remaining operations act on $\lambda$, $s$, the computational basis labels of $\sA$, and the $(Q+1)$-dimensional register $\sC$. See Fig.~\ref{fig:compression channel}. The finite-precision implementation of $V^\sC$ and the resulting gate count are established at the end of this proof, after the ideal-channel error estimate.


\begin{figure}
\adjustbox{width=180mm}{%
\begin{tikzpicture}[
    wire/.style={line width=1.6pt},
    cwire/.style={line width=0.6pt, double, double distance=2.2pt},
    box/.style={draw, line width=1.4pt, fill=white, font=\normalsize, align=center, minimum height=2.4cm},
    lab/.style={font=\small, fill=white, inner sep=2pt},
    ctrl/.style={fill=black, circle, inner sep=0pt, minimum size=5pt}
]
\node[box, minimum width=1.8cm] (sch) at (2.4,0.8) {Schur\\ transform};
\node[draw, line width=1.4pt, fill=white, minimum width=1.0cm, minimum height=0.7cm] (mlam) at (5.6,0.1) {};
\node[box, minimum width=1.8cm] (isch) at (8.4,0.8) {inverse\\ Schur\\ transform};
\node[box, minimum width=3.6cm] (sel) at (12.2,0.8) {on each copy:\\ $\ket{a}^\sA \mapsto \ket{a}^\sC$ for selected $a$,\\ $\ket{\!\!\perp}^\sC$ otherwise};
\node[box, minimum width=1.8cm] (haar) at (16.2,0.8) {\raisebox{-1.0cm}{$(V^\sC)^{\otimes K}$}};
\coordinate (P) at (-0.6,1.8);
\node[font=\large, anchor=east, fill=white, inner sep=2pt] at (-0.75,1.8) {\large{$\varrho^{\sA^K\sR}$}};
\draw[wire] (P) -- (0.4,2.8) -- (17.9,2.8);
\draw[wire] (P) -- (0.4,0.8) -- (sch.west);
\node[lab] at (0.7,3.2) {$\sR$};
\node[lab] at (0.7,0.4) {$\sA^{K}$};
\draw[wire] (sch.east |- 0,1.5) -- (4.8,1.5);
\draw[wire] (sch.east |- 0,0.8) -- (isch.west |- 0,0.8);
\draw[wire] (sch.east |- 0,0.1) -- (mlam.west);
\node[lab] at (4.05,1.8) {$\cW_\lambda(d_\sA)$};
\node[lab] at (4.0,1.05) {$\cS_\lambda$};
\node[lab] at (4.7,0.4) {$\lambda$};
\draw[line width=0.8pt] (5.32,-0.05) arc (180:0:0.28);
\draw[-Latex, line width=0.8pt] (5.6,-0.05) -- (5.88,0.37);
\draw[line width=1pt] (4.8,1.7) -- (4.8,1.3);
\draw[line width=1pt] (4.9,1.7) -- (4.9,1.3);
\node[lab, anchor=east] at (6.9,1.5) {$\ket{\mathrm{hw}_\lambda}$};
\draw[wire] (6.9,1.5) -- (isch.west |- 0,1.5);
\draw[wire] (isch.east) -- (sel.west);
\draw[wire] (sel.east) -- (haar.west);
\draw[wire] (haar.east) -- (17.9,0.8);
\coordinate (M) at (18.7,1.8);
\draw[wire] (17.9,2.8) -- (M);
\draw[wire] (17.9,0.8) -- (M);
\node[lab] at (17.6,3.2) {$\sR$};
\node[lab] at (17.6,0.4) {$\sC^{K}$};
\node[font=\large, anchor=west, fill=white, inner sep=2pt] at (18.85,1.8) {\large{$\big(\cC^{\sA^K\to\sC^K}\otimes\rid^\sR\big)\big(\varrho^{\sA^K\sR}\big)$}};
\draw[cwire] (mlam.south) -- (5.6,-1.2) -- (17.9,-1.2);
\draw[cwire] (8.4,-1.2) -- (isch.south);
\draw[cwire] (11.8,-1.2) -- (11.8,0 |- sel.south);
\node[ctrl] at (8.4,-1.2) {};
\node[ctrl] at (11.8,-1.2) {};
\node[lab] at (7.0,-1.2) {$\lambda$};
\node[lab, anchor=west] at (18.0,-1.2) {discard};
\draw[cwire] (0.4,-2.0) -- (17.9,-2.0);
\draw[cwire] (12.6,-2.0) -- (12.6,0 |- sel.south);
\node[ctrl] at (12.6,-2.0) {};
\node[lab] at (2.8,-2.0) {$s \sim \text{midpoints of } [w,2w]$};
\node[lab, anchor=west] at (18.0,-2.0) {discard};
\node[lab, align=center] at (16.1,1.2) {Haar\\ unitary};
\end{tikzpicture}}
    \caption{The channel $\cC^{\sA^K\to\sC^K}$ of \Cref{lem:compression}. The Schur transform separates $\lambda$, $\cW_\lambda(d_\sA)$, and $\cS_\lambda$; $\lambda$ is measured, $\cW_\lambda(d_\sA)$ is discarded and replaced by $\ket{\mathrm{hw}_\lambda}$, and the inverse Schur transform is applied. Each copy is then mapped into $\sC$ according to the row indices selected by $\lambda$ and $s$ in Eq.~\eqref{eq:selection rule}, followed by $V^\sC$. Double lines represent classical registers; $\sR$ is untouched.}
    \label{fig:compression channel}
\end{figure}
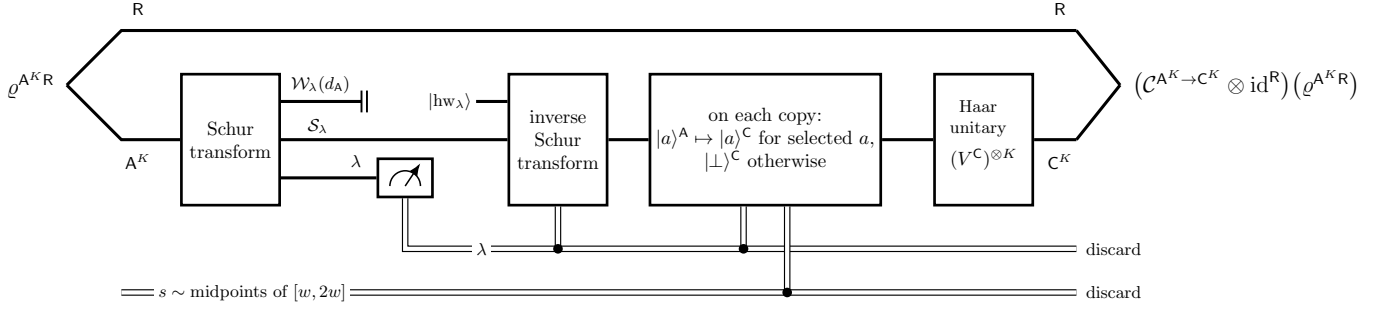


We next evaluate the error in Eq.~\eqref{eq:single compression error}.
The two channels $\cC^{\sA^K\to\sC^K}$ and $\cD_\tau^{\sA^K\to\sC^K}$ differ in the projector they measure on each copy of $\sA$: $\cD_\tau^{\sA^K\to\sC^K}$ measures $P_{\tau,s}^\sA$, which depends on $\tau$, while $\cC^{\sA^K\to\sC^K}$ measures the projector onto the indices selected from the measured $\lambda$.
We denote by $\mathbf{1}_{M \geq s}$ the spectral projector onto the eigenvalues of a Hermitian operator $M$ at least $s$, and by $\mathbf{1}_{\{\cdot\}}$ the indicator function with a corresponding condition. 
To compare the two projectors, we use $\mathbf{1}_{\bar{F}_i \geq s}$ for the average swap $\bar{F}_i \coloneqq \f{1}{K-1}\sum_{j \neq i} F_{ij}$, where $F_{ij}$ is the swap operator of the $i$-th and $j$-th copies of $\sA$.
This projector is close to $P_{\tau,s}^\sA$ on the $i$-th copy of $\sA$ under $\tau^{\otimes K}$, and it acts within each $\lambda$ sector since $\bar{F}_i$ lies in the group algebra $\bC[S_K]$ of $S_K$, which we identify with the span of the permutation operators of the $K$ copies when it acts on them. We thus bound its distance to each of the two projectors and combine the two bounds.

Let $\tau_i \coloneqq \bI^{\otimes(i-1)}\otimes\tau\otimes\bI^{\otimes(K-i)}$, so that $\mathbf{1}_{\tau_i \geq s}$ is $P_{\tau,s}^\sA$ on the $i$-th copy of $\sA$. Then, 
\begin{align}
\label{eq:input variance}
    \tr\big[\tau^{\otimes K}(\bar{F}_i - \tau_i)^2\big]
    = \f{1}{K-1} + \f{K-2}{K-1}\tr[\tau^3] - 2\tr[\tau^3] + \tr[\tau^3]
    = \f{1 - \tr[\tau^3]}{K-1}
    \leq \f{1}{K-1},
\end{align}
where we used $F_{ij}^2 = \bI$, $\tr\big[\tau^{\otimes K}F_{ij}F_{ij'}\big] = \tr[\tau^3]$ for $j \neq j'$, and $\tr\big[\tau^{\otimes K}\bar{F}_i\tau_i\big] = \tr\big[\tau^{\otimes K}\tau_i^2\big] = \tr[\tau^3]$.
Since $\tau^{\otimes K}$ commutes with the spectral projectors of the Hermitian operators $\bar{F}_i$ and $\tau_i$, the traces of $\tau^{\otimes K}$ against the products of their respective projectors define a joint probability distribution of a pair $(x, y)$ of eigenvalues of $\bar{F}_i$ and $\tau_i$.
As for the random boundaries in Refs.~\cite{hayashi2002universalvariablelengthsourcecoding, Bennett2006Universalquantumdatacompressionnondestructive}, a random threshold rarely separates close values: for any $x, y \in \bR$, the fraction of the $G$ midpoints $s$ lying between them is at most $|x - y|/w + 1/G$, as the $G$ subintervals of $[w, 2w]$ have width $w/G$. Using this joint probability distribution for $(x, y)$, we obtain
\begin{align}
\label{eq:input projectors}
    \bE_s\Big[\tr\big[\tau^{\otimes K}(\mathbf{1}_{\bar{F}_i \geq s} - \mathbf{1}_{\tau_i \geq s})^2\big]\Big]
    &= \bE_{x,y}\Big[\Pr_s\big[\min\{x,y\} < s \leq \max\{x,y\}\big]\Big] \\
    &\leq \f{1}{w}\bE_{x,y}\big[|x - y|\big] + \f{1}{G} \\
    &\leq \f{1}{w}\big(\bE_{x,y}\big[(x - y)^2\big]\big)^{1/2} + \f{1}{G} \\
    &\leq \f{1}{w\sqrt{K-1}} + \f{1}{G},
\end{align}
where the first equality exchanges the two averages, and we then use the above bound on the fraction of separating midpoints, the Cauchy--Schwarz inequality, and Eq.~\eqref{eq:input variance}.
This bounds $\mathbf{1}_{\bar{F}_i \geq s}$ against the projector $P_{\tau,s}^\sA$ of $\cD_\tau^{\sA^K\to\sC^K}$.

We turn to the corresponding bound against the projector in $\cC^{\sA^K\to\sC^K}$, which we evaluate on the state obtained from $\tau^{\otimes K}$ by the measurement of $\lambda$ and the replacement of the state on $\cW_\lambda(d_\sA)$ by $\ket{\mathrm{hw}_\lambda}$.
Let $\cM^{\sA^K\to\sA^K\sD}$ be the channel performing these two steps and storing the outcome $\lambda$ in a classical register $\sD$:
\begin{align}
\label{eq:carrier replacement}
    \cM(X) \coloneqq \sum_{\lambda \vdash K}\ketbra{\mathrm{hw}_\lambda}{\mathrm{hw}_\lambda}\otimes\tr_{\cW_\lambda(d_\sA)}\big[\Pi_\lambda X\Pi_\lambda\big]\otimes\ketbra{\lambda}{\lambda}.
\end{align}
As $\tau^{\otimes K}$ commutes with every permutation, Schur's lemma implies $\Pi_\lambda\tau^{\otimes K}\Pi_\lambda = X_\lambda\otimes\bI^{\cS_\lambda}$ with $X_\lambda \geq 0$, and hence
\begin{align}
\label{eq:replaced marginal}
    \cM(\tau^{\otimes K}) 
    = \sum_{\lambda \vdash K}p_\lambda \omega_\lambda \otimes \ketbra{\lambda}{\lambda}, \qquad 
    p_\lambda \coloneqq \tr\big[\Pi_\lambda\tau^{\otimes K}\big], \qquad 
    \omega_\lambda \coloneqq \ketbra{\mathrm{hw}_\lambda}{\mathrm{hw}_\lambda}\otimes\f{\bI^{\cS_\lambda}}{\dim\cS_\lambda}.
\end{align}
Note that each $\omega_\lambda$ commutes with every permutation.
Fix $\lambda$ with $p_\lambda > 0$ and a copy $i$, and put $c_a \coloneqq \lambda_a - a$, which is strictly decreasing in $a$ as $c_a - c_{a+1} = \lambda_a - \lambda_{a+1} + 1$. The channel $\cC^{\sA^K\to\sC^K}$ selects the row indices $a$ obeying Eq.~\eqref{eq:selection rule}, that is, those with $c_a \geq s(K-1)$. We define $R_{\lambda,s}^{(i)} \coloneqq \sum_{a: c_a \geq s(K-1)}\ketbra{a}{a}_i$, the projector onto them on the $i$-th copy, where $\ketbra{a}{a}_i \coloneqq \bI^{\otimes(i-1)}\otimes\ketbra{a}{a}\otimes\bI^{\otimes(K-i)}$.

We compare $R_{\lambda,s}^{(i)}$ with $\mathbf{1}_{\bar{F}_i \geq s}$ under $\omega_\lambda$.
Let $J_i \coloneqq (K-1)\bar{F}_i = \sum_{j \neq i}F_{ij}$, which is the sum of all transpositions of $S_K$ minus those of the subgroup $S_{K-1}$ fixing $i$.
Restricted to $S_{K-1}$, the space $\cS_\lambda$ is decomposed as $\bigoplus_r\cS_{\lambda - e_r}^{(i)}$ over the removable rows $r$ of $\lambda$, namely those with $\lambda_r > \lambda_{r+1}$, where $e_r$ is the $r$-th unit vector, so that $\lambda - e_r$ is a partition of $K-1$~\cite{goodman1998representations}.
Note that this decomposition depends on $i$, since $S_{K-1}$ fixing $i$ differs for each $i$.
Being central, the sum of all transpositions of $S_K$ acts on $\cS_\lambda$ as a scalar, $\f{1}{2}\sum_a\lambda_a(\lambda_a - 2a + 1)$, and that of $S_{K-1}$ acts on $\cS_{\lambda - e_r}^{(i)}$ as the same expression evaluated at $\lambda - e_r$.
Since $\lambda$ and $\lambda - e_r$ differ only in the $r$-th row, subtracting them yields that $J_i$ acts on $\cS_{\lambda - e_r}^{(i)}$ as $\f{1}{2}\lambda_r(\lambda_r - 2r + 1) - \f{1}{2}(\lambda_r - 1)(\lambda_r - 2r) = c_r$. Then, by writing $\Pi_{\lambda,r}^{(i)}$ for the projector onto $\cW_\lambda(d_\sA)\otimes\cS_{\lambda - e_r}^{(i)}$, we have $J_i = \sum_r c_r\Pi_{\lambda,r}^{(i)}$.
From $\bar{F}_i = J_i/(K-1)$, we obtain $\mathbf{1}_{\bar{F}_i \geq s} = \sum_{r: c_r \geq s(K-1)}\Pi_{\lambda,r}^{(i)}$ on the $\lambda$ sector.


Since $\omega_\lambda$ commutes with each $\Pi_{\lambda,r}^{(i)}$, we have 
\begin{align}
\label{eq:letter row law}
    \Pr[a, r] \coloneqq \tr\big[\omega_\lambda\ketbra{a}{a}_i\Pi_{\lambda,r}^{(i)}\big] = \tr\big[\Pi_{\lambda,r}^{(i)}\omega_\lambda\Pi_{\lambda,r}^{(i)}\ketbra{a}{a}_i\big] \geq 0.
\end{align}
Here, $\sum_a\ketbra{a}{a}_i = \bI$ and $\sum_r\Pi_{\lambda,r}^{(i)} = \Pi_\lambda$ make it a joint probability distribution of indices $(a, r)$, which satisfies
\begin{align}
\label{eq:highest weight identities}
    \Pr[a, r] = 0 \ \ \text{unless} \ \ a \leq r, \qquad 
    \Pr[a] = \f{\lambda_a}{K}, \qquad 
    \bE\big[(c_a - c_r)\mathbf{1}_{\{a = b\}}\big] = \f{1}{K}\sum_{j > b}\lambda_j \leq 1 
    \ \ \text{for every} \ \ b.
\end{align}
For the first, $\Pi_{\lambda,r}^{(i)}\omega_\lambda\Pi_{\lambda,r}^{(i)}$ is a mixture of $\ketbra{\mathrm{hw}_\lambda}{\mathrm{hw}_\lambda}\otimes\ketbra{v}{v}$ with $\ket{v} \in \cS_{\lambda - e_r}^{(i)}$, and $\ket{\mathrm{hw}_\lambda}\otimes\ket{v}$ lies in $\cH^\sA_i\otimes\cW_{\lambda - e_r}(d_\sA)\otimes\cS_{\lambda - e_r}^{(i)}$, the $S_{K-1}$-isotypic component of type $\lambda - e_r$ of the copies other than $i$. Applying $\ketbra{a}{a}_i$ leaves $\ket{a}_i\otimes\ket{u}$ with $\ket{u}$ of weight $\lambda - e_a$ in $\cW_{\lambda - e_r}(d_\sA)$, as the weight $\lambda$ under $\mathrm{diag}(t_1, \ldots, t_{d_\sA})^{\otimes K}$ is preserved. Every weight $\nu$ of $\cW_{\lambda - e_r}(d_\sA)$ obeys $\sum_{j \leq m}\nu_j \leq \sum_{j \leq m}(\lambda - e_r)_j$ for all $m$~\cite{goodman1998representations}, which at $\nu = \lambda - e_a$ fails at $m = r$ unless $a \leq r$; thus $\ket{u} = 0$ unless $a \leq r$.
For the other two, let $E_{ab} \coloneqq \sum_i\ketbra{a}{b}_i$, which commutes with every permutation and acts on the $\lambda$ sector as $\pi_\lambda(E_{ab})\otimes\bI^{\cS_\lambda}$, where $\pi_\lambda(E_{aa})\ket{\mathrm{hw}_\lambda} = \lambda_a\ket{\mathrm{hw}_\lambda}$ and $\pi_\lambda(E_{ab})\ket{\mathrm{hw}_\lambda} = 0$ for $a < b$. By the permutation invariance of $\omega_\lambda$, $\Pr[a] = \f{1}{K}\tr[\omega_\lambda E_{aa}] = \f{\lambda_a}{K}$, and
\begin{align}
    \bE\big[c_r\mathbf{1}_{\{a = b\}}\big]
    &= \tr\big[\omega_\lambda\ketbra{b}{b}_iJ_i\big] \\
    &= \f{1}{K}\tr\Big[\omega_\lambda\Big(\sum_{j}E_{bj}E_{jb} - d_\sA E_{bb}\Big)\Big] \\
    &= \f{1}{K}\Big(\lambda_b^2 + \sum_{j > b}(\lambda_b - \lambda_j) - d_\sA\lambda_b\Big) \\
    &= \f{1}{K}\Big(\lambda_bc_b - \sum_{j > b}\lambda_j\Big),
\end{align}
where the second equality uses $\sum_i\ketbra{b}{b}_iJ_i = \sum_jE_{bj}E_{jb} - d_\sA E_{bb}$, and the third uses $E_{bj}E_{jb} = E_{jb}E_{bj} + E_{bb} - E_{jj}$ at $j > b$ and $\pi_\lambda(E_{jb})\ket{\mathrm{hw}_\lambda} = 0$ at $j < b$. Subtracting this from $\bE[c_a\mathbf{1}_{\{a = b\}}] = c_b\lambda_b/K$ proves the third identity.

By expanding the square, $\tr[\omega_\lambda(R_{\lambda,s}^{(i)} - \mathbf{1}_{\bar{F}_i \geq s})^2]$ is the probability that exactly one of $c_a \geq s(K-1)$ and $c_r \geq s(K-1)$ holds. As $a \leq r$ on the support and $c_a$ is decreasing, this is the event $c_r < s(K-1) \leq c_a$, on which $\lambda_a \geq c_a \geq w(K-1)$; since $\sum_a\lambda_a = K$, at most $K/(w(K-1)) \leq 2/w \leq Q$ letters satisfy this, and hence $a \leq Q$. For fixed $(a, r)$, the fraction of the $G$ midpoints $s$ with $s(K-1) \in (c_r, c_a]$ is at most $(c_a - c_r)/(w(K-1)) + 1/G$, as in Eq.~\eqref{eq:input projectors}. Averaging over $(a, r)$ and using the third identity at each $b \leq Q$,
\begin{align}
\label{eq:output projectors}
    \bE_s\Big[\tr\big[\omega_\lambda\big(R_{\lambda,s}^{(i)} - \mathbf{1}_{\bar{F}_i \geq s}\big)^2\big]\Big]
    \leq \f{1}{w(K-1)}\sum_{b \leq Q}\bE\big[(c_a - c_r)\mathbf{1}_{\{a = b\}}\big] + \f{1}{G}
    \leq \f{Q}{w(K-1)} + \f{1}{G}.
\end{align}

We now derive Eq.~\eqref{eq:single compression error}. Both channels commute with the permutations of the copies, and the trace norm is monotone under the partial trace; we may thus take $\cI = \{h+1, \ldots, K\}$ and, enlarging $\sR$, $\varrho^{\sA^K\sR} = \ketbra{\psi}{\psi}$. We fix $s$, average over $V^\sC$, and bound the difference of the expectations of a fixed $L$ with $0 \leq L \leq \bI$ on $\sC^h\sR$ in the two outputs; the maximum over $L$ is the trace distance, and the average over $s$ is taken last.

On each of the $h$ copies, both channels measure a projector, on success embed its range isometrically into the span of $\{\ket{1}^\sC, \ldots, \ket{Q}^\sC\}$ and apply $V^\sC$, and on failure output $\ketbra{\!\!\perp}{\perp\!\!}^\sC$. For $\cD_\tau^{\sA^K\to\sC^K}$ the projector on the $i$-th copy is $\mathbf{1}_{\tau_i \geq s}$; for $\cC^{\sA^K\to\sC^K}$, the channel $\cM$ acts first and the projector is $R_{\lambda,s}^{(i)}$ at the measured $\lambda$, which we record in a classical register $\sD$ regarded, with the purifying system of $\cM$, as part of the reference. Let $\ket{\psi'}$ be the resulting pure state, whose marginal on $\sA^K\sR\sD$ is $(\cM\otimes\rid)(\psi)$ with $\lambda$ in $\sD$, and let $\hat{R}_i \coloneqq \sum_\lambda R_{\lambda,s}^{(i)}\otimes\ketbra{\lambda}{\lambda}^\sD$.

We first remove the isometries. For $b \in \{0,1\}^h$, let $J_b \coloneqq \{i : b_i = 1\}$, let $L_b$ be $L$ compressed by $\bra{\!\!\perp}^\sC$ on each copy $i \notin J_b$ and to the span of $\{\ket{1}^\sC, \ldots, \ket{Q}^\sC\}$ on each copy $i \in J_b$, and write $X^{\otimes J_b}$ for $X$ applied to the copies in $J_b$. Since $V^\sC$ fixes $\ket{\!\!\perp}^\sC$, the expectation of $L$ in the output of $\cD_\tau^{\sA^K\to\sC^K}$ is $\sum_b\bra{\psi}\Xi_bD_{\tau,s}^{\otimes J_b\dag}\bE_V\big[V^{\otimes J_b\dag}L_bV^{\otimes J_b}\big]D_{\tau,s}^{\otimes J_b}\Xi_b\ket{\psi}$, where $D_{\tau,s} \coloneqq V_{\tau, s}^{\sA\to\sC}P_{\tau,s}^\sA$ and $\Xi_b \coloneqq \prod_{i}(\mathbf{1}_{\tau_i \geq s})^{b_i}(\bI - \mathbf{1}_{\tau_i \geq s})^{1-b_i}$; for $\cC^{\sA^K\to\sC^K}$ on $\ket{\psi'}$, the same holds with $\hat{R}_i$ in place of $\mathbf{1}_{\tau_i \geq s}$ and the isometry $\ket{a}^\sA \mapsto \ket{a}^\sC$ on the range of $R_{\lambda,s}^{(i)}$, controlled on $\sD$, in place of $D_{\tau,s}$. The average over $V^\sC$ commutes with $U^{\otimes J_b}$ for every unitary $U$ on the span, and by Schur--Weyl duality it is the image, under the permutation representation on $|J_b|$ copies of the span, of an element $z_b \in \bC[S_{J_b}]\otimes\cB(\cH^\sR)$ that vanishes on the irreducible representations $\nu$ with more than $Q$ rows. Let $\widetilde{L}_b$ be the image of $z_b$ under the permutation representation on $(\cH^\sA)^{\otimes J_b}$, tensored with the identity on the other copies; it lies in $\bC[S_K]\otimes\cB(\cH^\sR)$, the span of the permutations of the $K$ copies tensored with $\cB(\cH^\sR)$. As $z_b$ has the same components in each $\cB(\cS_\nu)\otimes\cB(\cH^\sR)$ as the average, which lies between $0$ and $\bI$, also $0 \leq \widetilde{L}_b \leq \bI$. For any projector $P$ on $\cH^\sA$ of rank at most $Q$ and any isometry $D$ from its range into the span, extended by zero, $D^{\otimes J_b}$ commutes with the permutations of the copies and $D^{\otimes J_b\dag}D^{\otimes J_b} = P^{\otimes J_b}$; hence
\begin{align}
\label{eq:pullback}
    D^{\otimes J_b\dag}\,\bE_V\big[V^{\otimes J_b\dag}L_bV^{\otimes J_b}\big]\,D^{\otimes J_b}
    = P^{\otimes J_b}\,\widetilde{L}_b\,P^{\otimes J_b}.
\end{align}
As $\Xi_bP^{\otimes J_b} = \Xi_b$ at $P = P_{\tau,s}^\sA$, the $V^\sC$-averaged expectation for $\cD_\tau^{\sA^K\to\sC^K}$ is $\bra{\psi}\sum_b\Xi_b\widetilde{L}_b\Xi_b\ket{\psi}$, and that for $\cC^{\sA^K\to\sC^K}$ is $\bra{\psi'}\sum_b\hat{\Xi}_b\widetilde{L}_b\hat{\Xi}_b\ket{\psi'}$, where $\hat{\Xi}_b$ is $\Xi_b$ with $\hat{R}_i$ in place of $\mathbf{1}_{\tau_i \geq s}$.

We next replace the projectors by $\bar{\Pi}_i \coloneqq \mathbf{1}_{\bar{F}_i \geq s}$, one copy at a time. For a projector $P$, let $V_P \coloneqq P\otimes\ket{1} + (\bI - P)\otimes\ket{0}$, which records the outcome of $\{P, \bI - P\}$ in a fresh qubit. For projectors $P, P'$, a unit vector $\ket{\phi}$, an operator $0 \leq Y \leq \bI$, and unit vectors $\ket{u}, \ket{v}$,
\begin{align}
\label{eq:instrument distance}
    \big\|(V_P - V_{P'})\ket{\phi}\big\|^2 = 2\bra{\phi}(P - P')^2\ket{\phi},
    \qquad
    \big|\bra{u}Y\ket{u} - \bra{v}Y\ket{v}\big|
    \leq \f{1}{2}\big\|\ketbra{u}{u} - \ketbra{v}{v}\big\|_1
    \leq \big\|\ket{u} - \ket{v}\big\|.
\end{align}
Let $\hat{L} \coloneqq \sum_b\ketbra{b}{b}\otimes\widetilde{L}_b$ with $\ketbra{b}{b}$ on the $h$ outcome qubits, for which $0 \leq \hat{L} \leq \bI$, and let $W_0 \coloneqq V_{\mathbf{1}_{\tau_h \geq s}}\cdots V_{\mathbf{1}_{\tau_1 \geq s}}$ and $W_h \coloneqq V_{\bar{\Pi}_h}\cdots V_{\bar{\Pi}_1}$, the $i$-th factor writing into the $i$-th outcome qubit; then $\sum_b\Xi_b\widetilde{L}_b\Xi_b = W_0^\dag\hat{L}W_0$. Replacing the factors of $W_0$ by those of $W_h$ from the right, the $j$-th replacement acts on the unit vector $V_{\bar{\Pi}_{j-1}}\cdots V_{\bar{\Pi}_1}\ket{\psi}$, whose marginal on $\sA^K$ is $\tau^{\otimes K}$ because $\tau^{\otimes K}$ commutes with every $\bar{\Pi}_i$, while the remaining factors and $\hat{L}$ combine into an operator between $0$ and $\bI$. Equation~\eqref{eq:instrument distance} thus yields
\begin{align}
\label{eq:hybrid input}
    \big|\bra{\psi}W_0^\dag\hat{L}W_0\ket{\psi} - \bra{\psi}W_h^\dag\hat{L}W_h\ket{\psi}\big|
    \leq h\Big(2\tr\big[\tau^{\otimes K}\big(\mathbf{1}_{\tau_i \geq s} - \bar{\Pi}_i\big)^2\big]\Big)^{1/2},
\end{align}
the right-hand side being independent of $i$ by permutation invariance. The same argument on $\ket{\psi'}$, with $\hat{R}_i$ in place of $\mathbf{1}_{\tau_i \geq s}$ and the marginal $\sum_\lambda p_\lambda\omega_\lambda\otimes\ketbra{\lambda}{\lambda}^\sD$ on $\sA^K\sD$, which commutes with every $\bar{\Pi}_i$, bounds the difference between $\bra{\psi'}\sum_b\hat{\Xi}_b\widetilde{L}_b\hat{\Xi}_b\ket{\psi'}$ and $\bra{\psi'}W_h^\dag\hat{L}W_h\ket{\psi'}$ by $h\big(2\sum_\lambda p_\lambda\tr[\omega_\lambda(R_{\lambda,s}^{(i)} - \bar{\Pi}_i)^2]\big)^{1/2}$.

The two intermediate expectations coincide. Each $\bar{\Pi}_i$ is a polynomial in $\bar{F}_i$, so $W_h^\dag\hat{L}W_h$, a sum of products of the $\bar{\Pi}_i$, $\bI - \bar{\Pi}_i$, and $\widetilde{L}_b$, lies in $\bC[S_K]\otimes\cB(\cH^\sR)$. Such an operator $Z$ commutes with every $\Pi_\lambda$ and acts on $\cW_\lambda(d_\sA)\otimes\cS_\lambda\otimes\cH^\sR$ as $\bI^{\cW_\lambda(d_\sA)}\otimes Z_\lambda$; replacing the state on $\cW_\lambda(d_\sA)$ by $\ket{\mathrm{hw}_\lambda}$ therefore leaves its expectation unchanged, that is, $\tr[Z(\cM\otimes\rid)(\psi)] = \tr[Z\psi]$ by Eq.~\eqref{eq:carrier replacement}. As $\hat{L}$ and $W_h$ act trivially on $\sD$ and the purifying system, $\bra{\psi'}W_h^\dag\hat{L}W_h\ket{\psi'} = \bra{\psi}W_h^\dag\hat{L}W_h\ket{\psi}$.

By the triangle inequality, the difference of the two expectations at a fixed $s$ is at most the sum of the two bounds above. Averaging over $s$, using the concavity of the square root, Eqs.~\eqref{eq:input projectors} and~\eqref{eq:output projectors} at every $\lambda$, and $\sqrt{x + y} \leq \sqrt{x} + \sqrt{y}$, the left-hand side of Eq.~\eqref{eq:single compression error} is at most
\begin{align}
    h\bigg(\f{2}{w\sqrt{K-1}} + \f{2}{G}\bigg)^{1/2} + h\bigg(\f{2Q}{w(K-1)} + \f{2}{G}\bigg)^{1/2}
    &\leq \f{h}{\sqrt{w}}\bigg(\Big(\f{2}{\sqrt{K-1}}\Big)^{1/2} + \Big(\f{2Q}{K-1}\Big)^{1/2}\bigg) + \f{2\sqrt{2}h}{\sqrt{G}} \\
    &\leq \f{4h}{\sqrt{w}K^{1/4}} + \f{3h}{\sqrt{G}},
\end{align}
where the last step uses $K - 1 \geq K/2$ and $Q \leq \sqrt{K}/2$, which follow from $K \geq 4Q^2$. This is Eq.~\eqref{eq:single compression error}.

We complete the circuit implementation of the compression channel. Let $0 < \delta < 1$; the case $\delta \geq 1$ follows from any smaller tolerance. We pad $\sC$ to dimension $2^{\lceil\log{(Q+1)}\rceil}$ and apply \Cref{lem:haar implementation} with $q=Q$, tolerance $\delta/(4K)$, and $\zeta_\rH = \delta/4$. Its Haar subspace is the span of $\{\ket{1}^\sC,\ldots,\ket{Q}^\sC\}$, so the ideal unitary fixes $\ket{\!\!\perp}^\sC$ and the unused padding. We use the same returned circuit on every copy, each time with a work qubit initialized to $\ket{0}$ and discarded after the application. On the event that the lemma succeeds, the tensor products of the two isometries in Eq.~\eqref{eq:haar implementation} differ by at most $\delta/4$ in operator norm. Equation~\eqref{eq:rel of operator and diamond norm} and contractivity under discarding the work qubits therefore bound the diamond distance by $\delta/4$; the failure event adds at most $\zeta_\rH = \delta/4$. We implement the Schur transforms and the remaining operations within diamond distance $\delta/2$ following the conventions of \Cref{sec:numerics}. The total diamond distance is then at most $\delta$, and the gate count is as stated.

\end{proof}

The universal compression channel $\cC^{\sA^K\to\sC^K}$ of \Cref{lem:compression} acts on one side of the copies. In the two-party setting, where Alice and Bob share $(\rho^{\sA\sB})^{\otimes K}$, each party can apply the channel to its own system.
Let $z \coloneqq (s_\sA, s_\sB, V_\sA, V_\sB)$ be the two thresholds $s_\sA, s_\sB$ and the two Haar random unitaries $V_\sA, V_\sB$ drawn by the two channels. As in the construction of Eq.~\eqref{eq:truncation channel}, we set the state
\begin{align}
\label{eq:compact state}
    \sigma_z^{\sC_\sA\sC_\sB} \coloneqq \big(\cV_\sA^{\sC_\sA}\circ\cN_{\rho^\sA,s_\sA}^{\sA\to\sC_\sA}\big)\otimes\big(\cV_\sB^{\sC_\sB}\circ\cN_{\rho^\sB,s_\sB}^{\sB\to\sC_\sB}\big)\big(\rho^{\sA\sB}\big).
\end{align}
Then, we obtain the following corollary for two-sided compression.

\begin{corollary}[Universal collective compression on both sides]
\label{cor:compression}
For $\chi \in \{\sA, \sB\}$, let $w_\chi$, $Q_\chi$, and $G$ obey the conditions of \Cref{lem:compression} with $Q_\chi \geq \lceil 2/w_\chi \rceil$, and let $K \geq 4\max_\chi Q_\chi^2$ and $1 \leq h \leq K$. Suppose that the channel of \Cref{lem:compression} is applied to the systems $\sA^K$ and $\sB^K$ of $(\rho^{\sA\sB})^{\otimes K}$ at the parameters $(w_\sA, Q_\sA)$ and $(w_\sB, Q_\sB)$, respectively. Then, the trace distance between the output state on any $h$ selected copies and the mixture $\bE_z\big[(\sigma_z^{\sC_\sA\sC_\sB})^{\otimes h}\big]$ is bounded by 
\begin{align}
\label{eq:compression error}
    \varepsilon_{\mathrm{cp}}(h) \coloneqq \f{8h}{\sqrt{w_{\min}}K^{1/4}} + \f{6h}{\sqrt{G}},
\end{align}
where $w_{\min} \coloneqq \min_\chi w_\chi$, and the law of $z$ does not depend on the selected $h$ copies.
\end{corollary}

\begin{proof}[Proof of \Cref{cor:compression}]

We apply \Cref{lem:compression} first to $\sA^K$, keeping $\sB^K$ as part of the reference, and then to $\sB^K$; a trace-preserving operation on the other system leaves the marginal required by the lemma unchanged. Each of the two ideal channels uses a single threshold and a single unitary for all the copies, so their joint output on the selected $h$ copies is the mixture $\bE_z\big[\sigma_z^{\otimes h}\big]$. Summing the two errors of Eq.~\eqref{eq:single compression error} yields Eq.~\eqref{eq:compression error}.

\end{proof}

As we will see in \Cref{sec:achievability result}, our achievable rate in \Cref{thm:achievability by efficient protocol formal} is not stated in terms of the actual state $\rho^{\sA\sB}$. It concerns instead a state $\hat{\rho}^{\sA\sB}$ that is close to $\rho^{\sA\sB}$ and whose ranks are polynomially bounded. The compression, however, is applied to $\rho^{\sA\sB}$, and it is not immediate what becomes of $\hat{\rho}^{\sA\sB}$ in the compressed systems. The following lemma shows that each component $\sigma_z^{\sC_\sA\sC_\sB}$ of the mixture in \Cref{cor:compression} is close to $\hat{\rho}^{\sA\sB}$ up to local isometries depending on $z$, which leave the R\'enyi coherent information unchanged.

Let $\chi \in \{\sA,\sB\}$. For a tolerance parameter $\epsilon_\star \in (0, 1)$ and integers $\hat{r}, \hat{q}$ such that $\rank(\hat{\rho}^{\sA\sB}) \leq \hat{r}$ and $\rank(\hat{\rho}^\sA) \leq \hat{q}$, set
\begin{align}
\label{eq:compact dimensions}
    q_\sA &= \hat{q}, \qquad 
    q_\sB = \hat{r}\hat{q}, \qquad 
    w_\chi = \f{\epsilon_\star}{4q_\chi}, \qquad 
    Q_\chi = \bigg\lceil\f{2}{w_\chi}\bigg\rceil, \qquad 
    d_{\sC_\chi} = 2^{\lceil\log{(Q_\chi+1)}\rceil}.
\end{align}
Here, $q_\chi$ bounds the rank of $\hat{\rho}^\chi$, because $\rank(\hat{\rho}^\sB) \leq \rank(\hat{\rho}^{\sA\sB}) \rank(\hat{\rho}^\sA)$ holds.
From \Cref{lem:compression}, $Q_\chi$ is the dimension of the retained subspace, which leaves $\ket{\!\!\perp}^{\sC_\chi}$ and unused padding in $\sC_\chi$.

\begin{lemma}[Nearby state after truncation]
\label{lem:geometry}
Take the parameters of Eq.~\eqref{eq:compact dimensions}, and let $\f{1}{2}\|\rho^{\sA\sB} - \hat{\rho}^{\sA\sB}\|_1 \leq \epsilon_\star$. Then, for every $z$, there are isometries $W_\sA$ and $W_\sB$ from the supports of $\hat{\rho}^\sA$ and $\hat{\rho}^\sB$ into $\sC_\sA$ and $\sC_\sB$ such that the state $\hat{\sigma}_z^{\sC_\sA\sC_\sB} \coloneqq (W_\sA\otimes W_\sB)\hat{\rho}^{\sA\sB}(W_\sA\otimes W_\sB)^\dag$ satisfies 
\begin{align}
\label{eq:compact witness}
    \rF\big(\sigma_z^{\sC_\sA\sC_\sB}, \hat{\sigma}_z^{\sC_\sA\sC_\sB}\big) \geq 1 - 21\epsilon_\star.
\end{align}
\end{lemma}

As $\hat{\sigma}_z^{\sC_\sA\sC_\sB}$ is the image of $\hat{\rho}^{\sA\sB}$ under local isometries, $\widetilde{I}_\beta(\sC_\sA\rangle\sC_\sB)_{\hat{\sigma}_z} = \widetilde{I}_\beta(\sA\rangle\sB)_{\hat{\rho}}$ holds for every $\beta$. The lemma holds for every $z$, and this value does not depend on $z$, which is why a bound obtained for a fixed $z$ holds for the mixture over $z$ as well. Neither $\hat{\sigma}_z$ nor the isometries are computed by the protocol.

\begin{proof}[Proof of \Cref{lem:geometry}]

The statement is trivial for $\epsilon_\star \geq 1/21$, so we consider $\epsilon_\star \leq 1/21$.
We first confirm $\rank(\hat{\rho}^\sB) \leq \rank(\hat{\rho}^{\sA\sB}) \rank(\hat{\rho}^\sA)$, which was used in Eq.~\eqref{eq:compact dimensions}. Let $\hat{\rho}^{\sA\sB} = \sum_j p_j\ketbra{\psi_j}{\psi_j}^{\sA\sB}$ be its spectral decomposition, where the number of terms is $\rank(\hat{\rho}^{\sA\sB})$. As the support of $\psi_j^\sA$ is contained in that of $\hat{\rho}^\sA$, the Schmidt rank of $\ket{\psi_j}^{\sA\sB}$ across $\sA:\sB$ is at most $\rank(\hat{\rho}^\sA)$, and hence so is $\rank(\psi_j^\sB)$. As the support of $\hat{\rho}^\sB = \sum_j p_j\psi_j^\sB$ is the span of those of the $\psi_j^\sB$, we obtain the claim, and thus $q_\chi \geq \rank(\hat{\rho}^\chi)$ for $\chi \in \{\sA, \sB\}$.

We next bound the weight that the truncation discards from $\hat{\rho}^{\sA\sB}$ by $P_{\rho^\chi,s_\chi}$, which projects onto the eigenvalues of $\rho^\chi$ at least $s_\chi$. Let $\Pi_{\hat{\rho}}^\chi$ be the projector onto $\supp(\hat{\rho}^\chi)$, whose rank is at most $q_\chi$. Since $\tr[\Pi_{\hat{\rho}}^\chi\hat{\rho}^\chi] = 1$ and the trace distance between $\rho^\chi$ and $\hat{\rho}^\chi$ is at most $\epsilon_\star$, we have $\tr[\Pi_{\hat{\rho}}^\chi\rho^\chi] \geq 1 - \epsilon_\star$. The projector $\Pi_{\rho}^\chi$ onto the largest $q_\chi$ eigenvalues of $\rho^\chi$ maximizes this weight among the projectors of rank at most $q_\chi$, and hence $\tr[\Pi_{\rho}^\chi\rho^\chi] \geq 1 - \epsilon_\star$.
As $s_\chi \leq 2w_\chi$, the projector $P_{\rho^\chi,s_\chi}$ discards the eigenvalues below $2w_\chi$ from the range of $\Pi_\rho^\chi$. There are at most $q_\chi$ of them, and their total weight is at most $2q_\chi w_\chi = \epsilon_\star/2$. Thus $\tr[P_{\rho^\chi,s_\chi}\rho^\chi] \geq 1 - 3\epsilon_\star/2$, and the trace distance again gives $\tr[P_{\rho^\chi,s_\chi}\hat{\rho}^\chi] \geq 1 - 5\epsilon_\star/2$.

The first term of $\cV_\chi^{\sC_\chi}\circ\cN_{\rho^\chi,s_\chi}^{\chi\to\sC_\chi}$ in Eq.~\eqref{eq:compact state} has the Kraus operator $V_\chi V_{\rho^\chi,s_\chi}P_{\rho^\chi,s_\chi}$. Restricted to $\supp(\hat{\rho}^\chi)$, it becomes $Y_\chi \coloneqq V_\chi V_{\rho^\chi,s_\chi}P_{\rho^\chi,s_\chi}\Pi_{\hat{\rho}}^\chi$, and we write its polar decomposition as $Y_\chi = W_\chi\sqrt{H_\chi}$, where $H_\chi \coloneqq \Pi_{\hat{\rho}}^\chi P_{\rho^\chi,s_\chi}\Pi_{\hat{\rho}}^\chi$ obeys $0 \leq H_\chi \leq \bI$.
As $Q_\chi \geq q_\chi \geq \rank(\hat{\rho}^\chi)$ leaves room in $\sC_\chi$, we extend $W_\chi$ to an isometry on the whole of $\supp(\hat{\rho}^\chi)$, and take $W_\sA$ and $W_\sB$ in the statement to be these.
We denote by $\hat{\varsigma}_z^{\sC_\sA\sC_\sB}$ the state obtained by replacing $\rho^{\sA\sB}$ with $\hat{\rho}^{\sA\sB}$ in Eq.~\eqref{eq:compact state}.

Fix a purification $\ket{\hat{\rho}}^{\sA\sB\sE}$ of $\hat{\rho}^{\sA\sB}$. Since $\Pi_{\hat{\rho}}^\chi$ leaves it unchanged and $W_\chi^\dag W_\chi = \bI$, the overlap of $(W_\sA\otimes W_\sB)\ket{\hat{\rho}}^{\sA\sB\sE}$ and $(Y_\sA\otimes Y_\sB)\ket{\hat{\rho}}^{\sA\sB\sE}$ is
\begin{align}
    \bra{\hat{\rho}}\big(\sqrt{H_\sA}\otimes\sqrt{H_\sB}\big)\ket{\hat{\rho}}^{\sA\sB\sE}
    &= \tr\Big[\hat{\rho}^{\sA\sB}\big(\sqrt{H_\sA}\otimes\sqrt{H_\sB}\big)\Big] \\
    &\geq \tr\big[\hat{\rho}^{\sA\sB}(H_\sA \otimes H_\sB)\big] \\
    &\geq 1 - \tr\big[\hat{\rho}^\sA(\bI - H_\sA)\big] - \tr\big[\hat{\rho}^\sB(\bI - H_\sB)\big] \\
    &\geq 1 - 5\epsilon_\star,
\end{align}
where the first inequality uses $H_\chi \leq \sqrt{H_\chi}$ and the second $\bI - H_\sA\otimes H_\sB \leq (\bI - H_\sA)\otimes\bI + \bI\otimes(\bI - H_\sB)$.
The third inequality comes from $\tr[H_\chi\hat{\rho}^\chi] = \tr[P_{\rho^\chi,s_\chi}\hat{\rho}^\chi] \geq 1 - 5\epsilon_\star/2$, which holds as $\hat{\rho}^\chi$ is supported on the range of $\Pi_{\hat{\rho}}^\chi$.

Tracing out $\sE$, the state $(W_\sA\otimes W_\sB)\ket{\hat{\rho}}^{\sA\sB\sE}$ yields $\hat{\sigma}_z^{\sC_\sA\sC_\sB}$, while $(Y_\sA\otimes Y_\sB)\ket{\hat{\rho}}^{\sA\sB\sE}$ gives an operator, not a state, that is the part of $\hat{\varsigma}_z^{\sC_\sA\sC_\sB}$ orthogonal to $\ket{\!\!\perp}^{\sC_\sA}$ and $\ket{\!\!\perp}^{\sC_\sB}$. The fidelity $\rF(\hat{\sigma}_z^{\sC_\sA\sC_\sB}, \hat{\varsigma}_z^{\sC_\sA\sC_\sB})$ is thus at least the square of the overlap above, that is, $(1-5\epsilon_\star)^2 \geq 1 - 10\epsilon_\star$.
By the Fuchs--van de Graaf inequalities in Eq.~\eqref{eq:fuchsvandegraaf} and the assumption, we have $\sqrt{\rF(\rho^{\sA\sB}, \hat{\rho}^{\sA\sB})} \geq 1 - \epsilon_\star$, and hence $\rF(\rho^{\sA\sB}, \hat{\rho}^{\sA\sB}) \geq 1 - 2\epsilon_\star$. The channels in Eq.~\eqref{eq:compact state} cannot decrease the fidelity, which gives $\rF(\sigma_z^{\sC_\sA\sC_\sB}, \hat{\varsigma}_z^{\sC_\sA\sC_\sB}) \geq 1 - 2\epsilon_\star$.
Then, we obtain 
\begin{align}
    \sqrt{1-\rF\big(\sigma_z^{\sC_\sA\sC_\sB}, \hat{\sigma}_z^{\sC_\sA\sC_\sB}\big)}
    &\leq \sqrt{1-\rF\big(\sigma_z^{\sC_\sA\sC_\sB}, \hat{\varsigma}_z^{\sC_\sA\sC_\sB}\big)} + \sqrt{1-\rF\big(\hat{\varsigma}_z^{\sC_\sA\sC_\sB}, \hat{\sigma}_z^{\sC_\sA\sC_\sB}\big)} \\
    &\leq \sqrt{2\epsilon_\star} + \sqrt{10\epsilon_\star} \\
    &\leq \sqrt{21\epsilon_\star},
\end{align}
where we used the triangle inequality for the purified distance and $\sqrt{2}+\sqrt{10} \leq \sqrt{21}$.
This provides Eq.~\eqref{eq:compact witness}.

\end{proof}


\subsubsection{Alice's two-stage measurement}
\label{sec:two instruments}

Alice measures the compressed copies in two stages. The first acts on one copy at a time and reduces the dimension of her register while retaining most of the conditional R\'enyi entropy. The second applies a decoupling measurement to blocks of the resulting copies. Splitting the measurement in this way keeps the number of outcomes of the decoupling measurement small while preserving the target rate.

We here use another conditional R\'enyi entropy $\widetilde{H}_\alpha(\sA|\sE)_{\rho|\sigma} \coloneqq -\widetilde{D}_\alpha(\rho^{\sA\sE}\|\bI^\sA\otimes\sigma^\sE)$ with a reference state $\sigma^\sE$. By definition, $\widetilde{H}_\alpha(\sA|\sE)_\rho = \max_{\sigma^\sE}\widetilde{H}_\alpha(\sA|\sE)_{\rho|\sigma} \geq \widetilde{H}_\alpha(\sA|\sE)_{\rho|\rho}$. 
Set $p \coloneqq 1/\beta \in (1,2]$ and let $\alpha$ be the order dual to $\beta$, i.e., $1/\alpha + 1/\beta = 2$. Then, for a pure state $\ket{\rho}^{\sA\sB\sE}$, the following holds: 
\begin{align}
\label{eq:entropy conversion}
    \widetilde{I}_\beta(\sA\rangle\sB)_\rho
    = \widetilde{H}_\alpha(\sA|\sE)_\rho
    \leq \widetilde{H}_p(\sA|\sE)_{\rho|\rho},
\end{align}
where the equality is the duality in Eq.~\eqref{eq:duality} and the inequality follows from~{\cite[Corollary~4]{Tomamichel2014Relatingdifferentquantumgeneralizations}}.

Let $\ket{\psi}^{\sA\sB\sE}$ be a pure state with $d_\sA$ a power of two, and let $q < d_\sA$ also be a power of two. For a unitary $U^\sA$, let $\eta_U^{\sS\sB\sX\sE}$ be the state obtained from $\ket{\psi}^{\sA\sB\sE}$ by applying $U^\sA$ to $\sA$, measuring in the computational basis the label $x$ of the $d_\sA/q$ blocks of dimension $q$ and recording it in a register $\sX$, and retaining the $q$-dimensional register $\sS$ within the block. Let $\eta^{\sS\sB\sX\sU\sE}$ be the state in which $U^\sA$ is drawn from an exact unitary $2$-design and recorded in an additional register $\sU$. Then, we obtain the following lemmas.

\begin{lemma}[First-stage entropy retention]
\label{lem:first instrument}
Suppose that $\widetilde{H}_p(\sA|\sE)_{\psi|\psi} \geq J$. Then,
\begin{align}
\label{eq:first stage entropy}
    \widetilde{H}_p(\sS|\sX\sU\sE)_\eta
    \geq J - \f{1}{p-1}\log{\big(1 + 2^{-(p-1)(\log{q} - J)}\big)}.
\end{align}
The $2$-design can be taken to have fewer than $d_\sA^5$ elements, sampled and compiled into Clifford circuits of size $\poly(\log{d_\sA})$.
\end{lemma}

The loss in Eq.~\eqref{eq:first stage entropy} decreases with the margin $\log{q} - J$, and at $q = d_\sA$ we instead apply the identity, with no label, retaining the condition $\widetilde{H}_p(\sA|\sE)_{\psi|\psi} \geq J$.
The labels $\sX$ and $\sU$ produced at this stage, written $\sX_0$ and $\sU_0$ in \Cref{sec:decoder}, are retained as registers of the state on which the second stage acts; they are not outcomes that Bob's decoder for the decoupling, provided in the next subsection, has to reproduce.
The protocol in \Cref{sec:protocol} applies this lemma to $\sC_\sA$ with $q = 2^{m_0}$ in Eq.~\eqref{eq:instrument sizes}, where the margin is large enough that the output obeys $\widetilde{H}_p(\sS|\sX\sU\sE)_\eta \geq J - \xi/8$ for $\xi = \min\{\gamma, 1/4\}$, with $\sS$, $\sX$, and $\sU$ all of polynomial dimension. The circuit uses exact Clifford gates, with finite uniform sampling as in \Cref{sec:numerics}.

\begin{proof}[Proof of \Cref{lem:first instrument}]

Let $\Pi_x^\sA$ be the projector onto the $x$-th of the $d_\sA/q$ blocks of dimension $q$. Define the channels $\cP^\sA(X^\sA) \coloneqq \sum_x\Pi_x^\sA X^\sA\Pi_x^\sA$ and $\cF_U^\sA \coloneqq (\cU^{\sA})^\dag\circ\cP^\sA\circ\cU^\sA$ for a fixed $U$.
By the definition of an exact unitary $2$-design, the average of the latter channel over $U^\sA$ on an operator $Y^{\sA\sE}$ is given by
\begin{align}
\label{eq:pinching twirl}
    \bE_U\big[\big(\cF_U^\sA\otimes\rid^\sE\big)\big(Y^{\sA\sE}\big)\big] = \mu Y^{\sA\sE} + \nu\,\bI^\sA\otimes Y^\sE,
\end{align}
where $\mu = \f{d_\sA q-1}{d_\sA^2-1}$ and $\nu = \f{d_\sA-q}{d_\sA^2-1}$. 

A direct calculation implies
$\tr\big[\big(\cF_U(Y)\big)^p\big] = \tr\big[Y\big(\cF_U(Y)\big)^{p-1}\big]$ for $Y^{\sA\sE} \geq 0$.
Since $X \mapsto X^{p-1}$ is operator concave for $p \leq 2$, averaging this identity over $U$ yields
\begin{align}
    \bE_U\big[\tr\big[\big(\cF_U^\sA(Y^{\sA\sE})\big)^p\big]\big]
    \leq \tr\big[Y^{\sA\sE}\big(\mu Y^{\sA\sE} + \nu\,\bI^\sA\otimes Y^\sE\big)^{p-1}\big],
\end{align}
where we used Eq.~\eqref{eq:pinching twirl}. We then use the trace inequality~\cite{Dinh2019Twotraceinequalities}: for $A, B \geq 0$ and $1 < p \leq 2$, 
\begin{align}
\label{eq:trace subadditivity}
    \tr\big[A(A+B)^{p-1}\big] \leq \tr\big[A^p\big] + \tr\big[AB^{p-1}\big],
\end{align}
which is an equality at $p = 2$.
Using these bounds, we obtain 
\begin{align}
\label{eq:pinching moment}
    \bE_U\big[\tr\big[\big(\cF_U^\sA(Y^{\sA\sE})\big)^p\big]\big] \leq \mu^{p-1}\tr\big[\big(Y^{\sA\sE}\big)^p\big] + \nu^{p-1}\tr\big[\big(Y^\sE\big)^p\big].
\end{align}

We derive a lower bound on $\widetilde{H}_p(\sS|\sX\sU\sE)_\eta$, using the reference $\tau^{\sX\sU\sE} \coloneqq \pi^\sX\otimes\eta^\sU\otimes\psi^\sE$.
Take the above operator $Y^{\sA\sE}$ as $Y^{\sA\sE} = (\psi^\sE)^{-\f{p-1}{2p}}\psi^{\sA\sE}(\psi^\sE)^{-\f{p-1}{2p}}$. By the definition of $\widetilde{H}_p(\sA|\sE)_{\psi|\psi}$ and the condition of the lemma, we have 
\begin{align}
\label{eq:moment of Y}
    \tr\big[\big(Y^{\sA\sE}\big)^p\big] = 2^{-(p-1)\widetilde{H}_p(\sA|\sE)_{\psi|\psi}} \leq 2^{-(p-1)J}, \qquad
    \tr\big[\big(Y^\sE\big)^p\big] = 1,
\end{align}
where note that $Y^\sE = (\psi^\sE)^{1/p}$.
The post-measurement state $\eta$ is given by
\begin{align}
    \eta^{\sS\sX\sU\sE} = \sum_U\Pr[U]\ketbra{U}{U}^\sU\otimes\sum_x\ketbra{x}{x}^\sX\otimes \bra{x}^\sX U^\sA\psi^{\sA\sE}(U^\sA)^\dag\ket{x}^\sX,
\end{align}
where $\Pr[U]$ is the probability of drawing $U^\sA$ from the $2$-design, and $\ket{U}^\sU$ denotes the orthonormal basis state recording the description of $U^\sA$.
Then, a direct calculation provides
\begin{align}
    2^{-(p-1)\widetilde{H}_p(\sS|\sX\sU\sE)_{\eta|\tau}}
    = \Big(\f{d_\sA}{q}\Big)^{p-1}\bE_U\Big[\sum_x\tr\big[\big(\bra{x}^\sX U^\sA Y^{\sA\sE}(U^\sA)^\dag\ket{x}^\sX\big)^p\big]\Big]
    = \Big(\f{d_\sA}{q}\Big)^{p-1}\bE_U\big[\tr\big[\big(\cF_U^\sA(Y^{\sA\sE})\big)^p\big]\big].
\end{align}
Here, the weight $q/d_\sA$ of each $x$ in $\pi^\sX$ gives the factor $(d_\sA/q)^{p-1}$, and the second equality holds because the blocks are orthogonal. 
By Eqs.~\eqref{eq:pinching moment} and~\eqref{eq:moment of Y} with $\mu \leq q/d_\sA$ and $\nu \leq 1/d_\sA$, the right-hand side is at most $2^{-(p-1)J} + q^{-(p-1)}$. Since $\widetilde{H}_p(\sS|\sX\sU\sE)_\eta \geq \widetilde{H}_p(\sS|\sX\sU\sE)_{\eta|\tau}$, taking $-\f{1}{p-1}\log$ yields Eq.~\eqref{eq:first stage entropy}.

For an exact $2$-design, one convenient choice is the ensemble of Ref.~\cite{Cleve2016Nearlinearconstructionsexactunitary2designs}, which consists of fewer than $d_\sA^5$ Clifford unitaries on $\log{d_\sA}$ qubits, each sampled and compiled into a circuit of $\poly(\log{d_\sA})$ Clifford gates.

\end{proof}

The first-stage measurement described above acts on a single copy and is designed to reduce the dimension of Alice's register, retaining most of the conditional R\'enyi entropy. We next consider the second-stage measurement on multiple copies, whose goal is to achieve decoupling. The following lemma bounds its error in terms of the conditional R\'enyi entropy of the input and the output dimension of the measurement.

\begin{lemma}[Second-stage decoupling]
\label{lem:second instrument}
For $U^\sA$ drawn from an exact unitary $2$-design,
\begin{align}
\label{eq:decoupling}
    \bE_U\Big[1 - \rF\big(\eta_U^{\sS\sX\sE}, \pi^{\sS}\otimes\pi^\sX\otimes\psi^\sE\big)\Big]
    \leq 2^{1 - (1-\beta)\big(\widetilde{H}_p(\sA|\sE)_\psi - \log{q}\big)}.
\end{align}
\end{lemma}

The protocol draws $U$ for this stage from the uniform Clifford group on $\log{d_\sA}$ qubits, sampled and compiled in $\poly(\log{d_\sA})$ time~\cite{Koenig2014EfficientlyselectarbitraryClifford}. Finite sampling uses the capped rejection procedure of \Cref{sec:numerics}. Unlike the first stage, the seed enters Bob's decoder in \Cref{sec:decoder} only as a circuit, so the number of its values does not enter the dimension count.

\begin{proof}[Proof of \Cref{lem:second instrument}]

For any operator $Y^{\sA\sE}$, define the difference
\begin{align}
    \delta_U\big(Y^{\sA\sE}\big) \coloneqq \sum_x\ketbra{x}{x}^\sX\otimes \bra{x}^\sX U^\sA Y^{\sA\sE}(U^\sA)^\dag\ket{x}^\sX - \pi^\sS\otimes\pi^\sX\otimes Y^\sE.
\end{align}
From the triangle inequality and the monotonicity of the trace norm under channels, it follows that $\|\delta_U(Y)\|_1 \leq 2\|Y\|_1$. For the Hilbert--Schmidt norm, it satisfies
\begin{align}
    \|\delta_U(Y)\|_2^2
    &= \sum_x\tr\big[\big(\bra{x}^\sX U^\sA Y^{\sA\sE}(U^\sA)^\dag\ket{x}^\sX\big)^\dag\bra{x}^\sX U^\sA Y^{\sA\sE}(U^\sA)^\dag\ket{x}^\sX\big]
    - \f{1}{d_\sA}\tr\big[\big(Y^\sE\big)^\dag Y^\sE\big] \notag\\
    &= \tr\big[\big(Y^{\sA\sE}\big)^\dag\cF_U^\sA\big(Y^{\sA\sE}\big)\big]
    - \f{1}{d_\sA}\tr\big[\big(Y^\sE\big)^\dag Y^\sE\big],
\end{align}
where $\cF_U^\sA$ is the channel defined in the proof of \Cref{lem:first instrument}.
By Eq.~\eqref{eq:pinching twirl}, averaging it over $U$ gives
\begin{align}
    \bE_U\big[\|\delta_U(Y)\|_2^2\big]
    = \mu\tr\big[\big(Y^{\sA\sE}\big)^\dag Y^{\sA\sE}\big] + \Big(\nu - \f{1}{d_\sA}\Big)\tr\big[\big(Y^\sE\big)^\dag Y^\sE\big]
    \leq \f{q}{d_\sA}\|Y\|_2^2,
\end{align}
where we used $\mu \leq q/d_\sA$ and $\nu \leq 1/d_\sA$.

We regard $Y^{\sA\sE} \mapsto \bigoplus_U\delta_U(Y)$ as a single linear map, with the direct sum taken over $U$ of positive probability. Set $\Omega \coloneqq \bigoplus_U\Pr[U]\,\bI^{\sX\sS\sE}$ and equip the output with the weighted Schatten norms
\begin{align}
    \|Z\|_{s,\Omega} \coloneqq \big\|\Omega^{\f{1}{2s}}Z\Omega^{\f{1}{2s}}\big\|_s,
    \qquad 1 \leq s \leq 2.
\end{align}
For $Z = \bigoplus_U\delta_U(Y)$, this gives $\|Z\|_{s,\Omega} = \big(\bE_U[\|\delta_U(Y)\|_s^s]\big)^{1/s}$. Thus, the bounds for the trace norm and the Hilbert--Schmidt norm above give operator norms at most $2$ from $\|\cdot\|_1$ to $\|\cdot\|_{1,\Omega}$ and $(q/d_\sA)^{1/2}$ from $\|\cdot\|_2$ to $\|\cdot\|_{2,\Omega}$. Interpolating these two endpoints with the weights $(2\beta-1, 2-2\beta)$, which give $\f{1}{p} = (2\beta-1) + \f{2-2\beta}{2}$, the Riesz--Thorin theorem for weighted Schatten norms~\cite[Theorem~4]{Beigi2013Sandwicheddataprocessing} yields
\begin{align} 
    \label{eq:interpolation} \big(\bE_U\big[\|\delta_U(Y)\|_p^p\big]\big)^{1/p} \leq 2^{2\beta-1}\Big(\f{q}{d_\sA}\Big)^{1-\beta}\|Y\|_p. 
\end{align}

For a reference state $\tau^\sE$, we take $Y^{\sA\sE} = (\tau^\sE)^{-\f{1-\beta}{2}}\psi^{\sA\sE}(\tau^\sE)^{-\f{1-\beta}{2}}$. Then, we see that 
\begin{align}
    \delta_U(\psi) = \big(\bI^{\sX\sS}\otimes(\tau^\sE)^{\f{1-\beta}{2}}\big)\delta_U(Y)\big(\bI^{\sX\sS}\otimes(\tau^\sE)^{\f{1-\beta}{2}}\big).
\end{align}
Using H\"{o}lder's inequality with the exponents $\big(\f{2}{1-\beta}, p, \f{2}{1-\beta}\big)$, which satisfy $(1-\beta) + \f{1}{p} = 1$, gives $\|\delta_U(\psi)\|_1 \leq d_\sA^{1-\beta}\|\delta_U(Y)\|_p$, where note that $d_\sA$ coincides with the dimension of $\sX\sS$.
Then, noting that $\|Y\|_p = 2^{-(1-\beta)\widetilde{H}_p(\sA|\sE)_{\psi|\tau}}$, we obtain
\begin{align}
    \bE_U\big[\|\delta_U(\psi)\|_1\big]
    \leq 2q^{1-\beta} 2^{-(1-\beta)\widetilde{H}_p(\sA|\sE)_{\psi|\tau}},
\end{align}
where we used Eq.~\eqref{eq:interpolation} with $2^{2\beta-1} \leq 2$.
Using $1 - \rF(\rho,\sigma) \leq 2\big(1-\sqrt{\rF(\rho,\sigma)}\big) \leq \|\rho - \sigma\|_1$ by the Fuchs--van de Graaf inequality in Eq.~\eqref{eq:fuchsvandegraaf}, and then optimizing $\tau^\sE$, we arrive at Eq.~\eqref{eq:decoupling}.

\end{proof}


\subsubsection{Bob's decoder}
\label{sec:decoder}

After Alice's second-stage measurement, we construct Bob's decoder based on Uhlmann's theorem~\cite{uhlmann1976transition} to the canonical purification of the marginal on Alice's retained register and the environment. Decoupling makes this canonical purification close to a maximally entangled state tensored with a purification of the environment. The following lemma implements this transformation for states specified by a factored operator on a tensor power, with cost polynomial in the dimension of one tensor factor and the number of factors.

To provide the statement, we take $\sD_0$ to be a register of dimension $d_{\sD_0}$, a power of two, and set $\sD \coloneqq \sD_0^t$. We consider an operator on $\sD_0$ of the factored form
\begin{align}
\label{eq:one copy polar}
    C_0^{\sD_0} \coloneqq V_0^{\sD_0}\sqrt{\sigma_0^{\sD_0}},
    \qquad \sigma_0^{\sD_0} = W_0^{\sD_0}\,\mathrm{diag}(s_j^2)\,\big(W_0^{\sD_0}\big)^\dag,
\end{align}
where $V_0^{\sD_0}$ and $W_0^{\sD_0}$ are unitaries and $\sigma_0^{\sD_0}$ is a state, and set $C^\sD \coloneqq \big(C_0^{\sD_0}\big)^{\otimes t}$ and $\sigma^\sD \coloneqq \big(\sigma_0^{\sD_0}\big)^{\otimes t}$. The eigenvalues $\{s_j^2\}_j$ are rationals of total bit length $b_0$.

\begin{lemma}[Tensor polar circuit]
\label{lem:polar}
For any projector $P^\sD$ and any $\delta, a_\rp, \epsilon_\ru \in (0, 1/4)$, there is a unitary $\widehat{U}^{\sD\sW}$ that satisfies
\begin{align}
\label{eq:polar endpoint}
    \Big\|\big(\widehat{U}^{\sD\sW}\big)^\dag\big(C^\sD P^\sD\otimes\ket{0}^\sW\big) - \sqrt{P^\sD\sigma^\sD P^\sD}\otimes\ket{0}^\sW\Big\|_2 
    \leq 2\sqrt{2\delta} + \sqrt{2a_\rp} + \epsilon_\ru\big\|C^\sD P^\sD\big\|_2,
\end{align}
where the register $\sW$ consists of $g = \poly\big(d_{\sD_0}, t, \log{(1/\delta)}, 1/a_\rp, 1/\epsilon_\ru, b_0\big)$ qubits. Given circuits of size $\poly(g)$ for $V_0^{\sD_0}$, $W_0^{\sD_0}$, and a controlled application of $\bI^\sD - 2P^\sD$, together with the list $\{s_j^2\}_j$, the unitary $\widehat{U}^{\sD\sW}$ is implemented by a circuit of $\poly(g)$ one- and two-qubit gates, and its description is computed classically in $\poly(g)$ time.
\end{lemma}

Under the vectorization of Eq.~\eqref{eq:definition of vectorization}, the vectors $\rvec(C^\sD P^\sD)$ and $\rvec\big(\sqrt{P^\sD\sigma^\sD P^\sD}\big)$ lie on $\hat{\sD}\sD$ and have the same marginal on $\sD$, where $\hat{\sD}$ is a copy of $\sD$. The latter vector is proportional to the canonical purification of this marginal after normalization. \Cref{lem:polar} thus implements this change of purification approximately by $\big(\widehat{U}^{\hat{\sD}\sW}\big)^\dag$, acting only on the purifying register $\hat{\sD}$ and the initialized work register $\sW$. The circuit approximates an Uhlmann transformation from $\rvec(C^\sD P^\sD)$ to $\rvec\big(\sqrt{P^\sD\sigma^\sD P^\sD}\big)$.
In the protocol of \Cref{sec:protocol}, Bob holds $\hat{\sD}$, and the range of $P^\sD$ is identified with Alice's retained register $\sS$ together with the environment for each measurement outcome.

\begin{proof}[Proof of \Cref{lem:polar}]

Let $V^\sD \coloneqq \big(V_0^{\sD_0}\big)^{\otimes t}$, $D \coloneqq d_{\sD_0}^t$, and $Q^\sD \coloneqq \bI - P^\sD$. We regularize $\sigma$ as
\begin{align}
\label{eq:regularization}
    \sigma_\delta \coloneqq \f{\sigma + \delta\bI/D}{1+\delta},
    \qquad C_\delta \coloneqq V\sqrt{\sigma_\delta},
    \qquad F \coloneqq P + \sqrt{a_\rp}Q,
\end{align}
and write $C_\delta F = U_\rp R$ for the polar decomposition of $C_\delta F$, where $R = \sqrt{F\sigma_\delta F}$.
Both factors $C_\delta$ and $F$ are invertible.
The Powers--St{\o}rmer inequality~\cite{powers1970free} ensures $\|C_\delta-C\|_2 \leq \|\sigma_\delta - \sigma\|_1^{1/2} \leq \sqrt{2\delta}$.

We first derive the error bound in Eq.~\eqref{eq:polar endpoint}. In the block decomposition specified by projectors $P$ and $Q = \bI-P$, we write $R = \big(\begin{smallmatrix}R_{11} & R_{12}\\ R_{12}^\dag & R_{22}\end{smallmatrix}\big)$. 
Since $U_\rp^\dag C_\delta P = RP$ and the $P$ block of $|C_\delta P|$ is $\sqrt{R_{11}^2 + R_{12}R_{12}^\dag}$, we obtain
\begin{align}
    \big\|RP - |C_\delta P|\big\|_2^2 \leq 2\|R_{12}\|_2^2 \leq 2\tr\big[QR^2Q\big] = 2a_\rp\tr\big[Q\sigma_\delta\big] \leq 2a_\rp,
\end{align}
where we used $\sqrt{R_{11}^2+R_{12}R_{12}^\dag} \geq R_{11}$.
We next replace $C_\delta$ and $\sigma_\delta$ by $C$ and $\sigma$, respectively. The unitary invariance gives
\begin{align}
    \big\|U_\rp^\dag CP-U_\rp^\dag C_\delta P\big\|_2
    = \big\|(C-C_\delta)P\big\|_2
    \leq \sqrt{2\delta}.
\end{align}
Moreover, the Powers--St{\o}rmer inequality yields 
\begin{align}
    \Big\|\sqrt{P\sigma P}-\sqrt{P\sigma_\delta P}\Big\|_2
    \leq \big\|P(\sigma-\sigma_\delta)P\big\|_1^{1/2}
    \leq \sqrt{2\delta}.
\end{align}
Since $U_\rp^\dag C_\delta P=RP$ and $|C_\delta P|=\sqrt{P\sigma_\delta P}$, by combining these bounds, we obtain
\begin{align}
\label{eq:unitary error}
    \Big\|U_\rp^\dag CP-\sqrt{P\sigma P}\Big\|_2
    \leq \big\|(C-C_\delta)P\big\|_2
    + \big\|RP-|C_\delta P|\big\|_2 
    + \Big\|\sqrt{P\sigma_\delta P}-\sqrt{P\sigma P}\Big\|_2 
    \leq 2\sqrt{2\delta}+\sqrt{2a_\rp}.
\end{align}

It remains to account for the circuit approximation. Suppose that a unitary $\widehat{U}^{\sD\sW}$ satisfies
\begin{align}
    \Big\|\widehat{U}^{\sD\sW}\big(\bI^\sD\otimes\ket{0}^\sW\big)
    - U_\rp^\sD\otimes\ket{0}^\sW\Big\|_\infty
    \leq \epsilon_\ru.
\end{align}
Multiplying the difference by $(\widehat{U}^{\sD\sW})^\dag$ on the left and $(U_\rp^\sD)^\dag$ on the right gives the same bound for the inverse circuit. Hence, using $\|AB\|_2\leq\|A\|_\infty\|B\|_2$, we obtain
\begin{align}
\label{eq:block unitary error}
    \Big\|(\widehat{U}^{\sD\sW})^\dag
    \big(C^\sD P^\sD\otimes\ket{0}^\sW\big)
    - (U_\rp^\sD)^\dag C^\sD P^\sD\otimes\ket{0}^\sW\Big\|_2
    \leq \epsilon_\ru\big\|C^\sD P^\sD\big\|_2.
\end{align}
Adding this circuit error to the bound for $U_\rp^\dag CP$ in Eq.~\eqref{eq:unitary error} provides Eq.~\eqref{eq:polar endpoint}.

We next construct a circuit implementing $\widehat{U}^{\sD\sW}$ and evaluate its complexity. 
To do so, we express $U_\rp$ as an integral of products of operators that admit efficient block encodings. For $s > 0$, let
\begin{align}
    W_s \coloneqq \sqrt{\sigma_\delta}(\sigma_\delta+s^2\bI)^{-1/2}, \qquad 
    Z_s \coloneqq s(\sigma_\delta+s^2\bI)^{-1/2}, \qquad 
    A_s \coloneqq \bI+(a_\rp^{-1}-1)Z_sQZ_s.
\end{align}
The scalar integral $y^{-1/2}=\f{2}{\pi}\int_0^\infty(y+s^2)^{-1}\,\mathrm{d}s$, applied to $F\sigma_\delta F$, gives the first equality below~\cite{Higham2004Computingpolardecompositionmatrixsign}:
\begin{align}
\label{eq:polar integral}
    V^\dag U_\rp
    &= \f{2}{\pi}\int_0^\infty\sqrt{\sigma_\delta}F(F\sigma_\delta F+s^2\bI)^{-1}\,\mathrm{d}s
     = \f{2}{\pi}\int_{-\infty}^{\infty}W_s A_s^{-1} Z_sF^{-1}\,\mathrm{d}v,
     \qquad s=e^v,
\end{align}
where the second equality follows from $\sigma_\delta+s^2F^{-2}=(\sigma_\delta+s^2\bI)^{1/2}A_s(\sigma_\delta+s^2\bI)^{1/2}$.

We consider how to implement each factor in the integrand by block encoding.
The eigenvectors of $\sigma_\delta$ are $\big(W_0^{\sD_0}\big)^{\otimes t}\ket{a_1\cdots a_t}$, with corresponding eigenvalues
\begin{align}
    \lambda_{a_1,\ldots,a_t}
    \coloneqq \f{\prod_{j=1}^t s_{a_j}^2+\delta/D}{1+\delta},
\end{align}
where $a_i \in \{0,\ldots,d_{\sD_0}-1\}$ for $i = 1,\ldots,t$ are computational basis labels.
This eigenvalue can be computed reversibly from the label and the given list $\{s_j^2\}_j$.
For an eigenvalue $\lambda$ of $\sigma_\delta$, the corresponding eigenvalues of $W_s$ and $Z_s$ are $\sqrt{\lambda/(\lambda+s^2)}$ and $s/\sqrt{\lambda+s^2}$, respectively, both of which lie in $[0,1]$.
Then, we can construct block encodings of $W_s$ and $Z_s$ by the standard method for diagonal matrices~\cite[Chap.~6]{Lin2022LectureNotes}; briefly, apply $\big((W_0^{\sD_0})^\dag\big)^{\otimes t}$, reversibly compute the corresponding eigenvalue of $W_s$ or $Z_s$, and rotate a work qubit so that the amplitude of its $\ket{0}$ component equals that value. Uncomputing these calculations and returning to the original basis then gives a block encoding to any desired accuracy.
Similarly, using the given controlled reflection to distinguish the ranges of $P$ and $Q = \bI - P$, a conditional rotation followed by uncomputation gives a block encoding of $\sqrt{a_\rp}F^{-1}=\sqrt{a_\rp}P+Q$.

We express the inverse factor $A_s^{-1}$ by a Neumann series~\cite{Hiai2014Introductionmatrixanalysisapplications}. Let $H_s \coloneqq \bI-Z_sQZ_s$, so that $0 \leq H_s \leq\bI$. Then,
\begin{align}
\label{eq:inverse series}
    A_s^{-1} = a_\rp\sum_{j=0}^{\infty}(1-a_\rp)^jH_s^j,
    \qquad
    \Big\|A_s^{-1}-a_\rp\sum_{j=0}^{M-1}(1-a_\rp)^jH_s^j\Big\|_\infty
    \leq(1-a_\rp)^M.
\end{align}
Indeed, $\bI-(1-a_\rp)H_s=a_\rp A_s$, and the omitted coefficient sum is $(1-a_\rp)^M$.

We construct a block encoding of $A_s^{-1}$ using the truncated series of Eq.~\eqref{eq:inverse series}. The diagonal rotations above implement $G_s \coloneqq \begin{pmatrix} Z_s & -W_s \\ W_s & Z_s \end{pmatrix}$. The reflection $G_s^\dag(\bI-2\ketbra{0}{0}\otimes Q)G_s$ has $\bI-2Z_sQZ_s$ as its block between zero work states. Averaging this reflection with the identity gives a normalized block encoding of $H_s$.
The standard constructions for products and linear combinations of block-encoded matrices~\cite[Lemmas~29 and~30]{gilyen2019qsvt} then yield a block encoding of the truncated series. 
Its coefficients are nonnegative and sum to $1-(1-a_\rp)^M$; assigning the remaining weight $(1-a_\rp)^M$ to the zero operator keeps the normalization equal to one. The corresponding coefficient state is prepared on $M$ unary qubits by the rotation $\ket{0}\mapsto\sqrt{a_\rp}\ket{0}+\sqrt{1-a_\rp}\ket{1}$ on the first qubit, followed by the same rotation on each subsequent qubit controlled on the preceding qubit being in $\ket{1}$. Selecting the powers $H_s^j$ uses at most $M-1$ controlled applications of the block encoding of $H_s$, with separate work registers. To bound the total error by $\zeta$, choose $M=\lceil a_\rp^{-1}\ln(2/\zeta)\rceil$, so that the truncation error is at most $\zeta/2$. Implementing the circuit components so that their errors sum to at most $\zeta/2$ bounds the total error by $\zeta$, with additional gates and classical work polynomial in $M$ and $\log(1/\zeta)$.
Thus, we obtain a normalized block encoding of $A_s^{-1}$ with operator norm error at most $\zeta$.


We approximate the integral in Eq.~\eqref{eq:polar integral} by a finite sum. Set $\kappa\coloneqq\epsilon_\ru^2/128$ and $\sigma_{\min}\coloneqq\delta/[D(1+\delta)]$, so that $\sigma_\delta\geq\sigma_{\min}\bI$. The singular values of $\sqrt{\sigma_\delta}F$ lie in $[\sqrt{a_\rp\sigma_{\min}},1]$. Hence, the integrand in the first integral has norm at most $1/\sqrt{a_\rp\sigma_{\min}}$ and at most $1/s^2$. Choosing
\begin{align}
    s_{\mathrm{lo}}\coloneqq\f{\kappa\sqrt{a_\rp\sigma_{\min}}}{16},
    \qquad
    s_{\mathrm{hi}}\coloneqq\f{16}{\kappa}
\end{align}
bounds the sum of the two discarded tails, including the factor $2/\pi$, by $\kappa/(4\pi)<\kappa/8$. The retained interval in $v=\ln{s}$ has length $\ell\coloneqq\ln(s_{\mathrm{hi}}/s_{\mathrm{lo}})=\cO(\ln D+\ln(1/(\delta a_\rp\kappa)))$.

We discretize this interval by the midpoint rule. Since $\sqrt{\sigma_\delta}F=V^\dag U_\rp R$, the second integrand in Eq.~\eqref{eq:polar integral} is $V^\dag U_\rp e^vR(R^2+e^{2v}\bI)^{-1}$. For each eigenvalue $r>0$ of $R$,
\begin{align}
    f_r(v)\coloneqq\f{re^v}{r^2+e^{2v}}
    =\f{1}{2\cosh(v-\ln r)},
    \qquad |f_r'(v)|\leq\f12.
\end{align}
The derivative of the integrand therefore has norm at most $1/2$. A power-of-two number of equal intervals, each of width at most $\kappa/(4\ell)$, bounds the midpoint error by $\kappa/(16\pi)<\kappa/16$. Using the product and linear-combination constructions~\cite[Lemmas~29 and~30]{gilyen2019qsvt}, we combine the block encodings of $W_s$, $A_s^{-1}/2$, $Z_s$, and $\sqrt{a_\rp}F^{-1}$ over the midpoints and multiply by $V$. This produces a block encoding of a matrix $B$ with normalization $\Lambda\coloneqq4\ell/(\pi\sqrt{a_\rp})>1$. Taking each factor's block-encoding error at most $\kappa/(64\Lambda)$, and the combined error from controlled selection and scalar evaluation at most $\kappa/(16\Lambda)$, adds at most $\kappa/8$ to the error of $B$. Here the inverse construction above is used with $\zeta=\kappa/(64\Lambda)$ and followed by a fixed attenuation by $1/2$. Including the tail and midpoint errors, we obtain $\|B-U_\rp\|_\infty<\kappa/2$. Rounding the endpoints and midpoint locations is included in the scalar evaluation error using the same derivative bound.

Finally, we convert this block encoding into an approximate implementation of $U_\rp$. Choose the least odd $h\geq3$ such that $\Lambda\sin(\pi/(2h))\leq1$, and attenuate the block to $\sin(\pi/(2h))B$ with additional error at most $\kappa\sin(\pi/(2h))/8$. Its distance from $\sin(\pi/(2h))U_\rp$ is then at most $\kappa\sin(\pi/(2h))$. Robust oblivious amplitude amplification~\cite[Theorem~15]{gilyen2019qsvt}, including its phase correction, uses $h=\cO(\Lambda)$ calls and produces a unitary whose block between zero work states differs from $U_\rp$ by at most $2h\kappa\sin(\pi/(2h))\leq\pi\kappa$. By unitarity, expanding the squared norm of its output minus $U_\rp\ket{\psi}\otimes\ket{0}$ bounds this squared norm by $2\pi\kappa$ for every unit vector $\ket{\psi}$. Thus the error including all work registers is at most $\sqrt{2\pi\kappa}<\epsilon_\ru/2$. Finite-gate synthesis of the amplification circuit with error at most $\epsilon_\ru/2$ then supplies the required $\widehat{U}^{\sD\sW}$. All scalar evaluations use polynomially many bits in $t$, $b_0$, $\log D$, $\log(1/\delta)$, $\log(1/a_\rp)$, and $\log(1/\epsilon_\ru)$; in particular, the small denominators enter through $\log(1/\sigma_{\min})$. Since the number of midpoints is $\cO(\ell^2/\kappa)$, the work qubits, gate count, and classical computation time are polynomial in $d_{\sD_0}$, $t$, $\log(1/\delta)$, $1/a_\rp$, $1/\epsilon_\ru$, and $b_0$. Together with Eqs.~\eqref{eq:unitary error} and~\eqref{eq:block unitary error}, this proves Eq.~\eqref{eq:polar endpoint}.

\end{proof}


\begin{algorithm}[h]
\caption{Tensor polar circuit (In \Cref{lem:polar})}
\label{alg:tensor polar}
\SetKwInput{KwInput}{Input}
\SetKwInput{KwOutput}{Output}
\SetKwInput{KwParameters}{Parameters}
\SetKwComment{Comment}{$\triangleright$\ }{}
\SetCommentSty{textnormal}

\SetAlgoNoEnd
\SetAlgoNoLine
\KwInput{Circuits for $V_0^{\sD_0}$, $W_0^{\sD_0}$, and a controlled application of $\bI^\sD-2P^\sD$, together with the list $\{s_j^2\}_j$.}
\KwParameters{A positive integer $t$ and $\delta,a_\rp,\epsilon_\ru\in(0,1/4)$.}
\KwOutput{A unitary circuit $\widehat{U}^{\sD\sW}$ satisfying Eq.~\eqref{eq:polar endpoint}.}
\SetAlgoLined

Set $D \gets d_{\sD_0}^t$, $\kappa \gets \epsilon_\ru^2/128$, and $\sigma_{\min} \gets \delta/[D(1+\delta)]$. \\

Set $s_{\mathrm{lo}} \gets \kappa\sqrt{a_\rp\sigma_{\min}}/16$, $s_{\mathrm{hi}} \gets 16/\kappa$, and $\ell \gets \ln(s_{\mathrm{hi}}/s_{\mathrm{lo}})$. \\

Divide $[\ln s_{\mathrm{lo}},\ln s_{\mathrm{hi}}]$ into the smallest power-of-two number of equal intervals whose width is at most $\kappa/(4\ell)$, and denote their midpoints by $v_k$. \\

Set $\Lambda \gets 4\ell/(\pi\sqrt{a_\rp})$, $\zeta \gets \kappa/(64\Lambda)$, and $M \gets \lceil a_\rp^{-1}\ln(2/\zeta)\rceil$. \\

For $s=e^{v_k}$, construct normalized block encodings of $W_s$, $Z_s$, and $\sqrt{a_\rp}F^{-1}$ using reversible arithmetic, controlled rotations, and the given controlled reflection. \\

Construct a normalized block encoding of $H_s=\bI-Z_sQZ_s$, and combine its powers using the coefficients of the truncated series in Eq.~\eqref{eq:inverse series} to approximate $A_s^{-1}$. \\

Prepare a uniform superposition of the midpoint labels $k$, select the product block encoding of $W_s(A_s^{-1}/2)Z_s(\sqrt{a_\rp}F^{-1})$, and unprepare the labels. Apply $V=(V_0^{\sD_0})^{\otimes t}$ to obtain a block encoding of $B\approx U_\rp$ with normalization $\Lambda$. \\

Choose the least odd $h\geq3$ satisfying $\Lambda\sin(\pi/(2h))\leq1$. Attenuate the block to $\sin(\pi/(2h))B$ and apply robust oblivious amplitude amplification, including its phase correction~\cite[Theorem~15]{gilyen2019qsvt}. \\

Compile the resulting circuit using the component precisions specified in the proof of \Cref{lem:polar}. \\

Return the circuit $\widehat{U}^{\sD\sW}$.
\end{algorithm}


The following three lemmas of this subsection produce the input of \Cref{lem:polar}, namely the factored form of Eq.~\eqref{eq:one copy polar}, by local Pauli tomography of the shared state and subsequent classical computation. The first lemma is a diagonalization routine that the other two use. After presenting these lemmas and their proofs, we describe how we invoke them to construct Bob's decoder.

\begin{lemma}[Additive diagonalization]
\label{lem:diagonalization}
Let $n_0$ be a power of two, and let $H$ be an $n_0 \times n_0$ Hermitian matrix with $\|H\|_2 \leq M_0$ for some $M_0 \geq 1$, each entry of which is evaluated to any given accuracy $\epsilon$ in time polynomial in $n_0$ and $\log{(1/\epsilon)}$. Then, for every $\delta \in (0,1/2)$, a unitary circuit on a register of dimension $n_0$ and a list $\{\mu_j\}_j$ of rationals can be computed such that the unitary $\widehat{Q}$ implemented by the circuit satisfies
\begin{align}
\label{eq:diagonalization}
    \big\|H - \widehat{Q}\,\mathrm{diag}(\mu_j)\,\widehat{Q}^\dag\big\|_2 \leq \delta.
\end{align}
The circuit has $\poly\big(n_0, \log{M_0}, \log{(1/\delta)}\big)$ one- and two-qubit gates, and the classical running time is polynomial in these parameters.
\end{lemma}

\begin{proof}[Proof of \Cref{lem:diagonalization}]

We use the Jacobi eigenvalue algorithm~\cite{Golub2013Matrixcomputations}, which repeatedly annihilates a largest off-diagonal entry of the matrix by a rotation acting on the $2 \times 2$ principal block of its two indices. For $n_0 = 1$, no rotation is needed, and thus we take $n_0 \geq 2$ in the following. 
We perform at most $N_\rJ \coloneqq n_0^2\lceil\log{(8M_0/\delta)}\rceil$ rotations, allowing a rotation error $u_\rJ \coloneqq \delta/[128n_0^2(M_0+1)]$ and a matrix-update error $b_\rJ \coloneqq \delta/(64N_\rJ n_0^2)$ at each of them.
We first evaluate the entries of $H$ to form a rational working matrix within error $\delta/64$ in the Hilbert--Schmidt norm.
At each step, we stop if the off-diagonal part of the working matrix has Hilbert--Schmidt norm at most $\delta/8$. Otherwise, we choose a largest off-diagonal entry, take a rotation within operator norm $u_\rJ$ of one that diagonalizes its $2 \times 2$ principal block, record that rotation exactly, and update the Hermitian working matrix within Hilbert--Schmidt error $b_\rJ$.
As long as we have not stopped, the selected entry has modulus greater than $\delta/(8n_0)$, so the two eigenvalues of its principal block are separated by more than $\delta/(4n_0)$. The rotation is then evaluated with polynomially many bits, even when $H$ itself has repeated eigenvalues.
Let $a_j$ be the Hilbert--Schmidt norm of the off-diagonal part of the working matrix after $j$ steps.
An exact rotation at a largest off-diagonal entry contracts this norm by a factor of at most $1-1/n_0^2$~\cite{Golub2013Matrixcomputations,Oksa2017AsymptoticquadraticconvergenceblockJacobi}, and thus the two errors allowed at each step give
\begin{align}
    a_{j+1} \leq (1-1/n_0^2)a_j + 2u_\rJ(M_0+1) + b_\rJ
    \leq (1-1/n_0^2)a_j + \f{\delta}{32n_0^2}.
\end{align}
Then, after $N_\rJ$ steps, $a_{N_\rJ} \leq \delta/8 + \delta/16 = 3\delta/16$.

We take $\{\mu_j\}_j$ to be the diagonal of the final working matrix, and write $Q$ for the product of the recorded rotations. Following the conventions in \Cref{sec:numerics}, we compile $Q^\dag$ into a circuit on $\log{n_0}$ qubits within error $\delta/[32(M_0+1)]$ in the operator norm, and write $\widehat{Q}$ for the unitary that the circuit realizes. The bound $a_{N_\rJ} \leq 3\delta/16$, the accumulated update error $N_\rJ b_\rJ \leq \delta/64$, and the initial evaluation error $\delta/64$ place $QHQ^\dag$ within Hilbert--Schmidt distance $3\delta/16 + \delta/32$ of $\mathrm{diag}(\mu_j)$. Unitary invariance carries this bound to $\|H - Q^\dag\mathrm{diag}(\mu_j)Q\|_2$. Since $\|\mathrm{diag}(\mu_j)\|_2 \leq M_0 + 1$, replacing $Q^\dag$ by $\widehat{Q}$ adds at most $\delta/16$, which leaves the total error below $\delta$.

\end{proof}

\begin{lemma}[Tomography by local Pauli measurements; based on Ref.~\cite{Guta2020Faststatetomography}]
\label{lem:tomography}
Let $\cH^\sA$ and $\cH^\sB$ be Hilbert spaces of dimensions $d_\sA$ and $d_\sB$, both powers of two, and let $d \coloneqq d_\sA d_\sB$. For $\Delta, \zeta \in (0,1)$, set
\begin{align}
\label{eq:tomography count}
    N_\rtom(d,\Delta,\zeta) \coloneqq d^2\big\lceil 43d^2\Delta^{-2}\log{(d/\zeta)}\big\rceil.
\end{align}
Then, there is a protocol that acts on $N_\rtom(d,\Delta,\zeta)$ copies of a state $\sigma^{\sA\sB}$ and returns a circuit description and a list of rationals that determine a state $\theta^{\sA\sB}$. The state satisfies $\f{1}{2}\|\sigma^{\sA\sB} - \theta^{\sA\sB}\|_1 \leq \Delta$ with probability $1-\zeta$. The protocol measures one Pauli observable on $\sA$ and one on $\sB$ in each copy, on a schedule fixed in advance, and computes the output from the products of the two local outcomes. One-way classical communication transmitting the outcomes on $\sA$ to $\sB$ suffices.
\end{lemma}

\begin{proof}[Proof of \Cref{lem:tomography}]

For $\chi \in \{\sA, \sB\}$, let $\cP^\chi$ be the set of the $d_\chi^2$ tensor products of the single-qubit Pauli operators $\bI, X, Y, Z$ on the $\log{d_\chi}$ qubits of $\cH^\chi$. The $d^2$ operators $P = P^\sA \otimes P^\sB$ with $P^\sA \in \cP^\sA$ and $P^\sB \in \cP^\sB$ are then an orthogonal basis of the operators on $\cH^\sA\otimes\cH^\sB$. Each $P$ other than the identity is an observable with eigenvalues $\pm1$, and its two-outcome measurement is realized by measuring $P^\sA$ and $P^\sB$ locally and multiplying the outcomes. We assign a separate batch of $N_\rtom/d^2$ copies to each of the $d^2$ operators, take the empirical mean $\hat{a}_P$ of the products of the outcomes, and set $\hat{a}_\bI \coloneqq 1$ without measuring, which leaves that batch unused.

The least-squares estimator $\Xi \coloneqq d^{-1}\sum_P\hat{a}_PP$ is Hermitian with unit trace and rational entries, and obeys $\|\Xi\|_2 \leq \sqrt{d}$ by the orthogonality of the Pauli basis. Let $\Xi_\star$ be the state closest to $\Xi$ in the Hilbert--Schmidt norm, obtained by projecting $\Xi$ onto the set of states.
Then, Ref.~{\cite[Theorem~1]{Guta2020Faststatetomography}} with rank parameter $d$ gives $\f{1}{2}\|\sigma^{\sA\sB} - \Xi_\star\|_1 \leq \Delta/2$ except with probability $\zeta$, since the sample count of Eq.~\eqref{eq:tomography count} exceeds the one required in the reference.

We next compute a state close to $\Xi_\star$. The state $\Xi_\star$ has the eigenvectors of $\Xi$, with the eigenvalues replaced by their Euclidean projection onto the probability simplex~\cite{Smolin2012Efficientmethodcomputingmaximumlikelihood}. That projection is computed exactly in rational arithmetic by sorting the entries, determining the nonempty active set, subtracting its threshold $(\sum_{j \in \mathrm{active}}\lambda_j - 1)/|\mathrm{active}|$, and clipping at zero~\cite{wang2013projectionprobabilitysimplexefficient}.

We apply \Cref{lem:diagonalization} to $\Xi$ with $M_0 = d$ and $\delta = \Delta/(2\sqrt{d})$, and project the returned rational list as above. Since the projected list is nonnegative and sums to one, it and the returned circuit determine a state, which we take as the output $\theta^{\sA\sB}$.
By unitary invariance, $\theta^{\sA\sB}$ is the projection of the matrix determined by the circuit and the unprojected list.
As the projection onto the set of states is nonexpansive in the Hilbert--Schmidt norm and $\|X\|_1 \leq \sqrt{d}\|X\|_2$, we obtain $\f{1}{2}\|\theta^{\sA\sB} - \Xi_\star\|_1 \leq \Delta/4$. With the bound of the previous paragraph, this gives $\f{1}{2}\|\sigma^{\sA\sB} - \theta^{\sA\sB}\|_1 \leq 3\Delta/4 \leq \Delta$.

\end{proof}

\begin{lemma}[Factored form of a computable matrix]
\label{lem:factored model}
Let $d_0$ be a power of two, and let $C_\rin$ be a $d_0 \times d_0$ matrix with $\|C_\rin\|_2 = 1$, each entry of which is evaluated to any given accuracy $\epsilon$ in time $\poly\big(d_0, \log{(1/\epsilon)}\big)$. Then, for $\nu \in (0,1)$, one can compute circuits for two unitaries $V_0$ and $W_0$ on a register of dimension $2d_0$ and a list $\{s_j^2\}_j$ of nonnegative rationals summing to one, in time $\poly(d_0, 1/\nu)$. The matrix $C_0$ of Eq.~\eqref{eq:one copy polar} built from them obeys $\|C_0 - C_\rin\|_2 \leq \nu$, where $C_\rin$ is extended by zero rows and columns to dimension $2d_0$.
\end{lemma}

\begin{proof}[Proof of \Cref{lem:factored model}]

We obtain Eq.~\eqref{eq:one copy polar} from a singular value decomposition of $C_\rin$, computed with finite precision. We first evaluate the entries of $C_\rin$ to form a rational matrix $C$ with $\|C-C_\rin\|_2 \leq \nu/16$, so that $\big|\|C\|_2 - 1\big| \leq \nu/16$.
Let $\varpi \coloneqq \nu/(64\sqrt{d_0})$. We apply \Cref{lem:diagonalization} to $C^\dag C$, whose Hilbert--Schmidt norm is at most $2$, with $M_0 = 2$ and $\delta = \delta_\rG \leq \varpi^2\nu^2/(2^{16}d_0)$, and write $W_0$ and $\{\mu_j\}_j$ for the unitary and the list it returns.
We retain the indices $j$ such that $\mu_j \geq \varpi^2$, and put the square roots of those entries in the diagonal matrix $S$. Letting $W_{0,\rret}$ and $W_{0,\rdisc}$ be the columns of $W_0$ at the retained and the discarded indices, we set $A \coloneqq C W_{0,\rret}S^{-1}$, for which 
\begin{align}
    \|C W_{0,\rdisc}\|_2^2 \leq d_0(\varpi^2+\delta_\rG),
    \qquad \|A^\dag A-\bI\|_\infty \leq \f{\delta_\rG}{\varpi^2} < 1/4.
\end{align}
By the definitions of $\varpi$ and $\delta_\rG$, we see that $\|CW_{0,\rdisc}\|_2 \leq \nu/32$.

Set $\eta_\rG \coloneqq \delta_\rG/\varpi^2 < 1/4$, and let $B \coloneqq A(A^\dag A)^{-1/2}$, which is an isometry with $\|A-B\|_2 \leq \sqrt{d_0}\eta_\rG$ and $\|(B-A)S\|_2 \leq 4\eta_\rG$.
To synthesize a circuit for $B$ by \Cref{lem:isometry synthesis} with $V = B$ and $\delta = \nu/64$, we reorder the columns of $W_0$ and the entries of $\{\mu_j\}_j$ so that the retained indices come first.
The lemma takes as input a rational matrix within Hilbert--Schmidt error $\nu/8192$ of $B$, for which we use an evaluation of $A$ to Hilbert--Schmidt error $\nu/16384$.
This evaluation is well conditioned because every retained singular value is at least $\varpi$.
With $\|A - B\|_2 \leq \sqrt{d_0}\eta_\rG \leq \nu/16384$ for $\nu \leq 1$, the evaluated matrix is within $\nu/8192$ of $B$, as required.
The lemma returns a circuit for a unitary $U_0$ on a register of dimension $2d_0$, whose first columns approximate $B$ to error $\nu/64$ in the operator norm.
We extend both $C$ and $C_\rin$ by zero rows and columns to dimension $2d_0$, and pad the list $\{\mu_j\}_j$ with zeros.
On the same register, we extend $W_0$ to $W_0\oplus(\det W_0)^*\oplus\bI$, placing the compensating phase on one of the added coordinates. This extension has determinant one and agrees with the original $W_0$ on its first $d_0$ columns.

Following the conventions in \Cref{sec:numerics}, we compile this extension within error $\nu/64$ in the operator norm and retain the notation $W_0$ for the implemented unitary. We then normalize the retained entries of $\{\mu_j\}_j$ by their sum and set all remaining entries to zero. The resulting entries are nonnegative rationals summing to one, and we denote their list by $\{s_j^2\}_j$.
We set $V_0 \coloneqq U_0W_0^\dag$. The circuits for $V_0$ and $W_0$ and the normalized list then define the matrix $C_0$ of Eq.~\eqref{eq:one copy polar}. Writing $\widetilde{C}_0$ for the matrix obtained from the same circuits and the list before the normalization, we have
\begin{align}
    \big\|\widetilde{C}_0 - C\big\|_2 \leq \f{\nu}{32} + 4\eta_\rG + \f{2\sqrt{2}\nu}{64} 
    < \f{\nu}{8}.
\end{align}
Since $\big|\|C\|_2-1\big| \leq \nu/16$, the norm of $\widetilde{C}_0$ differs from one by less than $\nu/8 + \nu/16$, which bounds the change caused by the normalization. Adding the initial error $\nu/16$, we obtain $\|C_0 - C_\rin\|_2 < 7\nu/16 < \nu$.
The square roots $s_j$ are evaluated to the required precision when a circuit uses them. Finally, $\|C_0\|_2 = 1$ because the entries of $\{s_j^2\}_j$ sum to one.

\end{proof}

We finally construct Bob's decoder for the measurement of \Cref{sec:two instruments}. Suppose that the two parties share copies of the compressed state $\sigma_z^{\sC_\sA\sC_\sB}$ of Eq.~\eqref{eq:compact state}, that \Cref{lem:tomography} has supplied Bob with an estimate $\theta^{\sC_\sA\sC_\sB}$ of it, and that Alice applies the first stage of \Cref{lem:first instrument} to each copy and the second stage of \Cref{lem:second instrument} to each block of $t$ copies. We write $\sS_0$ for the register that the first stage retains, and $\sX_0$ and $\sU_0$ for the registers that record its outcome and the seed of the unitary it draws. Let $\eta_\theta^{\sS_0\sC_\sB\sX_0\sU_0}$ be the output of the first stage applied to $\sC_\sA$ of $\theta^{\sC_\sA\sC_\sB}$, and let $\ket{\eta_\theta}^{\sS_0\sC_\sB\sX_0\sU_0\sE}$ be its purification with a purifying system $\sE$. We obtain this purification from the circuit and the full eigenvalue list of $\theta^{\sC_\sA\sC_\sB}$, keeping a copy of each first-stage outcome and seed label in $\sE$. Choose a power of two $d_0$ no smaller than the dimensions of $\sC_\sB\sX_0\sU_0$ and $\sS_0\sE$, and enlarge $\sE$ by unused basis states so that $\dim\cH^{\sS_0\sE}=d_0$. We set $C_\rin \coloneqq \rvec^{-1}\big(\ket{\eta_\theta}\big)$ with the rows indexed by $\sC_\sB\sX_0\sU_0$ and padded with zeros to dimension $d_0$, and the columns indexed by $\sS_0\sE$. The entries of $C_\rin$ are evaluated to any given accuracy in polynomial time from the circuit and the list of $\theta^{\sC_\sA\sC_\sB}$. Hence \Cref{lem:factored model} applies at this $d_0$ and any $\nu \in (0,1)$, and returns $V_0^{\sD_0}$, $W_0^{\sD_0}$, and $\{s_j^2\}_j$ on a register $\sD_0$ of dimension $d_{\sD_0} \coloneqq 2d_0$. On the column side, we identify the additional coordinates with a further enlargement of $\sE$, so that the whole column space is $\cH^{\sS_0}\otimes\cH^\sE$ of dimension $d_{\sD_0}$. We also pad the row space to dimension $d_{\sD_0}$ and retain the notation $\ket{\eta_\theta}$ for the resulting zero extension. Since $\rvec$ preserves the norm, the matrix $C_0^{\sD_0}$ of Eq.~\eqref{eq:one copy polar} built from them obeys $\big\|\rvec\big(C_0^{\sD_0}\big) - \ket{\eta_\theta}\big\| \leq \nu$.

A block of $t$ copies carries $C^\sD = \big(C_0^{\sD_0}\big)^{\otimes t}$ and $\sigma^\sD$. On this block, the second stage acts on $\sS_0^t$, retains the register $\sS$, and records the outcome in $\sX$; Alice reports its seed $U$ and outcome $x$ to Bob. Since $\sS_0^t$ lies on the column side of $C^\sD$, the unitary $U^{\sS_0^t}$ multiplies $C^\sD$ from the right by $\big(U^{\sS_0^t}\big)^\top$ by Eq.~\eqref{eq:definition of vectorization}. We thus take
\begin{align}
\label{eq:decoder projector}
        J_x^{\sS\sE^t\to\sD} \coloneqq \big(U^{\sS_0^t}\big)^\top\big(\ket{x}^\sX\otimes\bI^{\sS}\big)\otimes\bI^{\sE^t},
    \qquad P^\sD \coloneqq J_x^{\sS\sE^t\to\sD}\big(J_x^{\sS\sE^t\to\sD}\big)^\dag,
\end{align}
for which $C^\sD J_x^{\sS\sE^t\to\sD}$ is the unnormalized amplitude of the branch $x$, and $\|C^\sD P^\sD\|_2^2$ is the probability $p_x$ of that branch at $C_0^{\sD_0}$. The ranges of the $J_x$ are mutually orthogonal and span $\cH^\sD$, so $\sum_xp_x = 1$. The reflection $\bI^\sD - 2P^\sD$ is the reflection about $\ket{x}^\sX$ conjugated by $\big(U^{\sS_0^t}\big)^\top$ and tensored with the identity on $\sE^t$, and Bob compiles it and its controlled version from the received $x$ and the Clifford circuit for $U$. All the inputs of \Cref{lem:polar} are thus available to Bob, and he obtains the circuit $\widehat{U}^{\sD\sW}$ for any $\delta, a_\rp, \epsilon_\ru \in (0, 1/4)$. His decoder applies the inverse of that circuit.


\subsection{Protocol}
\label{sec:protocol}

We fix the parameters and describe our entanglement distillation protocol. 
We here consider the protocol in terms of ideal operations, i.e., exact gates, exact finite random choices of thresholds, seeds, and codes, and Haar random unitaries. Their implementation by a finite gate set follows the conventions of \Cref{sec:numerics}, and the resulting errors are accounted for in the proof of \Cref{thm:achievability by efficient protocol formal}.

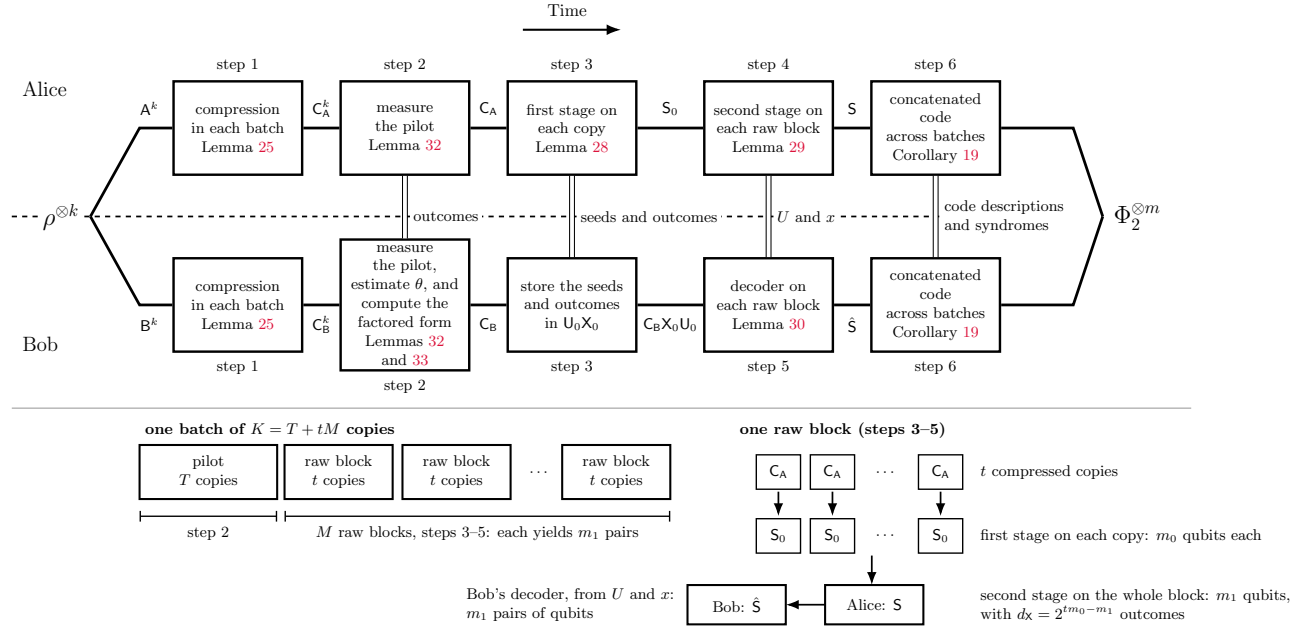
\begin{figure}
\adjustbox{width=170mm}{%
\begin{tikzpicture}[
    wire/.style={line width=1.6pt},
    cwire/.style={line width=0.6pt, double, double distance=2.2pt},
    box/.style={draw, line width=1.4pt, fill=white, font=\small, align=center, text width=2.4cm, minimum height=1.9cm},
    lab/.style={font=\small, fill=white, inner sep=2pt},
    party/.style={font=\large},
    seg/.style={draw, line width=1.2pt, fill=white, minimum height=1.1cm, align=center, font=\small},
    cop/.style={draw, line width=0.8pt, fill=white, minimum width=0.9cm, minimum height=0.7cm, font=\small},
    ttl/.style={font=\bfseries, anchor=west},
    arr/.style={-Latex, line width=1.0pt}
]
\draw[dashed, line width=1pt] (-2.2,0) -- (19.3,0);
\node[party, anchor=west] at (-2.1, 2.6) {Alice};
\node[party, anchor=west] at (-2.1,-2.6) {Bob};
\draw[-Latex, line width=1.2pt] (8.2,3.8) -- (10.2,3.8);
\node[font=\normalsize] at (9.1,4.2) {Time};
\coordinate (P) at (-0.6,0);
\node[font=\large, anchor=east, fill=white, inner sep=2pt] at (-0.75,0) {\Large{$\rho^{\otimes k}$}};
\node[box] (a1) at (2.4,1.8) {compression\\ in each batch\\ \Cref{lem:compression}};
\node[box] (a2) at (5.8,1.8) {measure\\ the pilot\\ \Cref{lem:tomography}};
\node[box] (a3) at (9.2,1.8) {first stage on\\ each copy\\ \Cref{lem:first instrument}};
\node[box] (a4) at (13.2,1.8) {second stage on\\ each raw block\\ \Cref{lem:second instrument}};
\node[box] (a5) at (16.6,1.8) {concatenated code\\ across batches\\ \Cref{cor:batch distillation}};
\node[box] (b1) at (2.4,-1.8) {compression\\ in each batch\\ \Cref{lem:compression}};
\node[box] (b2) at (5.8,-1.8) {measure the pilot,\\ estimate $\theta$, and\\ compute the factored form\\ \Cref{lem:tomography,lem:factored model}};
\node[box] (b3) at (9.2,-1.8) {store the seeds\\ and outcomes\\ in $\sU_0\sX_0$};
\node[box] (b4) at (13.2,-1.8) {decoder on\\ each raw block\\ \Cref{lem:polar}};
\node[box] (b5) at (16.6,-1.8) {concatenated code\\ across batches\\ \Cref{cor:batch distillation}};
\node[lab] at ([yshift=0.3cm]a1.north) {step 1};
\node[lab] at ([yshift=0.3cm]a2.north) {step 2};
\node[lab] at ([yshift=0.3cm]a3.north) {step 3};
\node[lab] at ([yshift=0.3cm]a4.north) {step 4};
\node[lab] at ([yshift=0.3cm]a5.north) {step 6};
\node[lab] at ([yshift=-0.3cm]b1.south) {step 1};
\node[lab] at ([yshift=-0.3cm]b2.south) {step 2};
\node[lab] at ([yshift=-0.3cm]b3.south) {step 3};
\node[lab] at ([yshift=-0.3cm]b4.south) {step 5};
\node[lab] at ([yshift=-0.3cm]b5.south) {step 6};
\draw[wire] (P) -- (0.4,1.8) -- (a1.west);
\draw[wire] (P) -- (0.4,-1.8) -- (b1.west);
\node[lab] at (0.6,2.2) {$\sA^{k}$};
\node[lab] at (0.6,-2.2) {$\sB^{k}$};
\draw[wire] (a1.east) -- (a2.west); \draw[wire] (a2.east) -- (a3.west); \draw[wire] (a3.east) -- (a4.west); \draw[wire] (a4.east) -- (a5.west);
\draw[wire] (b1.east) -- (b2.west); \draw[wire] (b2.east) -- (b3.west); \draw[wire] (b3.east) -- (b4.west); \draw[wire] (b4.east) -- (b5.west);
\node[lab] at (4.1,2.2) {$\sC_\sA^{k}$};
\node[lab] at (7.5,2.2) {$\sC_\sA$};
\node[lab] at (11.2,2.2) {$\sS_0$};
\node[lab] at (14.9,2.2) {$\sS$};
\node[lab] at (4.1,-2.2) {$\sC_\sB^{k}$};
\node[lab] at (7.5,-2.2) {$\sC_\sB$};
\node[lab] at (11.2,-2.2) {$\sC_\sB\sX_0\sU_0$};
\node[lab] at (14.9,-2.2) {$\hat{\sS}$};
\draw[cwire] (a2.south) -- (b2.north);
\draw[cwire] (a3.south) -- (b3.north);
\draw[cwire] (a4.south) -- (b4.north);
\draw[cwire] (a5.south) -- (b5.north);
\node[lab, anchor=west] at (5.9,0.0) {outcomes};
\node[lab, anchor=west] at (9.3,0.0) {seeds and outcomes};
\node[lab, anchor=west] at (13.3,0.0) {$U$ and $x$};
\node[lab, anchor=west, align=left] at (16.7,0.0) {code descriptions\\ and syndromes};
\coordinate (O) at (20.0,0);
\draw[wire] (a5.east) -- (19.4,1.8) -- (O);
\draw[wire] (b5.east) -- (19.4,-1.8) -- (O);
\node[font=\large, anchor=west, fill=white, inner sep=2pt] at (20.15,0) {\Large{$\Phi_2^{\otimes m}$}};
\draw[line width=0.6pt, black!40] (-2.2,-3.9) -- (21.8,-3.9);
\node[ttl] at (0.4,-4.35) {one batch of $K = T + tM$ copies};
\node[seg, minimum width=2.8cm] at (1.8,-5.2) {pilot\\ $T$ copies};
\node[seg, minimum width=2.2cm] at (4.45,-5.2) {raw block\\ $t$ copies};
\node[seg, minimum width=2.2cm] at (6.85,-5.2) {raw block\\ $t$ copies};
\node at (8.55,-5.2) {$\cdots$};
\node[seg, minimum width=2.2cm] at (10.1,-5.2) {raw block\\ $t$ copies};
\draw[|-|, line width=0.8pt] (0.4,-6.1) -- (3.2,-6.1);
\node[lab] at (1.8,-6.45) {step 2};
\draw[|-|, line width=0.8pt] (3.35,-6.1) -- (11.2,-6.1);
\node[lab] at (7.275,-6.45) {$M$ raw blocks, steps 3--5: each yields $m_1$ pairs};
\node[ttl] at (12.5,-4.35) {one raw block (steps 3--5)};
\foreach \x in {0,1,3}{
  \node[cop] at (13.4+\x*1.1,-5.2) {$\sC_\sA$};
}
\node at (13.4+2*1.1,-5.2) {$\cdots$};
\node[font=\small, anchor=west] at (17.4,-5.2) {$t$ compressed copies};
\foreach \x in {0,1,3}{
  \draw[arr] (13.4+\x*1.1,-5.6) -- (13.4+\x*1.1,-6.1);
  \node[cop] at (13.4+\x*1.1,-6.5) {$\sS_0$};
}
\node at (13.4+2*1.1,-6.5) {$\cdots$};
\node[font=\small, anchor=west] at (17.4,-6.5) {first stage on each copy: $m_0$ qubits each};
\draw[arr] (15.3,-6.9) -- (15.3,-7.5);
\node[seg, minimum width=2.0cm, minimum height=0.8cm] at (15.35,-7.9) {Alice: $\sS$};
\node[font=\small, anchor=west, align=left] at (17.4,-7.9) {second stage on the whole block: $m_1$ qubits,\\ with $d_\sX = 2^{tm_0-m_1}$ outcomes};
\draw[-Latex, line width=1.0pt] (14.35,-7.9) -- (13.55,-7.9);
\node[seg, minimum width=2.0cm, minimum height=0.8cm] at (12.55,-7.9) {Bob: $\hat{\sS}$};
\node[font=\small, anchor=east, align=left] at (11.4,-7.9) {Bob's decoder, from $U$ and $x$:\\ $m_1$ pairs of qubits};
\end{tikzpicture}}
    \caption{Structure of the protocol. Upper part: the operations of Alice and Bob in time order, labeled by the steps of \Cref{sec:protocol}, where double lines represent classical communication from Alice to Bob. Steps 1--5 are carried out independently in each of the $k_\rc$ batches. Lower left: each batch of $K$ copies consists of a pilot of $T$ copies and $M$ raw blocks of $t$ copies, and each raw block yields $m_1$ pairs of qubits. Lower right: the first stage acts on each compressed copy separately and the second stage on the whole raw block, which keeps the number of outcomes of the second stage polynomial, and Bob's decoder then produces $\hat{\sS}$ from the seed and the outcome that Alice reports. The code blocks that step~6 forms across the batches are shown in \Cref{fig:code blocks}.}
    \label{fig:diagram of ent distill}
\end{figure}

The $k$ input copies of the $\Theta(n)$-qubit state $\rho^{\sA\sB}$ are divided into $k_\rc$ independent batches of $K$ copies. In each batch, both parties, Alice and Bob, compress their registers using the compression channel in \Cref{sec:collective compression}, and the resulting copies are split into a pilot of $T$ copies and $M$ raw blocks of $t$ copies. The pilot is consumed by tomography, which supplies Bob with an estimate $\theta$ for his decoder. On each raw block, they obtain $m_1$ pairs of qubits by Alice's two-stage measurement of \Cref{sec:two instruments} and Bob's decoder of \Cref{sec:decoder}. \Cref{cor:batch distillation} then acts across batches and outputs $m = Mm_1m_\rc$ pairs of qubits. Thus, these parameters satisfy $k = Kk_\rc$ and $K = T + tM$. All communication is from Alice to Bob. See also Fig.~\ref{fig:diagram of ent distill}.

The protocol is constructed from integers $n, \hat{r}, \hat{q} \geq 1$ and rationals $J$, $\gamma$, and $\varepsilon$, together with $\beta \in [1/2, 1)$ and an integer $\ell_0 \geq 1/(1-\beta)$ fixed in advance. Here, $n$ indexes the family of states and determines the registers $\sA$ and $\sB$, and $J$, $\gamma$, and $\varepsilon$ are the target rate, the allowed rate loss, and the target error, respectively. The integers $\hat{r}$ and $\hat{q}$ fix the dimensions after the compression. We take $J \in (0, \log{\hat{q}}]$, $\gamma \in (0, J/4]$, and $\varepsilon \in (0,1)$. The quantities below are defined in order, each from these values and the quantities defined before it. None of them depends on $\rho^{\sA\sB}$.

We first fix the parameter $a$ that governs the loss in the rate and the threshold $u_\rc$ in Eq.~\eqref{eq:code parameters}; \Cref{cor:batch distillation} applies at this $a$ when each pair obtained from a raw block has infidelity at most $u_\rc$ with $\Phi_2$:
\begin{align}
\label{eq:rate parameters}
    a \coloneqq \f{\gamma}{8J}, \qquad 
    u_\rc = \f{a}{64\lceil\log{(16/a)}\rceil}.
\end{align}
As $\gamma \leq J/4$, we have $a \leq 1/32$, and the pilot will be chosen to occupy at most a fraction $a/2$ of each batch.

We next fix the entropy margins of Alice's two-stage measurement in \Cref{sec:two instruments}. 
Let $\xi \coloneqq \min\{\gamma, 1/4\}$. We set
\begin{align}
\label{eq:main parameters}
    B \coloneqq \ell_0\bigg\lceil\log{\bigg(\f{16\ell_0}{\xi}\bigg)}\bigg\rceil, \qquad 
    t \coloneqq \bigg\lceil\f{8\ell_0}{\xi}\bigg\lceil\log{\bigg(\f{32}{u_\rc}\bigg)}\bigg\rceil\bigg\rceil.
\end{align}
The first stage keeps up to $\lceil J + B\rceil$ qubits of each copy, and the margin $B$ over $J$ bounds the entropy lost in this stage by $\xi/8$. The second stage acts on each raw block of $t$ copies, and this $t$ bounds its decoupling error by $u_\rc/16$.

We then fix the dimensions after the compression and the sizes of the two-stage measurement. We take the tolerance parameter of \Cref{lem:geometry} to be
\begin{align}
\label{eq:epsilon star}
    \epsilon_\star \coloneqq \f{u_\rc}{2^{12}t}.
\end{align}
At this $\epsilon_\star$ and the given integers $\hat{r}$ and $\hat{q}$, Eq.~\eqref{eq:compact dimensions} fixes, for $\chi \in \{\sA, \sB\}$,
\begin{align}
\label{eq:compact dimensions protocol}
    w_\sA = \f{\epsilon_\star}{4\hat{q}}, \qquad 
    w_\sB = \f{\epsilon_\star}{4\hat{r}\hat{q}}, \qquad 
    Q_\chi = \bigg\lceil\f{2}{w_\chi}\bigg\rceil, \qquad 
    d_{\sC_\chi} = 2^{\lceil\log{(Q_\chi+1)}\rceil}.
\end{align}
Here, $[w_\chi, 2w_\chi]$ is the threshold interval of the compression channel of \Cref{lem:compression} applied to $\chi$, $Q_\chi$ is the dimension it retains, and $d_{\sC_\chi}$ is the dimension of $\sC_\chi$, the smallest power of two exceeding $Q_\chi$. We then set
\begin{align}
\label{eq:instrument sizes}
    m_0 \coloneqq \min\big\{\log{d_{\sC_\sA}},\lceil J+B\rceil\big\}, \qquad 
    m_1 \coloneqq \big\lfloor t(J-\xi/4)\big\rfloor, \qquad 
    d_\sX \coloneqq 2^{tm_0-m_1},
\end{align}
where $m_0$ is the number of qubits that the first stage keeps from each copy, $m_1$ is the number of qubits that the second stage keeps from each raw block, and $d_\sX$ is the number of outcomes of the second stage. Since $\lceil J+B\rceil > J$ and $d_{\sC_\sA} > Q_\sA \geq 8\hat{q}/\epsilon_\star > \hat{q}$ with $\log{\hat{q}} \geq J$, we have $m_0 > J$, and hence $m_1 < tm_0$. Moreover, $t \geq 8/\xi$ and $J \geq 4\gamma \geq 4\xi$ imply $m_1 \geq 1$. The number of outcomes obeys
\begin{align}
\label{eq:outcome count}
    \log{d_\sX} \leq t\Big(B + 1 + \f{\xi}{4}\Big) + 1.
\end{align}

We now fix the size of the pilot and the precisions of Bob's decoder and of the compression. We set
\begin{align}
\label{eq:outer budgets}
    \Delta \coloneqq \f{u_\rc}{2^{12}t}, \qquad
    \zeta \coloneqq \f{u_\rc}{8}, \qquad 
    T \coloneqq N_\rtom\big(d_{\sC_\sA}d_{\sC_\sB},\Delta,\zeta\big),
\end{align}
where $N_\rtom$ is the number of copies in Eq.~\eqref{eq:tomography count}. Applied to $T$ copies of a state on $\sC_\sA\sC_\sB$, the tomography of \Cref{lem:tomography} returns an estimate within trace distance $\Delta$ of that state, except with probability $\zeta$. For Bob's decoder, we set
\begin{align}
\label{eq:decoder errors}
    \nu \coloneqq \f{u_\rc}{2^{12}t}, \qquad 
    \delta \coloneqq \f{u_\rc^2}{2^{12}d_\sX}, \qquad 
    a_\rp \coloneqq \f{u_\rc^2}{2^{12}d_\sX}, \qquad
    \epsilon_\ru \coloneqq \f{u_\rc}{32},
\end{align}
where $\nu$ and $(\delta, a_\rp, \epsilon_\ru)$ are the parameters at which we apply \Cref{lem:factored model} and \Cref{lem:polar}, respectively. We then set
\begin{align}
\label{eq:grid size}
    h \coloneqq T + t, \qquad 
    G \coloneqq 2^{\lceil 2\log{(128h/u_\rc)}\rceil},
\end{align}
where $h$ is the number of copies in a batch to which we apply \Cref{cor:compression}, namely the pilot together with one raw block, and $G$ is the number of subintervals of $[w_\chi, 2w_\chi]$ in \Cref{lem:compression}. We take $G$ to be a power of two, so that the threshold is drawn exactly with uniform random bits.

We then fix the number of raw blocks and the batch size. We set
\begin{align}
\label{eq:allocation}
    M \coloneqq \bigg\lceil\f{1}{t}\max\bigg\{t, \, \f{2T}{a}, \, 4Q_\sB^2, \, \f{(128h/u_\rc)^4}{w_\sB^2}\bigg\}\bigg\rceil,
    \qquad K \coloneqq T + tM.
\end{align}
Note that $\hat{r} \geq 1$ gives $w_\sB \leq w_\sA$ and $Q_\sB \geq Q_\sA$ in Eq.~\eqref{eq:compact dimensions protocol}, and thus $w_{\min} = w_\sB$ and $\max_\chi Q_\chi = Q_\sB$ in \Cref{cor:compression}.
The first and third entries of the maximum in Eq.~\eqref{eq:allocation} give $K \geq h$ and $K \geq 4Q_\sB^2$. Together with $Q_\chi = \lceil 2/w_\chi\rceil$ and $w_\chi < 1/4$, which follows from $\epsilon_\star < 1$, these are the conditions of \Cref{cor:compression} for a batch of $K$ copies with $h = T + t$ selected copies. The second entry gives $T/K \leq a/2$, which bounds the fraction of each batch occupied by the pilot. The fourth entry and $G \geq (128h/u_\rc)^2$ bound the two terms of $\varepsilon_{\mathrm{cp}}(h)$ in Eq.~\eqref{eq:compression error} by $u_\rc/16$ and $3u_\rc/64$, respectively, and hence $\varepsilon_{\mathrm{cp}}(h) \leq u_\rc/8$.

We finally invoke \Cref{cor:batch distillation} at this $a$, at $m_\rb = Mm_1$ pairs per batch, and at the target error $\varepsilon_\rc \coloneqq \varepsilon/2$, which fixes $N$, $\zeta_\rc$, $L$, $k_\rc$, and $m_\rc$ through Eqs.~\eqref{eq:code target}, \eqref{eq:code parameters}, and~\eqref{eq:code lengths}. The remaining half of the target error is reserved for the finite-precision implementation. The protocol thus acts on $k \coloneqq Kk_\rc$ copies and outputs $m \coloneqq Mm_1m_\rc$ pairs of qubits.

The protocol proceeds through the following steps. All random choices in distinct batches are independent. Within each batch, the seeds of the first and the second stages of Alice's measurement are independent across distinct copies and distinct raw blocks, respectively.
\begin{enumerate}
    \item Partition the $k$ shared copies into $k_\rc$ batches of $K$ copies. In each batch, the party holding $\chi \in \{\sA, \sB\}$ applies the channel $\cC^{\chi^K\to\sC_\chi^K}$ of \Cref{lem:compression} to its $K$ copies of $\chi$ with the threshold interval $[w_\chi,2w_\chi]$ divided into $G$ subintervals and the retained dimension $Q_\chi$. Each copy of $\sC_\chi$ is then embedded into a larger space whose dimension $d_{\sC_\chi}$ is a power of two.
    \item In each batch, Alice and Bob measure the pilot, i.e., the first $T$ copies, by the fixed schedule of \Cref{lem:tomography}. Alice sends her outcomes, and Bob forms from their products an estimate $\theta^{\sC_\sA\sC_\sB}$ of the state of a single copy in that batch. From the circuit and the eigenvalue list of $\theta^{\sC_\sA\sC_\sB}$, Bob classically computes the matrix $C_\rin = \rvec^{-1}(\ket{\eta_\theta})$ of \Cref{sec:decoder}, where $\ket{\eta_\theta}^{\sS_0\sC_\sB\sX_0\sU_0\sE}$ is a purification of the output of the first stage of \Cref{lem:first instrument} at $q = 2^{m_0}$ applied to $\sC_\sA$ of $\theta^{\sC_\sA\sC_\sB}$. As this output contains every seed and outcome of the first stage in the registers $\sU_0$ and $\sX_0$, its computation does not use the values that Alice obtains in step 3. He applies \Cref{lem:factored model} to $C_\rin$ at $\nu$ of Eq.~\eqref{eq:decoder errors}, and obtains $V_0^{\sD_0}$, $W_0^{\sD_0}$, and $\{s_j^2\}_j$ of Eq.~\eqref{eq:one copy polar}.
    \item On $\sC_\sA$ of each of the remaining $tM$ copies in each batch, Alice performs the first stage of her measurement, given by \Cref{lem:first instrument} at $q = 2^{m_0}$, retains the register $\sS_0$ of $m_0$ qubits, and sends to Bob the seed of the unitary it draws and the measurement outcome. Bob stores them in the registers $\sU_0$ and $\sX_0$ as computational basis states and retains them together with $\sC_\sB$; these registers enter his decoder in step 5 as part of its input, and the decoder is not adapted to their values. At $m_0 = \log{d_{\sC_\sA}}$, this stage is the identity and has neither seed nor outcome.
    \item Partition the $tM$ copies in each batch into $M$ raw blocks of $t$ copies. On each raw block, Alice performs the second stage of her measurement, given by \Cref{lem:second instrument} at $q = 2^{m_1}$ with $\sS_0^t$ in place of $\sA$, retains $m_1$ qubits in $\sS$, and sends to Bob the seed of the unitary $U^{\sS_0^t}$ it draws and the outcome $x$.
    \item For each raw block, Bob forms from $U^{\sS_0^t}$ and $x$ the projector $P^\sD = (U^{\sS_0^t})^\top (\ketbra{x}{x}^\sX\otimes\bI^\sS)(U^{\sS_0^t})^*\otimes \bI^{\sE^t}$. Regarding his registers as $\hat{\sD}$, a copy of $\sD$ as in \Cref{sec:decoder}, he applies $(\widehat{U}^{\hat{\sD}\sW})^\dag$ of \Cref{lem:polar} at this $P$ and $\delta$, $a_\rp$, and $\epsilon_\ru$ of Eq.~\eqref{eq:decoder errors}, with $\sW$ initialized to $\ket{0}^\sW$. This approximately turns the state on $\hat{\sD}\sD$ into the canonical purification of its marginal on $\sD$. He then undoes Alice's unitary on the copy of $\sS_0^t$ by applying $(U^{\sS_0^t})^*$, keeps the resulting copy $\hat{\sS}$ of $\sS$, and discards the rest.
    \item Apply \Cref{cor:batch distillation} across the batches, and output the $m = Mm_1m_\rc$ pairs of qubits it produces.
\end{enumerate}


\subsection{Computationally efficient achievability result}
\label{sec:achievability result}

We now formally provide the performance that the protocol in \Cref{sec:protocol} achieves.
Let $\rho^{\sA\sB}$ be a $\Theta(n)$-qubit state.
The rate is stated through a further smoothing of the R\'{e}nyi coherent information, which restricts the rank of a marginal of the nearby state in addition to its global rank in Eq.~\eqref{eq:smooth coherent information global}. For $\hat{\epsilon} \in [0,1)$ and integers $\hat{r}, \hat{q} \geq 1$, we define
\begin{align}
\label{eq:smooth coherent information}
    \widetilde{I}_\beta^{\,\hat{\epsilon}, \hat{r}, \hat{q}}(\sA\rangle\sB)_\rho \coloneqq \sup\Big\{\widetilde{I}_\beta(\sA\rangle\sB)_{\hat{\rho}} : \f{1}{2}\big\|\rho^{\sA\sB} - \hat{\rho}^{\sA\sB}\big\|_1 \leq \hat{\epsilon}, \ \rank(\hat{\rho}^{\sA\sB}) \leq \hat{r}, \ \rank(\hat{\rho}^{\sA}) \leq \hat{q}\Big\},
\end{align}
where the supremum is over the states $\hat{\rho}^{\sA\sB}$ satisfying the conditions, and its value is $-\infty$ if no such state exists. Otherwise, the supremum is attained, since the set of such states is compact and $\widetilde{I}_\beta$ is continuous on it for $\beta \in [1/2, 1)\cup(1, \infty)$~\cite{Marwah2022UniformContinuitysandwichedrenyi}.
This quantity is at most $\widetilde{I}_\beta^{\,\hat{\epsilon}, \hat{r}}(\sA\rangle\sB)_\rho$ in Eq.~\eqref{eq:smooth coherent information global}, and both rank constraints are imposed on the nearby state $\hat{\rho}^{\sA\sB}$ alone, while $\rho^{\sA\sB}$ itself is arbitrary. At $\hat{\epsilon} = 0$, $\hat{r} = \rank(\rho^{\sA\sB})$, and $\hat{q} = \rank(\rho^\sA)$, it reduces to $\widetilde{I}_\beta(\sA\rangle\sB)_\rho$.
For every state $\hat{\rho}^{\sA\sB}$ with $\rank(\hat{\rho}^\sA) \leq \hat{q}$, the lower bound on the conditional R\'{e}nyi entropy by the marginal rank~\cite[Lemma~5.2]{Tomamichel2016QIPwithFiniteResources} yields
\begin{align}
\label{eq:target rate cap}
    \widetilde{I}_\beta(\sA\rangle\sB)_{\hat{\rho}}
    = -\widetilde{H}_\beta(\sA|\sB)_{\hat{\rho}}
    \leq \log{\rank(\hat{\rho}^\sA)}
    \leq \log{\hat{q}}.
\end{align}
Hence, $\widetilde{I}_\beta^{\,\hat{\epsilon}, \hat{r}, \hat{q}}(\sA\rangle\sB)_\rho \leq \log{\hat{q}}$.

\begin{theorem}[Efficient distillation for $k$ copies of a mixed state]
\label{thm:achievability by efficient protocol formal}
Fix a constant $\beta \in [1/2,1)$ and integers $\hat{r},\hat{q}=\poly(n)$. Let $J$, $\gamma$, and $\varepsilon$ be rationals of polynomial bit length with $J > 0$ and
\begin{align}
\label{eq:rate loss range}
    0 < \gamma \leq J/4, \qquad 
    1/\gamma = \cO\big(\log{n}/(\log\log n)^2\big), \qquad 
    0 < \varepsilon < 1, \qquad 
    \log(1/\varepsilon) = \poly(n).
\end{align}
Then, there is a rational $\epsilon_\star \in (0,1)$, which can be computed from $(J, \gamma)$ in $\poly(n)$ classical time and satisfies 
\begin{align}
\label{eq:smoothing radius}
    \epsilon_\star = \Theta\bigg(\f{\gamma\min\{\gamma,1\}}{J\log^2{(J/\gamma)}}\bigg),
\end{align}
such that the following holds.
For every $\hat{\epsilon} \in [0, \epsilon_\star]$ and every state $\rho^{\sA\sB}$ with $\widetilde{I}_\beta^{\,\hat{\epsilon}, \hat{r}, \hat{q}}(\sA\rangle\sB)_\rho \geq J$, the one-way LOCC protocol of \Cref{sec:protocol} acts on $k = \poly(n)$ copies and is implemented by a circuit of size $\poly(n)$. It distills Bell pairs from $\rho^{\sA\sB}$ at error $\varepsilon$ in the sense of \Cref{def:protocol}, with rate $R$ satisfying $J \geq R \geq J - \gamma$. The circuit is computed from $(n, \hat{r}, \hat{q}, J, \gamma, \varepsilon)$ in $\poly(n)$ classical time and is independent of $\hat{\epsilon}$ and $\rho^{\sA\sB}$.
\end{theorem}

\Cref{thm:achievability by efficient protocol formal} provides a universal protocol: a single circuit, computed from $(n, \hat{r}, \hat{q}, J, \gamma, \varepsilon)$ without any description of the state, distills Bell pairs at a rate at least $J - \gamma$ from every state $\rho^{\sA\sB}$ with $\widetilde{I}_\beta^{\,\hat{\epsilon}, \hat{r}, \hat{q}}(\sA\rangle\sB)_\rho \geq J$ at some $\hat{\epsilon} \in [0, \epsilon_\star]$. Since both rank conditions are imposed on $\hat{\rho}^{\sA\sB}$ alone, these states may have large rank, as do those obtained by adding a small amount of noise to $\hat{\rho}^{\sA\sB}$. The target rate carries no assumption of its own here, as Eq.~\eqref{eq:target rate cap} already bounds it by $\log{\hat{q}} = \cO(\log{n})$.

The universality is exactly what the state-agnostic model requires. For every $\hat{\epsilon} \in [0, \epsilon_\star]$, a single protocol works for the entire set $\scS$ of states with $\widetilde{I}_\beta^{\,\hat{\epsilon}, \hat{r}, \hat{q}}(\sA\rangle\sB)_\rho \geq J$, and its rate lower-bounds $E_{\rD, \ag}^{(k, \varepsilon)}(\sA:\sB)_\scS$. The circuit has size $\poly(n)$ and uses no description of the state. It thus lies in $\mathfrak{L}(s,k)$ at every state of $\scS$, and is also the output of a compiler that ignores its input. Hence the same rate lower-bounds $\widehat{E}_\rD^{(k,\varepsilon,s)}$ at every state of $\scS$, and $\widehat{E}_{\rD,\aw}^{(k,\varepsilon,0,s)}$ at every distribution over descriptions of states in $\scS$, at every large enough $s = \poly(n)$.

Case~(II) of \Cref{thm:achievability by efficient protocol} is obtained by choosing the parameters as follows. For a state $\hat{\rho}^{\sA\sB}$ in that theorem, we take $\hat{r} = \rank(\hat{\rho}^{\sA\sB})$ and $\hat{q} = \rank(\hat{\rho}^\sA)$, both of which are $\poly(n)$, together with $1/\gamma = \Theta(\log\log{n})$, a rational $J$ with $\widetilde{I}_\beta(\sA\rangle\sB)_{\hat{\rho}} - \gamma \leq J \leq \widetilde{I}_\beta(\sA\rangle\sB)_{\hat{\rho}}$, and an error $\varepsilon$ as in Eq.~\eqref{eq:rate loss range}. These obey Eq.~\eqref{eq:rate loss range} for large enough $n$, since $\widetilde{I}_\beta(\sA\rangle\sB)_{\hat{\rho}} = \Omega(1)$ implies $J \geq 4\gamma$, and Eq.~\eqref{eq:smoothing radius} yields $\epsilon_\star = \Omega\big(1/(\log{n}(\log\log{n})^4)\big)$, so that $\hat{\epsilon} = 1/\log^{1+c}{n}$ lies in $[0, \epsilon_\star]$ for every constant $c > 0$ and large enough $n$. Every state $\rho^{\sA\sB}$ within trace distance $1/\log^{1+c}{n}$ of $\hat{\rho}^{\sA\sB}$ obeys $\widetilde{I}_\beta^{\,\hat{\epsilon}, \hat{r}, \hat{q}}(\sA\rangle\sB)_\rho \geq \widetilde{I}_\beta(\sA\rangle\sB)_{\hat{\rho}} \geq J$, and the circuit, which depends on neither $\hat{\epsilon}$ nor $\rho^{\sA\sB}$, distills from each such state at a rate $R \geq J - \gamma \geq \widetilde{I}_\beta(\sA\rangle\sB)_{\hat{\rho}} - 2\gamma$, where $2\gamma = o(1)$.

\begin{proof}[Proof of \Cref{thm:achievability by efficient protocol formal}]

As $a \leq 1/32$ and $\ell_0$ is a constant integer such that $\ell_0 \geq 1/(1-\beta)$, Eqs.~\eqref{eq:rate parameters} and~\eqref{eq:main parameters} give $u_\rc = \Theta(a/\log(1/a))$ and $t = \Theta(\log(1/a)/\xi)$, and hence $\epsilon_\star$ defined in Eq.~\eqref{eq:epsilon star} satisfies
\begin{align}
    \epsilon_\star
    = \f{u_\rc}{2^{12}t}
    = \Theta\bigg(\f{a\xi}{\log^2(1/a)}\bigg)
    = \Theta\bigg(\f{\gamma\min\{\gamma,1\}}{J\log^2{(J/\gamma)}}\bigg),
\end{align}
where the last step uses $a = \gamma/(8J)$, $J/\gamma \geq 4$, and $\xi = \min\{\gamma, 1/4\}$. By the assumption $\widetilde{I}_\beta^{\,\hat{\epsilon}, \hat{r}, \hat{q}}(\sA\rangle\sB)_\rho \geq J$, we can fix a state $\hat{\rho}^{\sA\sB}$ attaining the supremum in Eq.~\eqref{eq:smooth coherent information}, which satisfies $\rank(\hat{\rho}^{\sA\sB}) \leq \hat{r}$, $\rank(\hat{\rho}^\sA) \leq \hat{q}$, $\f{1}{2}\|\rho^{\sA\sB}-\hat{\rho}^{\sA\sB}\|_1\leq\hat{\epsilon}\leq\epsilon_\star$, and $\widetilde{I}_\beta(\sA\rangle\sB)_{\hat{\rho}}\geq J$. Then Eq.~\eqref{eq:target rate cap} implies $J \leq \log{\hat{q}}$, and $J$, $\gamma$, and $\varepsilon$ lie in the ranges taken in \Cref{sec:protocol}. This state is used only in the analysis, while the protocol is constructed from $(n, \hat{r}, \hat{q}, J, \gamma, \varepsilon)$.

We first consider the protocol with ideal operations and show that, in every batch, the $m_1$ pairs obtained from each raw block in steps~1--5 of \Cref{sec:protocol} have infidelity at most $u_\rc$ with $\Phi_2^{\otimes m_1}$. To this end, using \Cref{cor:compression}, we analyze the pilot and one raw block after the compression through the mixture of the tensor powers of $\sigma_z^{\sC_\sA\sC_\sB}$ in Eq.~\eqref{eq:compact state}, and describe the decoder that Bob constructs from the pilot in step~2. We then show that Alice's measurement in steps~3 and~4 achieves decoupling at the state $\hat{\sigma}_z$ of \Cref{lem:geometry}, with little entropy lost in the first stage. This decoupling carries over to the pure state for which the decoder is built, and bounds the infidelity of the decoder output in step~5. \Cref{cor:batch distillation} in step~6 then yields the error and the rate of the ideal protocol. We then replace the exact gates, the exact finite random choices, and the ideal Haar unitaries of \Cref{lem:compression} with their implementations in \Cref{sec:numerics}, add the resulting errors, and verify the computational resources.

We begin with steps~1 and~2. As $\varepsilon_{\mathrm{cp}}(h)\leq u_\rc/8$ by Eq.~\eqref{eq:allocation}, where $h = T + t$, \Cref{cor:compression} places the state of the pilot and any one raw block of a batch after step~1 within trace distance $u_\rc/8$ of $\bE_z\big[(\sigma_z^{\sC_\sA\sC_\sB})^{\otimes(T+t)}\big]$. We analyze this mixture at a fixed $z$, and return to the actual state at the end. In step~2, the pilot consists of $T$ copies of $\sigma_z^{\sC_\sA\sC_\sB}$, and \Cref{lem:tomography} yields an estimate $\theta^{\sC_\sA\sC_\sB}$ satisfying $\f{1}{2}\|\theta^{\sC_\sA\sC_\sB}-\sigma_z^{\sC_\sA\sC_\sB}\|_1\leq\Delta$ except with probability at most $\zeta$. We call the pilot record successful when this holds, and condition on such a record. \Cref{lem:factored model} then supplies $C_0^{\sD_0}$ such that $\|\rvec(C_0) - \ket{\eta_\theta}\| \leq \nu$, where $\ket{\eta_\theta}^{\sS_0\sC_\sB\sX_0\sU_0\sE}$ is the purification of the first-stage output on $\theta^{\sC_\sA\sC_\sB}$ in \Cref{sec:decoder}. Bob's decoder is built for the pure state $\rvec(C^\sD)$ with $C^\sD = (C_0^{\sD_0})^{\otimes t}$, and is the inverse of the circuit of \Cref{lem:polar} at $\delta$, $a_\rp$, and $\epsilon_\ru$ of Eq.~\eqref{eq:decoder errors}. At the fixed $z$, the pilot and the raw block are in a product state, and the raw block thus remains $(\sigma_z^{\sC_\sA\sC_\sB})^{\otimes t}$ given the record.

We next analyze steps~3 and~4 at the fixed $z$. As no lower bound on the entropy of $\sigma_z^{\sC_\sA\sC_\sB}$ is available, we evaluate these steps on the state $\hat{\sigma}_z^{\sC_\sA\sC_\sB}$ of \Cref{lem:geometry}, which obeys Eq.~\eqref{eq:compact witness} and $\widetilde{I}_\beta(\sC_\sA\rangle\sC_\sB)_{\hat{\sigma}_z} = \widetilde{I}_\beta(\sA\rangle\sB)_{\hat{\rho}} \geq J$. By Eq.~\eqref{eq:entropy conversion}, a purification of $\hat{\sigma}_z^{\sC_\sA\sC_\sB}$ with a purifying system $\sE$ obeys $\widetilde{H}_p(\sC_\sA|\sE)_{\hat{\sigma}_z|\hat{\sigma}_z} \geq J$. Let $\ket{\eta_{\hat{\sigma}_z}}^{\sS_0\sC_\sB\sX_0\sU_0\sE}$ be the purification of the first-stage output on $\hat{\sigma}_z^{\sC_\sA\sC_\sB}$, constructed in the same way as $\ket{\eta_\theta}^{\sS_0\sC_\sB\sX_0\sU_0\sE}$ in \Cref{sec:decoder}, where $\sE$ keeps a copy of each outcome and seed of the first stage.
\Cref{lem:first instrument} with $q = 2^{m_0}$ yields $\widetilde{H}_p(\sS_0|\sE)_{\eta_{\hat{\sigma}_z}} \geq J - \xi/8$, as $\sE$ contains copies of $\sX_0$ and $\sU_0$. Indeed, if $m_0 = \log{d_{\sC_\sA}}$, the first stage is the identity and loses no entropy. Otherwise, $m_0 - J \geq B$, and hence $(p-1)(m_0 - J) \geq \log(16/((1-\beta)\xi))$ by $p-1 \geq 1-\beta$, $(1-\beta)\ell_0 \geq 1$, and Eq.~\eqref{eq:main parameters}; the loss in Eq.~\eqref{eq:first stage entropy} is then at most $\beta\xi/(16\ln{2}) < \xi/8$ by $\log(1+x) \leq x/\ln{2}$ and $1/(p-1) = \beta/(1-\beta)$. For $t$ copies, $\widetilde{H}_p(\sS_0^t|\sE^t)_{\eta_{\hat{\sigma}_z}^{\otimes t}} \geq t(J - \xi/8)$, since the tensor power of an optimal reference state for one copy is admissible for the $t$ copies. We apply the second stage in step~4, given by \Cref{lem:second instrument}, with $\sS_0^t$, $(\sC_\sB\sX_0\sU_0)^t$, and $\sE^t$ in place of $\sA$, $\sB$, and $\sE$, and with $q = 2^{m_1}$. Since $m_1 \leq t(J-\xi/4)$, the entropy exceeds $\log{q}$ by at least $t\xi/8$, and $(1-\beta)t\xi/8 \geq \lceil\log(32/u_\rc)\rceil$ by $(1-\beta)\ell_0 \geq 1$ and Eq.~\eqref{eq:main parameters}. Writing $\eta_U^{\sS(\sC_\sB\sX_0\sU_0)^t\sX\sE^t}$ for the state of \Cref{lem:second instrument} with $\ket{\eta_{\hat{\sigma}_z}}^{\otimes t}$ in place of $\ket{\psi}$, we thus obtain
\begin{align}
\label{eq:decoupling at nearby state}
    \bE_U\big[1 - \rF\big(\eta_U^{\sS\sX\sE^t}, \pi^{\sS}\otimes\pi^\sX\otimes(\eta_{\hat{\sigma}_z}^{\sE})^{\otimes t}\big)\big]
    \leq 2^{1-\lceil\log(32/u_\rc)\rceil}
    \leq \f{u_\rc}{16}.
\end{align}

We then carry Eq.~\eqref{eq:decoupling at nearby state} over to the pure state $\rvec(C^\sD)$ for which Bob's decoder is built in step~2. For a state $\tau^{\sC_\sA\sC_\sB}$, let $\omega_\tau^{\sS_0\sC_\sB\sX_0\sU_0}$ be the first-stage output on it, and let $\omega_0^{\sS_0\sC_\sB\sX_0\sU_0}$ be the marginal of $\rvec(C_0^{\sD_0})$ on $\sS_0\sC_\sB\sX_0\sU_0$, obtained by tracing out $\sE$. Then $\ket{\eta_{\hat{\sigma}_z}}^{\sS_0\sC_\sB\sX_0\sU_0\sE}$ and $\ket{\eta_\theta}^{\sS_0\sC_\sB\sX_0\sU_0\sE}$ purify $\omega_{\hat{\sigma}_z}^{\sS_0\sC_\sB\sX_0\sU_0}$ and $\omega_\theta^{\sS_0\sC_\sB\sX_0\sU_0}$, respectively. The purified distance obeys
\begin{align}
    \sqrt{1-\rF\big(\omega_{\hat{\sigma}_z}^{\otimes t}, \omega_0^{\otimes t}\big)}
    &\leq \sqrt{1-\rF\big(\omega_{\hat{\sigma}_z}^{\otimes t}, \omega_{\sigma_z}^{\otimes t}\big)} + \sqrt{1-\rF\big(\omega_{\sigma_z}^{\otimes t}, \omega_\theta^{\otimes t}\big)} + \sqrt{1-\rF\big(\omega_\theta^{\otimes t}, \omega_0^{\otimes t}\big)} \\
    &\leq \sqrt{21t\epsilon_\star} + \sqrt{2t\Delta} + \sqrt{t\nu^2} \\
    \label{eq:marginal chain}
    &\leq 5\sqrt{t\epsilon_\star} + \sqrt{2t\Delta} + \sqrt{2t\nu},
\end{align}
where the second line uses \Cref{lem:geometry} with the monotonicity of the fidelity, the Fuchs--van de Graaf inequality in Eq.~\eqref{eq:fuchsvandegraaf} with $\f{1}{2}\|\theta^{\sC_\sA\sC_\sB}-\sigma_z^{\sC_\sA\sC_\sB}\|_1\leq\Delta$, and $1 - |\braket{u}{v}|^2 \leq \|\ket{u} - \ket{v}\|^2$ for unit vectors with $\|\rvec(C_0^{\sD_0}) - \ket{\eta_\theta}\| \leq \nu$, each for $t$ copies. By Uhlmann's theorem, $(\omega_0^{\sS_0\sC_\sB\sX_0\sU_0})^{\otimes t}$ has a purification with the purifying system $\sE^t$, enlarged if necessary, whose purified distance from $\ket{\eta_{\hat{\sigma}_z}}^{\otimes t}$ is at most the right-hand side of Eq.~\eqref{eq:marginal chain}. 
Let $\check{\eta}_U^{\sS(\sC_\sB\sX_0\sU_0)^t\sX\sE^t}$ be the state of \Cref{lem:second instrument} computed from this purification in place of $\ket{\psi}$. As $\rvec(C^\sD)$ is another purification of $(\omega_0^{\sS_0\sC_\sB\sX_0\sU_0})^{\otimes t}$ with the purifying system $\sE^t$, computing $\check{\eta}_U$ from $\rvec(C^\sD)$ instead changes it only by an isometry on $\sE^t$, which leaves the fidelity below unchanged. At each $U$, both $\check{\eta}_U^{\sS\sX\sE^t}$ and $\pi^\sS\otimes\pi^\sX\otimes\check{\eta}_U^{\sE^t}$ are obtained from the purification given by Uhlmann's theorem through the same channels that yield $\eta_U^{\sS\sX\sE^t}$ and $\pi^\sS\otimes\pi^\sX\otimes(\eta_{\hat{\sigma}_z}^{\sE})^{\otimes t}$ from $\ket{\eta_{\hat{\sigma}_z}}^{\otimes t}$. The monotonicity and the triangle inequality of the purified distance thus yield
\begin{align}
    \bE_U\big[1 - \rF\big(\check{\eta}_U^{\sS\sX\sE^t}, \pi^{\sS}\otimes\pi^\sX\otimes\check{\eta}_U^{\sE^t}\big)\big]
    &\leq \bE_U\bigg[\Big(\sqrt{1 - \rF\big(\eta_U^{\sS\sX\sE^t}, \pi^{\sS}\otimes\pi^\sX\otimes(\eta_{\hat{\sigma}_z}^{\sE})^{\otimes t}\big)} + 2\Big(5\sqrt{t\epsilon_\star}+\sqrt{2t\Delta}+\sqrt{2t\nu}\Big)\Big)^2\bigg] \\
    &\leq \bigg(\Big(\bE_U\big[1 - \rF\big(\eta_U^{\sS\sX\sE^t}, \pi^{\sS}\otimes\pi^\sX\otimes(\eta_{\hat{\sigma}_z}^{\sE})^{\otimes t}\big)\big]\Big)^{1/2} + 2\Big(5\sqrt{t\epsilon_\star}+\sqrt{2t\Delta}+\sqrt{2t\nu}\Big)\bigg)^2 \\
    &\leq \bigg(\sqrt{\f{u_\rc}{16}} + 2\Big(5\sqrt{t\epsilon_\star}+\sqrt{2t\Delta}+\sqrt{2t\nu}\Big)\bigg)^2 \\
    &= \bigg(\f{1}{4} + \f{5+2\sqrt{2}}{32}\bigg)^2u_\rc \\
    \label{eq:decoupling at decoder target}
    &< \f{u_\rc}{4},
\end{align}
where the second line follows by expanding the square and applying Jensen's inequality, the third from Eq.~\eqref{eq:decoupling at nearby state}, and the fourth from $\epsilon_\star = \Delta = \nu = u_\rc/(2^{12}t)$ in Eqs.~\eqref{eq:epsilon star}, \eqref{eq:outer budgets}, and~\eqref{eq:decoder errors}.

We turn to Bob's decoder in step~5, and first consider the exact unitary that it approximates. Recall that a state $\eta$ has the canonical purification $\rvec\big(\sqrt{\eta^*}\big)$, whose marginal on the second tensor factor is $\eta$. For two states $\eta$ and $\tau$ on the same register, H\"{o}lder's inequality with the exponents $(4,2,4)$, applied to $\sqrt{\eta}\sqrt{\tau} = \eta^{1/4}(\eta^{1/4}\tau^{1/4})\tau^{1/4}$, shows $\rF(\eta, \tau) \leq \tr[\sqrt{\eta}\sqrt{\tau}]$. The right-hand side is the overlap of the canonical purifications of $\eta$ and $\tau$, whose fidelity is thus at least $\rF(\eta,\tau)^2$. If $\tau = \pi^\sS\otimes\tau^\sE$, the canonical purification of $\tau$ is the maximally entangled state of Schmidt rank $d_\sS$ between $\sS$ and a copy of $\sS$, tensored with the canonical purification of $\tau^\sE$.

Fix $U$ and write $\check{\eta}_U^{\sS\sX\sE^t} = \sum_x p_x\ketbra{x}{x}^\sX\otimes\check{\eta}_x^{\sS\sE^t}$, where $p_x = \|C^\sD P^\sD\|_2^2$ is the probability of the outcome $x$ as in \Cref{sec:decoder}, and $\check{\eta}_x^{\sS\sE^t}$ is an arbitrary state if $p_x = 0$. As the target in Eq.~\eqref{eq:decoupling at decoder target} is also block diagonal in $\sX$, the Cauchy--Schwarz inequality gives
\begin{align}
\label{eq:branch fidelity}
    \rF\big(\check{\eta}_U^{\sS\sX\sE^t}, \pi^\sS\otimes\pi^\sX\otimes\check{\eta}_U^{\sE^t}\big)
    = \bigg(\sum_x\sqrt{\f{p_x}{d_\sX}\rF_x}\bigg)^2
    \leq \sum_x p_x\rF_x,
\end{align}
where $\rF_x \coloneqq \rF\big(\check{\eta}_x^{\sS\sE^t}, \pi^\sS\otimes\check{\eta}_U^{\sE^t}\big)$. For each $x$, the adjoint of the unitary factor in the polar decomposition of $C^\sD P^\sD$, which the circuit of \Cref{lem:polar} approximates, maps $\rvec(C^\sD P^\sD)$ to $\rvec(|C^\sD P^\sD|)$. The latter is $\sqrt{p_x}$ times the canonical purification of the normalized marginal of $\rvec(C^\sD P^\sD)$ on $\sS\sE^t$, namely of $\check{\eta}_x^{\sS\sE^t}$. After $\sE^t$ and its copy are discarded, the output with the outcome $x$ thus has fidelity at least $\rF_x^2$ with $\Phi_2^{\otimes m_1}$ between $\sS$ and its copy $\hat{\sS}$, which Bob extracts in step~5 by applying $(U^{\sS_0^t})^*$. By Eq.~\eqref{eq:branch fidelity}, its infidelity averaged over $x$ is at most
\begin{align}
\label{eq:uhlmann infidelity}
    \sum_xp_x\big(1-\rF_x^2\big) \leq 2\sum_xp_x(1-\rF_x) \leq 2\big(1 - \rF\big(\check{\eta}_U^{\sS\sX\sE^t}, \pi^\sS\otimes\pi^\sX\otimes\check{\eta}_U^{\sE^t}\big)\big),
\end{align}
which is less than $u_\rc/2$ on average over $U$ by Eq.~\eqref{eq:decoupling at decoder target}.

Bob applies instead $(\widehat{U}^{\hat{\sD}\sW})^\dag$, the inverse of the circuit of \Cref{lem:polar} at $P^\sD$ of Eq.~\eqref{eq:decoder projector}, acting on his registers $\hat{\sD}$ and $\sW$. As $\|C^\sD P^\sD\|_2 = \sqrt{p_x}$ and $\sqrt{P^\sD\sigma^\sD P^\sD} = |C^\sD P^\sD|$, Eq.~\eqref{eq:polar endpoint} and the vectorization give, for each outcome $x$,
\begin{align}
\label{eq:decoder vector error}
    \Big\|\big(\widehat{U}^{\hat{\sD}\sW}\big)^\dag\big(\rvec(C^\sD P^\sD)\otimes\ket{0}^\sW\big) - \rvec(|C^\sD P^\sD|)\otimes\ket{0}^\sW\Big\|
    \leq 2\sqrt{2\delta}+\sqrt{2a_\rp}+\epsilon_\ru\sqrt{p_x}.
\end{align}
Both vectors have norm $\sqrt{p_x}$, and for a projector $\Pi$ and vectors $\ket{u}, \ket{v}$ of norm $\sqrt{p_x}$, we have $|\bra{u}\Pi\ket{u} - \bra{v}\Pi\ket{v}| \leq 2\sqrt{p_x}\|\ket{u}-\ket{v}\|$. Applying this with $\Pi$ the projector onto the target and summing over $x$, the additional infidelity is at most
\begin{align}
    \sum_x 2\sqrt{p_x}\big(2\sqrt{2\delta}+\sqrt{2a_\rp}+\epsilon_\ru\sqrt{p_x}\big)
    \leq 2\sqrt{d_\sX}\big(2\sqrt{2\delta}+\sqrt{2a_\rp}\big) + 2\epsilon_\ru
    = \f{(3\sqrt{2}+2)u_\rc}{32}
    < \f{7u_\rc}{32},
\end{align}
where we used $\sum_x\sqrt{p_x}\leq\sqrt{d_\sX}$, $\sum_xp_x = 1$, and Eq.~\eqref{eq:decoder errors}.
Moreover, steps~4 and~5 are applied to the first-stage outputs on $(\sigma_z^{\sC_\sA\sC_\sB})^{\otimes t}$ rather than to $\rvec(C^\sD)$. As these steps act only on Alice's and Bob's registers, this changes the fidelity by at most the trace distance between the two marginals on these registers, which is at most $t(\nu+\Delta)$ by $\|\rvec(C_0^{\sD_0}) - \ket{\eta_\theta}\| \leq \nu$ and $\f{1}{2}\|\theta^{\sC_\sA\sC_\sB}-\sigma_z^{\sC_\sA\sC_\sB}\|_1\leq\Delta$. Hence, for every successful pilot record, the state $\varphi^{\sS\hat{\sS}}$ of the $m_1$ pairs that Alice and Bob obtain from the raw block, averaged over $U$ and $x$, satisfies
\begin{align}
\label{eq:raw error}
    1 - \rF\big(\varphi^{\sS\hat{\sS}}, \Phi_2^{\otimes m_1}\big)
    \leq \f{u_\rc}{2} + \f{7u_\rc}{32} + t(\nu+\Delta)
    \leq \f{3u_\rc}{4},
\end{align}
where the last step uses $t(\nu+\Delta) = u_\rc/2^{11}$.

It remains to average over the pilot record and to return to the actual state. The bound of Eq.~\eqref{eq:raw error} holds at every $z$, and the fidelity with the pure target is linear in the state, so the same bound holds for the mixture over $z$. A pilot record fails with probability at most $\zeta = u_\rc/8$, in which case we bound the infidelity by one. Averaging over the record, the raw block of $\bE_z\big[(\sigma_z^{\sC_\sA\sC_\sB})^{\otimes(T+t)}\big]$ has infidelity at most $3u_\rc/4 + \zeta$. The protocol applies the same operations, including the decoder computed from the pilot, to the actual state after step~1, and the comparison with this mixture changes the fidelity by at most $\varepsilon_{\mathrm{cp}}(T+t) \leq u_\rc/8$. The $m_1$ pairs obtained from every raw block of every batch thus have infidelity at most $u_\rc$ with $\Phi_2^{\otimes m_1}$. By the monotonicity of the fidelity under the partial trace, each of these pairs has infidelity at most $u_\rc$ with $\Phi_2$.

The channel of steps~1--5 on one batch thus meets the conditions of \Cref{cor:batch distillation} at $m_\rb = Mm_1$, and step~6 invokes it at $\varepsilon_\rc = \varepsilon/2$. Its first statement bounds the infidelity of the output with $\Phi_2^{\otimes m}$ by $\varepsilon/2$, and its second statement gives $m_\rc/k_\rc \geq 1-4a$, which we use in the rate below.

We now implement the protocol with the conventions of \Cref{sec:numerics}. Let $G_0$ be an upper bound on the number of implementation calls other than the Haar unitaries over all $k_\rc$ batches and all code blocks, which is determined by $(n, \hat{r}, \hat{q}, J, \gamma, \varepsilon)$. We make each such call accurate to diamond distance $\varepsilon/[8(1+G_0)]$. Similarly, with an upper bound $R_0$ on the number of sampling calls, we bound the failure probability of each call by $\varepsilon/[8(1+R_0)]$. For the $k_\rc$ Haar unitaries in each laboratory, we use \Cref{lem:haar implementation} at $\varepsilon/(32k)$ with $\zeta_\rH=\varepsilon/(32k_\rc)$, and initialize and discard a work qubit at every application. In each laboratory, the $k$ applications, one on each copy, give tensor products of the implemented and the ideal isometries in Eq.~\eqref{eq:haar implementation}, which differ by at most $k\varepsilon/(32k) = \varepsilon/32$ in operator norm. Equation~\eqref{eq:rel of operator and diamond norm} and the contractivity under discarding the work qubits then bound the approximation error by $\varepsilon/16$ after summing over the two laboratories. The failure probability of the Haar unitaries is at most $2k_\rc\zeta_\rH = \varepsilon/16$. The failure events thus have total probability at most $3\varepsilon/16$, and all gate and arithmetic errors contribute at most $\varepsilon/8 + \varepsilon/16$.

The failure events above do not change the number $m$ of output pairs. Compared with the ideal protocol, the fidelity with $\Phi_2^{\otimes m}$ decreases by at most the diamond distance between the ideal and the implemented operations plus the total failure probability. The infidelity of the output is thus at most $\varepsilon/2 + 3\varepsilon/16 + 3\varepsilon/16 < \varepsilon$, which accounts for all the errors of the implemented protocol; namely, the protocol distills $m$ ebits from $\rho^{\sA\sB}$ at error $\varepsilon$.

We next verify the rate. Equation~\eqref{eq:allocation} implies $T/K\leq a/2$. Thus
\begin{align}
\label{eq:rate proof}
    R 
    = \f{m}{k}
    =\f{m_1}{t}\Big(1-\f{T}{K}\Big)\f{m_\rc}{k_\rc}
    \geq \f{m_1}{t}\Big(1-\f a2\Big)(1-4a)
    \geq \f{m_1}{t}\Big(1-\f{9a}{2}\Big),
\end{align}
where we used $m = Mm_1m_\rc$, $k = Kk_\rc$, and $K = T + tM$.
The factors are nonnegative because $a\leq1/32$. The floor in $m_1$ costs at most $1/t\leq\xi/8$, so $m_1/t\geq J-3\xi/8$. Since $aJ=\gamma/8$ and $\xi\leq\gamma$, Eq.~\eqref{eq:rate proof} yields $R\geq J-9\gamma/16-3\xi/8\geq J-15\gamma/16\geq J-\gamma$. Conversely, $m_1/t\leq J$ and the other factors are at most one, so $R\leq J$.

Finally, we show that all the resources are polynomial. As $J \leq \log{\hat{q}}$ and $\hat{q} = \poly(n)$, we have $J = \cO(\log{n})$. Since $1/\xi=\max\{1/\gamma,4\}$, Eqs.~\eqref{eq:rate parameters}, \eqref{eq:main parameters}, and~\eqref{eq:rate loss range} give
\begin{align}
    \f1a=\cO\Big(\f{(\log n)^2}{(\log\log n)^2}\Big),
    \qquad \f{1}{u_\rc}=\cO\Big(\f1a\log{\f1a}\Big),
    \qquad t=\cO\Big(\f{\log n}{\log\log n}\Big),
    \qquad B=\cO(\log\log n).
\end{align}
The inverses of $\epsilon_\star$, $w_\chi$, $\Delta$, $\zeta$, and $\nu$, the dimensions $d_{\sC_\chi}$ of Eq.~\eqref{eq:compact dimensions protocol}, and the dimension $d_{\sD_0}$ of $\sD_0$ in Eq.~\eqref{eq:one copy polar} are polynomial. Equation~\eqref{eq:outcome count} implies $d_\sX=\poly(n)$, and hence $1/\delta$, $1/a_\rp$, and $1/\epsilon_\ru$ in Eq.~\eqref{eq:decoder errors} are polynomial. \Cref{lem:factored model} then runs in polynomial time and returns a list $\{s_j^2\}_j$ of polynomial bit length $b_0$, and \Cref{lem:polar} yields a decoder of polynomial size, computed in polynomial time.

The quantities $T$, $h$, $G$, $M$, and $K$ in Eqs.~\eqref{eq:outer budgets}--\eqref{eq:allocation}, together with $N$, $L$, and $k_\rc$ of \Cref{cor:batch distillation} at $m_\rb = Mm_1$ and $\varepsilon_\rc = \varepsilon/2$, and hence $k = Kk_\rc$ and $m$, are polynomial by the bounds above and $\log(1/\varepsilon) = \poly(n)$ in Eq.~\eqref{eq:rate loss range}. The third statement of \Cref{lem:error reduction} makes the construction and the decoding of the two codes polynomial as well, since $\log{(1/a)}=\cO(\log\log{n})$ implies $\exp(\cO(\log^2{(1/a)}))=n^{o(1)}$. Tomography and the construction of the factored form are covered by \Cref{lem:diagonalization,lem:tomography,lem:factored model}, compression by \Cref{sec:collective compression}, Alice's two-stage measurement by \Cref{sec:two instruments}, and finite-precision compilation by \Cref{sec:numerics}. All the parameters of \Cref{sec:protocol} are rationals of polynomial bit length computed in $\poly(n)$ time, where the ceilings of their logarithms are computed by comparison with powers of two; in particular, $\epsilon_\star$ is computed from $(J, \gamma)$ at the fixed $\ell_0$. The implementation precisions and failure probabilities chosen above are $\varepsilon$ divided by polynomials in $n$, as $G_0$, $R_0$, and $k$ are polynomial, and the costs in \Cref{sec:numerics} depend on them only through their logarithms; these costs are thus polynomial under the same assumption on $\varepsilon$. All the classical messages, of polynomial total length, are sent from Alice to Bob, and Bob sends nothing, which proves the one-way property. As the whole circuit is computed from $(n, \hat{r}, \hat{q}, J, \gamma, \varepsilon)$ in $\poly(n)$ time, the circuit-generating algorithm is uniform.

\end{proof}


\section{Preparatory lemmas for converses}
\label{sec:Preparatory lemmas}

We collect here the lemmas needed to establish both converse bounds in \Cref{sec:Converse: State-agnostic setting,sec:Converse Description-aware setting}.
In \Cref{sec:flagged mixtures}, we show how the entropies and distillable entanglement of a state constructed from branches under a locally readable flag are determined by those of the branches. In \Cref{sec:spike state}, we construct a state whose conditional R\'{e}nyi entropy can be tuned through its order. In \Cref{sec:finite-copy converse}, we bound the rate of a distillation protocol at a finite number of copies by dominating the input state with a separable state.

\subsection{Flagged mixtures}
\label{sec:flagged mixtures}

Our states are constructed by mixing branches under a flag that both parties can read. The following lemmas characterize the entropies, the rank, and the distillable entanglement of such a mixture in terms of those of its branches.

\begin{lemma}[Conditional R\'{e}nyi entropy of a classical-quantum state]
\label{lem:conditional alpha entropystate}
Let $\alpha \in (1, \infty]$. For a state $\varrho^{\sA\sB\sC} = \sum_i p_i \varrho_i^{\sA\sB} \otimes \ketbra{i}{i}^\sC$, we have
\begin{align}
    \widetilde{H}_\alpha(\sA|\sB\sC)_\varrho 
    = - \frac{\alpha}{\alpha-1}\log{\sum_i p_i 2^{-\frac{\alpha-1}{\alpha}\widetilde{H}_\alpha(\sA|\sB)_{\varrho_i}}}.
\end{align}    
\end{lemma}

\begin{proof}[Proof of Lemma~\ref{lem:conditional alpha entropystate}]

By definition, $\widetilde{H}_\alpha(\sA|\sB\sC)_\varrho = -\min_{\sigma^{\sB\sC}} \widetilde{D}_\alpha(\varrho^{\sA\sB\sC} \|\bI^\sA \otimes \sigma^{\sB\sC})$.
The minimization is achieved by the state of the form $\sigma^{\sB\sC} = \sum_i q_i \sigma_i^\sB \otimes \ketbra{i}{i}^\sC$.
This is because, from the data-processing inequality, for the complete dephasing channel $\Delta^\sC$ defined by $\{\ket{i}^\sC\}_i$, it holds that $\widetilde{D}_\alpha(\varrho^{\sA\sB\sC} \| \bI^\sA \otimes \sigma^{\sB\sC}) \geq \widetilde{D}_\alpha(\Delta^\sC(\varrho^{\sA\sB\sC}) \| \bI^\sA \otimes \Delta^\sC(\sigma^{\sB\sC})) = \widetilde{D}_\alpha(\varrho^{\sA\sB\sC} \| \bI^\sA \otimes \Delta^\sC(\sigma^{\sB\sC}))$, where note that $\Delta^\sC(\varrho^{\sA\sB\sC}) = \varrho^{\sA\sB\sC}$.
The state $\sigma^{\sB\sC}$ of this form yields that, for $\alpha > 1$, 
\begin{align}
    \widetilde{H}_\alpha(\sA|\sB\sC)_\varrho
    &= -\min_{\{q_i, \sigma_i\}_i} \frac{1}{\alpha-1}\log{\sum_i p_i^\alpha q_i^{1-\alpha}\tr\big[(\sigma_i^{\frac{1-\alpha}{2\alpha}}\varrho_i\sigma_i^{\frac{1-\alpha}{2\alpha}})^\alpha\big]} \\
    &= -\frac{1}{\alpha-1}\log{\min_{\{q_i\}_i}\sum_i p_i^\alpha q_i^{1-\alpha} \min_{\{\sigma_i\}_i}\tr\big[(\sigma_i^{\frac{1-\alpha}{2\alpha}}\varrho_i\sigma_i^{\frac{1-\alpha}{2\alpha}})^\alpha\big]} \\
    &= -\frac{1}{\alpha-1}\log{\min_{\{q_i\}_i}\sum_i p_i^\alpha q_i^{1-\alpha} 
    2^{(1-\alpha)\widetilde{H}_\alpha(\sA|\sB)_{\varrho_i}}}.
\end{align}
Let $Q_{i, \alpha} = p_i^\alpha 2^{(1-\alpha)\widetilde{H}_\alpha(\sA|\sB)_{\varrho_i}}$.
We minimize $\sum_i q_i^{1-\alpha} Q_{i, \alpha}$ under the constraint $\sum_i q_i = 1$ using the method of Lagrange multipliers.
For $L = \sum_i q_i^{1-\alpha} Q_{i, \alpha} - \lambda(\sum_i q_i - 1)$, we obtain $\frac{\partial L}{\partial q_i} = (1-\alpha)q_i^{-\alpha}Q_{i, \alpha} - \lambda = 0$ and $\frac{\partial L}{\partial \lambda} = 1-\sum_i q_i = 0$; thus, $q_i = Q_{i, \alpha}^{1/\alpha}/\sum_j Q_{j, \alpha}^{1/\alpha}$.
Substituting this $q_i$, we arrive at
\begin{align}
    \widetilde{H}_\alpha(\sA|\sB\sC)_\varrho
    &= -\frac{1}{\alpha-1} \log{\sum_i \bigg(\frac{Q_{i, \alpha}^{1/\alpha}}{\sum_j Q_{j, \alpha}^{1/\alpha}}\bigg)^{1-\alpha}Q_{i, \alpha}} \\
    &= -\frac{\alpha}{\alpha-1} \log{\sum_i Q_{i, \alpha}^{1/\alpha}} \\
    &= -\frac{\alpha}{\alpha-1}\log{\sum_i p_i 2^{-\frac{\alpha-1}{\alpha}\widetilde{H}_\alpha(\sA|\sB)_{\varrho_i}}},
\end{align}
which establishes the claim for $\alpha \in (1, \infty)$. 

At $\alpha = \infty$, the identity follows by taking the limit $\alpha \to \infty$. Indeed, the left-hand side converges to $H_{\rm min}(\sA|\sB\sC)_\varrho$, and each term $\widetilde{H}_\alpha(\sA|\sB)_{\varrho_i}$ converges to $H_{\rm min}(\sA|\sB)_{\varrho_i}$. Since the sum is finite, the limit can be exchanged with the summation, and hence the identity extends to $\alpha = \infty$.

\end{proof}

\begin{lemma}[Flagged mixture of states]
\label{lem:flag mixture}
Let $J \geq 1$, $\{p_j\}_{j=1}^J$ be a probability distribution, and, for each $j$, $\ket{\chi_j}^{\sG_\sA^{(j)}\sG_\sB^{(j)}\sG_\sE^{(j)}}$ be a pure state. 
With flag registers $\sF_\sA, \sF_\sB, \sF_\sE$ of dimension $J$, define the pure state
\begin{align}
\label{eq:flag mixture}
    \ket{\Psi}^{\sA\sB\sE}
    \coloneqq \sum_{j=1}^{J}\sqrt{p_j}\,\ket{jjj}^{\sF_\sA\sF_\sB\sF_\sE}\ket{\chi_j}^{\sG_\sA^{(j)}\sG_\sB^{(j)}\sG_\sE^{(j)}}\bigotimes_{j'\neq j}\ket{0}^{\sG_\sA^{(j')}\sG_\sB^{(j')}\sG_\sE^{(j')}},
\end{align}
where $\sA = \sF_\sA\sG_\sA^{(1)}\cdots\sG_\sA^{(J)}$, and $\sB$ and $\sE$ are defined analogously. Then the following statements hold:
\begin{enumerate}[label=\textnormal{(\roman*)}]
    \item $\rank(\Psi^{\sA\sB}) = \sum_{j \,:\, p_j > 0}\rank\Big(\chi_j^{\sG_\sA^{(j)}\sG_\sB^{(j)}}\Big)$;
    \item $\widetilde{H}_\alpha(\sA|\sE)_\Psi
        = -\f{\alpha}{\alpha-1}\log{\bigg(\sum_{j=1}^{J} p_j 2^{-\f{\alpha-1}{\alpha}\widetilde{H}_\alpha(\sG_\sA^{(j)}|\sG_\sE^{(j)})_{\chi_j}}\bigg)}$ for every fixed $\alpha \in (1, \infty]$;
    \item $E_\rD(\sA:\sB)_\Psi \geq \sum_{j=1}^{J} p_j E_\rD\big(\sG_\sA^{(j)}:\sG_\sB^{(j)}\big)_{\chi_j}$;
    \item if $J = \poly(n)$, the distribution $\{p_j\}$ is efficiently computable, and each $\ket{\chi_j}^{\sG_\sA^{(j)}\sG_\sB^{(j)}\sG_\sE^{(j)}}$ is prepared by a circuit $C_j$ of size $\poly(n)$, then $\ket{\Psi}^{\sA\sB\sE}$ can be prepared with trace norm error $\negl(n)$ by a circuit of size $\poly(n)$ specified by $\{C_j\}_{j=1}^J$.
\end{enumerate}
Moreover, for every fixed $\alpha \in (1, \infty]$, if $p_1 = 1 - o(1)$ and $\widetilde{H}_\alpha(\sG_\sA^{(1)}|\sG_\sE^{(1)})_{\chi_1} \leq \widetilde{H}_\alpha(\sG_\sA^{(j)}|\sG_\sE^{(j)})_{\chi_j}$ for every $j \geq 2$, then (ii) reduces to
\begin{align}
\label{eq:flag mixture entropy dominant}
    \widetilde{H}_\alpha(\sA|\sE)_\Psi
    = \widetilde{H}_\alpha\big(\sG_\sA^{(1)}|\sG_\sE^{(1)}\big)_{\chi_1} + o(1).
\end{align}
\end{lemma}

\begin{proof}[Proof of \Cref{lem:flag mixture}]

Tracing out $\sE$ yields $\Psi^{\sA\sB} = \sum_j p_j \ketbra{jj}{jj}^{\sF_\sA\sF_\sB}\otimes\chi_j^{\sG_\sA^{(j)}\sG_\sB^{(j)}}\bigotimes_{j'\neq j}\ketbra{0}{0}^{\sG_\sA^{(j')}\sG_\sB^{(j')}}$, whose summands have mutually orthogonal supports; this gives (i). 
Since adjoining fixed pure product registers changes neither the conditional R\'{e}nyi entropy nor $E_\rD$, the $j$-th branch in Eq.~\eqref{eq:flag mixture} has the same values as $\ket{\chi_j}$.
Tracing out $\sB$ instead yields $\Psi^{\sA\sE} = \sum_j p_j \ketbra{j}{j}^{\sF_\sE}\otimes\varrho_j^{\sA\sG_\sE}$, where $\varrho_j^{\sA\sG_\sE} = \ketbra{j}{j}^{\sF_\sA}\otimes\chi_j^{\sG_\sA^{(j)}\sG_\sE^{(j)}}\bigotimes_{j'\neq j}\ketbra{0}{0}^{\sG_\sA^{(j')}\sG_\sE^{(j')}}$ and $\sG_\sE \coloneqq \sG_\sE^{(1)}\cdots\sG_\sE^{(J)}$. This is a classical-quantum state with classical register $\sF_\sE$; hence \Cref{lem:conditional alpha entropystate} gives (ii).
For (iii), the flag $\sF_\sA\sF_\sB$ is locally readable and identical on the two sides. Alice and Bob sort the input copies according to their value and distill each branch separately.
For (iv), the state $\sum_j\sqrt{p_j}\ket{jjj}^{\sF_\sA\sF_\sB\sF_\sE}$ can be prepared with trace norm error $\negl(n)$ using $\poly(n)$ gates, as $\{p_j\}$ is efficiently computable. Applying each $C_j$ controlled on the flag then takes $\poly(n)$ gates in total, and yields $\ket{\Psi}^{\sA\sB\sE}$ with trace norm error $\negl(n)$.

Finally, set $\Delta_j \coloneqq \widetilde{H}_\alpha(\sG_\sA^{(j)}|\sG_\sE^{(j)})_{\chi_j} - \widetilde{H}_\alpha(\sG_\sA^{(1)}|\sG_\sE^{(1)})_{\chi_1} \geq 0$, so that (ii) reads
\begin{align}
    \widetilde{H}_\alpha(\sA|\sE)_\Psi
    = \widetilde{H}_\alpha\big(\sG_\sA^{(1)}|\sG_\sE^{(1)}\big)_{\chi_1}
    - \f{\alpha}{\alpha-1}\log{\bigg(p_1 + \sum_{j \geq 2} p_j 2^{-\f{\alpha-1}{\alpha}\Delta_j}\bigg)}.
\end{align}
As $\Delta_j \geq 0$, every summand with $j \geq 2$ is at most $p_j$. The quantity inside the bracket is at most $1$, while it is at least $p_1 = 1 - o(1)$. Its logarithm is thus $-o(1)$, and thus the entire second term is $o(1)$ since $\f{\alpha}{\alpha-1} = \Theta(1)$ for fixed $\alpha \in (1, \infty]$, implying Eq.~\eqref{eq:flag mixture entropy dominant}.

\end{proof}

\begin{lemma}[Two-scale branch]
\label{lem:two scale branch}
Let $\sD_\sA$ and $\sD_\sB$ be single-qubit systems, and let $\ket{\hat{\phi}}^{\sM_\sA\sM_\sB}$ and $\ket{\phi}^{\sN_\sA\sN_\sB}$ be pure states.
For $\eta \in [0,1]$, define
\begin{align}
\label{eq:two scale branch}
    \ket{\zeta}^{\sL_\sA\sL_\sB}
    \coloneqq \sqrt{1 - \eta}\ket{00}^{\sD_\sA\sD_\sB}\ket{\hat{\phi}}^{\sM_\sA\sM_\sB}\ket{00}^{\sN_\sA\sN_\sB}
    + \sqrt{\eta}\ket{11}^{\sD_\sA\sD_\sB}\ket{00}^{\sM_\sA\sM_\sB}\ket{\phi}^{\sN_\sA\sN_\sB},
\end{align}
where $\sL_\sA \coloneqq \sD_\sA\sM_\sA\sN_\sA$ and $\sL_\sB \coloneqq \sD_\sB\sM_\sB\sN_\sB$.
Then the following statements hold.
\begin{enumerate}[label=\textnormal{(\roman*)}]
    \item For every fixed $\alpha \in (1, \infty]$,
    \begin{align}
    \label{eq:two scale renyi}
        (1-\eta)\,2^{-\f{\alpha-1}{\alpha}H_\alpha(\sM_\sA)_{\hat{\phi}}}
        \leq 2^{-\f{\alpha-1}{\alpha}H_\alpha(\sL_\sA)_{\zeta}}
        \leq (1-\eta)\,2^{-\f{\alpha-1}{\alpha}H_\alpha(\sM_\sA)_{\hat{\phi}}} + \eta\,2^{-\f{\alpha-1}{\alpha}H_\alpha(\sN_\sA)_{\phi}},
    \end{align}
    where the ratio $\f{\alpha-1}{\alpha}$ is understood to be $1$ when $\alpha = \infty$;
    \item $H_1(\sL_\sA)_{\zeta} \geq \eta H_1(\sN_\sA)_{\phi}$;
    \item if $\ket{\hat{\phi}}^{\sM_\sA\sM_\sB}$ and $\ket{\phi}^{\sN_\sA\sN_\sB}$ have Schmidt rank at most $2^{E_0}$ and $2^{E_\star}$ across $\sM_\sA : \sM_\sB$ and $\sN_\sA : \sN_\sB$, then the two terms in Eq.~\eqref{eq:two scale branch} are orthogonal due to the local flags $\sD_\sA\sD_\sB$ and have Schmidt rank at most $2^{E_0}$ and $2^{E_\star}$ across $\sL_\sA : \sL_\sB$, respectively.
\end{enumerate}
\end{lemma}

\begin{proof}[Proof of \Cref{lem:two scale branch}]

Tracing out $\sL_\sB$ gives $\zeta^{\sL_\sA} = (1-\eta)\ketbra{0}{0}^{\sD_\sA}\otimes\hat{\phi}^{\sM_\sA}\otimes\ketbra{0}{0}^{\sN_\sA} + \eta\ketbra{1}{1}^{\sD_\sA}\otimes\ketbra{0}{0}^{\sM_\sA}\otimes\phi^{\sN_\sA}$, whose two summands have orthogonal supports by the flag $\sD_\sA$.

For (i) at $\alpha \in (1,\infty)$, the orthogonality implies $\tr\big[(\zeta^{\sL_\sA})^\alpha\big] = (1-\eta)^\alpha\tr\big[(\hat{\phi}^{\sM_\sA})^\alpha\big] + \eta^\alpha\tr\big[(\phi^{\sN_\sA})^\alpha\big]$, and thus
\begin{align}
    (1-\eta)\big(\tr\big[(\hat{\phi}^{\sM_\sA})^\alpha\big]\big)^{1/\alpha}
    \leq \big(\tr\big[(\zeta^{\sL_\sA})^\alpha\big]\big)^{1/\alpha}
    \leq (1-\eta)\big(\tr\big[(\hat{\phi}^{\sM_\sA})^\alpha\big]\big)^{1/\alpha} 
    + \eta\big(\tr\big[(\phi^{\sN_\sA})^\alpha\big]\big)^{1/\alpha}.
\end{align}
Here the first inequality follows by discarding the second nonnegative term, and the second from the subadditivity of the function $x^{1/\alpha}$ for $\alpha > 1$.
As $\big(\tr[(\varrho^\sA)^{\alpha}]\big)^{1/\alpha} = 2^{-\f{\alpha-1}{\alpha}H_\alpha(\sA)_\varrho}$, we obtain Eq.~\eqref{eq:two scale renyi} for $\alpha \in (1, \infty)$.
At $\alpha = \infty$ the same orthogonality gives $\big\|\zeta^{\sL_\sA}\big\|_\infty = \max\big\{(1-\eta)\big\|\hat{\phi}^{\sM_\sA}\big\|_\infty, \, \eta\big\|\phi^{\sN_\sA}\big\|_\infty\big\}$, which again lies between the two sides.

For (ii), a state block-diagonalized by an orthonormal flag has entropy at least the average of the block entropies. Thus, $H_1(\sL_\sA)_{\zeta} \geq (1-\eta)H_1(\sM_\sA)_{\hat{\phi}} + \eta H_1(\sN_\sA)_{\phi} \geq \eta H_1(\sN_\sA)_{\phi}$.

Claim (iii) holds because the flags and the fixed vectors $\ket{0}$ are product across the cut, so that each term has the Schmidt rank of its remaining factor.

\end{proof}


\subsection{Spike state}
\label{sec:spike state}

We construct a state whose conditional R\'{e}nyi entropy varies with the order $\alpha$ at two prescribed points. The entropy stays at a fixed value up to $\alpha_\rc$, decreases beyond it, and changes sign at $\alpha_\rs$. The distillable entanglement remains at that fixed value. We call the state the \emph{spike state}.

\begin{lemma}[Spike state]
\label{lem:spikestate}
Let $\alpha_\rc, \alpha_\rs$ be arbitrary such that $1 < \alpha_\rc < \alpha_\rs \leq \infty$, and set
\begin{align}
\label{eq:spike exponent}
    \kappa \coloneqq \f{\alpha_\rs - \alpha_\rc}{\alpha_\rc(\alpha_\rs - 1)} \in (0, \, 1/\alpha_\rc],
    \ \ \ \ \ \ \ \ 
    g \coloneqq \big\lfloor r_\rs^{\kappa}\big\rfloor.
\end{align}
Then, for every $r_\rs \geq 2$, there exists a $\cO(\log{r_\rs})$-qubit pure state $\ket{\xi_{\mathrm{sp}}}^{\sG_\sA\sG_\sB\sG_\sE}$ with $\rank(\xi_{\mathrm{sp}}^{\sG_\sA\sG_\sB}) = r_\rs$, which can be prepared by a circuit of size $\poly(r_\rs)$.
For some separable state $\sigma_\xi^{\sG_\sA\sG_\sB}$, it satisfies
\begin{align}
\label{eq:alpha entropy of spike state}
    &\widetilde{H}_\alpha(\sG_\sA|\sG_\sE)_{\xi_{\mathrm{sp}}}
    = \log{g} - \f{\max\{\alpha, \alpha_\rc\} - \alpha_\rc}{\alpha_\rc\big(\max\{\alpha, \alpha_\rc\} - 1\big)}\log{r_\rs}
    - \f{1}{\alpha - 1}\log{\Big(1 + r_\rs^{-|\alpha-\alpha_\rc|/\alpha_\rc}\Big)} + o(1)
    \ \ \text{for every fixed} \ \alpha \in (1, \infty], \\
\label{eq:max entropy of spike state}
    &\xi_{\mathrm{sp}}^{\sG_\sA\sG_\sB} \leq g\,\sigma_\xi^{\sG_\sA\sG_\sB}, \\
\label{eq:IT distillable entanglement of spike state}
    &E_\rD(\sG_\sA:\sG_\sB)_{\xi_{\mathrm{sp}}} = \log{g},
\end{align}
where $o(1)$ vanishes as $r_\rs\to\infty$.
\end{lemma}

At constant $\alpha_\rc$ and $\alpha_\rs$ we have $\log{g} = \kappa\log{r_\rs} + o(1)$, and the coefficient of $\log{r_\rs}$ in Eq.~\eqref{eq:alpha entropy of spike state} decreases in $\alpha$ and vanishes at $\alpha = \alpha_\rs$. Thus, the state $\ket{\xi_{\mathrm{sp}}}$ satisfies that $\widetilde{H}_\alpha(\sG_\sA|\sG_\sE)_{\xi_{\mathrm{sp}}} = \Theta(\log{r_\rs}) > 0$ for each fixed $\alpha \in (1, \alpha_\rs)$ and $\widetilde{H}_\alpha(\sG_\sA|\sG_\sE)_{\xi_{\mathrm{sp}}} = -\Theta(\log{r_\rs}) < 0$ for $\alpha \in (\alpha_\rs, \infty]$ with $\alpha_\rs < \infty$. Moreover, $E_\rD(\sG_\sA:\sG_\sB)_{\xi_{\mathrm{sp}}} = \widetilde{H}_\alpha(\sG_\sA|\sG_\sE)_{\xi_{\mathrm{sp}}} + o(1)$ holds for $\alpha \in (1, \alpha_\rc)$. 
Note that the cases $\alpha = \infty$ and $\alpha_\rs = \infty$ are understood as the corresponding limits.

\begin{proof}[Proof of \Cref{lem:spikestate}]

We split $\sG_\sA = \sP_\sA\sQ_\sA$ and $\sG_\sB = \sP_\sB\sQ_\sB$, where $\sP_\sA$ and $\sP_\sB$ carry $\lceil\log{g}\rceil$ qubits each, $\sQ_\sA$ and $\sG_\sE$ carry $\lceil\log{r_\rs}\rceil$ qubits each, and $\sQ_\sB$ carries $\lceil\log{(r_\rs+1)}\rceil$ qubits. Set
\begin{align}
    q \coloneqq r_\rs^{1/\alpha_\rc - 1},
    \ \ \ \ \ \ 
    \lambda \coloneqq \f{qr_\rs - 1}{r_\rs - 1} \in [0, 1),
    \ \ \ \ \ \ 
    \ket{c_i}^{\sQ_\sB} \coloneqq \sqrt{\lambda}\ket{0} + \sqrt{1-\lambda}\ket{i},
\end{align}
where $\{\ket{i}\}_{i=0}^{r_\rs}$ are $r_\rs + 1$ orthonormal basis states of $\sQ_\sB$.
Each $\ket{c_i}$ is a unit vector, and $0 \leq \lambda < 1$ by $qr_\rs = r_\rs^{1/\alpha_\rc} \geq 1$ and $q < 1$.
We further define
\begin{align}
\label{eq:def of spike}
    \ket{\xi_{\mathrm{sp}}}^{\sG_\sA\sG_\sB\sG_\sE} \coloneqq \ket{\Phi_g}^{\sP_\sA\sP_\sB}\otimes\ket{\chi}^{\sQ_\sA\sQ_\sB\sG_\sE},
    \ \ \ \ \ 
    \ket{\Phi_g}^{\sP_\sA\sP_\sB} = \f{1}{\sqrt{g}}\sum_{i=1}^g \ket{i}^{\sP_\sA}\ket{i}^{\sP_\sB},
    \ \ \ \ \ 
    \ket{\chi}^{\sQ_\sA\sQ_\sB\sG_\sE} \coloneqq \f{1}{\sqrt{r_\rs}}\sum_{i=1}^{r_\rs}\ket{i}^{\sQ_\sA}\ket{c_i}^{\sQ_\sB}\ket{i}^{\sG_\sE}.
\end{align}

We first compute the reduced states. Tracing out $\sG_\sE$ gives $\chi^{\sQ_\sA\sQ_\sB} = \f{1}{r_\rs}\sum_i\ketbra{i}{i}^{\sQ_\sA}\otimes\ketbra{c_i}{c_i}^{\sQ_\sB}$, whose summands are mutually orthogonal, so $\xi_{\mathrm{sp}}^{\sG_\sA\sG_\sB} = \ketbra{\Phi_g}{\Phi_g}^{\sP_\sA\sP_\sB}\otimes\chi^{\sQ_\sA\sQ_\sB}$ is flat of rank $r_\rs$. Tracing out $\sQ_\sB$ instead, and using $\braket{c_{i'}}{c_i} = \lambda + (1-\lambda)\delta_{ii'}$,
\begin{align}
\label{inteq:spike AE}
    \chi^{\sQ_\sA\sG_\sE} = \f{1-\lambda}{r_\rs}\sum_i\ketbra{ii}{ii}^{\sQ_\sA\sG_\sE} + \lambda\ketbra{\Phi_{r_\rs}}{\Phi_{r_\rs}}^{\sQ_\sA\sG_\sE},
\end{align}
whose eigenvalues are $q$ once and $\f{1-q}{r_\rs-1}$ with multiplicity $r_\rs-1$.

Equation~\eqref{eq:max entropy of spike state} is immediate: we take $\sigma_\xi^{\sG_\sA\sG_\sB} \coloneqq \f{1}{g}\sum_{i=1}^{g}\ketbra{ii}{ii}^{\sP_\sA\sP_\sB}\otimes\chi^{\sQ_\sA\sQ_\sB}$, which is separable as $\chi^{\sQ_\sA\sQ_\sB}$ is. By the Cauchy--Schwarz inequality, $\ketbra{\Phi_g}{\Phi_g}^{\sP_\sA\sP_\sB} \leq \sum_{i=1}^{g}\ketbra{ii}{ii}^{\sP_\sA\sP_\sB}$ holds, and tensoring by $\chi^{\sQ_\sA\sQ_\sB} \geq 0$ preserves the ordering.
For Eq.~\eqref{eq:IT distillable entanglement of spike state}, discarding $\sQ_\sA\sQ_\sB$ leaves $\ket{\Phi_g}$, so $E_\rD(\sG_\sA:\sG_\sB)_{\xi_{\mathrm{sp}}} \geq \log{g}$. Conversely, the distillable entanglement is at most the relative entropy $\tr\big[\xi_{\mathrm{sp}}^{\sG_\sA\sG_\sB}\big(\log{\xi_{\mathrm{sp}}^{\sG_\sA\sG_\sB}} - \log{\sigma^{\sG_\sA\sG_\sB}}\big)\big]$ to any separable state $\sigma^{\sG_\sA\sG_\sB}$~\cite{khatri2024principlemodern}. 
Taking $\sigma = \sigma_\xi$, Eq.~\eqref{eq:max entropy of spike state} and the operator monotonicity of the logarithm give $\log{\xi_{\mathrm{sp}}^{\sG_\sA\sG_\sB}} \leq \log{g} + \log{\sigma_\xi^{\sG_\sA\sG_\sB}}$, and hence $E_\rD(\sG_\sA:\sG_\sB)_{\xi_{\mathrm{sp}}} \leq \log{g}$.

We turn to Eq.~\eqref{eq:alpha entropy of spike state}. Since $\chi^{\sQ_\sA\sG_\sE}$ is supported on two copies of the range of $\Pi \coloneqq \sum_{i=1}^{r_\rs}\ketbra{i}{i}$, the minimization below can be restricted to states $\sigma^{\sG_\sE}$ that satisfy $\sigma^{\sG_\sE} = \Pi\sigma^{\sG_\sE}\Pi$, with $\bI^{\sQ_\sA}$ replaced by $\Pi^{\sQ_\sA}$.
This is because the pinching by $\{\Pi, \bI - \Pi\}$ fixes $\chi^{\sQ_\sA\sG_\sE}$ and does not increase $\widetilde{D}_\alpha$, and neither does normalizing $\Pi\sigma^{\sG_\sE}\Pi$, by $\widetilde{D}_\alpha(\varrho\|c\,\sigma) = \widetilde{D}_\alpha(\varrho\|\sigma) - \log{c}$ for $c \leq 1$.

Let $W_{a,b}^{\sQ_\sA\sG_\sE} \coloneqq (X^aZ^b)^{\sQ_\sA}\otimes(X^aZ^{-b})^{\sG_\sE}$ with the Heisenberg--Weyl operators $X^s\ket{j} = \ket{j+s}$ mod $r_\rs$ and $Z^t\ket{j} = \omega^{jt}\ket{j}$, where $\omega = e^{i2\pi/r_\rs}$. 
Since both terms of Eq.~\eqref{inteq:spike AE} are invariant under $W_{a,b}^{\sQ_\sA\sG_\sE}$ and $\{X^aZ^{-b}\}_{a,b}$ forms a unitary $1$-design (see, e.g.,~\cite[Chap.~3]{khatri2024principlemodern}), the channel $\cW^{\sQ_\sA\sG_\sE}(\cdot) = \f{1}{r_\rs^2}\sum_{a,b}W_{a,b}^{\sQ_\sA\sG_\sE}(\cdot)(W_{a,b}^{\sQ_\sA\sG_\sE})^\dag$ fixes $\chi^{\sQ_\sA\sG_\sE}$ and maps $\Pi^{\sQ_\sA}\otimes\sigma^{\sG_\sE}$ such that $\sigma^{\sG_\sE} = \Pi\sigma^{\sG_\sE}\Pi$ to $\Pi^{\sQ_\sA}\otimes\Pi^{\sG_\sE}/r_\rs$. The data-processing inequality for $\widetilde{D}_\alpha$ therefore gives, for $\alpha \in (1,\infty)$,
\begin{align}
    \widetilde{H}_\alpha(\sQ_\sA|\sG_\sE)_\chi
    &= -\min_{\sigma^{\sG_\sE}}\widetilde{D}_\alpha\big(\chi^{\sQ_\sA\sG_\sE}\big\|\bI^{\sQ_\sA}\otimes\sigma^{\sG_\sE}\big) \\
    &= -\min_{\sigma^{\sG_\sE} = \Pi\sigma^{\sG_\sE}\Pi}\widetilde{D}_\alpha\big(\chi^{\sQ_\sA\sG_\sE}\big\|\Pi^{\sQ_\sA}\otimes\sigma^{\sG_\sE}\big) \\
    &\leq -\min_{\sigma^{\sG_\sE} = \Pi\sigma^{\sG_\sE}\Pi}\widetilde{D}_\alpha\big(\cW^{\sQ_\sA\sG_\sE}\big(\chi^{\sQ_\sA\sG_\sE}\big)\big\|\cW^{\sQ_\sA\sG_\sE}\big(\Pi^{\sQ_\sA}\otimes\sigma^{\sG_\sE}\big)\big) \\
    &= -\widetilde{D}_\alpha\big(\chi^{\sQ_\sA\sG_\sE}\big\|\Pi^{\sQ_\sA}\otimes\Pi^{\sG_\sE}/r_\rs\big).
\end{align}
Conversely, taking $\sigma^{\sG_\sE} = \Pi^{\sG_\sE}/r_\rs$ in the minimization above gives the reverse inequality. Hence, equality holds, and we have
\begin{align}
    \widetilde{H}_\alpha(\sQ_\sA|\sG_\sE)_\chi
    &= -\widetilde{D}_\alpha\big(\chi^{\sQ_\sA\sG_\sE}\big\|\Pi^{\sQ_\sA}\otimes\Pi^{\sG_\sE}/r_\rs\big) \\
    \label{inteq:spike renyi}
    &= \f{1}{1-\alpha}\log{\tr\big[(\chi^{\sQ_\sA\sG_\sE})^\alpha\big]} - \log{r_\rs}.
\end{align}
By the spectrum above, $\tr[(\chi^{\sQ_\sA\sG_\sE})^\alpha] = q^\alpha + (r_\rs-1)^{1-\alpha}(1-q)^\alpha = q^\alpha\big(1 + r_\rs^{\theta}\big)(1+o(1))$ with $\theta \coloneqq 1 - \alpha/\alpha_\rc$, since $q = o(1)$. Substituting this into Eq.~\eqref{inteq:spike renyi} and using $\log{(1+r_\rs^{\theta})} = \max\{0,\theta\}\log{r_\rs} + \log{(1+r_\rs^{-|\theta|})}$, we obtain 
\begin{align}
    \widetilde{H}_\alpha(\sQ_\sA|\sG_\sE)_\chi
    = \bigg(\f{\alpha(\alpha_\rc-1)}{\alpha_\rc(\alpha-1)} - 1 - \f{\max\{0,\theta\}}{\alpha-1}\bigg)\log{r_\rs}
    - \f{1}{\alpha-1}\log{\big(1+r_\rs^{-|\theta|}\big)} + o(1),
\end{align}
whose first coefficient equals $-\f{\alpha-\alpha_\rc}{\alpha_\rc(\alpha-1)}$ for $\alpha \geq \alpha_\rc$ and $0$ for $\alpha < \alpha_\rc$, that is, $-\f{\max\{\alpha,\alpha_\rc\}-\alpha_\rc}{\alpha_\rc(\max\{\alpha,\alpha_\rc\}-1)}$. 
Since $\ket{\xi_{\mathrm{sp}}}^{\sG_\sA\sG_\sB\sG_\sE}$ is the product of $\ket{\Phi_g}^{\sP_\sA\sP_\sB}$ with $\ket{\chi}^{\sQ_\sA\sQ_\sB\sG_\sE}$ and $\Phi_g^{\sP_\sA}$ is completely mixed, $\widetilde{H}_\alpha(\sG_\sA|\sG_\sE)_{\xi_{\mathrm{sp}}} = \log{g} + \widetilde{H}_\alpha(\sQ_\sA|\sG_\sE)_\chi$, which is Eq.~\eqref{eq:alpha entropy of spike state} for $\alpha \in (1,\infty)$.
For $\alpha = \infty$, taking the limit $\alpha \to \infty$ in Eq.~\eqref{inteq:spike renyi} shows $\widetilde{H}_\infty(\sQ_\sA|\sG_\sE)_\chi = -\log{\big(r_\rs\|\chi^{\sQ_\sA\sG_\sE}\|_\infty\big)} = -\f{\log{r_\rs}}{\alpha_\rc}$ by $\|\chi^{\sQ_\sA\sG_\sE}\|_\infty = q$, and adding $\log{g}$ as above gives Eq.~\eqref{eq:alpha entropy of spike state} at $\alpha = \infty$.

Finally, $\ket{\xi_{\mathrm{sp}}}^{\sG_\sA\sG_\sB\sG_\sE}$ is a pure state on $\cO(\log{r_\rs})$ qubits with an explicitly computable classical description, so it is prepared by a circuit of $\poly(r_\rs)$ one- and two-qubit gates~\cite[Sec.~4.5]{nielsen2010quantum}.

\end{proof}


\subsection{Finite-copy rate converse}
\label{sec:finite-copy converse}

The converse bounds cap the rate of a protocol acting on a finite number of copies. The following lemmas establish this by dominating the input with a separable state whose overlap with a maximally entangled state is small.

\begin{lemma}[Overlap of a separable state with a maximally entangled state]
\label{lem:sep overlap}
For $\ket{\Phi_d}^{\sA\sB} = \f{1}{\sqrt{d}}\sum_{i=1}^{d}\ket{ii}^{\sA\sB}$ and any separable state $\sigma^{\sA\sB}$, we have $\bra{\Phi_d}\sigma\ket{\Phi_d} \leq \f{1}{d}$.
\end{lemma}

\begin{proof}[Proof of \Cref{lem:sep overlap}]

A separable state $\sigma^{\sA\sB}$ is a convex combination of product states $\ketbra{a, b}{a, b}$, and it suffices to bound $|\braket{\Phi_d}{a, b}|^2$ for unit vectors $\ket{a}^{\sA} = \sum_{i=1}^{d}a_i\ket{i}^{\sA}$ and $\ket{b}^{\sB} = \sum_{i=1}^{d}b_i\ket{i}^{\sB}$, expanded in the bases used in the definition of $\ket{\Phi_d}^{\sA\sB}$.
We have $\braket{\Phi_d}{a, b} = \f{1}{\sqrt{d}}\sum_{i=1}^{d} a_ib_i$, whose absolute value is at most $\f{1}{\sqrt{d}}\big(\sum_i|a_i|^2\big)^{1/2}\big(\sum_i|b_i|^2\big)^{1/2} = \f{1}{\sqrt{d}}$ by the Cauchy--Schwarz inequality. Hence $|\braket{\Phi_d}{a, b}|^2 \leq \f{1}{d}$, and the claim follows.

\end{proof}

\begin{lemma}[Converse from separable domination]
\label{lem:sandwich converse}
Let $\tau^{\sA\sB}$ be a state, $k \geq 1$, and $\vartheta \in [0,1)$. Suppose that there exist $K \geq 0$, a positive semidefinite operator $\hat{\tau}$, and a separable state $\sigma$ on $(\sA\sB)^{\otimes k}$ such that $\big\|\tau^{\otimes k} - \hat{\tau}\big\|_1 \leq \vartheta$ and $\hat{\tau} \leq 2^{K}\sigma$.
Then any $k$-shot distillation protocol of \Cref{def:protocol} distilling $kR$ ebits from $\tau^{\sA\sB}$ at error $\varepsilon$ with $\varepsilon + \vartheta < 1$ obeys
\begin{align}
\label{eq:rateconverse sandwich}
    R \leq \f{K}{k} + \f{1}{k}\log{\f{1}{1 - \varepsilon - \vartheta}}.
\end{align}
Moreover, the same bound holds when the protocol acts on $\tau^{\otimes k}\otimes\varpi$ for a state $\varpi$ that is separable across $\sA:\sB$, provided the two conditions above hold for $\hat{\tau}\otimes\varpi$ and $\sigma\otimes\varpi$.
\end{lemma}

\begin{proof}[Proof of \Cref{lem:sandwich converse}]
For an LOCC $\Lambda$, let $\rF_\Lambda(\cdot) \coloneqq \tr\big[\big(\Phi_2^{\otimes kR}\otimes\bI\big)\Lambda(\cdot)\big]$ on the first $kR$ declared pairs, and $\rF_\Lambda \coloneqq 0$ if fewer are declared (see also Eq.~\eqref{eq:fidelity functional}). Set $Q \coloneqq \Lambda^\dag\big(\Phi_2^{\otimes kR}\otimes\bI\big)$, so that $0 \leq Q \leq \bI$ and $\tr[Q\tau^{\otimes k}] = \rF_\Lambda(\tau^{\otimes k}) \geq 1 - \varepsilon$.
Since $\|Q\|_\infty \leq 1$, H\"{o}lder's inequality implies that
\begin{align}
    \tr[Q\hat{\tau}] 
    &= \tr\big[Q\tau^{\otimes k}\big] - \tr\big[Q\big(\tau^{\otimes k} - \hat{\tau}\big)\big] \\
    &\geq \tr[Q\tau^{\otimes k}] - \big\|\tau^{\otimes k} - \hat{\tau}\big\|_1 \\
    &\geq 1 - \varepsilon - \vartheta,
\end{align}
and $Q \geq 0$ with $\hat{\tau} \leq 2^{K}\sigma$ gives $\tr[Q\sigma] \geq 2^{-K}(1 - \varepsilon - \vartheta)$. 
As $\Lambda$ maps separable inputs to separable outputs, and a partial trace preserves separability, $\tr[Q\sigma] = \rF_\Lambda(\sigma) \leq 2^{-kR}$ by \Cref{lem:sep overlap}.
Combining them and taking logarithms gives Eq.~\eqref{eq:rateconverse sandwich}. The last statement follows because $\varpi$ is separable, so $\sigma\otimes\varpi$ is, and tensoring a fixed state preserves the trace norm.

\end{proof}

\begin{lemma}[Separable bound by truncating light components]
\label{lem:truncation}
Let $\tau^{\sA\sB}$ be a state, $\vartheta \in (0,1)$, and $k \geq 1$. Then the conditions of \Cref{lem:sandwich converse} are satisfied in each of the following two cases.
\begin{enumerate}[label=\textnormal{(\arabic*)}]
    \item There are $J \geq 1$, positive semidefinite $\tau^{(0)}, \ldots, \tau^{(J)}$ with $\tau^{\sA\sB} = \sum_{i=0}^J\tau^{(i)}$, and separable positive semidefinite $\varsigma^{(0)}, \ldots, \varsigma^{(J)}$ with $\sum_{i=0}^J\tr[\varsigma^{(i)}] \leq 1$. They obey $\tau^{(i)} \leq 2^{E_i}\varsigma^{(i)}$ for every $i$ with some $E_0, \ldots, E_J \geq 0$, and $\tr[\tau^{(i)}] \leq w_i$ for every $i \geq 1$ with some $w_1, \ldots, w_J \in [0,1]$. Then $K = k\big(E_0 + \f{J}{\vartheta}\sum_{i=1}^J w_i E_i\big)$.
    \item There are $\eta \in [0,1]$ and $E_0, E_\star \geq 0$ such that $\tau^{\sA\sB} = \sum_{j} q_j\ketbra{\psi_j}{\psi_j}^{\sA\sB}$ for a probability distribution $\{q_j\}$ and pure states $\ket{\psi_j}^{\sA\sB} = \sqrt{1-\eta}\ket{\psi_{j,0}}^{\sA\sB} + \sqrt{\eta}\ket{\psi_{j,1}}^{\sA\sB}$,
    where $\ket{\psi_{j,0}}$ and $\ket{\psi_{j,1}}$ are orthonormal and have Schmidt rank at most $2^{E_0}$ and $2^{E_\star}$ across $\sA:\sB$. Then $K = k\big(E_0 + \f{9\eta}{\vartheta^2}\big(E_\star + \log{(k+1)}\big)\big) + \log{(k+1)}$.
\end{enumerate}
Both cases are closed under the following operation. Let $\{p_x\}$ be a probability distribution, and for each $x$, let $\tau_x^{\sA\sB}$ obey the conditions of the case under consideration with the same parameters. Let $\varsigma^{\sL_\sA\sL_\sB}(x)$ be a separable state, and let $U_x^{\sA\sL_\sA\sB\sL_\sB} = U_x^{\sA\sL_\sA}\otimes U_x^{\sB\sL_\sB}$ be a local unitary. Then the same $K$ is obtained for $\sum_{x}p_x U_x^{\sA\sL_\sA\sB\sL_\sB}\big(\tau_x^{\sA\sB}\otimes\varsigma^{\sL_\sA\sL_\sB}(x)\big)(U_x^{\sA\sL_\sA\sB\sL_\sB})^\dag$.
\end{lemma}

\begin{proof}[Proof of \Cref{lem:truncation}]
In case (1), for $c = c_1c_2 \cdots c_k \in \{0, \ldots, J\}^k$ let $|c|_i$ be the number of entries equal to $i$, set $t_i \coloneqq \lfloor Jkw_i/\vartheta \rfloor$ for $i \geq 1$, and let $\cG$ be the set of $c$ with $|c|_i \leq t_i$ for every $i \geq 1$. 
We have $\tau^{\otimes k} = \sum_{c}\bigotimes_{l=1}^k\tau^{(c_l)}$, and we take $\hat{\tau} \coloneqq \sum_{c \in \cG}\bigotimes_{l}\tau^{(c_l)}$. 
Every summand is positive semidefinite, so $\tau^{\otimes k} - \hat{\tau} \geq 0$ and $\|\tau^{\otimes k} - \hat{\tau}\|_1 = \sum_{c \notin \cG}\prod_{l}\tr[\tau^{(c_l)}]$. 
Since $\sum_{i=0}^J\tr[\tau^{(i)}] = 1$, the latter corresponds to $\Pr[c \notin \cG]$, the probability that $|c|_i > t_i$ for some $i \geq 1$, when the entries of $c$ are drawn independently with $\Pr[c_l = i] = \tr[\tau^{(i)}]$.
Markov's inequality applied to each $|c|_i$, whose mean is at most $kw_i$, and a union bound over $i$ therefore bound it by $\sum_{i \geq 1}\f{kw_i}{t_i+1} \leq \vartheta$, since $t_i = \lfloor Jkw_i/\vartheta \rfloor$ gives $\f{kw_i}{t_i+1} \leq \f{\vartheta}{J}$ for every $i$.
Hence, we obtain $\|\tau^{\otimes k} - \hat{\tau}\|_1 \leq \vartheta$.

Next, as $A\otimes B \leq A'\otimes B'$ holds for positive semidefinite operators satisfying $A \leq A'$ and $B \leq B'$, and $\tau^{(i)} \leq 2^{E_i}\varsigma^{(i)}$, every $c \in \cG$ obeys $\bigotimes_{l}\tau^{(c_l)} \leq 2^{\sum_{l}E_{c_l}}\bigotimes_l\varsigma^{(c_l)} = 2^{\sum_{i}|c|_iE_i}\bigotimes_{l}\varsigma^{(c_l)} \leq 2^{kE_0 + \sum_{i \geq 1}t_iE_i}\bigotimes_{l}\varsigma^{(c_l)}$.
Summing over $c \in \cG$ and then adjoining the terms with $c \notin \cG$, which are positive semidefinite, we obtain
\begin{align}
    \hat{\tau}
    \leq 2^{kE_0 + \sum_{i \geq 1}t_iE_i}\sum_{c \in \cG}\bigotimes_{l}\varsigma^{(c_l)}
    \leq 2^{kE_0 + \sum_{i \geq 1}t_iE_i}\sum_{c}\bigotimes_{l}\varsigma^{(c_l)}
    = 2^{kE_0 + \sum_{i \geq 1}t_iE_i}\,\bar{\sigma}^{\otimes k},
\end{align}
where $\bar{\sigma} \coloneqq \sum_{i=0}^J\varsigma^{(i)}$ is separable with $\tr[\bar{\sigma}] \leq 1$. Extending $\bar{\sigma}$ to unit trace by adding the missing weight on any product state yields a separable state $\sigma$ satisfying $\bar{\sigma}\leq\sigma$. The claim then follows from $t_i \leq Jkw_i/\vartheta$.

In case (2), for $c \in \{0,1\}^{k}$ let $|c|$ be the number of entries equal to $1$ and let $\cG$ be the set of $c$ with $|c| \leq t$, the value of $t$ being fixed below. Note that when the entries of $c$ are drawn independently with $\Pr[c_l = 1] = \eta$, then $\epsilon \coloneqq \sum_{c \notin \cG}\Pr[c] = \sum_{c \notin \cG}(1-\eta)^{k-|c|}\eta^{|c|} \leq \f{k\eta}{t+1}$ by Markov's inequality applied to $|c|$.
Expand $\bigotimes_{l=1}^k\ket{\psi_{j_l}}$ over $c$ and set 
\begin{align}
    \ket{u_{\vec{\jmath}}} \coloneqq \sum_{c \in \cG}\sqrt{(1-\eta)^{k-|c|}\eta^{|c|}}\bigotimes_{l=1}^k\ket{\psi_{j_l, c_l}},
\end{align}
for each $\vec{\jmath} = (j_1, \ldots, j_k)$, and $\hat{\tau} \coloneqq \sum_{\vec{\jmath}}q_{\vec{\jmath}}\ketbra{u_{\vec{\jmath}}}{u_{\vec{\jmath}}}$ with $q_{\vec{\jmath}} \coloneqq \prod_l q_{j_l}$. Writing $\bigotimes_l\ket{\psi_{j_l}} = \ket{u_{\vec{\jmath}}} + \ket{v_{\vec{\jmath}}}$, the vectors $\bigotimes_{l}\ket{\psi_{j_l, c_l}}$ being orthonormal across distinct $c$ gives $\|\ket{v_{\vec{\jmath}}}\|^2 = \epsilon$ for every $\vec{\jmath}$.
Since $\|\ketbra{a}{b}\|_1 = \|\ket{a}\|\|\ket{b}\|$ and $\|\ket{u_{\vec{\jmath}}}\| \leq 1$, the triangle inequality gives
\begin{align}
    \Big\|\bigotimes_l\ketbra{\psi_{j_l}}{\psi_{j_l}} - \ketbra{u_{\vec{\jmath}}}{u_{\vec{\jmath}}}\Big\|_1
    &= \big\|\ketbra{u_{\vec{\jmath}}}{v_{\vec{\jmath}}} + \ketbra{v_{\vec{\jmath}}}{u_{\vec{\jmath}}} + \ketbra{v_{\vec{\jmath}}}{v_{\vec{\jmath}}}\big\|_1 \\
    &\leq \big\|\ketbra{u_{\vec{\jmath}}}{v_{\vec{\jmath}}}\big\|_1 + \big\|\ketbra{v_{\vec{\jmath}}}{u_{\vec{\jmath}}}\big\|_1 + \big\|\ketbra{v_{\vec{\jmath}}}{v_{\vec{\jmath}}}\big\|_1 \\
    &\leq 2\sqrt{\epsilon} + \epsilon \\
    &\leq 3\sqrt{\epsilon},
\end{align}
the last step by $\epsilon \leq 1$. Averaging over $\vec{\jmath}$ with the weights $q_{\vec{\jmath}}$ gives $\|\tau^{\otimes k} - \hat{\tau}\|_1 \leq 3\sqrt{\epsilon}$, which is at most $\vartheta$ at $t \coloneqq \lfloor 9k\eta/\vartheta^2\rfloor$ by the Markov bound above.

The Schmidt rank of $\ket{u_{\vec{\jmath}}}$ across $\sA^{\otimes k}|\sB^{\otimes k}$ is at most $|\cG|2^{kE_0 + tE_\star}$, since it is multiplicative under tensor products and subadditive under sums. 
Moreover, a vector $\ket{u} = \sum_i\sqrt{\lambda_i}\ket{a_ib_i}$ of Schmidt rank $D$ obeys $\ketbra{u}{u} \leq D\sum_i\lambda_i\ketbra{a_ib_i}{a_ib_i}$ by the Cauchy--Schwarz inequality, where $\sum_i\lambda_i\ketbra{a_ib_i}{a_ib_i}$ is separable of trace $\|\ket{u}\|^2 \leq 1$.
Averaging over $\vec{\jmath}$ gives $\hat{\tau} \leq |\cG|\,2^{kE_0+tE_\star}\bar{\sigma}$ with $\bar{\sigma}$ separable of trace at most $1$. Extending $\bar{\sigma}$ to unit trace as in case (1) yields a separable state $\sigma$ satisfying $\bar{\sigma}\leq\sigma$. The claim then follows from $|\cG| \leq \sum_{i \leq t}\binom{k}{i} \leq (k+1)^{t+1}$ and $t \leq 9k\eta/\vartheta^2$.

The closure claims are immediate. In case (1), the operators $\sum_{x}p_x U_x\big(\tau_x^{(i)}\otimes\varsigma(x)\big)U_x^\dag$ and $\sum_{x}p_x U_x\big(\varsigma_x^{(i)}\otimes\varsigma(x)\big)U_x^\dag$ obey the conditions at the same $w_i$ and $E_i$, since tensoring by $\varsigma(x)$ and conjugating by a local unitary preserve the traces, the operator orderings, and separability. In case (2), expanding each $\varsigma(x)$ into pure states $\ket{\phi_\ell}$ product across the cut exhibits the resulting state as a mixture of $U_x\ket{\psi_j}\ket{\phi_\ell}$, whose two components remain orthonormal and keep their Schmidt ranks, since a local unitary and a factor product across the cut change neither.

\end{proof}

\section{Converse: Haar random construction and statistical indistinguishability}
\label{sec:Converse: State-agnostic setting}

We here show our converse bounds in the state-agnostic and the state-dependent models, for which no computational hardness assumption is used. We present the tools related to Haar randomness in \Cref{sec:haarrandomintegration}, construct the family of states from them in \Cref{sec:construction of the state family state agnostic}, and prove the bounds on it in \Cref{sec:statistical converse result}.

\subsection{Haar random tools}
\label{sec:haarrandomintegration}

In this section, we collect the tools related to Haar randomness that we use.

\begin{lemma}[Haar and Haar-subsystem states on $2m$ qubits; adapted from Ref.~{\cite[Definition~12, Corollary~1, and Lemma~9]{Leone2025Entanglementtheorywithlimited}}]
\label{lem:haar pair}
Let $m$ and $\mu$ obey $1 \leq \mu \leq 2m$, and let $\sA$ and $\sB$ be registers of $m$ qubits each. For $b \in \{0, 1\}$, let $\scE_b(m, \mu)$ be ensembles of pure states on $\sA\sB$: $\scE_0(m, \mu)$ consists of Haar random states on $2m$ qubits, and $\scE_1(m, \mu)$ of $U_\pi\ket{0^{2m-\mu}}\ket{\vartheta}$, where $\ket{\vartheta}$ is Haar random on $\mu$ qubits and $U_\pi$ is the unitary corresponding to a uniformly random permutation $\pi$ of the computational basis of $2m$ qubits. Then the following statements hold for $\ket{\varphi_b}^{\sA\sB} \sim \scE_b(m, \mu)$:
\begin{enumerate}[label=\textnormal{(\roman*)}]
    \item $\log{\rank(\varphi_1^{\sA})} \leq \mu$, and hence $H_1(\sA)_{\varphi_1} \leq \mu$;
    \item $\widetilde{H}_\alpha(\sA)_{\varphi_0} \geq \f{m}{4}$ for every $\alpha \in (0, \infty]$;
    \item $\big\|\bE_{\scE_0(m,\mu)}\big[\varphi_0^{\otimes k}\big] - \bE_{\scE_1(m,\mu)}\big[\varphi_1^{\otimes k}\big]\big\|_1 = \cO\big(k^2 2^{-\mu}\big)$ for every $k$; hence at $\mu = \omega(\log{n})$ the two ensembles are statistically indistinguishable at every $k = \poly(n)$.
\end{enumerate}
Statement (i) holds with certainty and (ii) with probability $1 - 2^{-\Omega(m)}$ over the choice of $\ket{\varphi_0}^{\sA\sB} \in \scE_0(m,\mu)$.
\end{lemma}

\begin{proof}[Proof of \Cref{lem:haar pair}]
For (i), a state of $\scE_1(m,\mu)$ is supported on $2^{\mu}$ computational basis states of $\sA\sB$, so its reduced state on $\sA$ has rank at most $2^{\mu}$ and $H_1(\sA)_{\varphi_1} \leq \log\rank(\varphi_1^{\sA}) \leq \mu$~{\cite[Corollary~1]{Leone2025Entanglementtheorywithlimited}}.
Regarding (ii), for a Haar random state $\ket{\varphi_0}$ on $2m$ qubits, $\bE\big[\tr\big[(\varphi_0^{\sA})^2\big]\big] \leq 2^{1-m}$~\cite{Lubkin1993AverageQuantalBehavior}. Markov's inequality gives $\tr\big[(\varphi_0^{\sA})^2\big] \leq 2^{-m/2}$ with probability $1 - 2^{1-m/2}$. As $\|\varrho\|_\infty \leq \|\varrho\|_2$ for any state $\varrho$, this yields $H_{\rm min}(\sA)_{\varphi_0} \geq \f{1}{2}H_2(\sA)_{\varphi_0} \geq \f{m}{4}$. The claim follows from $\widetilde{H}_\alpha(\sA)_{\varphi_0} \geq H_{\rm min}(\sA)_{\varphi_0}$ for $\alpha \in (0, \infty]$, which holds by the monotonicity of the R\'{e}nyi entropy in $\alpha$~\cite{muller2013quantumrenyinew}.
For (iii), this is Ref.~{\cite[Lemma~9]{Leone2025Entanglementtheorywithlimited}}.

\end{proof}

\begin{lemma}[Tilted pairs of ensembles; adapted from Ref.~{\cite[Lemma~12]{Leone2025Entanglementtheorywithlimited}}]
\label{lem:tilted pair}
For $b \in \{0,1\}$ and $i \in \{1, 2\}$, let $\scE_b^{(i)}$ be an ensemble of pure states on a register $\sR_i$, and let $\eta \in [0,1]$. Set
\begin{align}
\label{eq:tilted pair}
    \ket{\Psi_b}^{\sF\sR_1\sR_2}
    \coloneqq \sqrt{1-\eta}\ket{0}^{\sF}\ket{\psi_b^{(1)}}^{\sR_1}\ket{0}^{\sR_2}
    + \sqrt{\eta}\ket{1}^{\sF}\ket{0}^{\sR_1}\ket{\psi_b^{(2)}}^{\sR_2},
\end{align}
where $\ket{\psi_b^{(1)}}$ and $\ket{\psi_b^{(2)}}$ are drawn independently from $\scE_b^{(1)}$ and $\scE_b^{(2)}$, and suppose that at least one of the two pairs $\big(\scE_0^{(1)}, \scE_1^{(1)}\big)$ and $\big(\scE_0^{(2)}, \scE_1^{(2)}\big)$ of ensembles is invariant under multiplication by a global phase.
If $\varepsilon_1$ and $\varepsilon_2$ are non-decreasing functions such that 
\begin{align}
\label{inteq:20}
    \big\|\bE_{\scE_0^{(i)}}\big[\psi_0^{(i)\otimes k}\big] - \bE_{\scE_1^{(i)}}\big[\psi_1^{(i)\otimes k}\big]\big\|_1 \leq \varepsilon_i(k),
\end{align}
for every $k$ and every $i \in \{1,2\}$, then
\begin{align}
    \big\|\bE\big[\Psi_0^{\otimes k}\big] - \bE\big[\Psi_1^{\otimes k}\big]\big\|_1 \leq \varepsilon_1(k) + \varepsilon_2(k).
\end{align}
\end{lemma}

\begin{proof}[Proof of \Cref{lem:tilted pair}]

Let $\ket{\chi_b^{(1)}} \coloneqq \ket{\psi_b^{(1)}}^{\sR_1}\ket{0}^{\sR_2}$ and $\ket{\chi_b^{(2)}} \coloneqq \ket{0}^{\sR_1}\ket{\psi_b^{(2)}}^{\sR_2}$ on $\sR \coloneqq \sR_1\sR_2$. Attaching the fixed vector $\ket{0}$ does not change the trace distance, and hence the bound in Eq.~\eqref{inteq:20} holds for the $\ket{\chi_b^{(i)}}$ as well. 
By assumption, one of the two pairs is invariant under multiplication by a global phase; we take it to be $\big(\scE_0^{(1)}, \scE_1^{(1)}\big)$, as the other case is identical except that the phase appears in the second component. Then $\bE\big[\Psi_b^{\otimes k}\big] = \bE_\theta\bE\big[\Psi_{b,\theta}^{\otimes k}\big]$, where $\ket{\Psi_{b,\theta}} = \sqrt{1-\eta}\,e^{i\theta}\ket{0}^{\sF}\ket{\chi_b^{(1)}} + \sqrt{\eta}\ket{1}^{\sF}\ket{\chi_b^{(2)}}$ and $\theta$ is uniform on $[0, 2\pi)$.
Expanding the tensor power over subsets $T \subseteq \{1, \ldots, k\}$, where $T$ records which copies take the second component, we have 
\begin{align}
    \ket{\Psi_{b,\theta}}^{\otimes k}
    = \sum_T \sqrt{(1-\eta)^{k-|T|}\eta^{|T|}} e^{i\theta(k-|T|)}\ket{\Xi_T^b},
    \ \ \ \ \ 
    \ket{\Xi_T^{b}} \coloneqq \bigotimes_{l \in T}\ket{1}^{\sF}\ket{\chi_b^{(2)}}\bigotimes_{l \notin T}\ket{0}^{\sF}\ket{\chi_b^{(1)}},
\end{align}
where the vectors $\ket{\Xi_T^b}$ are orthonormal for distinct $T$ by the flags.
The average over $\theta$ annihilates every cross term with $|T| \neq |T'|$, leaving
\begin{align}
    \bE_\theta\big[\Psi_{b,\theta}^{\otimes k}\big]
    = \sum_{t=0}^{k}\binom{k}{t}(1-\eta)^{k-t}\eta^{t}\ketbra{\Xi_t^{b}}{\Xi_t^{b}},
    \ \ \ \ \ 
    \ket{\Xi_t^{b}} \coloneqq \binom{k}{t}^{-1/2}\sum_{T \,:\, |T| = t}\ket{\Xi_T^{b}}.
\end{align}

Let $U_\rP \coloneqq \sum_{f \in \{0,1\}^{k}}\ketbra{f}{f}^{\sF^{\otimes k}}\otimes T_{\pi_f}^{\sR^{\otimes k}}$, where $\pi_f$ is a permutation of $\{1, \ldots, k\}$ that moves the positions with $f_l = 1$ to the end and $T_{\pi_f}$ is its representation on $\sR^{\otimes k}$. This unitary is determined by the flags alone, and is thus the same for $b = 0$ and $b = 1$. Let $f_T \in \{0,1\}^k$ be the string with $(f_T)_l = 1$ exactly for $l \in T$, which is the content of the flag registers of $\ket{\Xi_T^b}$. At $|T| = t$, the unitary maps $\ket{\Xi_T^{b}} \mapsto \ket{f_T}^{\sF^{\otimes k}}\ket{\chi_b^{(1)}}^{\otimes(k-t)}\ket{\chi_b^{(2)}}^{\otimes t}$, whose factor on $\sR^{\otimes k}$ depends on $T$ only through $t$, and hence
\begin{align}
    U_\rP\ket{\Xi_t^{b}} = \ket{u_t}^{\sF^{\otimes k}}\ket{\chi_b^{(1)}}^{\otimes(k-t)}\ket{\chi_b^{(2)}}^{\otimes t},
    \ \ \ \ \ 
    \ket{u_t} \coloneqq \binom{k}{t}^{-1/2}\sum_{T \,:\, |T| = t}\ket{f_T},
\end{align}
with $\ket{u_t}$ independent of $b$ and of the drawn vectors.
As the two vectors are drawn independently, averaging gives
\begin{align}
    \bE\big[U_\rP\,\Xi_t^{b}\,U_\rP^\dag\big]
    = \ketbra{u_t}{u_t}^{\sF^{\otimes k}}\otimes\bE\big[\chi_b^{(1)\otimes(k-t)}\big]\otimes\bE\big[\chi_b^{(2)\otimes t}\big].
\end{align}

Combining this with the unitary invariance of the trace norm and the triangle inequality over $t$, we obtain
\begin{align}
    \big\|\bE\big[\Psi_0^{\otimes k}\big] - \bE\big[\Psi_1^{\otimes k}\big]\big\|_1
    \leq \sum_{t=0}^{k}\binom{k}{t}(1-\eta)^{k-t}\eta^{t}
    \Big\|\bE\big[\chi_0^{(1)\otimes(k-t)}\big]\otimes\bE\big[\chi_0^{(2)\otimes t}\big] - \bE\big[\chi_1^{(1)\otimes(k-t)}\big]\otimes\bE\big[\chi_1^{(2)\otimes t}\big]\Big\|_1.
\end{align}
Since $\|\rho\otimes\sigma - \rho'\otimes\sigma'\|_1 \leq \|\rho - \rho'\|_1 + \|\sigma - \sigma'\|_1$ for any states, each summand is at most $\varepsilon_1(k-t) + \varepsilon_2(t) \leq \varepsilon_1(k) + \varepsilon_2(k)$ by monotonicity, and the weights sum to $1$.

\end{proof}

\begin{lemma}[Scrambling of a labeled mixture]
\label{lem:maxmix}
Let $\{\sigma_j^{\sI_\sA\sI_\sB}\}_{j=1}^{M}$ be states on registers of dimensions $d_{\sI_\sA}$ and $d_{\sI_\sB}$, let $\sK_\sA$ and $\sK_\sB$ be registers of $\lceil\log{M}\rceil$ qubits, and set $\sA = \sI_\sA\sK_\sA$ and $\sB = \sI_\sB\sK_\sB$.
For unitaries $U_{1,j}$ on $\sI_\sA$ and $U_{2,j}$ on $\sI_\sB$, $j = 1, \ldots, M$, write $U_i = (U_{i,j})_{j=1}^{M}$ and define
\begin{align}
\label{eq:scrambled mixture}
    \varrho_M^{\sI_\sA\sI_\sB\sK_\sA\sK_\sB}
    \coloneqq \f{1}{M}\sum_{j=1}^{M}\big(U_{1,j}^{\sI_\sA}\otimes U_{2,j}^{\sI_\sB}\big)\sigma_j^{\sI_\sA\sI_\sB}\big(U_{1,j}^{\sI_\sA}\otimes U_{2,j}^{\sI_\sB}\big)^\dag\otimes\ketbra{jj}{jj}^{\sK_\sA\sK_\sB}.
\end{align}
Then the following statements hold.
\begin{enumerate}[label=\textnormal{(\roman*)}]
    \item For independent Haar random choices of all the unitaries and every $k \geq 1$,
    \begin{align}
        \big\|\bE_{U_1,U_2}\big[\varrho_M^{\otimes k}\big] - \tau_M^{\otimes k}\big\|_1
        \leq \f{k(k-1)}{M},
        \ \ \ \ \ 
        \tau_M^{\sI_\sA\sI_\sB\sK_\sA\sK_\sB} \coloneqq \f{\bI^{\sI_\sA\sI_\sB}}{d_{\sI_\sA}d_{\sI_\sB}}\otimes\f{1}{M}\sum_{j=1}^{M}\ketbra{jj}{jj}^{\sK_\sA\sK_\sB}.
    \end{align}
    \item Fix $U_2$, let the unitaries in $U_1$ be independent and Haar random, and write
    \begin{align}
    \label{eq:scrambled mixture mean}
        \bar{\varrho}_M^{\sI_\sA\sI_\sB\sK_\sA\sK_\sB}
        \coloneqq \bE_{U_1}\big[\varrho_M^{\sI_\sA\sI_\sB\sK_\sA\sK_\sB}\big]
        = \f{1}{M}\sum_{j=1}^{M}\f{\bI^{\sI_\sA}}{d_{\sI_\sA}}\otimes U_{2,j}^{\sI_\sB}\sigma_j^{\sI_\sB}(U_{2,j}^{\sI_\sB})^\dag\otimes\ketbra{jj}{jj}^{\sK_\sA\sK_\sB}.
    \end{align}
    For integers $k \geq 1$ and $t \geq 0$ with $k+t \leq d_{\sI_\sA}/2$,
    \begin{align}
    \label{eq:scrambled mixture domination}
        \bE_{U_1}\big[\varrho_M^{\otimes(k+t)}\big]
        \leq \f{(k+t)!}{t!}\exp\bigg(\f{3(k+t)^2}{d_{\sI_\sA}}\bigg)\bar{\varrho}_M^{\otimes k}\otimes\bE_{U_1}\big[\varrho_M^{\otimes t}\big],
    \end{align}
    where the last tensor factor is omitted at $t = 0$.
\end{enumerate}
\end{lemma}

In (i), at $M = 2^{\omega(\log{n})}$ and $k = \poly(n)$ the right-hand side is $\negl(n)$, so that $\varrho_M$ is statistically indistinguishable from the separable state $\tau_M$ using $\poly(n)$ copies. In (ii), $\bar{\varrho}_M$ is separable, and the bound holds without a lower bound on $M$.
The statements concern averages over the unitaries and not the state at fixed unitaries. A single choice of the unitaries must be selected through a quantity that is linear in the input, for which the average can be taken inside, and the lemma applied. This will be done with the success probability of a fixed protocol in the proof of \Cref{thm:agnostic no go}.

\begin{proof}[Proof of \Cref{lem:maxmix}]

For (i), expanding $\varrho_M^{\otimes k}$ over the label values $\vec{j} = (j_1, \ldots, j_k)$, the terms with distinct labels contain independent unitaries on every copy. For each such term, the single-copy Haar average~\cite{Mele2024IntroductionHaarMeasure} gives
\begin{align}
    \bE_{U_{1,j},U_{2,j}}\big[\big(U_{1,j}^{\sI_\sA}\otimes U_{2,j}^{\sI_\sB}\big)\sigma_j^{\sI_\sA\sI_\sB}\big(U_{1,j}^{\sI_\sA}\otimes U_{2,j}^{\sI_\sB}\big)^\dag\big]
    = \f{\bI^{\sI_\sA\sI_\sB}}{d_{\sI_\sA}d_{\sI_\sB}},
\end{align}
so the averaged term agrees with the corresponding term of $\tau_M^{\otimes k}$.
The label tuples with a repeated label carry total weight at most $\binom{k}{2}/M$ in both states, by a union bound over pairs of positions, and their terms differ by at most twice this weight in trace norm. Hence the distance is at most $2\binom{k}{2}/M = k(k-1)/M$.

For (ii), consider, for a dimension $d$, the Haar average channel $\cT_a(X) \coloneqq \bE_U[U^{\otimes a}X(U^\dag)^{\otimes a}]$ on $a$ copies of $\bC^d$; we apply the bounds below at $d = d_{\sI_\sA}$. Its normalized Choi operator is
\begin{align}
    \Omega_a \coloneqq d^{-a}\bE_U\big[(\rvec(U)\rvec(U)^\dag)^{\otimes a}\big].
\end{align}
Let $P_a$ be the projector onto the symmetric subspace of $(\bC^d\otimes\bC^d)^{\otimes a}$, and put $\Omega_0 = P_0 = 1$. Applying Ref.~{\cite[Lemmas~2.13 and~2.16]{tang2025conjugatequerieshelp}} to the maximally mixed state on $\bC^d$ shows that $\Omega_a$ has support $P_a$ and, on the sector indexed by $\lambda \vdash a$, its nonzero eigenvalue is $d_\lambda/(d^a\dim\cW_\lambda(d))$. The hook and Weyl dimension formulas~{\cite[Eq.~(C38)]{UQC}} give
\begin{align}
    \f{d_\lambda}{d^a\dim\cW_\lambda(d)}
    = \f{a!}{d^a\prod_{(i,j)\in\lambda}(d+j-i)}.
\end{align}
Consequently, for $a \in [1, d]$,
\begin{align}
    \f{a!}{d^a(d+a-1)^a}P_a
    \leq \Omega_a
    \leq \f{a!}{d^a(d-a+1)^a}P_a.
\end{align}
Write $q = k+t$. Since $P_q \leq \bI^{\otimes k}\otimes P_t$, where each identity acts on $\bC^d\otimes\bC^d$, these bounds imply
\begin{align}
    \Omega_q
    \leq \f{q!}{t!}\f{(1+(t-1)/d)^t}{(1-(q-1)/d)^q}\bigg(\f{\bI}{d^2}\bigg)^{\otimes k}\otimes\Omega_t
    \leq \f{q!}{t!}\exp\bigg(\f{3q^2}{d}\bigg)\bigg(\f{\bI}{d^2}\bigg)^{\otimes k}\otimes\Omega_t,
\end{align}
with the numerator interpreted as $1$ at $t = 0$. The last inequality uses $t \leq q \leq d/2$, $\ln(1+x) \leq x$, and $-\ln(1-x) \leq 2x$ for $x \in [0, 1/2]$. Let $\cD(X) \coloneqq \tr[X]\bI/d$ be the completely depolarizing channel. The last right-hand side is the normalized Choi operator of $\f{q!}{t!}\exp(3q^2/d)(\cD^{\otimes k}\otimes\cT_t)$. Thus the difference of this map and $\cT_q$ is completely positive, and its positivity is preserved after tensoring with the identity channel on any further registers. The same conclusion holds at $k = 0$.

We now expand $\varrho_M^{\otimes(k+t)}$ over its labels, keeping $U_2$ fixed. In each term, let $k_j$ and $t_j$ be the numbers of occurrences of label $j$ in the first $k$ and last $t$ copies, respectively. Apply the preceding channel inequality to the $k_j+t_j$ copies on which $U_{1,j}$ acts, with all other registers left untouched. The product of the factors is bounded by
\begin{align}
    \bigg(\prod_{j=1}^{M}\f{(k_j+t_j)!}{t_j!}\bigg)\exp\bigg(\f{3\sum_{j=1}^{M}(k_j+t_j)^2}{d}\bigg)
    \leq \f{(k+t)!}{t!}\exp\bigg(\f{3(k+t)^2}{d}\bigg).
\end{align}
For the factorial bound, add the first $k$ occurrences one at a time to the last $t$: at step $a$, the number of occurrences of the label being added is at most $t+a$. The exponential bound follows from $\sum_j(k_j+t_j) = k+t$. After depolarizing Alice's registers in the first $k$ copies, these copies are independent of $U_1$. Summing over the labels therefore gives $\bar{\varrho}_M^{\otimes k}\otimes\bE_{U_1}[\varrho_M^{\otimes t}]$, which proves Eq.~\eqref{eq:scrambled mixture domination}.

\end{proof}

We use the concentration inequality known as L\'{e}vy's lemma~\cite{levy1951problemes}. If each $\scM_i$ is a compact metric space with a probability measure satisfying a logarithmic Sobolev inequality with a constant $C_i$, then any Lipschitz function on the product space $\scM_1\times\cdots\times\scM_N$ concentrates at a rate governed by $\max_i C_i$~\cite{ledoux2001concentration, Meckes2019RandomMatrixTheoryClassicalCompact}. Since the unit sphere and the unitary group with their unitarily invariant measures satisfy these conditions with $C_i$ of order $1/d_i$, where $d_i$ is the dimension of the underlying Hilbert space, we have the following bound.

\begin{lemma}[L\'{e}vy's lemma on a product space; see, e.g., Ref.~{\cite[Chap.~5]{Meckes2019RandomMatrixTheoryClassicalCompact}}]
\label{lem:levy}
For $i = 1, \ldots, N$, let $\scM_i$ be the unit sphere or the unitary group of a $d_i$-dimensional Hilbert space, each equipped with the unitarily invariant probability measure and the metric $\|x_i - y_i\|_2$ induced by the inner product, where $x_i, y_i \in \scM_i$. Let $\scM \coloneqq \scM_1\times\cdots\times\scM_N$ carry the product measure and the metric $\|x - y\|_2 = \big(\sum_{i}\|x_i - y_i\|_2^2\big)^{1/2}$. There is a universal constant $c > 0$ such that, for every $L$-Lipschitz function $f : \scM \to \bR$ and every $t > 0$,
\begin{align}
\label{eq:levy}
    \Pr_{x}\Big[\big|f(x) - \bE[f]\big| \geq t\Big]
    \leq 2\exp\bigg(-\f{ct^2 \min_i{d_i}}{L^2}\bigg),
\end{align}
where $x$ is drawn from the product measure.
\end{lemma}

The Lipschitz constant on the product is obtained from those in the single arguments. If $f$ is $L_i$-Lipschitz in $x_i$ with the remaining arguments held fixed, then replacing the arguments one at a time and applying the triangle inequality yields $|f(x) - f(y)| \leq \sum_{i}L_i\|x_i - y_i\|_2 \leq \big(\sum_{i}L_i^2\big)^{1/2}\|x - y\|_2$ by the Cauchy--Schwarz inequality. Thus $L = \big(\sum_{i}L_i^2\big)^{1/2}$ is admissible.

We finally recall the random purification channel of Ref.~\cite{tang2025conjugatequerieshelp}, which maps $k$ copies of a state to the average of $k$ copies of a Haar random purification. The following definition and lemma give the properties of this channel that we will use.

\begin{definition}[Random purification channel; Ref.~{\cite[Lemma~2.11]{tang2025conjugatequerieshelp}}]
\label{def:random purification channel}
Let $\ket{\varrho}^{\sS\sR}$ be a fixed purification of a state $\varrho^\sS$ whose rank is at most $d_\sR = \dim\cH^\sR$, and let $\ket{\varrho_V}^{\sS\sR} = (\bI^\sS \otimes V^\sR)\ket{\varrho}^{\sS\sR}$ for a unitary $V^\sR$ on $\sR$. 
For every integer $k \geq 0$, the random purification channel $\cR^{(k)}$ is a channel from $\sS^{\otimes k}$ to $(\sS\sR)^{\otimes k}$, independent of $\varrho$, satisfying
\begin{align}
\label{eq:random purification channel}
    \cR^{(k)}\big((\varrho^\sS)^{\otimes k}\big) = \bE_{V}\big[(\ketbra{\varrho_V}{\varrho_V}^{\sS\sR})^{\otimes k}\big],
\end{align}
where $\bE_V$ averages over Haar random unitaries $V^\sR$, and $\cR^{(0)}$ acts as the identity on scalars.
\end{definition}

As any two purifications on $\sS\sR$ are related by a unitary acting on $\sR$, the distribution of $\ket{\varrho_V}$ for a Haar random unitary $V^\sR$ is independent of the choice of purification $\ket{\varrho}$.
Importantly, the $k$ copies in the output of $\cR^{(k)}$ all share the same unitary $V$. 
We use only the existence of the channel $\cR^{(k)}$, while Ref.~\cite{tang2025conjugatequerieshelp} provides an implementation with $d_\sR = \cO(d_\sS)$ and accuracy $\epsilon$ in diamond norm distance using $\poly(k, \log{d_\sS}, \log{(1/\epsilon)})$ gates.

\begin{lemma}[Statistical indistinguishability of random purifications]
\label{lem:random purification}
For every $k$ and any ensembles $\scE$ and $\scE'$ of states on $\sS$ of rank at most $d_\sR$, it holds that
\begin{align}
\label{eq:random purification hiding}
    \Big\|\bE_{\varrho\sim\scE}\bE_{V}\big[\ketbra{\varrho_V}{\varrho_V}^{\otimes k}\big]
    - \bE_{\sigma\sim\scE'}\bE_{V}\big[\ketbra{\sigma_V}{\sigma_V}^{\otimes k}\big]\Big\|_1
    = \Big\|\bE_{\varrho\sim\scE}\big[\varrho^{\otimes k}\big] - \bE_{\sigma\sim\scE'}\big[\sigma^{\otimes k}\big]\Big\|_1.
\end{align}
\end{lemma}

That is, two ensembles of mixed states are statistically indistinguishable if and only if the ensembles of their random purifications are.

\begin{proof}[Proof of \Cref{lem:random purification}]
As $\cR^{(k)}$ is linear, it commutes with the average over the ensemble, so the left-hand side of Eq.~\eqref{eq:random purification hiding} is $\big\|\cR^{(k)}\big(\bE_{\varrho\sim\scE}[\varrho^{\otimes k}]\big) - \cR^{(k)}\big(\bE_{\sigma\sim\scE'}[\sigma^{\otimes k}]\big)\big\|_1$ by Eq.~\eqref{eq:random purification channel}, which is at most the right-hand side by the monotonicity of the trace norm under channels.
Conversely, since $\tr_{\sR^{\otimes k}}\big[\bE_{V}[\ketbra{\varrho_V}{\varrho_V}^{\otimes k}]\big] = \varrho^{\otimes k}$, the same monotonicity of the trace norm under partial trace implies that the right-hand side is bounded above by the left-hand side.

\end{proof}


\subsection{Construction of the state family}
\label{sec:construction of the state family state agnostic}

We construct the family of states for our converse bounds in the state-agnostic and the state-dependent models. In \Cref{sec:parameters state agnostic}, we fix the parameters at which the ensembles of \Cref{lem:haar pair} are invoked; in \Cref{sec:construction state agnostic}, we build the states from them.

\subsubsection{Parameters}
\label{sec:parameters state agnostic}

Fix constants $\alpha_\rc, \alpha_\rs$ with $1 < \alpha_\rc < \alpha_\rs \leq \infty$ and a constant $\delta > 0$, and set
\begin{align}
\label{eq:agnostic params}
    \kappa = \f{\alpha_\rs - \alpha_\rc}{\alpha_\rc(\alpha_\rs - 1)} \in (0, \, 1/\alpha_\rc],
    \ \ \ \ \ \ \ \ 
    \eta = \log^{-(1+\delta)}{n},
    \ \ \ \ \ \ \ \ 
    \mu = \omega(\log{n}) \ \ \text{with} \ \ \eta\mu = o(1).
\end{align}
For instance, $\mu = \log{n}\log\log{n}$ meets the two conditions on $\mu$.
The parameter $\mu$ is the subsystem size parameter at which \Cref{lem:haar pair} is invoked below. We apply the lemma to registers of sizes $n$ and $\hat{m}$, where the latter case is considered only when $\hat{m} \geq \mu$. The condition $2m \geq \mu$ required by the lemma is satisfied in both cases: for size $n$ by $\mu = o(n)$, and for size $\hat{m}$ by $\hat{m} \geq \mu$.
The threshold $\mu = \omega(\log{n})$ serves two roles at once. It is the condition under which (iii) of \Cref{lem:haar pair} gives statistical indistinguishability at every $k \leq \poly(n)$, and it is what makes the selection argument for the state-dependent bound survive the union bound over the $2^{\poly(n)}$ efficient protocols, since the Haar component of the branch at $\hat{m} \geq \mu$ has dimension $2^{\Theta(\hat{m})}$.

\subsubsection{Construction}
\label{sec:construction state agnostic}

The family of states is built from three branches, mixed under a locally readable flag, replicated across a label, and then dressed by local unitaries.

The first branch is the spike state $\ket{\xi_\mathrm{sp}}^{\sG_\sA\sG_\sB\sG_\sE}$ of rank $r_\rs$ in \Cref{lem:spikestate}, where $r_\rs$ obeys either $r_\rs = 1$, in which case this branch carries zero weight and is absent, or $\omega(1) \leq r_\rs \leq \poly(n)$.
The second is $\ket{\phi_b}^{\sH_\sA\sH_\sB} \sim \scE_b(n, \mu)$ of \Cref{lem:haar pair} on $2n$ qubits, for which (i) and (ii) there lead to $H_1(\sH_\sA)_{\phi_1} \leq \mu$ and $\widetilde{H}_\alpha(\sH_\sA)_{\phi_0} \geq \f{n}{4}$.
The third is the two-scale state $\ket{\zeta_{b\hat{b}}}^{\sL_\sA\sL_\sB}$ of \Cref{lem:two scale branch}, parametrized by an integer $\hat{m}$ with $0 \leq \hat{m} \leq n$ and taken at the $\eta$ of Eq.~\eqref{eq:agnostic params}, where $\sL_\sA = \sD_\sA\hat{\sM}_\sA\sN_\sA$ and $\sL_\sB = \sD_\sB\hat{\sM}_\sB\sN_\sB$.
Its heavy component lives on $\hat{\sM}_\sA, \hat{\sM}_\sB$ of $\hat{m}$ qubits each, and is $\ket{\hat{\phi}_{\hat{b}}} \sim \scE_{\hat{b}}(\hat{m}, \mu)$ at $\hat{m} \geq \mu$ and $\ket{\hat{\phi}_0} = \ket{\hat{\phi}_1} = \ket{\Phi_2}^{\otimes\hat{m}}$ at $\hat{m} < \mu$, with empty registers at $\hat{m} = 0$, so that the branch is independent of $\hat{b}$ in the latter case. Its light component lives on $\sN_\sA, \sN_\sB$ of $n$ qubits each, and is the state $\ket{\phi_b}$ of the second branch, the entanglement amplifier.

We set their weights as
\begin{align}
\label{eq:agnostic weights}
    (\eta_0, \eta_1, \eta_2) = \begin{cases}
        (1 - \eta, \eta, 0) & \text{if} \ r_\rs \geq 2, \\
        (0, 0, 1) & \text{if} \ r_\rs = 1.
    \end{cases}
\end{align}
The two regimes are exclusive: in the first we take $\omega(1) \leq r_\rs \leq \poly(n)$ and $\hat{m} = 0$; in the second we set $r_\rs = 1$ and take any $0 \leq \hat{m} \leq n$. The first two branches then carry no weight. 
The second is not lost there, as it reappears as the light component of the third through Eq.~\eqref{eq:two scale branch}.
For $b, \hat{b} \in \{0,1\}$, we let $\ket{\Psi_{b\hat{b}}}^{\sI_\sA\sI_\sB\sI_\sE}$ be the flagged mixture of Eq.~\eqref{eq:flag mixture} built from these three branches and weights, where $\sI_\sA = \sF_\sA\sG_\sA\sH_\sA\sL_\sA$ and $\sI_\sB = \sF_\sB\sG_\sB\sH_\sB\sL_\sB$, while $\sI_\sE = \sF_\sE\sG_\sE$; that is,
\begin{align}
    \ket{\Psi_{b\hat{b}}}^{\sI_\sA\sI_\sB\sI_\sE}
    &= \sqrt{\eta_0} \ket{0}^{\sF_\sA\sF_\sB\sF_\sE}\ket{\xi_{\mathrm{sp}}}^{\sG_\sA\sG_\sB\sG_\sE}\ket{0}^{\sH_\sA\sH_\sB\sL_\sA\sL_\sB}
    + \sqrt{\eta_1}\ket{1}^{\sF_\sA\sF_\sB\sF_\sE}\ket{0}^{\sG_\sA\sG_\sB\sG_\sE}\ket{\phi_b}^{\sH_\sA\sH_\sB}\ket{0}^{\sL_\sA\sL_\sB}  \notag\\
    \label{eq:psi state agnostic}
    &\hspace{6pc} + \sqrt{\eta_2} \ket{2}^{\sF_\sA\sF_\sB\sF_\sE}\ket{0}^{\sG_\sA\sG_\sB\sG_\sE\sH_\sA\sH_\sB}\ket{\zeta_{b\hat{b}}}^{\sL_\sA\sL_\sB}.
\end{align}
Here, the state in the third branch is
\begin{align}
\label{eq:agnostic two scale}
    \ket{\zeta_{b\hat{b}}}^{\sL_\sA\sL_\sB}
    = \sqrt{1-\eta}\ket{0}^{\sD_\sA\sD_\sB}\ket{\hat{\phi}_{\hat{b}}}^{\hat{\sM}_\sA\hat{\sM}_\sB}\ket{0}^{\sN_\sA\sN_\sB}
    + \sqrt{\eta}\ket{1}^{\sD_\sA\sD_\sB}\ket{0}^{\hat{\sM}_\sA\hat{\sM}_\sB}\ket{\phi_b}^{\sN_\sA\sN_\sB}.
\end{align}
Since $\log{r_\rs} = \cO(\log{n})$ and $\hat{m} \leq n$, each of $\sI_\sA$ and $\sI_\sB$ carries $\Theta(n)$ qubits.

We then replicate the state by attaching a label and conjugate it by local unitaries, so that the rank is determined by the number of labels while the conditional R\'{e}nyi entropy and the information-theoretic distillable entanglement remain unchanged.
Let $r$ satisfy $r_\rs \leq r \leq 2^{\cO(n)}$, and set $M \coloneqq \big\lceil r/\rank(\Psi_{b\hat{b}}^{\sI_\sA\sI_\sB})\big\rceil$. Note that the rank of $\Psi_{b\hat{b}}^{\sI_\sA\sI_\sB}$ is independent of $b$ and $\hat{b}$.
By construction, $\rank(\Psi_{b\hat{b}}^{\sI_\sA\sI_\sB})M \in [r, \, r + \rank(\Psi_{b\hat{b}}^{\sI_\sA\sI_\sB})] \subseteq [r, \, 2r+1]$.
We use the same realization of $\ket{\Psi_{b\hat{b}}}^{\sI_\sA\sI_\sB\sI_\sE}$ for every label $j$ and every value of $M$.
With label registers $\sK_\sA, \sK_\sB, \sK_\sE$ of $\lceil\log{M}\rceil$ qubits and local unitaries $U_{1,j}$ on $\sI_\sA$ and $U_{2,j}$ on $\sI_\sB$, write $U_i = (U_{i,j})_{j=1}^{M}$ and set $\sA \coloneqq \sI_\sA\sK_\sA$ and $\sB \coloneqq \sI_\sB\sK_\sB$. We define
\begin{align}
\label{eq:agnostic purification}
    \ket{\Upsilon_{b\hat{b}}}^{\sA\sB\sE}
    = \f{1}{\sqrt{M}}\sum_{j=1}^{M}\big(U_{1,j}^{\sI_\sA}\otimes U_{2,j}^{\sI_\sB}\otimes\bI^{\sI_\sE}\big)\ket{\Psi_{b\hat{b}}}^{\sI_\sA\sI_\sB\sI_\sE}\ket{j}^{\sK_\sA}\ket{j}^{\sK_\sB}\ket{j}^{\sK_\sE},
\end{align}
where $\sE = \sI_\sE\sK_\sE$ is padded with ancillary qubits for later convenience so that $d_\sE = 2^{\Theta(n)} \geq d_\sA d_\sB$, with $d_\sA$ and $d_\sB$ the dimensions of $\sA$ and $\sB$. Tracing out $\sE$ gives the state
\begin{align}
\label{eq:agnostic mixed}
    \Upsilon_{b\hat{b}}^{\sA\sB}
    = \f{1}{M}\sum_{j=1}^{M}\big(U_{1,j}^{\sI_\sA}\otimes U_{2,j}^{\sI_\sB}\big)\Psi_{b\hat{b}}^{\sI_\sA\sI_\sB}\big(U_{1,j}^{\sI_\sA}\otimes U_{2,j}^{\sI_\sB}\big)^\dag\otimes\ketbra{jj}{jj}^{\sK_\sA\sK_\sB},
\end{align}
whose summands have mutually orthogonal supports. 
We write $\scF_{b\hat{b}}$ for the ensemble of states $\Upsilon_{b\hat{b}}^{\sA\sB}$ induced by the draw of $\ket{\Psi_{b\hat{b}}}$, at fixed $U_1$ and $U_2$.
By $\log{M} = \cO(n)$ we have $\dim{\cH^\sA}, \dim{\cH^\sB} = 2^{\Theta(n)}$.

Undoing the local unitaries conditioned on the labels gives $\Psi_{b\hat{b}}^{\sI_\sA\sI_\sB}\otimes\f{1}{M}\sum_j\ketbra{jj}{jj}^{\sK_\sA\sK_\sB}$. Appending or discarding this separable label is an LOCC operation, so the information-theoretic distillable entanglement equals that of $\Psi_{b\hat{b}}^{\sI_\sA\sI_\sB}$. Tracing out $\sB$ from the purification after undoing the local unitaries leaves identical blocks on $\sI_\sA\sI_\sE$, with the label copied to $\sK_\sA\sK_\sE$. Hence \Cref{lem:conditional alpha entropystate} also gives, for every $\alpha \in (1, \infty]$,
\begin{align}
\label{eq:agnostic invariance}
    \begin{aligned}
    E_\rD(\sA:\sB)_{\Upsilon_{b\hat{b}}}
    = E_\rD(\sI_\sA:\sI_\sB)_{\Psi_{b\hat{b}}}, \qquad
    \widetilde{H}_\alpha(\sA|\sE)_{\Upsilon_{b\hat{b}}}
    = \widetilde{H}_\alpha(\sI_\sA|\sI_\sE)_{\Psi_{b\hat{b}}}.
    \end{aligned}
\end{align}
These equalities hold for every fixed realization of $\ket{\Psi_{b\hat{b}}}$ and every $M$, regardless of the local unitaries.

The following lemma states that the ensembles $\scF_{b\hat{b}}$ are statistically indistinguishable across $b$ and across $\hat{b}$ from $\poly(n)$ copies. This is used in the proofs of \Cref{thm:agnostic unbounded,thm:agnostic no go} to transfer the success of a protocol on the high-entanglement side of the family to the low-entanglement one, where the cap is established.

\begin{lemma}[Statistical indistinguishability of the ensembles]
\label{lem:agnostic hiding}
Fix the parameters of the construction and the families of unitaries $U_1, U_2$, and let $\scF_{b\hat{b}}$ be the ensemble defined above. Then, for every $k$ and every $M$,
\begin{align}
    \big\|\bE\big[\Upsilon_{0\hat{b}}^{\otimes k}\big] - \bE\big[\Upsilon_{1\hat{b}}^{\otimes k}\big]\big\|_1
    = \cO\big(k^{2}2^{-\mu}\big),
    \ \ \ \ \ \ 
    \big\|\bE\big[\Upsilon_{b0}^{\otimes k}\big] - \bE\big[\Upsilon_{b1}^{\otimes k}\big]\big\|_1
    = \cO\big(k^{2}2^{-\mu}\big),
\end{align}
where each expectation is taken over $\scF_{b\hat{b}}$ at the corresponding pair of indices.
\end{lemma}

\begin{proof}[Proof of \Cref{lem:agnostic hiding}]

We first bound the distance at a single realization.
From Eqs.~\eqref{eq:agnostic weights} and~\eqref{eq:psi state agnostic}, at $r_\rs \geq 2$ the state $\ket{\Psi_{b\hat{b}}}$ is independent of $\hat{b}$ and is the flagged mixture $\sqrt{1-\eta}\ket{0}\ket{\xi_{\mathrm{sp}}} + \sqrt{\eta}\ket{1}\ket{\phi_b}$ of the fixed spike with $\ket{\phi_b}$; at $r_\rs = 1$ it is $\ket{2}\ket{0}\ket{\zeta_{b\hat{b}}}$ with $\ket{\zeta_{b\hat{b}}}$ of Eq.~\eqref{eq:agnostic two scale}. 
In either case, it is a superposition of two components carried by orthogonal flags, each of them either fixed or drawn from an ensemble of \Cref{lem:haar pair}, with $\hat{b}$ entering only the first and $b$ only the second.
These ensembles are invariant under multiplication by a global phase, so \Cref{lem:tilted pair} applies.
With $b$ fixed, the second component is the same at $\hat{b} = 0$ and $\hat{b} = 1$, and we can take $\varepsilon_2 = 0$. The first component is fixed at $\hat{m} < \mu$ and is a pair of ensembles of \Cref{lem:haar pair} at $\hat{m} \geq \mu$, whose (iii) provides $\varepsilon_1(k) = \cO(k^{2}2^{-\mu})$. Hence
\begin{align}
\label{inteq:agnostic single}
    \big\|\bE\big[\Psi_{b0}^{\otimes k}\big] - \bE\big[\Psi_{b1}^{\otimes k}\big]\big\|_1
    = \cO\big(k^{2}2^{-\mu}\big),
\end{align}
and likewise between $b = 0$ and $b = 1$, with the roles of the two components exchanged.

We next lift this to the mixture. The map that traces out $\sI_\sE$, appends the separable label $\f{1}{M}\sum_j\ketbra{jj}{jj}^{\sK_\sA\sK_\sB}$, and applies the fixed local unitaries conditioned on the labels is a channel independent of $b$ and $\hat{b}$. It maps $\Psi_{b\hat{b}}$ to $\Upsilon_{b\hat{b}}^{\sA\sB}$ by Eq.~\eqref{eq:agnostic mixed}. Applying its $k$-fold tensor power to Eq.~\eqref{inteq:agnostic single}, and likewise to the bound across $b$, gives the claim by the monotonicity of the trace norm under channels. The bound is uniform in $M$ and is $\negl(n)$ at $k = \poly(n)$ by $\mu = \omega(\log{n})$.

\end{proof}


\subsection{Statistical converse result}
\label{sec:statistical converse result}

We provide the converse bounds on the state-agnostic and the computational distillable entanglement of \Cref{def:agnostic distillable,def:comp distillable} using the family constructed in the previous subsection.
In \Cref{sec:statistical converse state agnostic}, we first consider a single protocol of unrestricted size that must succeed on the whole state set, as in model (M1).
In \Cref{sec:statistical converse state dependent}, we then consider model (M2), in which the protocol may depend on the state in any way but its circuit size is polynomial.
In this case, concentration and a union bound select one state against all such protocols.
Both models give the same cap.
In \Cref{sec:simplified statistical converse bound}, we provide a simplified statement of these bounds in terms of the R\'{e}nyi coherent information.
Each party may additionally receive $l = \poly(n)$ copies of a purification, each held entirely in that party's laboratory, where $l$ is a nonnegative integer.

\subsubsection{State-agnostic model}
\label{sec:statistical converse state agnostic}

In the state-agnostic setting, a single protocol must succeed on the whole set of states, with no restriction on its circuit size.
When additional copies of a purification are supplied, the protocol must also be agnostic to the choice of purification.

\begin{theorem}[No-go in the state-agnostic model]
\label{thm:agnostic unbounded}
Let $\alpha_\rc, \alpha_\rs$ be constants with $1 < \alpha_\rc < \alpha_\rs \leq \infty$ and let $\delta > 0$ be a constant. Take $\kappa$ and $\mu$ as in Eq.~\eqref{eq:agnostic params}. Let $r_\rs$ obey either $\omega(1) \leq r_\rs \leq \poly(n)$, in which case $\hat{m} = 0$, or $r_\rs = 1$, in which case $0 \leq \hat{m} \leq n$, and let $r$ obey $r_\rs \leq r \leq 2^{\cO(n)}$.
Let $\scS$ be the set of the states of Eq.~\eqref{eq:agnostic mixed}, namely
\begin{align}
\label{eq:agnostic family}
    \scS \coloneqq \Big\{\Upsilon_{00}^{\sA\sB}
    : \ket{\Psi_{00}}, \ U_1, \ U_2\Big\},
\end{align}
where the vector $\ket{\Psi_{00}}$ and the unitaries $U_{1,j}$, $U_{2,j}$, $j = 1, \ldots, M$, run over all realizations and all unitaries on the respective registers.
For a realization $\ket{\rho_0}^{\sA\sB\sE} = \ket{\Upsilon_{00}}^{\sA\sB\sE}$ of Eq.~\eqref{eq:agnostic purification}, the rank satisfies
\begin{align}
\label{eq:agnostic rank}
    &\rank(\rho_0^{\sA\sB})
    = \begin{cases}
        r & \ \text{if} \ \ r_\rs = 1, \\
        \Theta(r) & \ \text{if} \ \ r_\rs \geq 2,
    \end{cases}
\end{align}
and, with probability $1 - \negl(n)$ over the sampling, the same realization obeys
\begin{align}
\label{eq:agnostic renyi}
    &\widetilde{H}_\alpha(\sA|\sE)_{\rho_0}
    = \begin{cases}
        \hat{m} + o(1) & \ \text{if} \ \ r_\rs = 1 \ \text{and} \ \hat{m} < \mu, \\
        \Theta(\hat{m}) & \ \text{if} \ \ r_\rs = 1 \ \text{and} \ \hat{m} \geq \mu, \\
         \big(\kappa - \f{\max\{\alpha, \alpha_\rc\} - \alpha_\rc}{\alpha_\rc(\max\{\alpha, \alpha_\rc\} - 1)}\big) \log{r_\rs} + o(1) & \ \text{if} \ \ r_\rs \geq 2,
    \end{cases}
    \ \ \text{for every fixed} \ \ \alpha \in (1, \infty] \setminus\{\alpha_\rc\}; \\
    \label{eq:agnostic ed}
    &E_\rD(\sA:\sB)_{\rho_0} = \Omega\big(n\log^{-(1+\delta)}{n}\big).
\end{align}
For large enough $n$, the state-agnostic distillable entanglement satisfies
\begin{align}
\label{eq:agnostic cap unbounded}
    &E_{\rD, \ag}^{(k, l, \varepsilon)}(\sA:\sB)_\scS
    \leq \begin{cases}
        \min\{\hat{m}, \, \log{(k+2l)}\} + o(1) & \ \text{if} \ \ r_\rs = 1 \ \text{and} \ r = \poly(n), \\
        \min\{\kappa\log{r_\rs}, \, \log{(k+2l)}\} + o(1) & \ \text{if} \ \ r_\rs \geq 2 \ \text{and} \ r = \poly(n), \\
        0 & \ \text{if} \ \ r = 2^{\omega(\log{n})},
    \end{cases}
\end{align}
for any $k = \poly(n)$ with $k = \omega(1)$ and any nonnegative integer $l = \poly(n)$, where $\varepsilon \in [0, 1/2)$ is constant.
\end{theorem}

\begin{proof}[Proof of \Cref{thm:agnostic unbounded}]

We work with the family defined in Eq.~\eqref{eq:agnostic purification} for given $r_\rs$, $\hat{m}$, and $r$.
For a unitary $V$ acting on $\sE$, we define $\ket{\Upsilon_{b\hat{b},V}} \coloneqq (\bI^{\sA\sB}\otimes V^\sE)\ket{\Upsilon_{b\hat{b}}}$.
Since $V$ acts only on $\sE$, it leaves $\Upsilon_{00}^{\sA\sB}$ unchanged, and the rank, the conditional R\'{e}nyi entropy, and the information-theoretic distillable entanglement are independent of $V$.

We first compute the rank. By (i) of \Cref{lem:flag mixture}, $\rank(\Psi_{00}^{\sI_\sA\sI_\sB})$ is $r_\rs + 1$ at $r_\rs \geq 2$, since $\xi_{\mathrm{sp}}^{\sG_\sA\sG_\sB}$ is flat of rank $r_\rs$ and the branch of $\ket{\phi_0}^{\sH_\sA\sH_\sB}$ is pure, and is $1$ at $r_\rs = 1$, where the third branch alone carries positive weight and $\zeta_{00}^{\sL_\sA\sL_\sB}$ is pure.
The labels being locally orthogonal, the summands of Eq.~\eqref{eq:agnostic mixed} have mutually orthogonal supports, and conjugation by $U_{1,j}\otimes U_{2,j}$ preserves the rank of each block, so $\rank(\Upsilon_{00}^{\sA\sB}) = \rank(\Psi_{00}^{\sI_\sA\sI_\sB})M = \Theta(r)$ by the choice $M = \big\lceil r/\rank(\Psi_{00}^{\sI_\sA\sI_\sB})\big\rceil$. This is Eq.~\eqref{eq:agnostic rank}, which is $r$ itself at $r_\rs = 1$, where $\rank(\Psi_{00}^{\sI_\sA\sI_\sB}) = 1$ and $M = r$.

We next compute the conditional R\'{e}nyi entropy. By Eq.~\eqref{eq:agnostic invariance}, it equals $\widetilde{H}_\alpha(\sI_\sA|\sI_\sE)_{\Psi_{00}}$.
Statement (ii) of \Cref{lem:haar pair} gives $H_\alpha(\sH_\sA)_{\phi_0} \geq n/4$ for all $\alpha \in (0,\infty]$ with probability $1 - 2^{-\Omega(n)}$, and, when $\hat{m} \geq \mu$, also $H_\alpha(\hat{\sM}_\sA)_{\hat{\phi}_0} \geq \hat{m}/4$ with probability $1 - 2^{-\Omega(\hat{m})}$. Since $\hat{m} \geq \mu = \omega(\log{n})$ in the latter case, these events hold together with probability $1 - \negl(n)$ and cover all orders at once. We work on this event below.

At $r_\rs = 1$ the third branch alone has positive weight, so $\sI_\sE$ is in a pure state and $\widetilde{H}_\alpha(\sI_\sA|\sI_\sE)_{\Psi_{00}} = H_\alpha(\sL_\sA)_{\zeta_{00}}$.
At $\hat{m} < \mu$, the heavy component is $\ket{\Phi_2}^{\otimes\hat{m}}$, whereas the light component has $H_\alpha(\sN_\sA)_{\phi_0} \geq n/4$. The bounds in (i) of \Cref{lem:two scale branch} therefore give $H_\alpha(\sL_\sA)_{\zeta_{00}} = \hat{m} + o(1)$, since $\eta = o(1)$ and $\hat{m} < \mu = o(n)$.
At $\hat{m} \geq \mu$, the same bounds, together with $H_\alpha(\hat{\sM}_\sA)_{\hat{\phi}_0} \in [\hat{m}/4,\hat{m}]$ and $\hat{m} \leq n$, give $H_\alpha(\sL_\sA)_{\zeta_{00}} \in [\hat{m}/4 - o(1),\hat{m} + o(1)] = \Theta(\hat{m})$. These are the first two cases of Eq.~\eqref{eq:agnostic renyi}.

At $r_\rs \geq 2$, \Cref{lem:conditional alpha entropystate} applies to the flag $\sF_\sE$ of $\Psi_{00}$ and yields
\begin{align}
\label{inteq:agnostic spike mean}
    \widetilde{H}_\alpha(\sA|\sE)_{\Upsilon_{00}}
    = -\f{\alpha}{\alpha-1}\log\bigg((1-\eta)2^{-\f{\alpha-1}{\alpha}\widetilde{H}_\alpha(\sG_\sA|\sG_\sE)_{\xi_{\mathrm{sp}}}}
    + \eta 2^{-\f{\alpha-1}{\alpha}H_\alpha(\sH_\sA)_{\phi_0}}\bigg).
\end{align}
As $\widetilde{H}_\alpha(\sG_\sA|\sG_\sE)_{\xi_{\mathrm{sp}}} = \cO(\log{r_\rs}) = \cO(\log{n})$ by Eq.~\eqref{eq:alpha entropy of spike state}, the first term is at least $(1-\eta)n^{-\cO(1)}$, while the second is $\eta 2^{-\Omega(n)}$. Hence $\widetilde{H}_\alpha(\sA|\sE)_{\Upsilon_{00}} = \widetilde{H}_\alpha(\sG_\sA|\sG_\sE)_{\xi_{\mathrm{sp}}} + o(1)$, which is the third case of Eq.~\eqref{eq:agnostic renyi} by Eq.~\eqref{eq:alpha entropy of spike state}. At $\alpha = \infty$, the ratios involving $\alpha$ are understood by continuity.

We next bound the information-theoretic distillable entanglement from below. Equation~\eqref{eq:agnostic invariance} reduces this to the state $\Psi_{00}$.
At $r_\rs \geq 2$, applying (iii) of \Cref{lem:flag mixture} to the pure amplifier branch yields $E_\rD(\sI_\sA:\sI_\sB)_{\Psi_{00}} \geq \eta H_1(\sH_\sA)_{\phi_0}$.
At $r_\rs = 1$, the state $\Psi_{00}^{\sI_\sA\sI_\sB}$ is pure, so $E_\rD(\sI_\sA:\sI_\sB)_{\Psi_{00}} = H_1(\sL_\sA)_{\zeta_{00}} \geq \eta H_1(\sN_\sA)_{\phi_0}$ by (ii) of \Cref{lem:two scale branch}.
On the same event used above, $H_1(\sH_\sA)_{\phi_0} \geq n/4$, and likewise on $\sN_\sA$ in the second case. Thus $E_\rD(\sA:\sB)_{\Upsilon_{00}} \geq \eta n/4 = \Omega\big(n\log^{-(1+\delta)}{n}\big)$, proving Eq.~\eqref{eq:agnostic ed}.

We now bound the state-agnostic distillable entanglement.
Let $\Lambda$ be a single protocol of unrestricted size that, for every state in $\scS$ and every purification of it on $\sA\sB\sE$, distills at rate $R$ and error $\varepsilon$ from $k$ shared copies and $l = \poly(n)$ local copies of that purification per party.
Let $\rF_\Lambda(\cdot) \coloneqq \tr\big[\big(\Phi_2^{\otimes kR}\otimes\bI\big)\Lambda(\cdot)\big]$ on the first $kR$ declared pairs, and $\rF_\Lambda \coloneqq 0$ if fewer are declared, as in Eq.~\eqref{eq:fidelity functional}.

We first bound the rate in terms of the number of copies. Fix a realization of $\ket{\Psi_{00}}$. Let $\cR^{(2l)}$ be the random purification channel of \Cref{def:random purification channel}, so that
\begin{align}
\label{inteq:agnostic sample block}
    \big(\rid^{\otimes k} \otimes \cR^{(2l)}\big)\big(\Upsilon_{00}^{\sA\sB \otimes(k+2l)}\big)
    = \Upsilon_{00}^{\sA\sB \otimes k} \otimes \bE_{V^\sE}\big[\ketbra{\Upsilon_{00,V}}{\Upsilon_{00,V}}^{\sA\sB\sE \otimes 2l}\big].
\end{align}
The channel acts only on the last $2l$ copies. It need not be an LOCC, as it is used to express the averaged input and not as part of the protocol.
For fixed $U_2$, apply (ii) of \Cref{lem:maxmix} with $t = 2l$ and every $\sigma_j = \Psi_{00}^{\sI_\sA\sI_\sB}$, and then apply $\rid^{\otimes k}\otimes\cR^{(2l)}$ to the operator inequality. The first $k$ copies on its right-hand side are in the separable state $\bE_{U_1}[\Upsilon_{00}]$. The remaining factor becomes an average of $\ketbra{\Upsilon_{00,V}}{\Upsilon_{00,V}}^{\otimes 2l}$, with $l$ full copies in each laboratory, and is also separable across the two laboratories. Thus the entire comparison state is separable. By the linearity and positivity of $\rF_\Lambda$, and \Cref{lem:sep overlap},
\begin{align}
\label{inteq:agnostic logarithmic fidelity}
    \bE_{U_1,U_2,V^\sE}\big[\rF_\Lambda\big(\Upsilon_{00}^{\otimes k}\otimes\ketbra{\Upsilon_{00,V}}{\Upsilon_{00,V}}^{\otimes 2l}\big)\big]
    \leq \f{(k+2l)!}{(2l)!}\exp\bigg(\f{3(k+2l)^2}{d_{\sI_\sA}}\bigg)2^{-kR}
    = \f{(k+2l)!}{(2l)!}\big(1+\negl(n)\big)2^{-kR},
\end{align}
where all unitaries are independent and Haar random. The last equality uses $d_{\sI_\sA} = 2^{\Theta(n)}$ and $k,l = \poly(n)$. This estimate holds for every realization of $\ket{\Psi_{00}}$.
Since $\Lambda$ succeeds at every choice of the state and its purification, the left-hand side is at least $1-\varepsilon$. Taking logarithms gives
\begin{align}
\label{inteq:agnostic logarithmic rate}
    R \leq \f{1}{k}\log{\f{(k+2l)!}{(2l)!}} + \cO(1/k) + \negl(n)
    \leq \log{(k+2l)} + o(1),
\end{align}
where we used $(k+2l)!/(2l)! = \prod_{i=1}^{k}(2l+i) \leq (k+2l)^k$ and $k = \omega(1)$.

At $r = \poly(n)$, fix arbitrary $(U_1,U_2)$.
By the assumption on $\Lambda$, we have $\rF_{\Lambda}\big(\Upsilon_{00}^{\otimes k}\otimes\ketbra{\Upsilon_{00,V}}{\Upsilon_{00,V}}^{\otimes 2l}\big) \geq 1 - \varepsilon$ at every realization $\Upsilon_{00} \in \scF_{00}$ and every $V^\sE$, and hence
\begin{align}
    \bE_{\scF_{00}, V^\sE}\big[\rF_{\Lambda}\big(\Upsilon_{00}^{\otimes k}\otimes\ketbra{\Upsilon_{00,V}}{\Upsilon_{00,V}}^{\otimes 2l}\big)\big] \geq 1 - \varepsilon.
\end{align}
Since the input is obtained from $\ketbra{\Upsilon_{00,V}}{\Upsilon_{00,V}}^{\otimes(k+2l)}$ by tracing out $\sE$ on the first $k$ copies, we have
\begin{align}
\label{inteq:agnostic purification comparison}
    &\Big\|\tr_{\sE^{\otimes k}}\Big[\bE_{\scF_{00}, V^\sE}\big[\ketbra{\Upsilon_{00,V}}{\Upsilon_{00,V}}^{\otimes(k+2l)}\big]\Big] - \tr_{\sE^{\otimes k}}\Big[\bE_{\scF_{11}, V^\sE}\big[\ketbra{\Upsilon_{11,V}}{\Upsilon_{11,V}}^{\otimes(k+2l)}\big]\Big]\Big\|_1 \notag\\
    &\leq \Big\|\bE_{\scF_{00}, V^\sE}\big[\ketbra{\Upsilon_{00,V}}{\Upsilon_{00,V}}^{\otimes(k+2l)}\big] - \bE_{\scF_{11}, V^\sE}\big[\ketbra{\Upsilon_{11,V}}{\Upsilon_{11,V}}^{\otimes(k+2l)}\big]\Big\|_1 \\
    &= \Big\|\bE_{\scF_{00}}\big[\Upsilon_{00}^{\otimes(k+2l)}\big] - \bE_{\scF_{11}}\big[\Upsilon_{11}^{\otimes(k+2l)}\big]\Big\|_1 \\
    &= \cO\big((k+2l)^{2}2^{-\mu}\big) \\
    &= \negl(n),
\end{align}
where the first line uses the monotonicity of the trace norm under partial trace, the second \Cref{lem:random purification}, and the third \Cref{lem:agnostic hiding} along the hybrid $\scF_{00} \to \scF_{01} \to \scF_{11}$, whose first hop is trivial at $\hat{m} < \mu$.
Hence, we obtain 
\begin{align}
    1 - \varepsilon
    \leq \bE_{\scF_{00}, V^\sE}\big[\rF_{\Lambda}\big(\Upsilon_{00}^{\otimes k}\otimes\ketbra{\Upsilon_{00,V}}{\Upsilon_{00,V}}^{\otimes 2l}\big)\big]
    \leq \max_{\Upsilon_{11} \in \scF_{11},\, V^\sE}\rF_{\Lambda}\big(\Upsilon_{11}^{\otimes k}\otimes\ketbra{\Upsilon_{11,V}}{\Upsilon_{11,V}}^{\otimes 2l}\big) + \negl(n),
\end{align}
which implies that $\Lambda$ distills at the same rate $R$ with fidelity $\geq 1 - \varepsilon - \negl(n)$ from some realization $\Upsilon_{11}^{\sA\sB} \in \scF_{11}$, supplied with the purification $\ket{\Upsilon_{11,V}}$ at some $V^\sE$.

We bound $R$ on that realization at error $\varepsilon + \negl(n)$, taking $\vartheta \coloneqq (1-\varepsilon)/2$, which obeys $\varepsilon + \negl(n) + \vartheta < 1$ for large $n$; the term $\f{1}{k}\log\f{2}{1-\varepsilon}$ below changes by $o(1)$ under this error.
The $2l$ local samples are pure states held entirely by one party, hence separable across the cut, and the last statement of \Cref{lem:sandwich converse} applies.
To apply \Cref{lem:truncation} to $\Upsilon_{11}^{\sA\sB}$, note that undoing its local unitaries gives $\Psi_{11}^{\sI_\sA\sI_\sB}$ tensored with the separable label, under which its conditions are preserved; it thus suffices to check the conditions of \Cref{lem:truncation} for $\Psi_{11}^{\sI_\sA\sI_\sB}$.

At $r_\rs \geq 2$ the state $\Psi_{11}^{\sI_\sA\sI_\sB}$ is the flagged mixture of $\xi_{\mathrm{sp}}^{\sG_\sA\sG_\sB}$ with weight $1-\eta$ and the pure state $\phi_1$ with weight $\eta$.
We apply (1) of \Cref{lem:truncation} at $J = 1$, taking $\tau^{(0)}$ and $\tau^{(1)}$ to be the two flag components of $\Psi_{11}^{\sI_\sA\sI_\sB}$.
Equation~\eqref{eq:max entropy of spike state} yields $E_0 = \log{g}$, and the branch weight of Eq.~\eqref{eq:agnostic weights} yields $w_1 = \eta$. By (i) of \Cref{lem:haar pair}, which holds with certainty, the Schmidt rank of $\ket{\phi_1}$ is at most $2^\mu$, and hence $E_1 = \mu$.
Hence $K = k\big(\log{g} + \f{\eta\mu}{\vartheta}\big)$, and by Eqs.~\eqref{eq:rateconverse sandwich} and~\eqref{eq:spike exponent},
\begin{align}
\label{inteq:agnostic comparison rate spike}
    R
    \leq \log{g} + \f{\eta\mu}{\vartheta} + \f{1}{k}\log{\f{2}{1-\varepsilon}} 
    = \kappa\log{r_\rs} + o(1),
\end{align}
where we used $\eta\mu = o(1)$ and $k = \omega(1)$.

At $r_\rs = 1$, the state $\Psi_{11}^{\sI_\sA\sI_\sB}$ is the pure state $\zeta_{11}$ of Eq.~\eqref{eq:agnostic two scale}. We apply (2) of \Cref{lem:truncation} with every $\ket{\psi_j} = \ket{\zeta_{11}}$ at the $E_0$ and $E_\star$ of (iii) of \Cref{lem:two scale branch}.
Hence $K = k\big(E_0 + \f{9\eta}{\vartheta^2}\big(E_\star + \log{(k+1)}\big)\big) + \log{(k+1)}$, and by Eq.~\eqref{eq:rateconverse sandwich},
\begin{align}
\label{inteq:agnostic comparison rate pure}
    R \leq E_0 + \f{9\eta}{\vartheta^2}\big(E_\star + \log{(k+1)}\big) + \f{\log{(k+1)}}{k} + \f{1}{k}\log{\f{2}{1-\varepsilon}}.
\end{align}
Since the heavy component is $\ket{\Phi_2}^{\otimes\hat{m}}$ at $\hat{m} < \mu$, $E_0 = \hat{m}$, while $E_\star = \mu$ by (i) of \Cref{lem:haar pair}. At $\hat{m} \geq \mu$, the same statement (i) gives $E_0 = E_\star = \mu$. The correction is $o(1)$ by $\eta\mu = o(1)$, $\eta\log{k} = \cO(\log^{-\delta}{n}) = o(1)$ for $k = \poly(n)$, and $k = \omega(1)$. Therefore, $R \leq \hat{m} + o(1)$ at $\hat{m} < \mu$ and $R \leq \mu + o(1)$ at $\hat{m} \geq \mu$. Combining these bounds and Eq.~\eqref{inteq:agnostic comparison rate spike} with Eq.~\eqref{inteq:agnostic logarithmic rate} gives the first two cases of Eq.~\eqref{eq:agnostic cap unbounded}, since $\log{(k+2l)} < \mu$ for large enough $n$.

We turn to the case of $r = 2^{\omega(\log{n})}$, where $M \geq r/\poly(n) = 2^{\omega(\log{n})}$ by $\rank(\Psi_{00}^{\sI_\sA\sI_\sB}) \leq r_\rs + 1 = \poly(n)$. Statement (i) of \Cref{lem:maxmix} applies with every $\sigma_j = \Psi_{00}^{\sI_\sA\sI_\sB}$ and gives trace-norm error at most $(k+2l)(k+2l-1)/M = \negl(n)$.
Fix any realization of $\ket{\Psi_{00}}$. The assumption on $\Lambda$ gives output fidelity at least $1 - \varepsilon$ at every $(U_1,U_2,V^\sE)$, and hence also after averaging over independent Haar random choices of all their unitaries.
As $\rF_\Lambda$ is linear in its input and $\cR^{(2l)}$ is independent of $(U_1,  U_2)$, we obtain 
\begin{align}
\label{inteq:agnostic average fidelity}
    \bE_{U_1,  U_2, V^\sE}\big[\rF_\Lambda\big(\Upsilon_{00}^{\otimes k}\otimes\ketbra{\Upsilon_{00,V}}{\Upsilon_{00,V}}^{\otimes 2l}\big)\big]
    = \rF_{\Lambda}\Big(\big(\rid^{\otimes k}\otimes\cR^{(2l)}\big)\Big(\bE_{U_1,  U_2}\big[\Upsilon_{00}^{\otimes (k+2l)}\big]\Big)\Big)
    \leq 2^{-kR} + \negl(n),
\end{align}
where the inequality is obtained as follows.
Statement (i) of \Cref{lem:maxmix} at $k+2l$ copies replaces the average by $\tau_M^{\otimes(k+2l)}$ up to $\negl(n)$. Its image under $\rid^{\otimes k}\otimes\cR^{(2l)}$ is separable across the two laboratories. The first $k$ copies remain in the separable state $\tau_M$, while the sample block is $\bE_{V^\sE}\big[\ketbra{\tau_{M,V}}{\tau_{M,V}}^{\otimes 2l}\big]$ by Eq.~\eqref{eq:random purification channel} at $d_\sE \geq d_\sA d_\sB$, a mixture of states product across the copies, each of which lies entirely in one laboratory. An LOCC maps separable inputs to separable outputs, and \Cref{lem:sep overlap} bounds the overlap with $\Phi_2^{\otimes kR}$ by $2^{-kR}$.

It follows that $2^{-kR} + \negl(n) \geq 1 - \varepsilon > 1/2$, and hence $kR = 0$ for large enough $n$, since $kR$ is a nonnegative integer.
This proves the last case of Eq.~\eqref{eq:agnostic cap unbounded}.
Neither concentration nor a union bound is needed, and the circuit size of $\Lambda$ is unrestricted.

\end{proof}

\subsubsection{Computationally efficient state-dependent model}
\label{sec:statistical converse state dependent}

We next allow the protocol to depend on the input state, as in model (M2), while restricting its circuit size to be polynomial.
The average fidelity bounds in the proof of \Cref{thm:agnostic unbounded} do not by themselves select a single hard instance against all such protocols.
We use concentration and a union bound to make this selection, together with a purification whose additional local copies do not invalidate the bound.

\begin{theorem}[No-go for efficient protocols]
\label{thm:agnostic no go}
Fix the parameters and the state set $\scS$ of \Cref{thm:agnostic unbounded}.
Then, for large enough $n$, there exists a $\Theta(n)$-qubit pure state $\ket{\rho_0}^{\sA\sB\sE}$ with $\rho_0^{\sA\sB} \in \scS$ satisfying Eqs.~\eqref{eq:agnostic rank}--\eqref{eq:agnostic ed} and
\begin{align}
\label{eq:agnostic cap}
    &\widehat{E}_\rD^{(k, l, \varepsilon, s)}(\sA:\sB)_{\rho_0}
    \leq \begin{cases}
        \min\{\hat{m}, \, \log{(k+2l)}\} + o(1) & \ \text{if} \ \ r_\rs = 1 \ \text{and} \ r = \poly(n), \\
        \min\{\kappa\log{r_\rs}, \, \log{(k+2l)}\} + o(1) & \ \text{if} \ \ r_\rs \geq 2 \ \text{and} \ r = \poly(n), \\
        0 & \ \text{if} \ \ r = 2^{\omega(\log{n})},
    \end{cases}
\end{align}
for any $k = \poly(n)$ with $k = \omega(1)$, any nonnegative integer $l = \poly(n)$, and every $s = \poly(n)$, where $\varepsilon \in [0, 1/2)$ is constant.
The state $\ket{\rho_0}^{\sA\sB\sE}$ is obtained from a realization of the ensemble of Eq.~\eqref{eq:agnostic purification} at $b = \hat{b} = 0$ by a unitary on $\sE$, and the realization of $\ket{\Psi_{00}}$ can be chosen independently of $r$.
\end{theorem}

\begin{proof}[Proof of \Cref{thm:agnostic no go}]

We use the states $\ket{\Upsilon_{b\hat{b},V}}$ and the output fidelity $\rF_\Lambda$ defined in the proof of \Cref{thm:agnostic unbounded}.
The state appearing in the statement, $\ket{\rho_0}^{\sA\sB\sE} = \ket{\Upsilon_{00,V}}^{\sA\sB\sE}$, is obtained by selecting a realization of $\ket{\Psi_{00}}$ and unitaries $(U_1,U_2,V^\sE)$.
The rank is given by Eq.~\eqref{eq:agnostic rank}, and Eqs.~\eqref{eq:agnostic renyi} and~\eqref{eq:agnostic ed} hold with probability $1 - \negl(n)$ over the sampling of $\ket{\Psi_{00}}$, as proved in \Cref{thm:agnostic unbounded}.
These properties are independent of $U_1$, $U_2$, and $V^\sE$, and the same realization of $\ket{\Psi_{00}}$ obeys them at every $M$.

For a protocol $\Lambda \in \mathfrak{L}(s, k, l)$ declaring $kR$ pairs, we use $\rF_\Lambda$ to evaluate their fidelity.
We select the vectors of $\ket{\Psi_{00}}$ and the unitaries $(U_1,U_2,V^\sE)$ at which the fidelity stays near its average for every protocol at once, by L\'{e}vy's lemma and a union bound over the class of efficient protocols. To this end, we regard $\rF_\Lambda\big(\Upsilon_{00}^{\otimes k}\otimes\ketbra{\Upsilon_{00,V}}{\Upsilon_{00,V}}^{\otimes 2l}\big)$ as a function of all these vectors and unitaries, and record its Lipschitz constants.

For any state $\varrho$ and unitaries $U, U'$, we have $\|U\varrho U^\dag - U'\varrho U'^\dag\|_1 \leq 2\|U - U'\|_\infty \leq 2\|U - U'\|_2$, while $\rF_\Lambda$ moves by at most half the trace-norm distance of its input, since it is of the form $\tr[Q\,\cdot\,]$ with $0 \leq Q \leq \bI$, to which the variational form of the trace distance applies. The map attaching the coherent label and applying its controlled unitaries is an isometry, so replacing $\ket{\Psi_{00}}$ with $\ket{\Psi_{00}'}$ moves $\ket{\Upsilon_{00,V}}$ by $\big\|\ket{\Psi_{00}} - \ket{\Psi_{00}'}\big\|$. Within $\ket{\Psi_{00}}$, the vector $\ket{\phi_0}$ carries amplitude $\sqrt{\eta}$ and the vector $\ket{\hat{\phi}_0}$ amplitude $\sqrt{1-\eta}$. Thus,
\begin{align}
    &\big|\rF_\Lambda(\Upsilon_{00}^{\otimes k}\otimes\ketbra{\Upsilon_{00,V}}{\Upsilon_{00,V}}^{\otimes 2l}) - \rF_\Lambda(\Upsilon_{00}'^{\otimes k}\otimes\ketbra{\Upsilon_{00,V}'}{\Upsilon_{00,V}'}^{\otimes 2l})\big| \notag\\
    &\leq \f{k}{2}\big\|\Upsilon_{00} - \Upsilon_{00}'\big\|_1 + l\big\|\ketbra{\Upsilon_{00,V}}{\Upsilon_{00,V}} - \ketbra{\Upsilon_{00,V}'}{\Upsilon_{00,V}'}\big\|_1 \\
    &\leq (k+2l)\sqrt{\eta}\big\|\ket{\phi_0} - \ket{\phi_0'}\big\|,
\end{align}
where $\Upsilon_{00}'$ and $\ket{\Upsilon_{00,V}'}$ are obtained from Eqs.~\eqref{eq:agnostic mixed} and~\eqref{eq:agnostic purification} by replacing $\ket{\phi_0}$ with $\ket{\phi_0'}$.
We here used $\|\rho^{\otimes k} - \varrho^{\otimes k}\|_1 \leq k\|\rho - \varrho\|_1$ and $\|\ketbra{u}{u} - \ketbra{v}{v}\|_1 \leq 2\|\ket{u} - \ket{v}\|$.
The same bound holds for $\ket{\hat{\phi}_0}$, with $\sqrt{\eta}\big\|\ket{\phi_0} - \ket{\phi_0'}\big\|$ replaced by $\sqrt{1-\eta}\big\|\ket{\hat{\phi}_0} - \ket{\hat{\phi}_0'}\big\| \leq \big\|\ket{\hat{\phi}_0} - \ket{\hat{\phi}_0'}\big\|$.
Thus $\rF_\Lambda$ is $(k+2l)\sqrt{\eta}$-Lipschitz in $\ket{\phi_0}$, which is Haar random on $2n$ qubits, and $(k+2l)$-Lipschitz in $\ket{\hat{\phi}_0}$, which is Haar random on $2\hat{m}$ qubits and present only at $\hat{m} \geq \mu$.

By $\|U\varrho U^\dag - U'\varrho U'^\dag\|_1 \leq 2\|U - U'\|_2$, replacing $U_{1,j}$ by $U_{1,j}'$ moves $\Upsilon_{00}^{\sA\sB}$ by at most $\f{2}{M}\|U_{1,j}-U_{1,j}'\|_2$ in trace norm and $\ket{\Upsilon_{00,V}}$ by at most $\f{1}{\sqrt{M}}\|U_{1,j}-U_{1,j}'\|_2$ in Euclidean norm, and the same holds for $U_{2,j}$, while replacing $V^\sE$ by $V'^\sE$ moves each of the $2l$ sample copies by at most $2\|V^\sE - V'^\sE\|_2$ in trace norm.
Since $\rF_\Lambda$ moves by at most half the trace-norm distance of its input, it is $\big(\f{k}{M}+\f{2l}{\sqrt{M}}\big)$-Lipschitz in each of the $2M$ unitaries $U_{1,j},U_{2,j}$ and $2l$-Lipschitz in $V^\sE$ with respect to the Hilbert--Schmidt norm.

On the product of the spheres and unitary groups, the Lipschitz constant is at most $\big((k+2l)^2\eta + (k+2l)^2 + 2M(\f{k}{M}+\f{2l}{\sqrt{M}})^2 + 4l^2\big)^{1/2} \leq 3(k+2l)$, uniformly in $M$.
The concentration is governed by the smallest factor of dimension $\min\{2^{2n}, 2^{2\hat{m}}, d_{\sI_\sA}, d_{\sI_\sB}, d_\sE\}$, with $2^{2\hat{m}}$ omitted at $\hat{m} < \mu$, where $\ket{\hat{\phi}_0}$ is fixed.
L\'{e}vy's lemma in \Cref{lem:levy} with the threshold $t = \min\{2^{2n}, 2^{2\hat{m}}, d_{\sI_\sA}, d_{\sI_\sB}, d_\sE\}^{-1/4} = \negl(n)$, with the same omission, implies that
\begin{align}
\label{inteq:agnostic levy poly}
    \Pr\Big[\big|\rF_\Lambda - \bE_{\scF_{00}, U_1, U_2, V^\sE}[\rF_\Lambda]\big| \geq t\Big]
    \leq 2\exp\bigg(-\Omega\bigg(\f{\min\{2^{2n}, 2^{2\hat{m}}, d_{\sI_\sA}, d_{\sI_\sB}, d_\sE\}^{1/2}}{(k+2l)^{2}}\bigg)\bigg),
\end{align}
whose exponent is at least $2^{3\mu/4 - \cO(1)}$ at every $s, k, l \leq 2^{\mu/8}$, by $2^{2\hat{m}} \geq 2^{2\mu}$ at $\hat{m} \geq \mu$, $d_{\sI_\sA}, d_{\sI_\sB}, d_\sE = 2^{\Theta(n)}$, and $\mu = o(n)$.
At $\hat{m} < \mu$, the minimum runs over the remaining factors and the exponent is $2^{\Omega(n)}$.
A union bound over the protocols of $\mathfrak{L}(s, k, l)$ at every such $s, k, l$ jointly, of which there are $2^{\cO(2^{\mu/8}\mu)}$ by Eq.~\eqref{eq:protocol count}, leaves a failure probability $\negl(n)$.

The realization of $\ket{\Psi_{00}}$ can be chosen independently of $M$ as follows. Fix any constant $C > 0$ and sample its vectors once, while sampling $(U_1,U_2,V^\sE)$ separately for each $M \in [1, 2^{Cn}]$. The preceding bound is uniform in $M$, so a further union bound over these $2^{Cn}$ choices still leaves a failure probability $\negl(n)$, since $\mu = \omega(\log{n})$.
The events of Eqs.~\eqref{eq:agnostic renyi} and~\eqref{eq:agnostic ed} depend only on $\ket{\Psi_{00}}$ and each fails with probability $\negl(n)$; they need not be counted again for each $M$.
Consequently, there exist one realization of $\ket{\Psi_{00}}$ and, for each $M$, a choice of $(U_1,U_2,V^\sE)$ determining $\ket{\rho_0}^{\sA\sB\sE} = \ket{\Upsilon_{00,V}}^{\sA\sB\sE}$, such that Eqs.~\eqref{eq:agnostic renyi} and~\eqref{eq:agnostic ed} hold and
\begin{align}
\label{inteq:agnostic selection}
    \rF_\Lambda\big(\rho_0^{\otimes k}\otimes\ketbra{\rho_0}{\rho_0}^{\otimes 2l}\big)
    \leq \bE_{\scF_{00}, U_1, U_2, V^\sE}\big[\rF_\Lambda\big(\Upsilon_{00}^{\otimes k}\otimes\ketbra{\Upsilon_{00,V}}{\Upsilon_{00,V}}^{\otimes 2l}\big)\big] + \negl(n),
\end{align}
for every $\Lambda \in \mathfrak{L}(s,k,l)$ and all $s,k,l \leq 2^{\mu/8}$.
The selection is made before $s$, $k$, and $l$ are quantified, and, as $\mu = \omega(\log{n})$, every $s, k, l = \poly(n)$ falls below $2^{\mu/8}$ at large enough $n$.
Taking $C$ large enough covers the range $r \leq 2^{\cO(n)}$ fixed in the construction.

Suppose now that a protocol $\Lambda$ of $\mathfrak{L}(s, k, l)$ distills at a rate $R$ from the $k$ copies of $\rho_0^{\sA\sB}$ and the $2l$ local samples of its purification at error $\varepsilon$, so that $\rF_{\Lambda}\big(\rho_0^{\sA\sB \otimes k}\otimes\ketbra{\rho_0}{\rho_0}^{\otimes 2l}\big) \geq 1 - \varepsilon$.
The bound in Eq.~\eqref{inteq:agnostic logarithmic fidelity} is uniform in the realization of $\ket{\Psi_{00}}$ and in $U_2$, and hence also bounds the average in Eq.~\eqref{inteq:agnostic selection}. We obtain
\begin{align}
    1 - \varepsilon - \negl(n)
    \leq \f{(k+2l)!}{(2l)!}\big(1+\negl(n)\big)2^{-kR},
\end{align}
and therefore $R \leq \log{(k+2l)} + o(1)$ by the same calculation as in Eq.~\eqref{inteq:agnostic logarithmic rate}.

At $r = \poly(n)$, Eq.~\eqref{inteq:agnostic purification comparison} bounds the trace-norm distance between the averaged inputs from $\scF_{00}$ and $\scF_{11}$ by $\negl(n)$, including the additional local copies of the purification, for every fixed $(U_1,U_2)$.
Together with Eq.~\eqref{inteq:agnostic selection} and the linearity of $\rF_\Lambda$, averaging also over these unitaries gives
\begin{align}
    1 - \varepsilon
    &\leq \rF_{\Lambda}\big(\rho_0^{\otimes k}\otimes\ketbra{\rho_0}{\rho_0}^{\otimes 2l}\big) \\
    &\leq \bE_{\scF_{11}, U_1, U_2, V^\sE}\big[\rF_{\Lambda}\big(\Upsilon_{11}^{\otimes k}\otimes\ketbra{\Upsilon_{11,V}}{\Upsilon_{11,V}}^{\otimes 2l}\big)\big] + \negl(n) \\
    &\leq \max_{\Upsilon_{11},\, U_1,\, U_2,\, V^\sE}\rF_{\Lambda}\big(\Upsilon_{11}^{\otimes k}\otimes\ketbra{\Upsilon_{11,V}}{\Upsilon_{11,V}}^{\otimes 2l}\big) + \negl(n),
\end{align}
where the maximum runs over all realizations of $\scF_{11}$ and all the displayed unitaries.
Thus, $\Lambda$ distills at the same rate $R$ with fidelity $\geq 1 - \varepsilon - \negl(n)$ from some realization $\Upsilon_{11}^{\sA\sB}$, supplied with the purification $\ket{\Upsilon_{11,V}}$ at some $V^\sE$.
The conditions of \Cref{lem:truncation} were verified for every such realization in the proof of \Cref{thm:agnostic unbounded}.
Equations~\eqref{inteq:agnostic comparison rate spike} and~\eqref{inteq:agnostic comparison rate pure}, at error $\varepsilon + \negl(n)$, therefore give $R \leq \kappa\log{r_\rs} + o(1)$ at $r_\rs \geq 2$ and $R \leq \hat{m} + o(1)$ at $r_\rs = 1$ and $\hat{m} < \mu$.
At $r_\rs = 1$ and $\hat{m} \geq \mu$, we have $\log{(k+2l)} = \cO(\log{n}) < \mu \leq \hat{m}$ for large enough $n$.
Combining these bounds gives the first two cases of Eq.~\eqref{eq:agnostic cap}.

At $r = 2^{\omega(\log{n})}$, Eq.~\eqref{inteq:agnostic average fidelity} bounds the average over $(U_1,U_2,V^\sE)$ by $2^{-kR} + \negl(n)$ for every fixed realization of $\ket{\Psi_{00}}$, and hence also after averaging over that realization.
Together with Eq.~\eqref{inteq:agnostic selection}, this gives $2^{-kR} + \negl(n) \geq 1 - \varepsilon > 1/2$, and hence $kR = 0$ as $kR$ is a nonnegative integer.
This is the last case of Eq.~\eqref{eq:agnostic cap}.

\end{proof}

Model (M2) places no constraint on how a protocol depends on the state, so the class over which the union bound is taken already contains the protocols that use the rank and the conditional R\'{e}nyi entropy; the cap therefore also applies to the protocol of \Cref{thm:achievability by efficient protocol formal}, which is constructed from them, whenever it acts on $k = \omega(1)$ copies.

The result is statistical in origin, and efficient protocols are considered only through the union bound by which a single realization, together with its supplied purification, is selected.
The circuit-size restriction cannot be dispensed with for a single known state: at $r = 2^{\omega(\log{n})}$, parties who know $U_1$ and $U_2$ and have unbounded computational power can read the label, undo the corresponding unitaries, and distill at the rate of Eq.~\eqref{eq:agnostic ed}.

\subsubsection{A simplified converse bound}
\label{sec:simplified statistical converse bound}

The rank in Eq.~\eqref{eq:agnostic rank} is set by $r$ and ranges up to $2^{\Theta(n)}$, while $E_\rD(\sA:\sB)_{\rho_0}$ and $\widetilde{H}_\alpha(\sA|\sE)_{\rho_0}$ remain unchanged when $r$ is varied with the realization of $\ket{\Psi_{00}}$ fixed.
The exact $0$ of the last case of Eq.~\eqref{eq:agnostic cap} rests on the indistinguishability from the separable state $\tau_M$ of \Cref{lem:maxmix}, which cannot hold at polynomial rank, since a state of rank $2^{\cO(\log{n})}$ has purity at least $1/\poly(n)$ and a SWAP test on two copies distinguishes it from that state, whose purity is $1/(M d_{\sI_\sA}d_{\sI_\sB}) = 2^{-\Theta(n)}$.
The indistinguishability from that state is what \Cref{lem:maxmix} supplies at superpolynomial rank.
At $r_\rs = 1$ and $r = 1$ the state $\rho_0^{\sA\sB}$ is pure, and $\hat{m}$ sweeps $\widetilde{H}_\alpha(\sA|\sE)_{\rho_0}$ from $o(1)$ to $\Theta(n)$ while Eq.~\eqref{eq:agnostic ed} is unchanged.

Specializing to a fixed $\alpha \in (1, \alpha_\rc)$, we consider the polynomial-rank cases of Eq.~\eqref{eq:agnostic cap}.
At $r_\rs \geq 2$, Eq.~\eqref{eq:agnostic renyi} gives $\widetilde{H}_\alpha(\sA|\sE)_{\rho_0} = \kappa\log{r_\rs} + o(1)$, since the term subtracted from $\kappa$ is $0$ for $\alpha < \alpha_\rc$.
At $r_\rs = 1$ and $\hat{m} < \mu$, it gives $\widetilde{H}_\alpha(\sA|\sE)_{\rho_0} = \hat{m} + o(1)$.
At $r_\rs = 1$ and $\hat{m} \geq \mu$, the entropy is $\Theta(\hat{m}) = \omega(\log{n})$, while $\log{(k+2l)} = \cO(\log{n})$ for polynomially many copies.
Hence, in all these cases the cap is $\min\{\widetilde{H}_\alpha(\sA|\sE)_{\rho_0}, \, \log{(k+2l)}\} + o(1)$.
Since the rank of Eq.~\eqref{eq:agnostic rank} is $\Theta(r)$ throughout, the last case of Eq.~\eqref{eq:agnostic cap} applies at every $r = 2^{\omega(\log{n})}$, with the same conditional R\'{e}nyi entropy and information-theoretic distillable entanglement.
The duality in Eq.~\eqref{eq:duality} at $1/\alpha + 1/\beta = 2$ carries $\alpha \in (1, \alpha_\rc)$ onto $\beta \in (\beta_\rc, 1)$ with $\beta_\rc \coloneqq \f{\alpha_\rc}{2\alpha_\rc - 1}$, and gives the following corollary.

\begin{corollary}[Simplified statistical no-go; see \Cref{thm:agnostic unbounded,thm:agnostic no go}]
\label{cor:agnostic no go informal}
There exist a constant $\beta_\rc\in(1/2,1)$ and a family of mixed states $\{\rho^{\sA\sB}\}_n$, selected from state sets $\scS$ of \Cref{thm:agnostic unbounded}, together with purifications $\ket{\rho}^{\sA\sB\sE}$, such that the following holds for any $k = \poly(n)$ with $k = \omega(1)$, any nonnegative integer $l = \poly(n)$, and any constant $\varepsilon \in [0, 1/2)$.
Each protocol acts on $k$ shared copies and $l$ additional local copies of a purification per party.
For every constant $\beta \in (\beta_\rc, 1)$, no LOCC protocol in the state-agnostic model required to succeed on every state in $\scS$ and every purification of it on $\sA\sB\sE$, whatever its circuit size, distills at a rate $R$ larger than
\begin{align}
    \begin{cases}
        \min\{\widetilde{I}_{\beta}(\sA\rangle\sB)_\rho, \, \log{(k+2l)}\} + o(1),
        & \text{if } \rank(\rho^{\sA\sB}) = \poly(n), \\
        0,
        & \text{if } \rank(\rho^{\sA\sB}) = 2^{\omega(\log{n})}.
    \end{cases}
\end{align}
The same bound holds for every computationally efficient LOCC protocol acting on $\rho^{\sA\sB}$ with the selected purification $\ket{\rho}^{\sA\sB\sE}$, even when the protocol is chosen specifically for that state.
The information-theoretic distillable entanglement satisfies $E_\rD(\sA:\sB)_\rho = \widetilde{\Omega}(n)$.
\end{corollary}

The threshold $\varepsilon < 1/2$ in \Cref{thm:agnostic no go} is what makes the last case of Eq.~\eqref{eq:agnostic cap} exactly $0$, and it cannot be relaxed. The value $1/2$ is the maximal overlap of a single Bell pair with a separable state, and a fidelity of exactly $1/2$ is attained without touching the input. At $\varepsilon \geq 1/2$, the LOCC that discards the input and locally prepares $\big(\f{1}{2}(\ketbra{00}{00} + \ketbra{11}{11})\big)^{\otimes t}$ with $t = \lfloor\log\f{1}{1-\varepsilon}\rfloor \geq 1$ declares $t$ Bell pairs with fidelity $2^{-t} \geq 1-\varepsilon$ and probability $1$. Beyond the threshold, only the exact $0$ is lost: for any constant $\varepsilon < 1$, the chain leading to $2^{-kR} + \negl(n) \geq 1 - \varepsilon$ still applies, and yields $\widehat{E}_\rD^{(k, l, \varepsilon, s)}(\sA:\sB)_{\rho_0} = \cO(1/k)$, which vanishes for $k = \omega(1)$.
However, this does not mean that some entanglement remains to be distilled: this protocol never touches the input, and the same rate is declared on every state.

The restriction $k = \omega(1)$ is of the same origin. At constant $k$, a protocol may declare more pairs than the state carries by leaving them in a larger space: at the spike branch, retaining $\Phi_g^{\otimes k} = \Phi_{g^k}$ and declaring $m$ pairs attains fidelity $g^{k}2^{-m}$, which admits $m = \lfloor k\log{g} + \log\f{1}{1-\varepsilon}\rfloor$ at error $\varepsilon$, provided $2^m \geq g^k$. The excess is a constant number of pairs at any $k$, which the term $\f{1}{k}\log\f{2}{1-\varepsilon}$ of Eq.~\eqref{eq:rateconverse sandwich} records and $k = \omega(1)$ makes vanish.

\section{Converse: Cryptographic construction and efficiently preparable states}
\label{sec:Converse Description-aware setting}

We provide our converse bounds in the description-aware case, under conjectured cryptographic hardness assumptions. We introduce the cryptographic tools in \Cref{sec:cryptographic tools} and the public-key pseudoentanglement built from them in \Cref{sec:public key pseudoentanglement}. We then construct the family of states in \Cref{sec:construction of the state family state aware} and prove the bounds on it in \Cref{sec:description-aware converse result}.

\subsection{Cryptographic tools}
\label{sec:cryptographic tools}

Our converse bounds in the description-aware case rely on two cryptographic ingredients---a lossy function family and a labeled generator---both of which we instantiate from the conjectured hardness of the learning with errors (LWE) decision problem~\cite{regev2009Onlatticeslearningwitherrors} with binary secrets. We introduce the LWE assumptions in \Cref{sec:lwe} and the two ingredients in \Cref{sec:lf pke prg}.

\subsubsection{Learning with errors}
\label{sec:lwe}

Below, $\ell \geq 1$ denotes the LWE dimension, which plays the role of the security parameter, $m$ the number of noisy linear equations, $q$ the modulus, and $\sigma$ the standard deviation of the error distribution. We reserve $n$ for the number of qubits under consideration.

\begin{definition}[LWE decision problem with binary secrets~\cite{Micciancio2018OntheHardnessofLWEwithBinarySecrets}]
\label{def:lwe}
Let $\bU_q^{m\times\ell}$ denote the uniform distribution over $\bZ_q^{m\times\ell}$, and let $\bD_{q,\sigma}^{m\times\ell}$ denote the distribution over $\bZ_q^{m\times\ell}$ in which every entry is independently sampled from the discretized Gaussian distribution with standard deviation $\sigma$. We write $\bU_q^m \coloneqq \bU_q^{m \times 1}$ and $\bD_{q,\sigma}^m \coloneqq \bD_{q,\sigma}^{m \times 1}$, and regard $\{0,1\}^\ell$ as a subset of $\bZ_q^\ell$.
The $\mathrm{LWE}_{\ell, m, q, \sigma}^{\rm bin}$ decision problem is the problem of distinguishing the following two distributions over $\bZ_q^{m\times\ell}\times\bZ_q^{m}$:
\begin{enumerate}
    \item $(A, u) \sim \bU_q^{m\times\ell}\times \bU_q^m$,
    \item $(A, As + \delta)$ with $A \sim \bU_q^{m\times\ell}$, \ $s$ uniform on $\{0,1\}^\ell$, \ $\delta \sim \bD_{q,\sigma}^m$.
\end{enumerate}
\end{definition}

The advantage of an algorithm in this problem is the absolute difference of its acceptance probabilities on the two distributions.
We say that the $\mathrm{LWE}_{\ell, m, q, \sigma}^{\rm bin}$ decision problem is $(T, \epsilon)$-hard if every family of quantum circuits of size at most $T(\ell)$ has advantage at most $\epsilon(\ell)$ for all sufficiently large $\ell$. We quantify over circuits rather than over uniform algorithms, as is standard when the size and advantage bounds are stated as explicit functions of the security parameter; this is the stronger requirement of the two, and it matches the class of protocols against which our converse bounds are proved.

We consider the $\mathrm{LWE}_{\ell, m, q, \sigma}^{\rm bin}$ decision problem in the regime
\begin{align}
\label{eq:lwe params}
    q = \poly(\ell + m) \text{: prime},
    \ \ \ \  \ell^{1/2}\log{\ell} \leq \sigma < q,
\end{align}
and invoke its hardness against circuits of superpolynomial size at two numbers of equations $m$, polynomially and subexponentially many in $\ell$. These are the usual parameter choices for LWE~\cite{regev2010Thelearningwitherrorsproblem}, except that the lower bound on $\sigma$ is raised to a noise level at which the reduction from standard LWE to binary LWE applies~\cite{Micciancio2018OntheHardnessofLWEwithBinarySecrets}.

\begin{assumption}[Hardness assumptions for LWE with binary secrets]
\label{assumpt:lwe assumptions}
We consider the following two assumptions on the quantum hardness of the $\mathrm{LWE}_{\ell,m,q,\sigma}^{\rm bin}$ decision problem, each required to hold for every $(m, q, \sigma)$ obeying Eq.~\eqref{eq:lwe params} with $m$ in the range stated in the item.
\begin{enumerate}[label=\textnormal{(H\arabic*)}, leftmargin=4em]
    \item Hardness at polynomially many equations: there exist constants $c_1, c_2 > 0$ such that, for every $m = \poly(\ell)$, the problem is $\big(2^{c_1\ell/\log(2 + m\log{q}/\ell)}, \ 2^{-c_2\ell/\log(2 + m\log{q}/\ell)}\big)$-hard.
    \item Hardness at subexponentially many equations: there exist constants $1 > \gamma_1 > \gamma_2 > 0$ such that, for every $m = 2^{\cO(\ell^{\gamma_2})}$, the problem is $\big(T, 2^{-\Omega(\ell^{\gamma_1})}\big)$-hard for every $T = 2^{\cO(\ell^{\gamma_2})}$.
\end{enumerate}
\end{assumption}

Assumption (H1) is used for the first application of \Cref{lem:small register pke} in \Cref{sec:parameters state aware} and for the labeled generator of \Cref{lem:labeled generator}, and (H2) for the second application there and, in one parameter regime, for the third. The second application makes the information-theoretic distillable entanglement of our family nearly extensive at a subexponential number of equations. At $m = \poly(\ell)$, the exponent in (H1) is $\Omega(\ell/\log{\ell})$, and (H1) implies the bound of (H2) there; (H2) adds content only at $m$ superpolynomial in $\ell$, where (H1) asserts nothing.

We state both assumptions directly rather than deriving them from the worst-case hardness of lattice problems. The known reduction from standard LWE to LWE with binary secrets~\cite{Micciancio2018OntheHardnessofLWEwithBinarySecrets} maps dimension $k$ to dimension $\Theta(k\log{q})$ and increases the noise, and thus does not carry the exponent of (H1) as a function of $\ell$. We state the size and advantage bounds explicitly because we shall take $\ell$ polylogarithmic in $n$, against adversaries of size $\poly(n)$, where hardness polynomial in $\ell$ and an advantage negligible only in $\ell$ are insufficient.

The dependence of (H1) on $m$ accommodates known subexponential attacks. For example, for $m=10\ell$, $q=\Theta(\ell^3)$ prime, and
$\sigma=\ell$, combining the small-secret BKW algorithm of
Ref.~\cite[Theorem~4]{Kirchner2015AnImprovedBKWAlgorithmforLWE}
with the sample amplification of
Ref.~\cite[Lemma~2]{Herold2018OnasymptoticcomplexitysolvingLWE}
gives an algorithm for our decision problem with running time
$2^{\cO(\ell/\log\log{\ell})}$. The exponent in (H1) is $\Theta(\ell/\log\log{\ell})$ for this parameter family; a hardness bound $2^{\Omega(\ell)}$ uniform over all $m=\poly(\ell)$ would therefore be false.
At $m = \Theta(\ell/\log{\ell})$ and $q = \poly(\ell)$, the parameters of the labeled generator below, the exponent is $\Theta(\ell)$, and the explicit dependence on $m$ also covers the larger polynomial numbers of equations used by the lossy function family of \Cref{lem:small modulus lossy}.

The subexponential scaling of the number of equations and of the circuit size in (H2) also appears in Ref.~{\cite[Assumption~2.19]{bouland2024PublicKeyPseudoentanglement}} and Ref.~{\cite[Assumption~3.5]{bouland2024hardnesslearninggroundstate}}; the secret, modulus, and noise regimes there differ from ours, and we state (H2) directly in the form used here. We take $\gamma_2 < 1/2$ without loss of generality, as (H2) at a constant $\gamma_2$ implies (H2) at every smaller one. Its circuit size bound $2^{\cO(\ell^{\gamma_2})}$ coincides with $\poly(m)$ at $m = 2^{\Theta(\ell^{\gamma_2})}$ and exceeds it at fewer equations, where we use it as well.

\subsubsection{Lossy function and labeled generator}
\label{sec:lf pke prg}

For a positive integer $p$, we partition $\bZ_q$ into $p$ consecutive bins, each containing $\lceil q/p\rceil$ or $\lfloor q/p\rfloor$ elements, and write $\lfloor\cdot\rfloor_p : \bZ_q \to \bZ_p$ for the map sending an element to the index of the bin containing it. 
We extend this notation coordinatewise to vectors.

\begin{definition}[Lossy function family~\cite{Peikert2011LossyTrapdoorFunctions}]
\label{def:lossy function}
Let $\mu(n)$ be an input length and $\tau(n) \geq 0$ a leakage parameter. A \emph{lossy function family} at $(\mu, \tau)$ is a family $\{g_k\}_k$ of functions on $\{0,1\}^{\mu}$ computable by circuits of size $\poly(n)$, indexed by a key $k$ drawn from either of two distributions $\cK_n^0$ and $\cK_n^1$ sampleable in time $\poly(n)$, such that the following holds.
\begin{enumerate}[label=\textnormal{(L\arabic*)}, leftmargin=4em]
    \item For $k \leftarrow \cK_n^{0}$, the function $g_k$ is injective with probability $1 - \negl(n)$.
    \item For $k \leftarrow \cK_n^{1}$, it holds that $|\rIm(g_k)| \leq 2^{\tau}$ with probability $1 - \negl(n)$.
\end{enumerate}
For parameters $T$ and $\epsilon$, the family is \emph{$(T, \epsilon)$-mode-indistinguishable} if every family of quantum circuits of size at most $T(n)$ has advantage at most $\epsilon(n)$ in distinguishing $\cK_n^{0}$ from $\cK_n^{1}$ given the key.
\end{definition}

\Cref{def:lossy function} corresponds to the notion of an almost-always lossy (nontrapdoor) function family in Ref.~\cite{Peikert2011LossyTrapdoorFunctions}. The following lemma gives such a function family from LWE with binary secrets, on inputs encoded as concatenations of $a$ one-hot blocks of length $L$; the encoding keeps the leakage at $\ell\log(2(a+1))$, independent of the input length $a\log{L}$. Statements (I) and (II) supply (L1) and (L2), respectively, while (III) transfers the $(T,\epsilon)$ of the $\mathrm{LWE}_{\ell,m,q,\sigma}^{\rm bin}$ decision problem to mode indistinguishability, up to a factor of $aL$ in the advantage and a $\poly(n)$ overhead in the circuit size.

\begin{lemma}[Lossy functions on block-one-hot inputs at a polynomial modulus; adapted from Ref.~{\cite[Definitions~2.16 and~3.2, and Theorem~3.3]{bouland2024PublicKeyPseudoentanglement}}]
\label{lem:small modulus lossy}
Let $\ell$, $a$, and $L$ be efficiently computable functions of $n$, with $a \geq 1$ an integer and $L \geq 2$ a power of two, obeying $\ell = \omega(\log{n})$ and $\ell, aL = \poly(n)$. Set
\begin{align}
\label{eq:lossy params}
    p \coloneqq 2^{\lceil\log{\ell}\rceil},
    \ \ \ \ m \coloneqq \bigg\lceil \f{\ell + 2a\log{L}}{\log{p}} \bigg\rceil,
    \ \ \ \ \sigma \coloneqq 3\ell^{1/2}\log{\ell},
    \ \ \ \ q \coloneqq \text{the least prime with } q \geq 30mpa\ell\log{\ell},
\end{align}
which obey Eq.~\eqref{eq:lwe params}. We identify $x \in \{0,1\}^{a\log{L}}$ with $(x_1, \ldots, x_a) \in \{0, \ldots, L-1\}^a$, and let $\iota(x) \in \{0,1\}^{aL}$ be the vector whose $j$-th block of $L$ coordinates has a one at position $x_j$ and zeros elsewhere.
For $b \in \{0, 1\}$, let $\cK_n^{b}$ be the efficiently sampleable distribution over keys $k = (A, t)$ in which $t \sim \bU_q^m$ and
\begin{align}
    A \sim
    \begin{cases}
        \bU_q^{m\times aL} & \ b = 0, \\
        \text{the law of} \ B^\top C + E, \ \text{where} \ B \sim \bU_q^{\ell\times m}, \ C \ \text{uniform on} \ \{0,1\}^{\ell\times aL}, \ E \sim \bD_{q,\sigma}^{m \times aL} & \ b = 1,
    \end{cases}
\end{align}
independently. A key $k$ determines a function $g_k : \{0,1\}^{a\log{L}} \to \bZ_p^m$ given by $g_k(x) \coloneqq \lfloor A\iota(x) + t\rfloor_p$. Then the following statements hold.
\begin{enumerate}[label=\textnormal{(\Roman*)}]
    \item For $k \leftarrow \cK_n^{0}$, the function $g_k$ is injective with probability $1 - \cO(2^{-\ell})$.
    \item For $k \leftarrow \cK_n^{1}$, it holds that $|\rIm(g_k)| \leq (2(a+1))^{\ell}$ with probability $1 - \cO(2^{-\ell})$.
    \item Any quantum circuit that, given the key $k$, distinguishes $\cK_n^{0}$ from $\cK_n^{1}$ with advantage $\epsilon$ yields a circuit solving the $\mathrm{LWE}_{\ell, m, q, \sigma}^{\rm bin}$ decision problem with advantage $\epsilon/(aL)$, with a $\poly(n)$ overhead in size.
\end{enumerate}
In particular, $\cK_n^0$ and $\cK_n^1$ determine a lossy function family at $(a\log{L}, \ \ell\log(2(a+1)))$.
\end{lemma}

\begin{proof}[Proof of \Cref{lem:small modulus lossy}]

We first check Eq.~\eqref{eq:lwe params}. By Bertrand's postulate~\cite{Hardy2008AnIntroductiontotheTheoryofNumbers}, $q \leq 60mpa\ell\log{\ell}$, and $a \leq m\log{p}$ by the choice of $m$, so that $q = \poly(\ell + m)$; also $\ell^{1/2}\log{\ell} \leq \sigma < q$.

(I) We follow the approach of Ref.~{\cite[Lemma~3.4 and Corollary~3.5]{bouland2024PublicKeyPseudoentanglement}}.
Let $x \neq x' \in \{0,1\}^{a\log{L}}$. Since $\iota(x)$ and $\iota(x')$ are distinct block-one-hot vectors, there is a coordinate $j$ with $\iota(x)_j = 1$ and $\iota(x')_j = 0$. Condition on $t$ and on all columns of $A$ except the $j$-th one, $a_j$. Then $A\iota(x')+t$ is fixed, whereas $A\iota(x)+t$ is uniform on $\bZ_q^m$ as it includes the independent uniform summand $a_j$.
Thus, each coordinate of $A\iota(x)+t$ is independent and uniform on $\bZ_q$, and falls into the bin containing the corresponding coordinate of $A\iota(x')+t$ with probability at most $\lceil q/p\rceil/q \leq p^{-1}(1+p/q)$.
Multiplying over the coordinates and using $mp/q \leq 1$, we obtain $\Pr[g_k(x) = g_k(x')] \leq p^{-m}e^{mp/q} \leq e p^{-m}$.
A union bound over the fewer than $2^{2a\log{L}}$ pairs of distinct inputs gives a failure probability at most $e 2^{2a\log{L} - m\log{p}} \leq e2^{-\ell}$ by the choice of $m$.

(II) We follow the proof of Ref.~{\cite[Lemma~3.8]{bouland2024PublicKeyPseudoentanglement}}. Represent the entries of $E$ by their centered integer lifts, and let $L_E$ be the event that some entry has absolute value at least $5\ell\log{\ell}$. The Gaussian tail bound and a union bound over the $maL$ entries give $\Pr[L_E] \leq 2maL e^{-25\ell/18} \leq 2^{-\ell}$ for large $\ell$, as $maL = \poly(n) = 2^{o(\ell)}$. On $\lnot L_E$, since $\iota(x)$ has Hamming weight $a$, every coordinate of $E\iota(x)$ has absolute value at most $5a\ell\log{\ell}$.
As $q > a$, the vector $C\iota(x)$ can be identified with an element of $\{0, \ldots, a\}^\ell$, so the map $x \mapsto C\iota(x)$ has at most $(a+1)^\ell$ nonempty fibers $\cX_y \coloneqq \{x : C\iota(x) = y\}$. For $x \in \cX_y$, we have $A\iota(x) + t = w_y + E\iota(x)$ with $w_y \coloneqq B^\top y + t$ independent of $x$.
Every bin has at least $\lfloor q/p\rfloor > 10a\ell\log{\ell} + 1$ elements. Hence, on $\lnot L_E$, the $i$-th coordinate of $g_k(x)$ takes at most two values as $x$ ranges over $\cX_y$, and more than one only if $(w_y)_i$ lies within distance $5a\ell\log{\ell}$ of one of the $p$ bin boundaries; let $\mathbf{1}_{y,i}$ indicate this event, so that $|g_k(\cX_y)| \leq 2^{\sum_i \mathbf{1}_{y,i}}$. As $t$ is uniform and independent of $(B, C, E)$, the coordinates of $w_y$ are independent and uniform on $\bZ_q$, and the boundary neighborhoods contain at most $p(10a\ell\log{\ell}+1)$ points of $\bZ_q$, so $\Pr[\mathbf{1}_{y,i} = 1] \leq p(10a\ell\log{\ell}+1)/q \leq 1/(2m)$. Then $\bE\big[2^{\sum_i \mathbf{1}_{y,i}}\big] \leq (1 + 1/(2m))^m \leq e^{1/2}$. Summing over the fibers gives $\bE[|\rIm(g_k)|] \leq e^{1/2}(a+1)^\ell$ on $\lnot L_E$, and Markov's inequality at the threshold $(2(a+1))^\ell$ gives the claim together with $\Pr[L_E] \leq 2^{-\ell}$.
In Ref.~\cite{bouland2024PublicKeyPseudoentanglement}, the fiber of $y = 0$ is bounded by its cardinality, requiring an exponential modulus; the shift $t$ avoids this, as does a half-bin shift of the rounding~{\cite[Lemma~3.8]{bouland2024hardnesslearninggroundstate}}.

(III) The reduction follows the proof of Ref.~{\cite[Lemma~2.20]{bouland2024PublicKeyPseudoentanglement}}. Under $\cK_n^1$, the $j$-th column of $B^\top C + E$ is $B^\top c_j + \delta_j$ with matrix $B^\top \sim \bU_q^{m\times\ell}$, secret $c_j$ uniform on $\{0,1\}^\ell$, and error $\delta_j \sim \bD_{q,\sigma}^m$, whereas under $\cK_n^0$ every column is uniform. Replacing the columns one at a time gives $aL+1$ hybrids interpolating between $\cK_n^{1}$ and $\cK_n^0$. Thus, a distinguisher with advantage $\epsilon$ has advantage at least $\epsilon/(aL)$ on some pair of adjacent hybrids, which differ in a single column corresponding to a single LWE instance with a binary secret. Given the instance, the reduction inserts it as that column, samples the other columns with the same $B^\top$ and $t$ itself, and runs the distinguisher, with a $\poly(n)$ overhead.

\end{proof}

For the large-rank regime, we use a generator whose output is hidden from efficient circuits given its label, while an unbounded party recovers the output from the label. The generator hashes LWE samples with binary secrets; its seed consists of the secret and of the randomness of the following error sampler, and is thus linear in the LWE dimension.

\begin{lemma}[Gaussian sampler with a short seed]
\label{lem:gaussian sampler}
Let $m = \lceil 4\ell/\log{\ell}\rceil$, $\sigma = \ell$, and $\ell^{10} \leq q \leq 2\ell^{10}$. Then there is a function $\mathrm{Samp} : \{0,1\}^{8\ell} \to \bZ_q^m$ computable by a circuit of size $\poly(\ell)$ such that, for all sufficiently large $\ell$, $|\rIm(\mathrm{Samp})| \leq 2^{7\ell}$ and the output on a uniform input is within total variation distance $2^{-\ell}$ of $\bD_{q,\sigma}^m$.
\end{lemma}

\begin{proof}[Proof of \Cref{lem:gaussian sampler}]

Set $R \coloneqq \lceil 4\ell^{3/2}\rceil < q/2$, and identify the centered interval $[-R, R]$ with its reduction modulo $q$. The Gaussian tail bound and a union bound over the $m$ coordinates give $\Pr_{\delta \sim \bD_{q,\sigma}^m}[\delta \notin [-R, R]^m] \leq 2me^{-8\ell} \leq 2^{-2\ell}$. The box has $(2R+1)^m = 2^{(6+o(1))\ell} \leq 2^{7\ell}$ points, as $\log(2R+1) = \f{3}{2}\log{\ell} + \cO(1)$.
Let $\bD'$ be $\bD_{q,\sigma}^m$ conditioned on the box, a product distribution. Its one-coordinate probabilities can be rounded in time $\poly(\ell)$ to nonnegative multiples of $2^{-4\ell}$ summing to one, each changing by at most $2^{-4\ell}$. Assign to the points of the box, ordered lexicographically, consecutive subintervals of $[0,1)$ of lengths given by the product of the rounded one-coordinate probabilities, and let $\mathrm{Samp}$ map $r \in \{0,1\}^{8\ell}$ to the point whose subinterval contains $r2^{-8\ell}$. By the product structure, the point is found coordinate by coordinate from the cumulative one-coordinate probabilities, with $\poly(\ell)$ operations on $\poly(\ell)$-bit integers and without enumerating the box. The rounding contributes at most $m(2R+1)2^{-4\ell} \leq 2^{-2\ell}$ to the total variation distance, the grid of spacing $2^{-8\ell}$ contributes at most $(2R+1)^m2^{-8\ell} = 2^{-(2-o(1))\ell}$, and the tail contributes at most $2^{-2\ell}$. The total variation distance is thus at most $2^{-\ell}$ for all sufficiently large $\ell$.

\end{proof}

For bitstrings $u$ and $u'$, let $u\|u'$ denote their concatenation. A family of functions $\{h_k : \cX \to \cY\}_{k \in \cK}$ is \emph{$r$-wise independent} if, for every $r$ distinct $x_1, \ldots, x_r \in \cX$ and a uniformly sampled $k \in \cK$, the values $h_k(x_1), \ldots, h_k(x_r)$ are independent and uniformly distributed over $\cY$. 
For every $r$ and all $\cX, \cY$ whose cardinalities are powers of two, there exist such families with keys of length $r\max\{\log|\cX|, \log|\cY|\}$ and evaluation time $\poly(\log|\cX|, \log|\cY|, r)$~{\cite[Corollary~3.34]{Vadhan2012Pseudorandomness}}.

Fix $\ell$, and let $m$, $\sigma$, and $q$ be as in \Cref{lem:gaussian sampler}, with $q$ the least prime obeying $q \geq \ell^{10}$, so that $q \leq 2\ell^{10}$ by Bertrand's postulate~\cite{Hardy2008AnIntroductiontotheTheoryofNumbers} and $(\ell, m, q, \sigma)$ obey Eq.~\eqref{eq:lwe params}. Set
\begin{align}
\label{eq:generator lengths}
    d \coloneqq 9\ell,
    \ \ \ \ d_V \coloneqq 18\ell,
    \ \ \ \ d_z \coloneqq \lfloor m\log{q}\rfloor - 2\ell - d_V - d,
\end{align}
where $d_z \geq \ell$ for large $\ell$ as $m\log{q} \geq 40\ell$. Identifying $\bZ_q^m$ injectively with $\{0,1\}^{\lceil m\log{q}\rceil}$, let $\{h_{k^{\rm hash}} : \bZ_q^m \to \{0,1\}^{d_V + d + d_z}\}_{k^{\rm hash}}$ be a pairwise independent function family.
For an output length $d_x$, set $N \coloneqq \lceil d_x/d_z\rceil$, sample $A_i \sim \bU_q^{m\times\ell}$ and $k_i^{\rm hash}$ uniformly for $i = 1, \ldots, N$, independently, and set $k^{\rm gen} \coloneqq (A_i, k_i^{\rm hash})_{i=1}^N$.
For a seed $\omega_1 \in \{0,1\}^{d}$, write $\omega_i = s_i\|r_i$ with $s_i \in \{0,1\}^{\ell}$ and $r_i \in \{0,1\}^{8\ell}$, and recursively
\begin{align}
\label{eq:generator stage}
    h_{k_i^{\rm hash}}\big(A_is_i + \mathrm{Samp}(r_i)\big) = V_i\|\omega_{i+1}\|z_i,
    \ \ \ \ V_i \in \{0,1\}^{d_V}, \ \omega_{i+1} \in \{0,1\}^{d}, \ z_i \in \{0,1\}^{d_z},
\end{align}
for $i = 1, \ldots, N$. The label and the output of the generator are $V_{k^{\rm gen}}(\omega_1) \coloneqq V_1$ and $x_{k^{\rm gen}}(\omega_1) \coloneqq$ the first $d_x$ bits of $z_1\|\cdots\|z_N$.

\begin{lemma}[Labeled generator from LWE with binary secrets]
\label{lem:labeled generator}
The functions $V_{k^{\rm gen}}$ and $x_{k^{\rm gen}}$ above satisfy the following statements.
\begin{enumerate}[label=\textnormal{(\Roman*)}]
    \item The description $k^{\rm gen}$ is sampleable, and $V_{k^{\rm gen}}$ and $x_{k^{\rm gen}}$ are computable, by circuits of size $\poly(d_x, \ell)$.
    \item Except with probability $2^{-\ell}$ over $k^{\rm gen}$, the function $V_{k^{\rm gen}}$ takes at least $2^\ell$ values, and there is a function $\mathrm{Rec}_{k^{\rm gen}}$ such that $\mathrm{Rec}_{k^{\rm gen}}(V_{k^{\rm gen}}(\omega)) = x_{k^{\rm gen}}(\omega)$ for every $\omega \in \{0,1\}^d$.
    \item Let $k \geq 1$, let $\omega^{(1)}, \ldots, \omega^{(k)}$ be independent and uniform on $\{0,1\}^{d}$, and let $u_1, \ldots, u_k$ be independent and uniform on $\{0,1\}^{d_x}$. Any quantum circuit that, given $k^{\rm gen}$, distinguishes $\big(V_{k^{\rm gen}}(\omega^{(j)}), x_{k^{\rm gen}}(\omega^{(j)})\big)_{j=1}^k$ from $\big(V_{k^{\rm gen}}(\omega^{(j)}), u_j\big)_{j=1}^k$ with advantage $\epsilon$ yields a circuit solving the $\mathrm{LWE}_{\ell, m, q, \sigma}^{\rm bin}$ decision problem with advantage $\epsilon/(2kN) - 2^{1-\ell}$, with a $\poly(k, d_x, \ell)$ overhead in size.
\end{enumerate}
\end{lemma}

\begin{proof}[Proof of \Cref{lem:labeled generator}]

(I) Each of the $N = \cO(1 + d_x/\ell)$ stages evaluates $\mathrm{Samp}$, a matrix-vector product over $\bZ_q$, and a pairwise independent function, all of size $\poly(\ell)$.

(II) For a matrix $A$, let $\cR_A \coloneqq \{As + \mathrm{Samp}(r) : s \in \{0,1\}^\ell, \ r \in \{0,1\}^{8\ell}\}$, whose cardinality is at most $2^{\ell}2^{7\ell} = 2^{8\ell}$ by \Cref{lem:gaussian sampler}. Let $h^V_{k^{\rm hash}}$ denote the first $d_V$ bits of $h_{k^{\rm hash}}$. By pairwise independence and a union bound over the pairs, $h^V_{k_1^{\rm hash}}$ is injective on $\cR_{A_1}$ except with probability $|\cR_{A_1}|^2 2^{-d_V} \leq 2^{-2\ell}$. Likewise, $A_1$ is injective on $\{0,1\}^\ell$ except with probability $2^{2\ell}q^{-m} \leq 2^{-38\ell}$.
On these events, $V_{k^{\rm gen}}$ takes at least $2^\ell$ values, since $A_1s + \mathrm{Samp}(r)$ at a fixed $r$ takes $2^\ell$ values in $\cR_{A_1}$. Given $V_1$, the function $\mathrm{Rec}_{k^{\rm gen}}$ finds the unique $y_1 \in \cR_{A_1}$ with $h^V_{k_1^{\rm hash}}(y_1) = V_1$, reads $\omega_2$ and $z_1$ off $h_{k_1^{\rm hash}}(y_1)$, and evaluates the remaining stages from $\omega_2$. Seeds producing the same $y_1$ produce the same label and output, so the seed itself need not be recovered.

(III) We first consider one stage on a uniform seed $\omega_i$, jointly with uniformly sampled $(A_i, k_i^{\rm hash})$. Replacing $\mathrm{Samp}(r_i)$ by a sample from $\bD_{q,\sigma}^m$ changes the distribution by at most $2^{-\ell}$ by \Cref{lem:gaussian sampler}; the pair $(A_i, A_is_i + \delta)$ is then an LWE instance with binary secret, so replacing $A_is_i + \delta$ by a uniform $u \in \bZ_q^m$ is detected with advantage at most that of the resulting LWE distinguisher; the leftover hash lemma with pairwise independence and $\log|\bZ_q^m| \geq d_V + d + d_z + 2\ell$ then bounds the distance of $(k_i^{\rm hash}, h_{k_i^{\rm hash}}(u))$ from uniform by $2^{-\ell}$. Hence, the output $V_i\|\omega_{i+1}\|z_i$ of the stage is indistinguishable from a uniform string up to the LWE advantage and $2^{1-\ell}$.
We replace the stages by uniform strings in their forward order within each of the $k$ copies, giving $kN$ switches. Once the stages before $i$ in a copy have been replaced, its $\omega_i$ is uniform and independent of everything else, so the one-stage argument applies; the same $(A_i, k_i^{\rm hash})$ serves all copies, since the reduction for one switch samples the stage-$i$ values of the other copies itself, with fresh secrets and errors under the same $A_i$. After the LWE replacement, the uniform input of the switched copy is independent of the other copies conditioned on $(A_i, k_i^{\rm hash})$, and the leftover hash bound is unchanged in the presence of their values. Thus, $\big(V_{k^{\rm gen}}(\omega^{(j)}), x_{k^{\rm gen}}(\omega^{(j)})\big)_{j=1}^k$ is indistinguishable from all labels and outputs uniform, up to $kN$ times the LWE advantage plus $kN2^{1-\ell}$. Discarding the outputs shows the same for the labels alone, and appending independent uniform outputs to them and applying the triangle inequality doubles the bound. A distinguisher with advantage $\epsilon$ therefore has advantage at least $\epsilon/(2kN) - 2^{1-\ell}$ against some switch, which the reduction above turns into an LWE distinguisher with a $\poly(k, d_x, \ell)$ overhead.

\end{proof}


\subsection{Public-key pseudoentanglement}
\label{sec:public key pseudoentanglement}

We describe the public-key pseudoentanglement construction of Ref.~\cite{bouland2024PublicKeyPseudoentanglement} in the form we use: a pair of phase-state families built from a lossy function family of \Cref{def:lossy function}.
The following lemma gives the public-key pseudoentanglement result for a bipartite pure state on $2\mu$ qubits, split evenly across the two registers.

\begin{lemma}[Public-key pseudoentanglement on $2\mu$ qubits; adapted from Ref.~{\cite[Definition~3.12, Theorem~3.13, and Lemma~3.14]{bouland2024PublicKeyPseudoentanglement}}]
\label{lem:small register pke}
Let $\mu = \poly(n)$ obey $\mu = \omega(\log{n})$, and let $\{g_k\}_k$ be a lossy function family of \Cref{def:lossy function} at $(\mu, \tau)$ that is $(\poly(n), \negl(n))$-mode-indistinguishable. Let $\sA$ and $\sB$ be registers of $\mu$ qubits each. Then, for $b \in \{0, 1\}$, there exist efficiently sampleable ensembles $\scC_b$ of $\poly(n)$-size quantum circuits on $\sA\sB$, each supplied with a classical description, such that the following statements hold for $\ket{\varphi_b}^{\sA\sB} \coloneqq C_b^{\sA\sB}\ket{0^{2\mu}}^{\sA\sB}$, where $C_b^{\sA\sB} \sim \scC_b$:
\begin{enumerate}[label=\textnormal{(\roman*)}]
    \item $\log{\rank(\varphi_1^{\sA})} \leq \tau$, and hence $H_1(\sA)_{\varphi_1} \leq \tau$;
    \item $\widetilde{H}_\alpha(\sA)_{\varphi_0} \geq \f{\mu}{4}$ for every $\alpha \in (0, \infty]$;
    \item The ensembles $\scC_0$ and $\scC_1$ are computationally indistinguishable to quantum circuits of size $\poly(n)$, even when their classical descriptions are provided as input.
\end{enumerate}
Statements \textnormal{(i)} and \textnormal{(ii)} hold with probability $1 - \negl(n)$ over the choice of $C_b^{\sA\sB} \sim \scC_b$.
\end{lemma}

\begin{proof}[Proof of \Cref{lem:small register pke}]

Encode the outputs of the functions $g_k$ as bit strings of a common length $\poly(n)$ under a fixed injective encoding. Sample $k^{\rm fin}$ uniformly from the key set of a $4$-wise independent function family $\{h_{k^{\rm fin}}\}$ from the pairs $(y, j)$ of such a string $y$ and $j \in \{0,1\}^{\mu}$ to $\{0,1\}$, sample $k^{\rm rep} \leftarrow \cK_n^b$, and set $\hat{k} \coloneqq (k^{\rm rep}, k^{\rm fin})$.
Let $C_b^{\sA\sB}$ be the circuit that applies a Hadamard gate to each qubit of $\ket{0^{2\mu}}^{\sA\sB}$ and then the phase oracle of $s_{\hat{k}}$, preparing
\begin{align}
    \ket{\varphi_b}^{\sA\sB} = \f{1}{2^\mu} \sum_{i, j \in \{0,1\}^{\mu}} (-1)^{s_{\hat{k}}(i\|j)}\ket{i}^{\sA}\ket{j}^{\sB},
    \ \ \ \ \ \ 
    s_{\hat{k}}(i\|j) \coloneqq h_{k^{\rm fin}}\big(g_{k^{\rm rep}}(i)\big\|j\big).
\end{align}
Since all component functions are computable by circuits of size $\poly(n)$ and $\mu = \poly(n)$, the circuit $C_b^{\sA\sB}$ has size $\poly(n)$ and is efficiently described by $\hat{k}$.
Let $T_{\hat{k}} \in \{\pm 1\}^{2^\mu \times 2^\mu}$ have entries $(T_{\hat{k}})_{ij} \coloneqq (-1)^{s_{\hat{k}}(i\|j)}$, so that $\varphi_b^{\sA} = T_{\hat{k}}T_{\hat{k}}^{\top}/2^{2\mu}$.

(i) The $i$-th row of $T_{\hat{k}}$ depends on $i$ only through $g_{k^{\rm rep}}(i)$, so $T_{\hat{k}}$ has at most $|\rIm(g_{k^{\rm rep}})|$ distinct rows.
As $\varphi_1^{\sA} = T_{\hat{k}}T_{\hat{k}}^{\top}/2^{2\mu}$, this gives $\rank(\varphi_1^{\sA}) = \rank(T_{\hat{k}}) \leq |\rIm(g_{k^{\rm rep}})|$, and hence $H_1(\sA)_{\varphi_1} \leq \log\rank(\varphi_1^{\sA}) \leq \log|\rIm(g_{k^{\rm rep}})|$, which (L2) of \Cref{def:lossy function} bounds by $\tau$ with probability $1-\negl(n)$.

(ii) We follow the fourth-moment computation of Ref.~{\cite[Lemma~3.14]{bouland2024PublicKeyPseudoentanglement}}. Condition on a $k^{\rm rep} \leftarrow \cK_n^0$ at which $g_{k^{\rm rep}}$ is injective, which by (L1) fails with probability $\negl(n)$. The strings $g_{k^{\rm rep}}(i)\|j$ are then pairwise distinct over $(i, j)$, so the entries of $T_{\hat{k}}$ are $4$-wise independent uniform signs over the choice of $k^{\rm fin}$. Writing $D \coloneqq 2^\mu$, we obtain
\begin{align}
    \bE_{k^{\rm fin}}\big[\tr\big((\varphi_0^{\sA})^2\big)\big]
    = \f{1}{D^4}\sum_{i, i', j, j' \in \{0,1\}^{\mu}} \bE_{k^{\rm fin}}\big[(T_{\hat{k}})_{ij}(T_{\hat{k}})_{i'j}(T_{\hat{k}})_{i'j'}(T_{\hat{k}})_{ij'}\big]
    = \f{2D^3 - D^2}{D^4}
    \leq \f{2}{D},
\end{align}
where the expectation vanishes unless $i = i'$ or $j = j'$, in which case it equals $1$. Markov's inequality gives $\tr((\varphi_0^{\sA})^2) \leq D^{-1/2}$ with probability $1 - 2D^{-1/2}$, that is, $-\log\|\varphi_0^{\sA}\|_2 \geq \f{\mu}{4}$ with probability $1 - 2^{1-\mu/2} = 1 - \negl(n)$ by $\mu = \omega(\log{n})$. As $\|\rho\|_\infty \leq \|\rho\|_2$ for any state $\rho$, we obtain $H_{\rm min}(\sA)_{\varphi_0} = -\log\|\varphi_0^{\sA}\|_\infty \geq \f{\mu}{4}$. Claim (ii) follows from $\widetilde{H}_\alpha(\sA)_{\varphi_0} \geq H_{\rm min}(\sA)_{\varphi_0}$, which holds by the monotonicity of the R\'{e}nyi entropy in $\alpha$.

(iii) The two descriptions differ only in $k^{\rm rep}$, and $k^{\rm fin}$ is sampled identically in the two ensembles. A distinguisher of size $\poly(n)$ for the descriptions therefore yields one of size $\poly(n)$ for $\cK_n^0$ and $\cK_n^1$ with the same advantage, which is $\negl(n)$ by the mode indistinguishability assumed.

\end{proof}

Under (H1), whose scope is $m = \poly(\ell)$, placing the pair of \Cref{lem:small register pke} on $\Theta(n)$ qubits with the family of \Cref{lem:small modulus lossy}, whose number of equations is $m = \Theta((\mu + \ell)/\log{\ell})$ at input length $\mu$, forces $\ell = n^{\Omega(1)}$, and the low-entanglement side is then $\tau = \ell\log(2(a+1)) = n^{\Omega(1)}$. Under (H2), where $m$ may be subexponential in $\ell$, the same register size admits a polylogarithmic $\ell$.


\subsection{Construction of the state family}
\label{sec:construction of the state family state aware}

We construct the family of states for our description-aware converse bounds, using the cryptographic primitives introduced so far. In \Cref{sec:parameters state aware}, we specify the parameters for invoking the cryptographic tools of \Cref{sec:cryptographic tools}; in \Cref{sec:construction state aware}, we use these tools to construct the states.

\subsubsection{Parameters}
\label{sec:parameters state aware}

Throughout, we assume (H1) and (H2) of \Cref{assumpt:lwe assumptions}. We invoke \Cref{lem:small register pke} at three register sizes, $\mu$, $\mu'$, and $\hat{\mu}$, with the lossy function family of \Cref{lem:small modulus lossy} at $(\ell, a, L)$, $(\ell', a', L')$, and $(\hat{\ell}, \hat{a}, \hat{L})$, respectively; the first application uses (H1), the second (H2), and the third reuses one of them. We then fix the range of the LWE dimension $\lambda$ of the labeled generator of \Cref{lem:labeled generator}.

Let $f$ be an efficiently computable function with $f = \omega(1)$ and $\lceil f^2\rceil\lceil\log{n}\rceil \leq n$, and set
\begin{align}
\label{eq:instantiations}
    a = \lceil f^2\rceil,
    \ \ \ \ L = 2^{\lceil\log{n}\rceil},
    \ \ \ \ \ell = \bigg\lceil \f{f\log{n}}{\log(2(a+1))} \bigg\rceil,
    \ \ \ \ \mu = a\log{L},
\end{align}
with $p$, $m$, $\sigma$, and $q$ as in Eq.~\eqref{eq:lossy params}. These meet the conditions of \Cref{lem:small modulus lossy}: $\ell = \Theta(f\log{n}/\log(2+f)) = \omega(\log{n})$, and $\ell, aL = \poly(n)$. They also place the $\mathrm{LWE}_{\ell, m, q, \sigma}^{\rm bin}$ decision problem in the scope of (H1): $m, q = \poly(\ell)$ by $f, \log{n} \leq \ell$ for large $n$, and $m\log{q}/\ell = \Theta(f\log(2+f))$ by $m = \Theta(f^2\log{n}/\log{\ell})$ and $\log{q} = \Theta(\log{\ell})$, so that the exponent in (H1) is $\Theta(f\log{n}/\log^2(2+f)) = \omega(\log{n})$. Every quantum circuit of size $\poly(n)$ thus has advantage $\negl(n)$ in the problem, and (III) of \Cref{lem:small modulus lossy} makes the family $(\poly(n), \negl(n))$-mode-indistinguishable at
\begin{align}
\label{eq:first leakage}
    (\mu, \tau) = \big(a\log{L}, \ \ell\log(2(a+1))\big),
    \ \ \ \ \tau = (1+o(1))f\log{n} = o(\mu),
\end{align}
where $\mu = \Theta(f^2\log{n}) = \omega(\log{n})$ and $\mu \leq n$. Then \Cref{lem:small register pke} yields two computationally indistinguishable ensembles of circuits on $2\mu$ qubits, one satisfying $\widetilde{H}_\alpha \geq \mu/4$ for every $\alpha \in (0, \infty]$ and the other satisfying $H_1 \leq \tau$.

For the second application, let $\gamma_1 > \gamma_2$ be the constants of (H2), and set
\begin{align}
\label{eq:amplifier params}
    \ell' = \big\lceil \log^{1/\gamma_2}{n} \big\rceil,
    \ \ \ \ a' = n,
    \ \ \ \ L' = 2,
    \ \ \ \ \mu' = n,
\end{align}
with $p'$, $m'$, $\sigma'$, and $q'$ as in Eq.~\eqref{eq:lossy params}. The conditions of \Cref{lem:small modulus lossy} hold by $\ell' = \omega(\log{n})$, which follows from $\gamma_2 < 1/2$, and $\ell', a'L' = \poly(n)$. The problem $\mathrm{LWE}_{\ell', m', q', \sigma'}^{\rm bin}$ lies in the scope of (H2): $m' = \Theta(n/\log{\ell'}) = 2^{\cO({\ell'}^{\gamma_2})}$ by ${\ell'}^{\gamma_2} \geq \log{n}$, and every quantum circuit of size $\poly(n) = 2^{\cO({\ell'}^{\gamma_2})}$ has advantage at most $2^{-\Omega({\ell'}^{\gamma_1})} = \negl(n)$ by ${\ell'}^{\gamma_1} \geq \log^{\gamma_1/\gamma_2}{n} = \omega(\log{n})$. The family is thus $(\poly(n), \negl(n))$-mode-indistinguishable at $(\mu', \tau')$ with
\begin{align}
\label{eq:amplifier leakage}
    \tau' = \ell'\log(2(n+1)) = \cO\big(\log^{1/\gamma_2+1}{n}\big),
\end{align}
and \Cref{lem:small register pke} yields two computationally indistinguishable ensembles of circuits on $2n$ qubits, one satisfying $\widetilde{H}_\alpha \geq n/4$ for every $\alpha \in (0, \infty]$ and the other satisfying $H_1 \leq \tau'$.

For the third application, let $\hat{\mu}$ be an integer with $0 \leq \hat{\mu} \leq n$, and set
\begin{align}
\label{eq:heavy params}
    (\hat{\ell}, \hat{a}, \hat{L})
    = \begin{cases}
        (\ell, a, L) & \ \text{if} \ \ \hat{\mu} = \mu, \\
        (\ell', \hat{\mu}, 2) & \ \text{otherwise},
    \end{cases}
\end{align}
so that $\hat{a}\log{\hat{L}} = \hat{\mu}$ in either case. This application is used only at $\hat{\mu} \geq \hat{\ell}$, where we take $\hat{p}$, $\hat{m}$, $\hat{\sigma}$, and $\hat{q}$ as in Eq.~\eqref{eq:lossy params}. In the first case, it is the first application, and in the second case it lies in the scope of (H2) as the second application does, by $\hat{m} \leq m'$ and $\hat{a}\hat{L} = 2\hat{\mu} \leq 2n$. Hence, at $\hat{\mu} \geq \hat{\ell}$, \Cref{lem:small register pke} yields two computationally indistinguishable ensembles of circuits on $2\hat{\mu}$ qubits, one satisfying $\widetilde{H}_\alpha \geq \hat{\mu}/4$ for every $\alpha \in (0, \infty]$ and the other satisfying $H_1 \leq \hat{\tau}$, where
\begin{align}
\label{eq:heavy leakage}
    \hat{\tau} = \hat{\ell}\log(2(\hat{a}+1))
    = \begin{cases}
        \tau & \ \text{if} \ \ \hat{\mu} = \mu, \\
        \ell'\log(2(\hat{\mu}+1)) & \ \text{otherwise}.
    \end{cases}
\end{align}

Finally, the labeled generator of \Cref{lem:labeled generator} is specified by its LWE dimension $\lambda$, which either equals $0$, in which case the generator is not used, or satisfies
\begin{align}
\label{eq:lambda range}
    \lambda = \omega(\log{n}),
    \ \ \ \ \lambda \leq n.
\end{align}
At $\lambda > 0$, the $\mathrm{LWE}^{\rm bin}$ problem of the generator of \Cref{lem:labeled generator} lies in the scope of (H1): its number of equations is $\lceil 4\lambda/\log{\lambda}\rceil$, its standard deviation is $\lambda$, and its modulus lies between $\lambda^{10}$ and $2\lambda^{10}$, at which the argument of the logarithm in (H1) is $\Theta(1)$ and the exponent is $\Theta(\lambda)$. Hence, the problem is $\big(2^{\Omega(\lambda)}, \ 2^{-\Omega(\lambda)}\big)$-hard. Applying (III) of \Cref{lem:labeled generator} at an output length $d_x = \poly(n)$ and $k = \poly(n)$ copies, where $N = \poly(n)$, every quantum circuit of size $\poly(n) = 2^{o(\lambda)}$ has advantage at most $2kN\big(2^{-\Omega(\lambda)} + 2^{1-\lambda}\big) = \negl(n)$ in distinguishing the labeled outputs from the labels with uniform strings.


\subsubsection{Construction}
\label{sec:construction state aware}

Fix constants $\alpha_\rc, \alpha_\rs$ with $1 < \alpha_\rc < \alpha_\rs \leq \infty$, take the parameters of \Cref{sec:parameters state aware}, and set
\begin{align}
\label{eq:master params}
    \kappa = \f{\alpha_\rs - \alpha_\rc}{\alpha_\rc(\alpha_\rs - 1)} \in (0, \, 1/\alpha_\rc].
\end{align}

The family of states is built from three branches, mixed under a locally readable flag and then dressed by a labeled Heisenberg--Weyl displacement. The first branch is the spike state $\ket{\xi_{\mathrm{sp}}}^{\sG_\sA\sG_\sB\sG_\sE}$ of \Cref{lem:spikestate} at rank $r_\rs$, where $r_\rs$ obeys either $r_\rs = 1$, in which case this branch carries zero weight and is absent, or $\omega(1) \leq r_\rs \leq \poly(n)$. The second is $\ket{\phi_b}^{\sH_\sA\sH_\sB} \coloneqq C'_b\ket{0}$ on $2n$ qubits, with $C'_b$ drawn from the ensembles $\scC'_b$ of \Cref{lem:small register pke} at the parameters of Eq.~\eqref{eq:amplifier params}, with $\mu' = n$.
The third is the two-scale state $\ket{\zeta_{b\hat{b}}}^{\sL_\sA\sL_\sB}$ of \Cref{lem:two scale branch} at the $\eta$ of Eq.~\eqref{eq:master weights} below.
Its heavy component is $\ket{\hat{\phi}_{\hat{b}}}^{\hat{\sM}_\sA\hat{\sM}_\sB} \coloneqq \hat{C}_{\hat{b}}^{\hat{\sM}_\sA\hat{\sM}_\sB}\ket{0^{2\hat{\mu}}}^{\hat{\sM}_\sA\hat{\sM}_\sB}$ on registers $\hat{\sM}_\sA, \hat{\sM}_\sB$ of $\hat{\mu}$ qubits each.
At $\hat{\mu} \geq \hat{\ell}$, the circuit $\hat{C}_{\hat{b}}$ is drawn from the ensembles $\hat{\scC}_{\hat{b}}$ of \Cref{lem:small register pke} at the parameters of Eq.~\eqref{eq:heavy params}.
At $\hat{\mu} < \hat{\ell}$, we take $\hat{C}_0 = \hat{C}_1$ to be the fixed circuit preparing $\ket{\Phi_2}^{\otimes\hat{\mu}}$, the registers being empty at $\hat{\mu} = 0$; the branch is then independent of $\hat{b}$.
Its light component is the state $\ket{\phi_b}^{\sN_\sA\sN_\sB}$ prepared by the same circuit $C'_b$ as the second branch, on registers of $n$ qubits each.

We set their weights as
\begin{align}
\label{eq:master weights}
    (\eta_0, \eta_1, \eta_2)
    = \begin{cases}
        (1 - \eta, \ \eta, \ 0) & \ \text{if} \ \ r_\rs \geq 2, \\
        (0, \ 0, \ 1) & \ \text{if} \ \ r_\rs = 1,
    \end{cases}
    \ \ \ \ \ \ \ \ \ 
    \eta = \log^{-(1/\gamma_2+2)}{n}.
\end{align}
In the first case, the third branch carries no weight and is absent. In the second case, the first two branches carry no weight; the second is not lost, as it reappears as the light component of the third through Eq.~\eqref{eq:master two scale}.
For $b, \hat{b} \in \{0,1\}$, let $\ket{\Psi_{b\hat{b}}}^{\sI_\sA\sI_\sB\sI_\sE}$ be the flagged mixture built from these three branches and weights, where $\sI_\sA = \sF_\sA\sG_\sA\sH_\sA\sL_\sA$ and $\sI_\sB = \sF_\sB\sG_\sB\sH_\sB\sL_\sB$, while $\sI_\sE = \sF_\sE\sG_\sE$; that is,
\begin{align}
    \ket{\Psi_{b\hat{b}}}^{\sI_\sA\sI_\sB\sI_\sE}
    &= \sqrt{\eta_0} \ket{0}^{\sF_\sA\sF_\sB\sF_\sE}\ket{\xi_{\mathrm{sp}}}^{\sG_\sA\sG_\sB\sG_\sE}\ket{0}^{\sH_\sA\sH_\sB\sL_\sA\sL_\sB}
    + \sqrt{\eta_1}\ket{1}^{\sF_\sA\sF_\sB\sF_\sE}\ket{0}^{\sG_\sA\sG_\sB\sG_\sE}\ket{\phi_b}^{\sH_\sA\sH_\sB}\ket{0}^{\sL_\sA\sL_\sB}  \notag\\
    &\hspace{6pc} + \sqrt{\eta_2} \ket{2}^{\sF_\sA\sF_\sB\sF_\sE}\ket{0}^{\sG_\sA\sG_\sB\sG_\sE\sH_\sA\sH_\sB}\ket{\zeta_{b\hat{b}}}^{\sL_\sA\sL_\sB}.
\end{align}
Here the state in the third branch is
\begin{align}
\label{eq:master two scale}
    \ket{\zeta_{b\hat{b}}}^{\sL_\sA\sL_\sB}
    = \sqrt{1-\eta}\ket{0}^{\sD_\sA\sD_\sB}\ket{\hat{\phi}_{\hat{b}}}^{\hat{\sM}_\sA\hat{\sM}_\sB}\ket{0}^{\sN_\sA\sN_\sB}
    + \sqrt{\eta}\ket{1}^{\sD_\sA\sD_\sB}\ket{0}^{\hat{\sM}_\sA\hat{\sM}_\sB}\ket{\phi_b}^{\sN_\sA\sN_\sB}.
\end{align}
Since $\log{r_\rs} = \cO(\log{n})$ and $\hat{\mu} \leq n$, each of $\sI_\sA$ and $\sI_\sB$ carries $\Theta(n)$ qubits, and we pad $\sI_\sA$ with ancillas $\ket{0}$ so that both carry the same number $n' = \Theta(n)$ of qubits for simplicity.

For a bitstring $z = (z_1, \ldots, z_{n'}) \in \{0,1\}^{n'}$, let $X(z)=\bigotimes_{i=1}^{n'}X_i^{z_i}$ and $Z(z)=\bigotimes_{i=1}^{n'}Z_i^{z_i}$ be the qubit Heisenberg--Weyl operators on $n'$ qubits, acting as $X(z)\ket{y}=\ket{y\oplus z}$ and $Z(z)\ket{y}=(-1)^{z\cdot y}\ket{y}$ for $y\in\{0,1\}^{n'}$. 
Then, a displacement $x = (x_A, x_B) \in \{0,1\}^{4n'}$ with $x_A = (x_A^X, x_A^Z)$ and $x_B = (x_B^X, x_B^Z)$, where $x_A^X, x_A^Z, x_B^X, x_B^Z \in \{0,1\}^{n'}$, defines
\begin{align}
    \label{eq:def of w}
    W^{\sI_\sA\sI_\sB}(x) = W^{\sI_\sA}(x_A) \otimes W^{\sI_\sB}(x_B),
\end{align}
with the local factors $W^{\sI_\sA}(x_A) = \big(X(x_A^X)Z(x_A^Z)\big)^{\sI_\sA}$ and $W^{\sI_\sB}(x_B) = \big(X(x_B^X)Z(x_B^Z)\big)^{\sI_\sB}$.

We then dress the state with a labeled displacement. Recall from Eq.~\eqref{eq:lambda range} that $\lambda$ is either $0$ or obeys $\omega(\log{n}) = \lambda \leq n$; here we further require $\lambda \leq n'$. At $\lambda = 0$, the label is the displacement itself, in plaintext. At $\lambda > 0$, the displacement is the output of the labeled generator of \Cref{lem:labeled generator} at LWE dimension $\lambda$ and output length $d_x = 4n'$, and the label is its label, so that the number of distinct displacements is at most $2^{d} = 2^{9\lambda}$, set by the seed length rather than by the displacement length $4n'$. We sample the description $k^{\rm gen}$ of the generator and write $x(\omega) \coloneqq x_{k^{\rm gen}}(\omega) \in \{0,1\}^{4n'}$ and $V(\omega) \coloneqq V_{k^{\rm gen}}(\omega)$ for $\omega \in \{0,1\}^{d}$.

Sampling $\omega \in \{0,1\}^{d}$ uniformly, we define
\begin{align}
\label{eq:master purification}
    \ket{\Upsilon_{b\hat{b}}}^{\sA\sB\sE}
    = \f{1}{\sqrt{2^{d}}}\sum_{\omega \in \{0,1\}^{d}} W^{\sI_\sA\sI_\sB}(x(\omega))\ket{\Psi_{b\hat{b}}}^{\sI_\sA\sI_\sB\sI_\sE}\ket{V(\omega)}^{\sK_\sA}\ket{V(\omega)}^{\sK_\sB}\ket{\omega}^{\sJ_\sE},
\end{align}
where $\sA = \sI_\sA\sK_\sA$, $\sB = \sI_\sB\sK_\sB$, and $\sE = \sI_\sE\sJ_\sE$. 
Let $\cV \coloneqq \rIm(V)$, and, for $v \in \cV$, let $p_v \coloneqq 2^{-d}\big|\{\omega : V(\omega) = v\}\big|$. Except with probability $2^{-\lambda}$ over $k^{\rm gen}$, (II) of \Cref{lem:labeled generator} gives $|\cV| \geq 2^{\lambda}$ and a function $\mathrm{Rec}_{k^{\rm gen}}$ with $\mathrm{Rec}_{k^{\rm gen}}(V(\omega)) = x(\omega)$ for every $\omega$. On this event, let $x_v \coloneqq \mathrm{Rec}_{k^{\rm gen}}(v)$; tracing out $\sE$ gives the state 
\begin{align}
\label{eq:master mixed}
    \Upsilon_{b\hat{b}}^{\sA\sB}
    = \sum_{v \in \cV} p_v \big(W(x_v)\Psi_{b\hat{b}}W(x_v)^\dag\big)^{\sI_\sA\sI_\sB}\otimes\ketbra{v}{v}^{\sK_\sA}\otimes\ketbra{v}{v}^{\sK_\sB}.
\end{align}
The state $\Upsilon_{b\hat{b}}^{\sA\sB}$ is fully specified by the classical description $(C'_b, \hat{C}_{\hat{b}}, k^{\rm gen})$. An unbounded party recovers the displacement $x_v$ from either label $v$ by $\mathrm{Rec}_{k^{\rm gen}}$, while by (III) of \Cref{lem:labeled generator} and the paragraph after Eq.~\eqref{eq:lambda range}, no circuit of size $\poly(n)$ distinguishes the labels with the displacements from the labels with uniform strings, even from $\poly(n)$ copies.
When $\lambda = 0$, for a prescribed integer $M$ such that $1 \leq M \leq \poly(n)$, we instead let $\cX_M$ consist of the lexicographically first $M$ strings of $\{0,1\}^{4n'}$ and replace the sum of Eq.~\eqref{eq:master purification} by the uniform sum over $x \in \cX_M$, with $\ket{x}^{\sK_\sA}\ket{x}^{\sK_\sB}\ket{x}^{\sJ_\sE}$ in place of $\ket{V(\omega)}^{\sK_\sA}\ket{V(\omega)}^{\sK_\sB}\ket{\omega}^{\sJ_\sE}$. 
Then, Eq.~\eqref{eq:master mixed} holds with $\cV = \cX_M$, $p_v = 1/M$, and $x_v = v$. At $M = 1$ the displacement is the fixed $x = 0^{4n'}$, for which $W(x) = \bI$, and $\ket{\Upsilon_{b\hat{b}}}$ is $\ket{\Psi_{b\hat{b}}}$ tensored with a fixed pure state.

Hereafter, for an LOCC protocol $\Lambda$ and an integer $kR \geq 0$, we write $\rF_\Lambda(\cdot) \coloneqq \tr\big[\big(\Phi_2^{\otimes kR}\otimes\bI\big)\Lambda(\cdot)\big]$ on the first $kR$ declared pairs, and $\rF_\Lambda \coloneqq 0$ if fewer are declared, as in Eq.~\eqref{eq:fidelity functional}.


\subsection{Description-aware converse result}
\label{sec:description-aware converse result}

We provide the converse bound on the description-aware distillable entanglement of \Cref{def:aware distillable} using the family constructed in the previous subsection.  The bound comes in two regimes: at a polynomial rank, the cap is set by the spike and the weakly entangled counterpart of the heavy component of the two-scale branch, while at a superpolynomial rank, the label hides the displacement and no entanglement is distilled at all. The information-theoretic distillable entanglement is nearly extensive throughout.

\begin{theorem}[Description-aware no-go from LWE with binary secrets]
\label{thm:lwe mixed state no go}
Let $\alpha_\rc, \alpha_\rs$ be constants with $1 < \alpha_\rc < \alpha_\rs \leq \infty$ and take $\kappa$ as in Eq.~\eqref{eq:master params}. Assume (H1) and (H2) of \Cref{assumpt:lwe assumptions}, and take the parameters of \Cref{sec:parameters state aware} at any admissible $f$ and $\hat{\mu}$.
Let $r_\rs$ obey either $r_\rs = 1$ or $\omega(1) \leq r_\rs \leq \poly(n)$, and let $r$ obey either $r_\rs \leq r \leq \poly(n)$ or $\omega(\log{n}) = \log{r} \leq \cO(n)$.
Then there is an efficiently sampleable ensemble of efficiently preparable $\Theta(n)$-qubit pure states $\ket{\rho_0}^{\sA\sB\sE}$, each supplied with a description of its preparation circuit.
Let $\scD$ denote the distribution over these circuit descriptions, as in \Cref{def:aware distillable}.
The rank satisfies
\begin{align}
    \label{eq:no go rank}
    &\rank(\rho_0^{\sA\sB})
    = \begin{cases}
        r & \ \text{if} \ \ r_\rs = 1 \ \text{and} \ r \leq \poly(n), \\
        \Theta(r) & \ \text{if} \ \ r_\rs \geq 2 \ \text{and} \ r \leq \poly(n), \\
        2^{\Theta(\log{r})} & \ \text{if} \ \ \log{r} = \omega(\log{n}),
    \end{cases}
\end{align}
with certainty when $r \leq \poly(n)$ and with probability $1 - \negl(n)$ over the sampling when $\log{r} = \omega(\log{n})$.
With probability $1 - \negl(n)$ over the sampling, the same state obeys
\begin{align}
    \label{eq:no go renyi}
    &\widetilde{H}_\alpha(\sA|\sE)_{\rho_0}
    = \begin{cases}
        \hat{\mu} + o(1) & \ \text{if} \ \ r_\rs = 1 \ \text{and} \ \hat{\mu} < \hat{\ell}, \\
        \Theta(\hat{\mu}) & \ \text{if} \ \ r_\rs = 1 \ \text{and} \ \hat{\mu} \geq \hat{\ell}, \\
        \big(\kappa - \f{\max\{\alpha, \alpha_\rc\} - \alpha_\rc}{\alpha_\rc(\max\{\alpha, \alpha_\rc\} - 1)}\big) \log{r_\rs} + o(1) & \ \text{if} \ \ r_\rs \geq 2,
    \end{cases}
    \ \ \text{for every fixed} \ \ \alpha \in (1, \infty] \setminus\{\alpha_\rc\}; \\
    \label{eq:no go ed}
    &E_\rD(\sA:\sB)_{\rho_0}
    = \widetilde{\Omega}(n).
\end{align}
For large enough $n$, the description-aware distillable entanglement satisfies
\begin{align}
    \label{eq:no go cap}
    &\widehat{E}_{\rD, \aw}^{(k, \varepsilon, \nu, s)}(\sA:\sB)_{\scD}
    \leq \begin{cases}
        \hat{\mu} + o(1) & \ \text{if} \ \ r_\rs = 1, \ \hat{\mu} < \hat{\ell}, \ \text{and} \ r \leq \poly(n), \\
        \hat{\tau} + o(1) & \ \text{if} \ \ r_\rs = 1, \ \hat{\mu} \geq \hat{\ell}, \ \text{and} \ r \leq \poly(n), \\
        \kappa\log{r_\rs} + o(1) & \ \text{if} \ \ r_\rs \geq 2 \ \text{and} \ r \leq \poly(n), \\
        0 & \ \text{if} \ \ \log{r} = \omega(\log{n}),
    \end{cases}
\end{align}
for any $\omega(1) \leq k \leq \poly(n)$, every $s = \poly(n)$, and every $\nu = \negl(n)$, where $\varepsilon$ is any constant with $\varepsilon < 1/2$.
\end{theorem}

The same rate bound holds when each party additionally receives polynomially many local copies of $\ket{\rho_0}^{\sA\sB\sE}$, since these can be prepared from the given circuit description with polynomial overhead.
The rank in Eq.~\eqref{eq:no go rank} is set by $r$ independently of the entropy parameters $\hat{\mu}$ and $r_\rs$, and its cases exhaust the two regimes $\poly(n)$ and $2^{\omega(\log{n})}$.

At $r_\rs = 1$ the parameter $\hat{\mu}$ sweeps $\widetilde{H}_\alpha(\sA|\sE)_{\rho_0}$ from $o(1)$ to $\Theta(n)$ while Eq.~\eqref{eq:no go ed} is unchanged, with $\rho_0^{\sA\sB}$ pure at $r = 1$. At polynomial rank, the cap of Eq.~\eqref{eq:no go cap} equals this entropy up to $o(1)$ at $\hat{\mu} < \hat{\ell}$ and stays $\hat{\tau} + o(1)$ beyond it. 
The exact $0$ of the last case comes from the computational indistinguishability of $\rho_0^{\sA\sB}$ from a separable state whose marginal on $\sI_\sA\sI_\sB$ is maximally mixed, and this indistinguishability cannot hold at polynomial rank, since a state of rank $2^{\cO(\log{n})}$ has purity at least $1/\poly(n)$ and a SWAP test on two copies distinguishes it from that state, whose purity is $2^{-\Theta(n)}$. The condition $\varepsilon < 1/2$ for that case is optimal in both directions for the same reasons as after \Cref{thm:agnostic no go}, where the upper bound is supplied here by (III) of \Cref{lem:labeled generator}. Past the threshold, the bound degrades gracefully: for any constant $\varepsilon < 1$ the argument for that case still gives $2^{-kR} \geq 1-\varepsilon-\negl(n)$, hence $\widehat{E}_{\rD, \aw}^{(k, \varepsilon, \nu, s)}(\sA:\sB)_{\scD} = \cO(1/k)$, and this level is attained in this error regime as well.

Specializing to a fixed $\alpha \in (1, \alpha_\rc)$, at which Eq.~\eqref{eq:no go renyi} reduces to $\kappa\log{r_\rs} + o(1)$ at $r_\rs \geq 2$, the cases of Eq.~\eqref{eq:no go cap} at $r \leq \poly(n)$ become the two entries of a single minimum: at $r_\rs \geq 2$ the cap is $\widetilde{H}_\alpha(\sA|\sE)_{\rho_0} + o(1)$, the smaller entry by $r_\rs \leq \poly(n)$, while at $r_\rs = 1$ and $\hat{\mu} = \mu$ it is $\tau + o(1) = (1+o(1))f\log{n}$ by Eq.~\eqref{eq:first leakage}, the smaller entry by $\tau = o(\mu)$; for a function $\theta = \omega(\log{n})$, taking $f = \min\{\theta, \sqrt{n}\}/(2\log{n})$ gives $\tau \leq \theta$ for large $n$, so that the two entries are at most $\min\{\widetilde{H}_\alpha(\sA|\sE)_{\rho_0}, \theta\} + o(1)$.
The third case of Eq.~\eqref{eq:no go rank} applies at every $r$ with $\log{r} = \omega(\log{n})$, while Eq.~\eqref{eq:no go renyi} is unchanged. 
The duality in Eq.~\eqref{eq:duality} maps $\alpha \in (1, \alpha_\rc)$ to $\beta \in (\beta_\rc, 1)$, where $\beta_\rc \coloneqq \f{\alpha_\rc}{2\alpha_\rc - 1}$, yielding \Cref{thm:lwe mixed state no go informal}.

\begin{corollary}[Simplified description-aware no-go from LWE with binary secrets]
\label{thm:lwe mixed state no go informal}
Assume that the LWE decision problem with binary secrets is hard for quantum algorithms as in (H1) and (H2) of \Cref{assumpt:lwe assumptions}, and let $\theta$ be any efficiently computable function with $\theta = \omega(\log{n})$.
Then, there exist a constant $\beta_\rc \in (1/2, 1)$ and an efficiently sampleable ensemble of efficiently preparable mixed states $\rho^{\sA\sB}$, each supplied with a description of an efficient circuit that prepares a purification $\ket{\rho}^{\sA\sB\sE}$, such that, for every constant $\beta \in (\beta_\rc, 1)$, no LOCC protocol compiled from the description by an efficient circuit fixed beforehand distills, on all but a $\negl(n)$ fraction of the ensemble, from $k$ copies at error $\varepsilon$ and at a rate $R$ larger than
\begin{align}
    \begin{cases}
        \min\{\widetilde{I}_{\beta}(\sA\rangle\sB)_\rho, \ \theta\} + o(1),
        & \text{if } \rank(\rho^{\sA\sB}) = \poly(n), \\
        0,
        & \text{if } \rank(\rho^{\sA\sB}) = 2^{\omega(\log{n})}.
    \end{cases}
\end{align}
This bound holds for any $\omega(1) \leq k \leq \poly(n)$ and any constant $\varepsilon < 1/2$.
The information-theoretic distillable entanglement satisfies $E_\rD(\sA:\sB)_\rho = \widetilde{\Omega}(n)$ with probability $1-\negl(n)$ over the sampling.
\end{corollary}

The second entry $\theta$ is any prescribed function $\omega(\log{n})$. As shown later, \Cref{prop:indistinguishability rank} gives conditions under which an indistinguishable ensemble of polynomial global rank cannot yield a $\cO(\log{n})$ information-theoretic rate cap uniformly over polynomial numbers of copies.


We now prove \Cref{thm:lwe mixed state no go}.

\begin{proof}[Proof of \Cref{thm:lwe mixed state no go}]

We take $\ket{\rho_0}^{\sA\sB\sE}$ to be $\ket{\Upsilon_{00}}^{\sA\sB\sE}$ of the construction above at the given $r_\rs$ and $\hat{\mu}$. The parameter $r$ fixes the remaining freedom: at $r \leq \poly(n)$ we take $\lambda = 0$ and $M = \big\lceil r/\rank(\Psi_{00}^{\sI_\sA\sI_\sB})\big\rceil$, and at $\log{r} = \omega(\log{n})$ we take $\lambda = \Theta(\log{r})$, with the constant chosen so that $\lambda \leq n$, which obeys Eq.~\eqref{eq:lambda range}.

We first check efficient preparability. Every branch is efficiently preparable: the spike by \Cref{lem:spikestate} with $r_\rs \leq \poly(n)$ and the others by \Cref{lem:small register pke}; thus (iv) of \Cref{lem:flag mixture} makes each $\ket{\Psi_{b\hat{b}}}$ efficiently preparable and specified by the descriptions of $\hat{C}_{\hat{b}}$ and $C'_b$.
The displacement and the label written into $\sK_\sA\sK_\sB$ are efficiently computable, both being $x$ itself at $\lambda = 0$ and being $x(\omega)$ and $V(\omega)$ at $\lambda > 0$ by (I) of \Cref{lem:labeled generator}, so the same holds for $\ket{\Upsilon_{b\hat{b}}}$.

We next compute the rank. By (i) of \Cref{lem:flag mixture}, $\rank(\Psi_{00}^{\sI_\sA\sI_\sB})$ is $1$ at $r_\rs = 1$, where the third branch alone has positive weight and $\zeta_{00}^{\sL_\sA\sL_\sB}$ is pure, and is $r_\rs + 1$ at $r_\rs \geq 2$, since $\xi_{\mathrm{sp}}^{\sG_\sA\sG_\sB}$ is flat of rank $r_\rs$ and the second branch is pure. The summands of Eq.~\eqref{eq:master mixed} have mutually orthogonal supports across distinct labels $v$ and each has rank $\rank(\Psi_{00}^{\sI_\sA\sI_\sB})$, implying $\rank(\Upsilon_{00}^{\sA\sB}) = \rank(\Psi_{00}^{\sI_\sA\sI_\sB})|\cV|$. 
At $\lambda = 0$ this is $\rank(\Psi_{00}^{\sI_\sA\sI_\sB})M$, which the choice of $M$ above makes $r$ at $r_\rs = 1$ and $\Theta(r)$ at $r_\rs \geq 2$.
At $r_\rs \geq 2$, the product $\rank(\Psi_{00}^{\sI_\sA\sI_\sB})M$ lies between $r$ and $2r + 1$ by $r_\rs \leq r$. At $\lambda > 0$, on the event of (II) of \Cref{lem:labeled generator}, the two factors obey $\rank(\Psi_{00}^{\sI_\sA\sI_\sB}) \leq r_\rs + 1 = 2^{o(\lambda)}$ by $r_\rs \leq \poly(n)$ and $2^{\lambda} \leq |\cV| \leq 2^{d} = 2^{9\lambda}$; hence, their product is $2^{\Theta(\lambda)} = 2^{\Theta(\log{r})}$.
This is Eq.~\eqref{eq:no go rank}.

We now compute the conditional R\'{e}nyi entropy. Tracing out $\sB$ from Eq.~\eqref{eq:master purification}, the distinctness of the labels removes all off-diagonal terms between seeds with distinct $V(\omega)$, and grouping the seeds with $V(\omega) = v$, on which $x(\omega) = x_v$ on the event of (II) of \Cref{lem:labeled generator}, gives
\begin{align}
    \Upsilon_{00}^{\sA\sE}
    = \sum_{v \in \cV} p_v W^{\sI_\sA}((x_v)_A)\Psi_{00}^{\sI_\sA\sI_\sE}W^{\sI_\sA}((x_v)_A)^\dag\otimes\ketbra{v}{v}^{\sK_\sA}\otimes\ketbra{S_v}{S_v}^{\sJ_\sE},
\end{align}
where $\ket{S_v} \coloneqq (2^{d}p_v)^{-1/2}\sum_{\omega : V(\omega) = v}\ket{\omega}$. The vectors $\{\ket{S_v}\}_{v}$ are orthonormal as the sets $\{\omega : V(\omega) = v\}$ are disjoint; at $\lambda = 0$ the same form holds with $\ket{v}^{\sJ_\sE}$ in place of $\ket{S_v}$.
This is a classical--quantum state with classical register $\sJ_\sE$, and every block is $\Psi_{00}^{\sI_\sA\sI_\sE}$ up to a local unitary on $\sI_\sA$ and a tensored pure state on $\sK_\sA$, under both of which $\widetilde{H}_\alpha$ is invariant. Then, \Cref{lem:conditional alpha entropystate} implies $\widetilde{H}_\alpha(\sA|\sE)_{\Upsilon_{00}} = \widetilde{H}_\alpha(\sI_\sA|\sI_\sE)_{\Psi_{00}}$. We evaluate the right-hand side below.

At $r_\rs = 1$, $\sI_\sE$ is in a pure state and $\widetilde{H}_\alpha(\sI_\sA|\sI_\sE)_{\Psi_{00}} = H_\alpha(\sI_\sA)_{\Psi_{00}} = H_\alpha(\sL_\sA)_{\zeta_{00}}$. Applying (i) of \Cref{lem:two scale branch} with $\ket{\hat{\phi}} = \ket{\hat{\phi}_0}$ and $\ket{\phi} = \ket{\phi_0}$, and bounding the light component by $H_\alpha(\sN_\sA)_{\phi_0} \geq \f{n}{4}$ using (ii) of \Cref{lem:small register pke}, we obtain
\begin{align}
    (1-\eta)\,2^{-\f{\alpha-1}{\alpha}H_\alpha(\hat{\sM}_\sA)_{\hat{\phi}_0}}
    \leq 2^{-\f{\alpha-1}{\alpha}\widetilde{H}_\alpha(\sI_\sA|\sI_\sE)_{\Psi_{00}}}
    \leq (1-\eta)\,2^{-\f{\alpha-1}{\alpha}H_\alpha(\hat{\sM}_\sA)_{\hat{\phi}_0}} + \eta\,2^{-\f{\alpha-1}{\alpha}\f{n}{4}}.
\end{align}
At $\hat{\mu} < \hat{\ell}$ we have $H_\alpha(\hat{\sM}_\sA)_{\hat{\phi}_0} = \hat{\mu}$, and thus $\widetilde{H}_\alpha(\sI_\sA|\sI_\sE)_{\Psi_{00}} = \hat{\mu} + o(1)$ by $\hat{\mu} \leq \hat{\ell} = o(n)$ and $\eta = o(1)$; at $\hat{\mu} \geq \hat{\ell}$, $H_\alpha(\hat{\sM}_\sA)_{\hat{\phi}_0}$ lies between $\hat{\mu}/4$ and $\hat{\mu}$, the lower bound by (ii) of \Cref{lem:small register pke} at the parameters of Eq.~\eqref{eq:heavy params} and the upper one by the size of $\hat{\sM}_\sA$. Hence, $\widetilde{H}_\alpha(\sI_\sA|\sI_\sE)_{\Psi_{00}} \in [\hat{\mu}/4 - \cO(1), \ \hat{\mu} + o(1)] = \Theta(\hat{\mu})$. 
These are the first two cases of Eq.~\eqref{eq:no go renyi}.
At $r_\rs \geq 2$ we evaluate $\widetilde{H}_\alpha(\sI_\sA|\sI_\sE)_{\Psi_{00}}$ through (ii) of \Cref{lem:flag mixture}. The spike carries the smallest branch entropy: the second branch has $\widetilde{H}_\alpha \geq n/4$ by (ii) of \Cref{lem:small register pke}, whereas $\widetilde{H}_\alpha(\sG_\sA|\sG_\sE)_{\xi_{\mathrm{sp}}} = \cO(\log{r_\rs}) = o(n)$ by Eq.~\eqref{eq:alpha entropy of spike state}. As $\eta_1 = o(1)$, Eq.~\eqref{eq:flag mixture entropy dominant} returns this spike entropy, whose third term is $o(1)$ for every fixed $\alpha \neq \alpha_\rc$. This is the third case of Eq.~\eqref{eq:no go renyi}.

We next bound the information-theoretic distillable entanglement. An unbounded party recovers the displacement from $\sK_\sA$ or $\sK_\sB$, by reading the plaintext label at $\lambda = 0$ and by applying $\mathrm{Rec}_{k^{\rm gen}}$ to the label at $\lambda > 0$, on the event of (II) of \Cref{lem:labeled generator}. Undoing $W(x_v)$ locally maps every copy to $\Psi_{00}^{\sI_\sA\sI_\sB}$. Conversely, $\Upsilon_{00}^{\sA\sB}$ is obtained from $\Psi_{00}^{\sI_\sA\sI_\sB}$ by an LOCC operation in which Alice samples $v$ with probability $p_v$ and sends it to Bob, and both apply the local factors of $W(x_v)$ and write $v$ into $\sK_\sA\sK_\sB$. Hence $E_\rD(\sA:\sB)_{\Upsilon_{00}} = E_\rD(\sI_\sA:\sI_\sB)_{\Psi_{00}}$. At $r_\rs \geq 2$, the $\ket{\phi_0}$ branch is pure with $H_1(\sH_\sA)_{\phi_0} \geq n/4$ by (ii) of \Cref{lem:small register pke}. Applying (iii) of \Cref{lem:flag mixture} to this branch yields $E_\rD(\sI_\sA:\sI_\sB)_{\Psi_{00}} \geq \eta n/4$. At $r_\rs = 1$ the state $\Psi_{00}^{\sI_\sA\sI_\sB}$ is pure, so $E_\rD(\sI_\sA:\sI_\sB)_{\Psi_{00}} = H_1(\sL_\sA)_{\zeta_{00}} \geq \eta H_1(\sN_\sA)_{\phi_0} \geq \eta n/4$ by (ii) of \Cref{lem:two scale branch}. Either bound is $\Omega\big(n/\log^{1/\gamma_2+2}{n}\big) = \widetilde{\Omega}(n)$ by Eq.~\eqref{eq:master weights}. This is Eq.~\eqref{eq:no go ed}.

We finally bound the computationally distillable entanglement, first for $\lambda=0$, with $\vartheta \coloneqq (1-\varepsilon)/2$.
The state $\Upsilon_{11}^{\sA\sB}$ of Eq.~\eqref{eq:master mixed} is a mixture over $v$ of $\Psi_{11}^{\sI_\sA\sI_\sB}$ dressed by the local unitary $W(x_v)$ and tensored with the separable classical state $\ketbra{v}{v}^{\sK_\sA}\otimes\ketbra{v}{v}^{\sK_\sB}$. Since none of these three operations affects the conditions of \Cref{lem:truncation}, it suffices to verify the conditions for $\Psi_{11}^{\sI_\sA\sI_\sB}$ when we utilize the lemma.

At $r_\rs \geq 2$ the state $\Psi_{11}^{\sI_\sA\sI_\sB}$ is the classical flagged mixture of $\xi_{\mathrm{sp}}^{\sG_\sA\sG_\sB}$ and the pure state $\phi_1^{\sH_\sA\sH_\sB}$ with the weights in Eq.~\eqref{eq:master weights}.
Then, (1) of \Cref{lem:truncation} applies at $J = 1$.
The spike is the heavy component, with $E_0 = \log{g}$ by Eq.~\eqref{eq:max entropy of spike state}, and the branch of $\phi_1$ is light, of weight $w_1 = \eta$ and, by (i) of \Cref{lem:small register pke} at the parameters of Eq.~\eqref{eq:amplifier params}, of $E_1 = \tau'$. The correction $\f{1}{\vartheta}w_1E_1$ is $\cO(1/\log{n}) = o(1)$ by Eqs.~\eqref{eq:amplifier leakage} and~\eqref{eq:master weights}.
Hence $K = k\big(\log{g} + \f{1}{\vartheta}w_1E_1\big)$, and \Cref{lem:sandwich converse} gives
\begin{align}
    R \leq \log{g} + \f{1}{\vartheta}w_1E_1 + \f{1}{k}\log{\f{2}{1-\varepsilon}}
    = \kappa\log{r_\rs} + o(1),
\end{align}
at $k = \omega(1)$. 
At $r_\rs = 1$ the state $\Psi_{11}^{\sI_\sA\sI_\sB}$ is essentially the pure state $\zeta_{11}$ in Eq.~\eqref{eq:master two scale}; thus (2) of \Cref{lem:truncation} applies at the $E_0$ and $E_\star$ of (iii) of \Cref{lem:two scale branch}, with $E_0 = \hat{\mu}$ at $\hat{\mu} < \hat{\ell}$ and $E_0 = \hat{\tau}$ at $\hat{\mu} \geq \hat{\ell}$ by (i) of \Cref{lem:small register pke}, and $E_\star = \tau'$ for the light $\ket{\phi_1}$ on $\sN_\sA\sN_\sB$. Hence $K = k\big(E_0 + \f{9\eta}{\vartheta^2}\big(E_\star + \log{(k+1)}\big)\big) + \log{(k+1)}$, and \Cref{lem:sandwich converse} gives
\begin{align}
    R \leq E_0 + \f{9\eta}{\vartheta^2}\big(E_\star + \log{(k+1)}\big) + \f{\log{(k+1)}}{k} + \f{1}{k}\log{\f{2}{1-\varepsilon}}
    = E_0 + o(1),
\end{align}
where the correction term is $o(1)$ by $\eta E_\star = o(1)$ as above, $\eta\log{k} = o(1)$, and $k = \omega(1)$. The two values of $E_0$ above make the right-hand side $\hat{\mu} + o(1)$ at $\hat{\mu} < \hat{\ell}$ and $\hat{\tau} + o(1)$ at $\hat{\mu} \geq \hat{\ell}$. These are the first three cases of Eq.~\eqref{eq:no go cap}.

We transfer this cap for $\Upsilon_{11}^{\sA\sB}$ to $\Upsilon_{00}^{\sA\sB}$ along the hybrid $\Upsilon_{00}\to\Upsilon_{01}\to\Upsilon_{11}$. The first hop switches $\hat{C}_0 \to \hat{C}_1$ and the second $C'_0 \to C'_1$. The descriptions at the ends of each hop are computationally indistinguishable to quantum circuits of size $\poly(n)$ by (iii) of \Cref{lem:small register pke} applied to $(\hat{\scC}_0,\hat{\scC}_1)$ and to $(\scC'_0,\scC'_1)$; the first hop is trivial at $\hat{\mu} < \hat{\ell}$, where $\hat{C}_0 = \hat{C}_1$. The spike and the labels are identical throughout.

Suppose a protocol distills from $\Upsilon_{00}^{\sA\sB}$ at a rate $R$ with error $\varepsilon$, where the protocol is the output of a circuit of polynomial size that is fixed independently of the sample and receives the sampled description as input. For $b\hat{b} \in \{00, 01, 11\}$, consider the efficient test that prepares $(\Upsilon_{b\hat{b}}^{\sA\sB})^{\otimes k}$ from the description, runs the protocol $\Lambda$ that the same circuit outputs on this description, and measures the first $kR$ declared pairs in the Bell basis, rejecting outright if fewer are declared.\footnote{Here, we analyze the tests with ideal state preparations. Since each test uses only polynomially many copies, the negligible preparation errors change its acceptance probability by at most $\negl(n)$.} Its acceptance probability is $\rF_\Lambda\big(\Upsilon_{b\hat{b}}^{\otimes k}\big)$, which at $b\hat{b} = 00$ is at least $1 - \varepsilon$ for all but a $\negl(n)$ fraction of the descriptions, so that its average over the sampling is at least $1 - \varepsilon - \negl(n)$. Since the two hops above change this average only by $\negl(n)$ each, at $b\hat{b} = 11$ it is at least $1 - \varepsilon - \negl(n)$. The cap above applies to a description only if the bounds on $E_0$, $E_1$, and $E_\star$ hold there, which by (i) of \Cref{lem:small register pke} happens with probability $1 - \negl(n)$ over the sampling. The descriptions violating any of these bounds contribute at most $\negl(n)$ to the average, and the acceptance probability at $b\hat{b} = 11$ is thus at least $1 - \varepsilon - \negl(n)$ for some description obeying all of them. Consequently, $\rF_\Lambda$ is positive there, so at least $kR$ pairs are declared, and the output on the first $kR$ of them has fidelity $\geq 1 - \varepsilon - \negl(n)$ with $\Phi_2^{\otimes kR}$ for large enough $n$. Since discarding the remaining declared pairs is a local operation and $\varepsilon + \negl(n) + \vartheta < 1$ for large $n$, \Cref{lem:sandwich converse} applies to this protocol at error $\varepsilon + \negl(n)$ and gives the same cap on $R$ up to $o(1)$.

At $\lambda > 0$ we instead exploit the hidden displacement. For the protocol supposed above, $\rF_{\Lambda}(\Upsilon_{00}^{\otimes k}) \geq 1 - \varepsilon$ at all but a $\negl(n)$ fraction of the descriptions, so that $\bE\big[\rF_{\Lambda}(\Upsilon_{00}^{\otimes k})\big] \geq 1 - \varepsilon - \negl(n)$, the expectation being over the sampling of $(C'_0, \hat{C}_0, k^{\rm gen})$.

Now consider the algorithm that, given $k^{\rm gen}$ and $k$ pairs $(v_1, u_1), \ldots, (v_k, u_k)$ of a label $v_j \in \{0,1\}^{d_V}$ and a string $u_j \in \{0,1\}^{4n'}$, samples $C'_0$ and $\hat{C}_0$ internally, computes $\Lambda$ from the description $(C'_0, \hat{C}_0, k^{\rm gen})$, prepares $\bigotimes_{j=1}^{k}\Big(\big(W(u_j)\Psi_{00} W(u_j)^\dag\big)^{\sI_\sA\sI_\sB}\otimes\ketbra{v_j}{v_j}^{\sK_\sA}\otimes\ketbra{v_j}{v_j}^{\sK_\sB}\Big)$, runs $\Lambda$, and measures $\{\Phi_2^{\otimes kR}, \bI - \Phi_2^{\otimes kR}\}$ on the first $kR$ declared pairs, accepting on the first outcome and rejecting outright when fewer are declared. 
When $(v_j, u_j) = (V(\omega_j), x(\omega_j))$ for independent uniform seeds $\omega_1, \ldots, \omega_k$, the prepared state averaged over the seeds is $(\Upsilon_{00}^{\sA\sB})^{\otimes k}$ by Eq.~\eqref{eq:master purification}, so by the linearity of $\rF_\Lambda$ the acceptance probability is $\bE\big[\rF_\Lambda(\Upsilon_{00}^{\otimes k})\big]$, where the expectation is taken over the description it samples.
On the other hand, when $u_1, \ldots, u_k$ are independent uniform strings and $v_j = V(\omega_j)$, the average over each $u_j$ acts on the corresponding copy as the full twirl $2^{-4n'}\sum_u W(u)(\cdot)W(u)^\dag$, which maps every state on $\sI_\sA\sI_\sB$ to $\pi^{\sI_\sA\sI_\sB}$ by the $1$-design property of the Heisenberg--Weyl operators~\cite[Chap.~3]{khatri2024principlemodern}, while the average over $\omega_j$ turns the label into $\varsigma^{\sK_\sA\sK_\sB} \coloneqq \bE_\omega\big[\ketbra{V(\omega)}{V(\omega)}^{\sK_\sA}\otimes\ketbra{V(\omega)}{V(\omega)}^{\sK_\sB}\big]$. The acceptance probability is then $\bE\big[\rF_\Lambda\big((\pi\otimes\varsigma)^{\otimes k}\big)\big]$.

The algorithm has size $\poly(n)$, so the two acceptance probabilities differ by at most $\negl(n)$ by (III) of \Cref{lem:labeled generator}, as shown after Eq.~\eqref{eq:lambda range}. Therefore, $\bE\big[\rF_{\Lambda}\big((\pi\otimes\varsigma)^{\otimes k}\big)\big] \geq 1 - \varepsilon - \negl(n)$.
The state $\pi^{\sI_\sA\sI_\sB}\otimes\varsigma^{\sK_\sA\sK_\sB}$ is separable across $\sA:\sB$, and an LOCC maps separable inputs to separable outputs, which overlap $\Phi_2^{\otimes kR}$ by at most $2^{-kR}$ at every description, and hence on average.
Hence $2^{-kR} \geq 1 - \varepsilon - \negl(n) > 1/2$ for constant $\varepsilon < 1/2$, and $kR = 0$ as $kR$ is a nonnegative integer.

\end{proof}

We have taken (H1) and (H2) at full strength to state the theorem uniformly over the parameters and to cover the widest ranges of rank and entropy. Neither the construction nor the proof depends on that strength, which only sets the ranges. Weakening (H1) to $(2^{\ell^{c}}, 2^{-\ell^{c}})$-hardness for a constant $c \in (0,1)$ at $m = \poly(\ell)$ forces $\ell^{c}, \lambda^{c} = \omega(\log{n})$, which confines the second entry $\theta$ of \Cref{thm:lwe mixed state no go informal} to $\theta = \widetilde{\Omega}(\log^{1/c}{n})$ and pushes the zero-rate case of Eq.~\eqref{eq:no go cap} out to $\log{r} = \omega(\log^{1/c}{n})$, leaving the ranks in between uncovered. Under polynomial hardness alone, $\ell$ and $\lambda$ are instead $n^{\Omega(1)}$, and both bounds become polynomial in $n$. Assumption (H2) acts elsewhere, entering only through the second application of \Cref{lem:small register pke} and, at $\hat{\mu} \neq \mu$, the third, both at the LWE dimension $\ell'$. Dropping it and running these applications under (H1) at $\ell' = n^{\delta/2}$ for a constant $\delta \in (0,1)$, with $\eta = n^{-\delta}$ to keep $\eta\tau' = o(1)$, turns Eq.~\eqref{eq:no go ed} into $\Omega(n^{1-\delta})$. In every case, at a small enough exponent for $\ell$ and $\lambda$, Eq.~\eqref{eq:no go ed} stays polynomially above Eq.~\eqref{eq:no go cap}; the assumptions determine the range of the separation, not its existence.

We choose binary-secret LWE on the same ground. Binary secrets keep the leakage at $\ell\log(2(a+1))$ in \Cref{lem:small modulus lossy} and the seed length at $9\ell$ in \Cref{lem:labeled generator}, avoiding the additional $\log{q}$ factor in the length of a secret in $\bZ_q^\ell$. The known reduction from standard LWE increases the dimension by a factor of order $\log{q}$~\cite{Micciancio2018OntheHardnessofLWEwithBinarySecrets}, which carries the same factor into the leakage and the seed length. Under hardness assumptions for standard LWE, this separation between efficient and information-theoretic distillation would still be obtained by adjusting the parameters, with narrower guaranteed ranges of rank and entropy.


\section{Tightness of the bounds via the achievability protocols and the converse families}
\label{sec:achievability on converse families}

We compare the achievability results of \Cref{sec:sample efficient result,sec:achievability result} with the converse bounds of \Cref{sec:Converse: State-agnostic setting,sec:Converse Description-aware setting}. Case~(I) of \Cref{thm:achievability by efficient protocol} requires only the global rank of the nearby state to be polynomial, whereas Case~(II) requires both its global and marginal ranks to be polynomial and gives an efficient protocol. By Eq.~\eqref{eq:target rate cap}, the marginal-rank condition bounds the R\'{e}nyi coherent information by $\cO(\log{n})$. We take a constant error $\varepsilon \in (0,1/2)$ throughout the comparison.

We first show that the R\'{e}nyi coherent information is attained from polynomially many copies on the families of both converse results at $r_\rs = \omega(1)$ and $r = \poly(n)$ in \Cref{thm:agnostic no go,thm:lwe mixed state no go}. In this regime, each block $\Psi_{00}^{\sI_\sA\sI_\sB}$ is the flagged mixture of the spike state with weight $1-\eta$ and the second branch with weight $\eta$, where $\eta$ is given in Eqs.~\eqref{eq:agnostic params} and~\eqref{eq:master weights}, respectively. We delete the second branch and set
\begin{align}
\label{eq:spike block}
    \ket{\hat{\Psi}}^{\sI_\sA\sI_\sB\sI_\sE}
    \coloneqq \ket{0}^{\sF_\sA\sF_\sB\sF_\sE}\ket{\xi_{\mathrm{sp}}}^{\sG_\sA\sG_\sB\sG_\sE}\ket{0}^{\sH_\sA\sH_\sB\sL_\sA\sL_\sB}.
\end{align}
Let $\hat{\rho}^{\sA\sB}$ be the state obtained from $\rho_0^{\sA\sB}$ by replacing the block $\Psi_{00}^{\sI_\sA\sI_\sB}$ in Eq.~\eqref{eq:agnostic mixed} or Eq.~\eqref{eq:master mixed} with $\hat{\Psi}^{\sI_\sA\sI_\sB}$, while keeping the labels, the local unitaries $U_{1,j}\otimes U_{2,j}$, and the displacements $W(x_v)$. It is defined on the same registers as $\rho_0^{\sA\sB}$. Their trace distance is at most $\eta$: the blocks differ by this amount because their branches have orthogonal flag values, and applying the same local unitaries and taking the mixture over the labels cannot increase the distance.

The spike state has $\rank(\xi_{\mathrm{sp}}^{\sG_\sA\sG_\sB}) = r_\rs$, and $\xi_{\mathrm{sp}}^{\sG_\sA} = \Phi_g^{\sP_\sA}\otimes\chi^{\sQ_\sA}$ has rank $gr_\rs$, as both factors in Eq.~\eqref{eq:def of spike} are completely mixed on their supports. The labels have mutually orthogonal supports on each side, and local unitaries preserve the rank. Hence, with $M$ the number of labels,
\begin{align}
    \rank(\hat{\rho}^{\sA\sB}) = Mr_\rs,
    \qquad
    \rank(\hat{\rho}^{\sA}) = Mgr_\rs,
\end{align}
both of which are $\poly(n)$ by $M \leq r = \poly(n)$ and $g \leq r_\rs = \poly(n)$. Case~(I) uses the first of the two, and Case~(II) uses both.

The conditional R\'{e}nyi entropy of $\hat{\rho}^{\sA\sB}$ coincides with that of the spike state. Indeed, Eq.~\eqref{eq:agnostic invariance} and the corresponding computation in the proof of \Cref{thm:lwe mixed state no go} apply to the purification obtained by the same replacement in Eqs.~\eqref{eq:agnostic purification} and~\eqref{eq:master purification}. As every block is now $\hat{\Psi}$, they give $\widetilde{H}_\alpha(\sA|\sE)_{\hat{\rho}} = \widetilde{H}_\alpha(\sG_\sA|\sG_\sE)_{\xi_{\mathrm{sp}}}$ for every fixed $\alpha \in (1, \infty]$. At a fixed $\alpha \in (1, \alpha_\rc)$, Eq.~\eqref{eq:alpha entropy of spike state} reduces this to $\log{g} + o(1) = \kappa\log{r_\rs} + o(1)$, since $r_\rs = \omega(1)$. By the duality in Eq.~\eqref{eq:duality}, we obtain
\begin{align}
\label{eq:spike block entropy}
    \widetilde{I}_\beta(\sA\rangle\sB)_{\hat{\rho}} = \kappa\log{r_\rs} + o(1),
\end{align}
for every constant $\beta \in (\beta_\rc, 1)$. This agrees with $\widetilde{I}_\beta(\sA\rangle\sB)_{\rho_0}$ up to $o(1)$ by the third cases of Eqs.~\eqref{eq:agnostic renyi} and~\eqref{eq:no go renyi}. The same quantity $\kappa\log{r_\rs} + o(1)$ bounds the distillation rate from above by the second cases of Eqs.~\eqref{eq:agnostic cap unbounded} and~\eqref{eq:agnostic cap}, and the third case of Eq.~\eqref{eq:no go cap}.

Both cases of \Cref{thm:achievability by efficient protocol} thus apply to $\hat{\rho}^{\sA\sB}$. In Case~(I), we take a constant $C$ large enough that $C\log{n}$ exceeds the value in Eq.~\eqref{eq:spike block entropy} for large $n$, which is possible as $r_\rs = \poly(n)$. The trace distance $\eta$ is at most $1/\log^{1+c}{n}$ for $c \in (0, \delta]$ in \Cref{thm:agnostic no go}, with $\delta$ the constant in Eq.~\eqref{eq:agnostic params}, and for $c \in (0, 1/\gamma_2 + 1]$ in \Cref{thm:lwe mixed state no go}, with $\gamma_2$ the constant of (H2) in \Cref{assumpt:lwe assumptions}. Each protocol therefore distills from $\rho_0^{\sA\sB}$ at a rate $\kappa\log{r_\rs} - o(1)$.

At the parameters used to obtain Cases~(I) and~(II), $J = \Omega(1)$ and $1/\gamma = \Theta(\log\log{n})$. Both protocols invoke \Cref{cor:batch distillation} at $a = \gamma/(8J)$ by Eqs.~\eqref{eq:threshold regularization} and~\eqref{eq:rate parameters}, and their numbers of copies obey
\begin{align}
    k = \poly(n), \qquad k \geq N \geq 4/a = 32J/\gamma = \omega(1).
\end{align}
They therefore lie in the range of the converse bounds. Neither protocol uses additional copies of a purification.

The ranks and Eq.~\eqref{eq:spike block entropy} depend neither on the realizations nor on the local unitaries and the displacements. A single protocol of each case thus serves the whole set $\scS$ of \Cref{thm:agnostic unbounded} and every description sampled in \Cref{thm:lwe mixed state no go}. The rate of Case~(II) is achieved in all three computational models, as shown after \Cref{thm:achievability by efficient protocol formal}, and that of Case~(I) lower-bounds $E_{\rD, \ag}^{(k, \varepsilon)}(\sA:\sB)_\scS$ with no restriction on the circuit size, as shown after \Cref{thm:achievability by schur coding}.

At the numbers of copies used by these protocols, the respective converse bounds therefore characterize the optimal rate by $\widetilde{I}_\beta(\sA\rangle\sB)_{\rho_0}$ up to $o(1)$ for every constant $\beta \in (\beta_\rc, 1)$: for the set $\scS$ in model~(M1), for the selected state in model~(M2), and for the distribution $\scD$ of descriptions in model~(M3). Varying $r_\rs = \omega(1)$ within $\poly(n)$ at a fixed $\kappa$ gives this characterization throughout the range from $\omega(1)$ to $\cO(\log{n})$.

The same argument applies at $r_\rs = 1$ when $1 \leq \hat{m} = \cO(\log{n})$ in \Cref{thm:agnostic no go}, or $1 \leq \hat{\mu} = \cO(\log{n})$ in \Cref{thm:lwe mixed state no go}. Deleting the light component of the two-scale branch leaves the blocks $\ket{\Phi_2}^{\otimes\hat{m}}$ or $\ket{\Phi_2}^{\otimes\hat{\mu}}$, which give polynomial global and marginal ranks and $\widetilde{I}_\beta(\sA\rangle\sB)_{\hat{\rho}} = \hat{m}$ or $\hat{\mu}$, agreeing up to $o(1)$ with the entries $\hat{m}$ and $\hat{\mu}$ in the first cases of Eqs.~\eqref{eq:agnostic cap} and~\eqref{eq:no go cap}, respectively.

Since the two components of the two-scale branch are superposed coherently, the trace distance is now $\sqrt{\eta}$. This is at most $1/\log^{1+c}{n}$ at $c = 1/(2\gamma_2)$ in \Cref{thm:lwe mixed state no go}, whereas in \Cref{thm:agnostic no go} it is so at $\delta > 1$, with $c = (\delta - 1)/2$.

At $r_\rs = 1$ and $\omega(\log{n}) = \hat{m} < \mu$ in \Cref{thm:agnostic no go}, at the same $\delta > 1$, the blocks $\ket{\Phi_2}^{\otimes\hat{m}}$ have superpolynomial marginal rank and Case~(II) does not apply. Their labeled mixture still has $\rank(\hat{\rho}^{\sA\sB}) = M \leq r = \poly(n)$, so Case~(I) applies. To compare its rate with the logarithmic converse bound, fix any constant $B \in (0,1/2)$ and take
\begin{align}
    J \coloneqq \f{4B\log{M} + \log{n}}{1-2B}, \qquad
    1/\gamma = \Theta(\log\log{n}).
\end{align}
Since $J = \Theta(\log{n}) < \hat{m}$ for large $n$, \Cref{thm:achievability by schur coding} with $\hat{r} = M$ gives a single protocol for $\scS$ at rate $R \geq J-\gamma$. Equation~\eqref{eq:schur copy count} then gives $R-B\log{k} \geq \log{n} - \cO(\log\log{n}) > 0$ for large $n$. Combining this with Eq.~\eqref{eq:agnostic cap unbounded} at $l = 0$, we obtain
\begin{align}
    B\log{k}
    \leq E_{\rD, \ag}^{(k, \varepsilon)}(\sA:\sB)_\scS
    \leq \log{k} + o(1)
\end{align}
for the number $k = \poly(n)$ of copies used by this protocol. Thus, in this regime, the Schur-coding protocol attains any fixed fraction $B < 1/2$ of the logarithmic upper bound, with $B$ chosen independently of the rank bound. The required number of copies may depend on both.

We finally examine whether the second cap $\theta$ in the description-aware converse can be reduced to $\cO(\log{n})$ by comparison with a computationally indistinguishable ensemble of polynomial global rank whose distillation rate is capped at this order uniformly over polynomial numbers of copies, with no restriction on circuit size. Related restrictions on mixed-state pseudoentanglement were obtained in terms of logarithmic negativity and fidelity of separability~\cite{Tarabunga2026quantifyingmixed}. Here we combine the SWAP-test argument of Refs.~\cite{Ji2018PseudorandomQuantumStates,aaronson2023quantumpseudoentanglement,bouland2024PublicKeyPseudoentanglement} with \Cref{thm:achievability by schur coding} to bound the distillation rate of the comparison ensemble from below.

\begin{proposition}[Distillation from indistinguishable ensembles]
\label{prop:indistinguishability rank}
Let $\scE_0$ and $\scE_1$ be ensembles of states on the same registers $\sA\sB$ of $\Theta(n)$ qubits that are computationally indistinguishable given polynomially many copies. Suppose that, for some constant $\beta \in (1/2,1)$ and some function $h = \omega(\log{n})$, $\widetilde{I}_\beta(\sA\rangle\sB)_\rho \geq h$ with probability $1-\negl(n)$ over $\rho \sim \scE_0$, and that every $\sigma \sim \scE_1$ has $\rank(\sigma^{\sA\sB}) \leq \hat{r} = \poly(n)$. Then, for every constant $C > 0$ and every constant $\varepsilon \in (0,1)$, there is a single one-way LOCC protocol using $k = \poly(n)$ copies that distills from $\sigma$ at rate at least $C\log{n} - o(1)$ and error $\varepsilon$, with probability $1-\negl(n)$ over $\sigma \sim \scE_1$. No restriction is imposed on its circuit size.
\end{proposition}

\begin{proof}[Proof of \Cref{prop:indistinguishability rank}]

We first derive two entropy inequalities for an arbitrary state $\varrho^{\sA\sE}$ with a purification $\ket{\varrho}^{\sA\sB\sE}$. For $\alpha > 1$, the data-processing inequality and $\tr[(\varrho^\sA)^\alpha] \geq \|\varrho^\sA\|_\infty^\alpha$ give
\begin{align}
    \label{inteq:1}
    \widetilde{H}_\alpha(\sA|\sE)_\varrho
    \leq \widetilde{H}_\alpha(\sA)_\varrho
    \leq \f{\alpha}{\alpha-1}H_{\rm min}(\sA)_\varrho.
\end{align}
The chain rule and the data-processing inequality for the min-entropy also give
\begin{align}
    H_{\rm min}(\sA)_\varrho
    &= H_{\rm min}(\sB\sE)_\varrho \\
    &\leq H_{\rm min}(\sB|\sE)_\varrho + \log{\rank(\varrho^\sE)} \\
    &\leq H_{\rm min}(\sB)_\varrho + \log{\rank(\varrho^\sE)} \\
    &= H_{\rm min}(\sA\sE)_\varrho + \log{\rank(\varrho^\sE)} \\
    \label{inteq:2}
    &\leq H_{\rm min}(\sA|\sE)_\varrho + 2\log{\rank(\varrho^\sE)},
\end{align}
where the equalities use the purity of $\ket{\varrho}^{\sA\sB\sE}$.

By the duality in Eq.~\eqref{eq:duality}, Eq.~\eqref{inteq:1} at $\alpha = \beta/(2\beta-1)$, and $\tr[(\rho^\sA)^2] \leq \|\rho^\sA\|_\infty$, we have
\begin{align}
    \bE_{\rho \sim \scE_0}\tr[(\rho^\sA)^2]
    \leq 2^{-\f{1-\beta}{\beta} h} + \negl(n)
    = \negl(n).
\end{align}
The SWAP test on two copies of $\rho^\sA$ has acceptance probability $(1+\tr[(\rho^\sA)^2])/2$ and is efficient. Computational indistinguishability therefore gives $\bE_{\sigma \sim \scE_1}\tr[(\sigma^\sA)^2] = \negl(n)$.

For a purification of $\sigma^{\sA\sB}$, the rank bound gives $\rank(\sigma^\sE) \leq \hat{r}$. Equation~\eqref{inteq:2} and $\|\sigma^\sA\|_\infty^2 \leq \tr[(\sigma^\sA)^2]$ then imply
\begin{align}
    \widetilde{I}_{1/2}(\sA\rangle\sB)_\sigma
    = H_{\rm min}(\sA|\sE)_\sigma 
    \geq H_{\rm min}(\sA)_\sigma - 2\log{\hat{r}} 
    \geq -\f{1}{2}\log{\tr[(\sigma^\sA)^2]} - 2\log{\hat{r}}.
\end{align}
For $J = C\log{n}$, Markov's inequality thus yields
\begin{align}
    \Pr_{\sigma \sim \scE_1}\big[\widetilde{I}_{1/2}(\sA\rangle\sB)_\sigma < J\big]
    \leq 2^{2J}\hat{r}^4 \bE_{\sigma \sim \scE_1}\tr[(\sigma^\sA)^2]
    = \negl(n).
\end{align}
Applying \Cref{thm:achievability by schur coding} with rank bound $\hat{r}$, order $\beta = 1/2$, $\hat{\epsilon} = 0$, and $\gamma = \Theta(1/\log\log n)$ gives a single protocol with $k = \poly(n)$ and rate $R \geq J-\gamma = C\log{n}-o(1)$ at error $\varepsilon$.

\end{proof}

Consequently, under these assumptions, the ensemble $\scE_1$ admits no $\cO(\log{n})$ information-theoretic rate cap that is uniform over polynomial numbers of copies. This does not rule out bounding the rate of a particular protocol, or the rate at a specified number of copies, by choosing a comparison ensemble for that purpose.\footnote{Under the additional assumption that quantum-secure extremely lossy functions~\cite{Zhandry2019TheMagicofELFs} exist, it may be possible to obtain an $\cO(\log{n})$ cap for a single family, with the implicit constant depending on the protocol. The comparison would require the distinguishing advantage to be bounded only by a sufficiently small constant, for a circuit-size bound determined by the given protocol.}

\section{Separation: Conditional entropies and distillable entanglement}
\label{Separations for conditional entropies and converse bounds}

This section proves the three results. We first prove the rank bound on the separation of R\'{e}nyi orders below. We then read the converse theorems at the parameters that exhibit the separations, at fixed orders in \Cref{sec:proof of separation} and at growing ones in \Cref{sec:proof of nearly maximal separation}.

\proprenyiminentropies*

\begin{proof}[Proof of \Cref{prop:renyi min relation}]

Let $\ket{\varrho}^{\sA\sB\sE}$ be a purification of $\varrho^{\sA\sB}$, and let $\alpha \in (1, \infty)$ be the dual order of $\beta$, with $1/\alpha + 1/\beta = 2$. By the duality in Eq.~\eqref{eq:duality}, together with $\alpha/(\alpha-1) = \beta/(1-\beta)$ and $\rank(\varrho^\sE) = \rank(\varrho^{\sA\sB})$, the claim is equivalent to
\begin{align}
    H_{\rm min}(\sA|\sE)_\varrho
    \leq \widetilde{H}_\alpha(\sA|\sE)_\varrho
    \leq \f{\alpha}{\alpha-1}\big(H_{\rm min}(\sA|\sE)_\varrho + 2\log{\rank(\varrho^\sE)}\big).
\end{align}
The first inequality follows from the monotonicity of the conditional R\'{e}nyi entropy in its order. Combining Eqs.~\eqref{inteq:1} and~\eqref{inteq:2}, derived in the proof of \Cref{prop:indistinguishability rank}, gives the second inequality.

\end{proof}


\subsection{Proof of \texorpdfstring{\Cref{prop:separation}}{Theorem~\ref{prop:separation}}}
\label{sec:proof of separation}

We read \Cref{thm:agnostic no go,thm:agnostic unbounded,thm:lwe mixed state no go} at the parameters where the spike is present, namely $r_\rs \geq 2$ and $\hat{m} = 0$, or $\hat{\mu} = 0$ in \Cref{thm:lwe mixed state no go}. At these parameters, the conditional R\'{e}nyi entropy of the family is the spike entropy of \Cref{lem:spikestate} and bounds the distillation rate from above at the orders $\alpha \in (1, \alpha_\rc)$.
Two gaps appear there. The first is between the information-theoretic and the computational distillable entanglement, where the former is polynomially large and the latter is $\cO(\log{r_\rs})$.
The second is inside the conditional R\'{e}nyi entropies themselves, which change sign at $\alpha = \alpha_\rs$: the orders governing the cap are positive, while the min-entropy, the order $\alpha = \infty$, is negative. \Cref{prop:separation} exhibits both gaps for a family of any rank $\omega(1) \leq r \leq \poly(n)$, where the rank is set by $r_\rs$ itself at these parameters. Through the duality in Eq.~\eqref{eq:duality}, the parameters $\alpha_\rc$ and $\alpha_\rs$ of these theorems correspond to $\beta_\rc = \alpha_\rc/(2\alpha_\rc - 1)$ and $\beta_\rs = \alpha_\rs/(2\alpha_\rs - 1)$ in \Cref{prop:separation}, which we restate below.

\threegapsseparations*

\begin{proof}[Proof of \Cref{prop:separation}]

Given constants $\beta_\rs, \beta_\rc$ such that $1/2 < \beta_\rs < \beta_\rc < 1$ and $r$ satisfying $\omega(1) \leq r \leq \poly(n)$, let $\alpha_\rc \coloneqq \beta_\rc/(2\beta_\rc - 1)$ and $\alpha_\rs \coloneqq \beta_\rs/(2\beta_\rs - 1)$, which satisfy $1 < \alpha_\rc < \alpha_\rs < \infty$. By the duality in Eq.~\eqref{eq:duality}, statements (1) and (2) are equivalent to the following: $\widetilde{H}_\alpha(\sA|\sE)_\rho$ is $+\Theta(\log{r})$ for every fixed $\alpha \in (1, \alpha_\rs)$ and $-\Theta(\log{r})$ for every fixed $\alpha \in (\alpha_\rs, \infty]$, and the rate is at most $\widetilde{H}_\alpha(\sA|\sE)_\rho + o(1)$ for any fixed $\alpha \in (1, \alpha_\rc)$.
We take the family of \Cref{thm:agnostic no go} at $\hat{m} = 0$ and $r_\rs = r - 1$, which obeys $\omega(1) \leq r_\rs \leq \poly(n)$ and $\log{r_\rs} = \Theta(\log{r})$.
Then $\rank(\Psi_{00}^{\sI_\sA\sI_\sB}) = r_\rs + 1 = r$ and $M = 1$. We take $\rho^{\sA\sB}$ to be the state $\rho_0^{\sA\sB}$ provided by \Cref{thm:agnostic no go} at these parameters; it is $\Psi_{00}^{\sI_\sA\sI_\sB}$ up to a local unitary, and $\rank(\rho^{\sA\sB}) = r$.

At these parameters, Eq.~\eqref{eq:agnostic renyi} reads $\widetilde{H}_\alpha(\sA|\sE)_{\rho} = \big(\kappa - \f{\max\{\alpha, \alpha_\rc\} - \alpha_\rc}{\alpha_\rc(\max\{\alpha, \alpha_\rc\} - 1)}\big)\log{r_\rs} + o(1)$ for every fixed $\alpha \in (1, \infty]\setminus\{\alpha_\rc\}$, whose coefficient is $\kappa > 0$ at $\alpha \leq \alpha_\rc$, decreases in $\alpha$, and vanishes at $\alpha = \alpha_\rs$, as recorded after \Cref{lem:spikestate}. With $\log{r_\rs} = \Theta(\log{r})$, the entropy is thus $+\Theta(\log{r})$ for every fixed $\alpha \in (1, \alpha_\rs)$ and $-\Theta(\log{r})$ for $\alpha \in (\alpha_\rs, \infty]$. At $\alpha = \alpha_\rc$, which Eq.~\eqref{eq:agnostic renyi} excludes, Eq.~\eqref{eq:alpha entropy of spike state} still applies, where its third term reads $-\f{1}{\alpha_\rc-1}\log{2}$ instead of $o(1)$, and the value is $\kappa\log{r_\rs} + \cO(1) = \Theta(\log{r})$ all the same, since the constant term is absorbed by $\log{r} = \omega(1)$. This is (1).

For (2), Eq.~\eqref{eq:agnostic ed} gives $E_\rD(\sA:\sB)_{\rho} = \widetilde{\Omega}(n) = n^{\Omega(1)}$, while $r = \poly(n)$ and $r_\rs \geq 2$ give the upper bound $\kappa\log{r_\rs} + o(1)$ by Eq.~\eqref{eq:agnostic cap}. At every fixed $\alpha \in (1, \alpha_\rc)$ the coefficient subtracted from $\kappa$ above vanishes, so this upper bound is $\widetilde{H}_\alpha(\sA|\sE)_{\rho} + o(1)$.

The construction uses no computational hardness assumption, and \Cref{thm:agnostic unbounded} caps a protocol required to succeed on the whole set by the same bound with no restriction on its size. Under \Cref{assumpt:lwe assumptions}, the same reading of \Cref{thm:lwe mixed state no go} at $\hat{\mu} = 0$, $r_\rs = r - 1$, and $\lambda = 0$ yields an efficiently preparable family supplied with a description of its preparation circuit, whose state $\rho^{\sA\sB}$, taken in the same way, again has rank $r$ as $\ket{\Psi_{00}}$ is there tensored with a fixed pure state. At these parameters, Eqs.~\eqref{eq:no go renyi} and~\eqref{eq:no go cap} coincide with the evaluations above and Eq.~\eqref{eq:no go ed} gives $E_\rD(\sA:\sB)_{\rho} = \widetilde{\Omega}(n) = n^{\Omega(1)}$. Both gaps therefore persist for $\widehat{E}_{\rD, \aw}^{(k, \varepsilon, \nu, s)}$.

\end{proof}

Fix $\alpha \in (1,\alpha_\rc)$ and let $\beta$ be its dual order, with $1/\alpha + 1/\beta = 2$. In either construction, let $\hat{\rho}^{\sA\sB}$ be the normalized spike branch of $\rho^{\sA\sB}$, including the same local unitaries and ancillary registers. By \Cref{lem:spikestate}, its global rank is $r_\rs$ and its marginal rank on Alice's side is polynomial. The orthogonal flags give $\big\|\rho^{\sA\sB} - \hat{\rho}^{\sA\sB}\big\|_1 = 2\eta$, and the entropy evaluation above gives
\begin{align}
    \widetilde{I}_\beta(\sA\rangle\sB)_{\hat{\rho}}
    = \widetilde{H}_\alpha(\sG_\sA|\sG_\sE)_{\xi_{\mathrm{sp}}}
    = \widetilde{H}_\alpha(\sA|\sE)_{\rho} + o(1)
    = \Theta(\log{r}).
\end{align}
Choose rationals $\gamma$ and $J$ of polynomial bit length with $\gamma = \Theta(1/\log\log n)$ and
$\widetilde{I}_\beta(\sA\rangle\sB)_{\hat{\rho}} - \gamma \leq J \leq \widetilde{I}_\beta(\sA\rangle\sB)_{\hat{\rho}}$.
These choices satisfy Eq.~\eqref{eq:rate loss range} for large enough $n$, and $J = \cO(\log{n})$ gives
\begin{align}
    \epsilon_\star
    = \Omega\bigg(\f{1}{\log{n}(\log\log n)^4}\bigg),
    \qquad
    2\eta = o(\epsilon_\star),
\end{align}
where the second relation follows from Eq.~\eqref{eq:agnostic params} in the statistical construction and Eq.~\eqref{eq:master weights} in the LWE construction.
Thus \Cref{thm:achievability by efficient protocol formal} applies and supplies an efficient one-way LOCC protocol using a polynomial number of copies, with
\begin{align}
    R \geq J - \gamma
    \geq \widetilde{I}_\beta(\sA\rangle\sB)_{\hat{\rho}} - 2\gamma
    = \widetilde{H}_\alpha(\sA|\sE)_{\rho} - o(1).
\end{align}
The rank bounds and the R\'{e}nyi coherent information of the nearby state are independent of the realization, so the same protocol works for the whole set and can also be used in the description-aware model (M3).
This matches the cap in (2) of \Cref{prop:separation} up to an additive $o(1)$. Efficient distillation from these states is therefore characterized by the conditional R\'{e}nyi entropy at the orders $\alpha \in (1,\alpha_\rc)$, which is positive, while it is negative at the orders $\alpha \in (\alpha_\rs, \infty]$, including the conditional min-entropy. Since $\alpha_\rc$ and $\alpha_\rs$ can be chosen freely, no single fixed order characterizes the efficient rate for all these families, a phenomenon specific to mixed states. The state $\rho^{\sA\sB}$ has rank $r$, and at $r = n^{\Theta(1)}$ the values $\pm\Theta(\log{r})$ are as far apart as \Cref{prop:renyi min relation} permits at that rank, leading to $\pm\Theta(\log{n})$.


\subsection{Proof of \texorpdfstring{\Cref{prop:nearly maximal separation}}{Proposition~\ref{prop:nearly maximal separation}}}
\label{sec:proof of nearly maximal separation}

We prove \Cref{prop:nearly maximal separation} by showing its formal version, \Cref{prop:degenerate spike}, using the families in \Cref{thm:agnostic no go,thm:lwe mixed state no go} with $\hat{m} = 0$ and $\hat{\mu} = 0$, respectively, and growing orders $\alpha_\rc,\alpha_\rs = \omega(\log{n})$, which correspond to $\beta_\rc - 1/2 = o(1/\log{n})$ in \Cref{sec:overview of main results}.
In this case, $\kappa$ obeys $\kappa \leq 1/\alpha_\rc = o(1/\log{n})$, and $r_\rs \leq \poly(n)$ gives $g = 1$ in \Cref{lem:spikestate}.
The spike state then carries no entanglement, and its conditional R\'{e}nyi entropy is $o(1)$ at every fixed order $\alpha \in (1,\infty]$. The converse bound on the computational distillable entanglement is also $o(1)$, while the amplifier branch is untouched and keeps the information-theoretic rate polynomially large.
The first gap of \Cref{prop:separation} thus widens to $\widetilde{\Omega}(n)$ against $o(1)$, while the second closes, since both entropies at its two ends are now $o(1)$. The rank reaches $1$ as well, at $r_\rs = 1$ and $r = 1$, where the state is pure.

\begin{proposition}[Nearly maximal separation at a degenerate spike]
\label{prop:degenerate spike}
Let $1 < \alpha_\rc < \alpha_\rs \leq \infty$ with $\alpha_\rc = \omega(\log{n})$, and consider either of the following.
\begin{enumerate}[label=\textnormal{(\Roman*)}]
    \item The family of Eq.~\eqref{eq:agnostic purification} at $\hat{m} = 0$, at any $r_\rs$ and any $r \leq \poly(n)$ admissible in \Cref{thm:agnostic no go}, and at the $\eta$ and $\mu$ of Eq.~\eqref{eq:agnostic params}.
    \item The family of Eq.~\eqref{eq:master purification} at $\hat{\mu} = 0$ and $\lambda = 0$, at any $r_\rs$ and any $r \leq \poly(n)$ admissible in \Cref{thm:lwe mixed state no go}, under \Cref{assumpt:lwe assumptions}.
\end{enumerate}
In case (II), the orders are chosen to be efficiently computable.
Then, for large enough $n$, there exists a $\Theta(n)$-qubit pure state $\ket{\rho_0}^{\sA\sB\sE}$ whose reduced state has rank $\Theta(r) \leq \poly(n)$, equal to $r$ at $r_\rs = 1$, and satisfies
\begin{align}
    \label{eq:degenerate renyi}
    &\widetilde{H}_\alpha(\sA|\sE)_{\rho_0} = o(1) \ \ \text{for every fixed} \ \ \alpha \in (1, \infty]; \\
    \label{eq:degenerate ed}
    &E_\rD(\sA:\sB)_{\rho_0} = \widetilde{\Omega}(n); \\
    \label{eq:degenerate cap}
    &\widehat{E}_\rD^{(k, \varepsilon, s)}(\sA:\sB)_{\rho_0} = o(1) \ \ \text{in case (I)},
    \ \text{for any constant} \ \varepsilon \ \text{with} \ \varepsilon < 1 \ \text{and any} \ s = \poly(n), \ \omega(1) \leq k \leq \poly(n).
\end{align}
In case (I), the same cap holds for $E_{\rD, \ag}^{(k, \varepsilon)}(\sA:\sB)_\scS$, where $\scS$ is the set defined in \Cref{thm:agnostic unbounded} at the present parameters, with no restriction on the size of the single protocol.
In case (II), the family is efficiently preparable, and the cap of Eq.~\eqref{eq:degenerate cap} holds for $\widehat{E}_{\rD, \aw}^{(k, \varepsilon, \nu, s)}(\sA:\sB)_{\scD}$ at every $\nu = \negl(n)$ and the distribution $\scD$ of descriptions of its preparation circuits.
\end{proposition}

Taking $r_\rs = 1$ in either case gives $\rank(\rho_0^{\sA\sB}) = r$ at every $1 \leq r \leq \poly(n)$, with $E_\rD(\sA:\sB)_{\rho_0} = \widetilde{\Omega}(n)$ by Eq.~\eqref{eq:degenerate ed}, while the corresponding caps on $\widehat{E}_\rD^{(k, \varepsilon, s)}$, $E_{\rD, \ag}^{(k, \varepsilon)}$, and $\widehat{E}_{\rD, \aw}^{(k, \varepsilon, \nu, s)}$ are all $o(1)$. These provide \Cref{prop:nearly maximal separation}, with $\rho^{\sA\sB}$ taken to be $\rho_0^{\sA\sB}$. Note that restricting instead to $\varepsilon < 1/2$ extends the two caps of case (I) to every $r$ with $\omega(\log{n}) = \log{r} \leq \cO(n)$, where they are exactly $0$ rather than $o(1)$ by the last case of Eq.~\eqref{eq:agnostic cap}. In case (II), the same holds for $\widehat{E}_{\rD, \aw}^{(k, \varepsilon, \nu, s)}$ at $\lambda = \Theta(\log{r})$ and $\omega(\log{n}) = \log{r} \leq \cO(n)$, where the rank is $2^{\Theta(\log{r})}$, by the last case of Eq.~\eqref{eq:no go cap}.

Although \Cref{thm:agnostic no go,thm:agnostic unbounded,thm:lwe mixed state no go} are stated at constants $\alpha_\rc, \alpha_\rs$, the construction remains available at $\alpha_\rc, \alpha_\rs = \omega(\log{n})$, since \Cref{lem:spikestate} holds at arbitrary $1 < \alpha_\rc < \alpha_\rs \leq \infty$.
We reevaluate Eqs.~\eqref{eq:spike exponent} and~\eqref{eq:alpha entropy of spike state} at these growing orders in the following proof.

\begin{proof}[Proof of \Cref{prop:degenerate spike}]

We first evaluate the spike state at $\alpha_\rc = \omega(\log{n})$, and take $r_\rs \geq 2$.
As $\alpha_\rc < \alpha_\rs$ forces $\alpha_\rs = \omega(\log{n})$, the exponent $\kappa$ in Eq.~\eqref{eq:spike exponent} obeys $\kappa \leq \f{1}{\alpha_\rc} = o(1/\log{n})$. Hence $\kappa\log{r_\rs} = o(1)$ by $r_\rs \leq \poly(n)$, and $g = \lfloor r_\rs^\kappa\rfloor = 1$ for sufficiently large $n$.
We substitute this into Eq.~\eqref{eq:alpha entropy of spike state}; the first term is $\log{g} = 0$, the second is $o(1)$ for every fixed finite $\alpha > 1$ since $\alpha < \alpha_\rc$ for large $n$, and the third is $o(1)$.
Collecting these evaluations, and using Eq.~\eqref{eq:max entropy of spike state} at $g = 1$, we obtain
\begin{align}
\label{inteq:degenerate spike}
    \widetilde{H}_\alpha(\sG_\sA|\sG_\sE)_{\xi_\mathrm{sp}} = o(1) \ \ \text{for every fixed} \ \alpha \in (1,\infty],
    \ \ \ \ \ \ \ \ \ \
    \xi_\mathrm{sp}^{\sG_\sA\sG_\sB} \leq \sigma_\xi^{\sG_\sA\sG_\sB},
\end{align}
where we note that $H_{\min}(\sG_\sA|\sG_\sE)_{\xi_\mathrm{sp}} = -\f{\log{r_\rs}}{\alpha_\rc} = o(1)$ at $\alpha = \infty$.

The proofs of \Cref{thm:agnostic unbounded,thm:lwe mixed state no go} invoke the spike through the statements of \Cref{lem:spikestate}, which we have just evaluated at growing orders. We can therefore run the proofs for the values given in Eq.~\eqref{inteq:degenerate spike}. In case (II), the efficient choice of the orders also preserves the efficient preparation of the spike.

We compute the conditional R\'{e}nyi entropy, for which the spike still carries the smallest branch entropy and the weight $1-\eta = 1 - o(1)$. In case (I), Eq.~\eqref{inteq:agnostic spike mean} holds verbatim at the weights of Eq.~\eqref{eq:agnostic weights}. Its first term is $(1-\eta)2^{-o(1)} = 1 - o(1)$ by Eq.~\eqref{inteq:degenerate spike} and $\f{\alpha}{\alpha-1} = \Theta(1)$, with the ratio read as $1$ at $\alpha = \infty$, while its second is at most $\eta\,2^{-\Omega(n)}$ with probability $1 - \negl(n)$, as in the proof of \Cref{thm:agnostic unbounded}. In case (II), the branch of $\ket{\phi_0}$ obeys $\widetilde{H}_\alpha \geq n/4$ by (ii) of \Cref{lem:small register pke}, so that Eq.~\eqref{eq:flag mixture entropy dominant} returns the spike entropy at the weights of Eq.~\eqref{eq:master weights}. In either case $\widetilde{H}_\alpha(\sA|\sE)_{\rho_0} = \widetilde{H}_\alpha(\sG_\sA|\sG_\sE)_{\xi_\mathrm{sp}} + o(1) = o(1)$, which is Eq.~\eqref{eq:degenerate renyi}.

We next bound the information-theoretic distillable entanglement. Both proofs derive the bound from the branch of weight $\eta$ that carries the high-entanglement state $\ket{\phi_0}$, which does not involve the spike. Equation~\eqref{eq:degenerate ed} therefore restates Eqs.~\eqref{eq:agnostic ed} and~\eqref{eq:no go ed}.

We finally bound the computational distillable entanglement. The selection and transfer arguments in the proofs of \Cref{thm:agnostic no go,thm:lwe mixed state no go} are unchanged and yield a state $\rho_1^{\sA\sB}$ of the family at $b = \hat{b} = 1$ that is distilled at the same rate $R$ with error $\varepsilon + \negl(n)$. The conditions of (1) of \Cref{lem:truncation}, verified there, now hold at $E_0 = \log{g} = 0$ by Eq.~\eqref{inteq:degenerate spike}, while the low-entanglement state $\ket{\phi_1}$ keeps its weight $\eta$ and its $E_1$, which comes from (i) of \Cref{lem:haar pair} in case (I) and from (i) of \Cref{lem:small register pke} in case (II). Hence \Cref{lem:sandwich converse} at $\vartheta = (1-\varepsilon)/2$ gives $R = \cO(\eta\mu + 1/k) = o(1)$ in case (I) and $R = \cO(\eta\tau' + 1/k) = o(1)$ in case (II), for every $\omega(1) \leq k \leq \poly(n)$ and every constant $\varepsilon < 1$. This is Eq.~\eqref{eq:degenerate cap} in case (I), and the same bound holds for $\widehat{E}_{\rD, \aw}^{(k, \varepsilon, \nu, s)}$ in case (II), where the transfer runs through the descriptions. In case (I), the same bound holds for $E_{\rD, \ag}^{(k, \varepsilon)}(\sA:\sB)_\scS$ at the set $\scS$ specified in the statement, by the argument of \Cref{thm:agnostic unbounded}, which needs no selection and hence no restriction on the size of the protocol.

It remains to treat the case $r_\rs = 1$, where the spike branch is absent, and the family is carried by the two-scale branch with a product state as its heavy component. For the conditional R\'{e}nyi entropy and the information-theoretic distillable entanglement, Eq.~\eqref{eq:degenerate renyi} is the first case of Eqs.~\eqref{eq:agnostic renyi} and~\eqref{eq:no go renyi} at $\hat{m} = 0$ and $\hat{\mu} = 0$, respectively, and Eq.~\eqref{eq:degenerate ed} restates Eqs.~\eqref{eq:agnostic ed} and~\eqref{eq:no go ed}, which do not depend on $r_\rs$. For the cap of the computational distillable entanglement, we apply (2) of \Cref{lem:truncation} at $E_0 = 0$, where the remaining terms are $o(1)$ as in the proofs of \Cref{thm:agnostic unbounded,thm:lwe mixed state no go}.

The rank is $\Theta(r)$ by Eq.~\eqref{eq:agnostic rank} in case (I) and by Eq.~\eqref{eq:no go rank} in case (II), and equals $r$ at $r_\rs = 1$, where it reaches $1$ at $r = 1$.

Note that the restriction $\varepsilon < 1/2$ of \Cref{thm:agnostic no go,thm:lwe mixed state no go} is not imposed here, since it serves only the case of a superpolynomial rank in both, which is reached through \Cref{lem:maxmix} in the former and through the hidden displacement in the latter. That case is absent at $r \leq \poly(n)$ and $\lambda = 0$.

\end{proof}

The separation is nearly maximal, up to a polylogarithmic factor. Our bound on the computational rate is $o(1)$ when $\eta\mu = o(1)$ in case (I) and $\eta\tau' = o(1)$ in case (II). Since $\mu = \omega(\log{n})$ by (iii) of \Cref{lem:haar pair} and $\tau' = \omega(\log{n})$ by Eq.~\eqref{eq:amplifier leakage}, these conditions require $\eta = o(1/\log{n})$. The information-theoretic lower bound used in Eq.~\eqref{eq:degenerate ed} is proportional to $\eta n$, which is $\widetilde{\Omega}(n)$ at the weights of Eqs.~\eqref{eq:agnostic params} and~\eqref{eq:master weights}.

For the fixed orders of \Cref{prop:separation}, the optimal computational rate is captured by the conditional R\'{e}nyi entropy at an intermediate order $\alpha \in (1, \alpha_\rc)$, despite the negative conditional min-entropy. When $\alpha_\rc = \omega(\log{n})$, the conditional R\'{e}nyi entropy tends to zero at every fixed order $\alpha \in (1,\infty]$, including the conditional min-entropy. Thus, the construction covers both the regime in which an intermediate order governs the optimal rate, specific to mixed states, and the regime of the pure-state results of Ref.~\cite{Leone2025Entanglementtheorywithlimited} (Theorems~23 and~27 for the statistical and LWE settings, respectively), while also allowing genuinely mixed states of polynomial rank.


\end{document}